\documentclass[modern]{aastex63}
\usepackage{graphicx}
\usepackage{mathtools}
\graphicspath{{./images_revised/}}
\usepackage{amsmath}

\newcommand{\Msolar}{M$_{\odot}$}
\newcommand{\Rsolar}{$R_{\odot}$}
\newcommand{\kms}{km s$^{-1}$}

\newcommand{\PRV}{$P_\mathrm{RV}$}
\newcommand{\PPM}{$P_{\mu}$}

\newcommand{\logg}{log$\,g$}
\newcommand{\teff}{$T_\mathrm{eff}$}
\newcommand{\gming}{$G_{BP} - G_{RP}$}
\newcommand{\gaia}{\textit{Gaia} DR2}

\newcommand{\ClusAmpCass}{82.5}
\newcommand{\ClusAmpErrCass}{3.3}
\newcommand{\ClusAvgRVCass}{-53.5}
\newcommand{\ClusAvgRVerrCass}{0.1}
\newcommand{\ClusSigCass}{1.5}
\newcommand{\ClusSigErrCass}{0.1}
\newcommand{\FldAmpCass}{5.7}
\newcommand{\FldAmpErrCass}{0.8}
\newcommand{\FldAvgRVCass}{-36.7}
\newcommand{\FldAvgRVerrCass}{4.5}
\newcommand{\FldSigCass}{25.7}
\newcommand{\FldSigErrCass}{4.7}

\newcommand{\numcatSMCass}{459}
\newcommand{\numcatSNCass}{436}
\newcommand{\numcatBMCass}{83}
\newcommand{\numcatBNCass}{30}
\newcommand{\numcatBLMCass}{22}
\newcommand{\numcatBLNCass}{32}
\newcommand{\numcatBUCass}{8}

\newcommand{\numcatVRRCass}{113}

\newcommand{\ClusAmpPM}{57.0}
\newcommand{\ClusAmpPMerr}{1.3}
\newcommand{\ClusAvgPMRA}{-0.959}
\newcommand{\ClusAvgPMRAerr}{0.003}
\newcommand{\ClusAvgPMDEC}{-1.986}
\newcommand{\ClusAvgPMDECerr}{0.003}
\newcommand{\ClusPMRASig}{0.131}
\newcommand{\ClusPMRASigerr}{0.003}
\newcommand{\ClusPMDECSig}{0.140}
\newcommand{\ClusPMDECSigerr}{0.004}

\newcommand{\FldAmpPM}{0.2}
\newcommand{\FldAmpPMerr}{0.1}
\newcommand{\FldAvgPMRA}{-1.8}
\newcommand{\FldAvgPMRAerr}{1.0}
\newcommand{\FldAvgPMDEC}{-1.6}
\newcommand{\FldAvgPMDECerr}{0.9}
\newcommand{\FldPMRASig}{1.5}
\newcommand{\FldPMRASigerr}{2.0}
\newcommand{\FldPMDECSig}{1.9}
\newcommand{\FldPMDECSigerr}{2.5}

\newcommand{\vsini}{$v\,$sin$\,i$}

\newcommand{\PcircCass}{8.22}
\newcommand{\cperrupCass}{+3.51}
\newcommand{\cperrdnCass}{-1.35}

\shorttitle{Radial Velocities and Binary Orbits in NGC 7789}
\shortauthors{Nine et al.}

\begin{document}

\title{WIYN Open Cluster Study. LXXVII. Radial-Velocity Measurements and Spectroscopic Binary Orbits in the Open Cluster NGC 7789}

\author{Andrew C. Nine}
\affiliation{Department of Astronomy, University of Wisconsin-Madison, \\ 475 N Charter St, Madison, WI 53706, USA}
\email{anine@astro.wisc.edu}

\author{Katelyn E. Milliman}
\affiliation{Department of Space Studies, American Public University, \\ 111 W Congress St, Charles Town, WV 25414, USA}
\affiliation{Department of Astronomy, University of Wisconsin-Madison, \\ 475 N Charter St, Madison, WI 53706, USA}

\author{Robert D. Mathieu}
\affiliation{Department of Astronomy, University of Wisconsin-Madison, \\ 475 N Charter St, Madison, WI 53706, USA}

\author{Aaron M. Geller}
\affiliation{Center for Interdisciplinary Exploration and Research in Astrophysics (CIERA) \\ and Department of Physics and Astronomy, Northwestern University, \\ 1800 Sherman Ave., Evanston, IL 60208, USA}
\affiliation{Adler Planetarium, Department of Astronomy, \\ 1300 S. Lake Shore Drive, Chicago, IL 60605, USA}

\author{Emily M. Leiner}
\affiliation{Center for Interdisciplinary Exploration and Research in Astrophysics (CIERA) \\ and Department of Physics and Astronomy, Northwestern University, \\ 1800 Sherman Ave., Evanston, IL 60208, USA}
\affiliation{Department of Astronomy, University of Wisconsin-Madison, \\ 475 N Charter St, Madison, WI 53706, USA}

\author{Imants Platais}
\affiliation{Department of Physics and Astronomy, Johns Hopkins University, \\ 3400 N Charles St, Baltimore, MD 21218, USA}

\author{Benjamin M. Tofflemire}
\affiliation{Department of Astronomy, University of Texas at Austin, \\ 2515 Speedway, Stop C1400, Austin, TX 78712, USA}
\affiliation{Department of Astronomy, University of Wisconsin-Madison, \\ 475 N Charter St, Madison, WI 53706, USA}

\begin{abstract}
We introduce the stellar sample of the WIYN Open Cluster Study radial-velocity survey for the rich open cluster NGC 7789 (1.6 Gyr, [Fe/H] = +0.02). This sample lies within an 18$'$ circular radius from the cluster center (10 pc in projection, or about 2 core radii), and includes giants, red-clump stars, blue stragglers, red stragglers, sub-subgiants, and main-sequence stars down to one mag below the turnoff. Our survey began in 2005 and comprises more than 9,000 radial-velocity measurements  from the Hydra Multi-Object Spectrograph on the WIYN 3.5m telescope. We identify 564 likely cluster members and present the orbital solutions for 83 cluster binary stars with periods between 1.45$\,$d and 4200$\,$d. From the main-sequence binary solutions we fit a circularization period of~\PcircCass$^{\cperrupCass}_{\cperrdnCass}\,$~d. We calculate an incompleteness-corrected main-sequence binary frequency of 31\% $\pm$ 3\% for binaries with periods less than 10$^{4}$ days, similar to other WOCS open clusters of all ages. We detect a blue straggler binary frequency of 31\% $\pm$ 15\%, consistent with the similarly aged cluster NGC 6819.
\vspace{1cm}
\end{abstract}

\section{Introduction}
The WIYN Open Cluster Study (WOCS; \citealt{Mathieu2000}) seeks to probe a wide variety of questions in stellar astrophysics by acquiring comprehensive photometric, astrometric, and spectroscopic data on a selected set of rich, nearby open clusters that span age and metallicity (100 Myr $\lesssim$ $t$ $\lesssim$ 8 Gyr; -0.4 $\lesssim$ [Fe/H] $\lesssim$ +0.3). Within WOCS, extensive data have been combined and presented for M35 (150 Myr; \citealt{ Geller2010, Thompson2014, Leiner2015}), NGC 6819 (2.5 Gyr; \citealt{Platais2013, Yang2013, Milliman2014}), M67 (4 Gyr; Geller et al. 2020, and references therein, in prep.), NGC 188 (7 Gyr; \citealt{Sarajedini1999, Platais2003, Geller2012}) and NGC 6791 (8 Gyr; \citealt{Platais2013, Tofflemire2014}). These data have been the foundation for a wide array of scientific discoveries including insights into blue straggler (BSS) formation (\citealt{GellerNature2011, Milliman2015, Gosnell2015, MathieuLeiner2019ASEP, Leiner2019BL}), an age-rotation relationship for cool stars (\citealt{Meibom2015}), open cluster sub-subgiants (\citealt{Milliman2016, Geller2017a, Geller2017b, Leiner2017}) and more (\citealt{Sandquist2018, Deliyannis2019}).

In this paper we introduce the stellar sample of the WOCS radial-velocity (RV) survey of NGC 7789 ($\alpha$ = 23$^h$57$^m$21$^s$.6, $\delta$ = $+$56$^{\circ}$43$'$22$''$  J2000). NGC 7789 is a well-populated, 1.6 Gyr (\citealt{Gim1998}) open cluster at a distance of 2.0 kpc with [Fe/H] = +0.02 $\pm$ 0.04 determined by \cite{Jacobson2011}. \cite{Wu2009} measure a core radius of 8.82$'$ $\pm$ 0.91$'$ and determine a cluster mass between 5150 \Msolar~and 7710 \Msolar~depending on the measurement method used. 

Photometric studies of NGC 7789 begin with \cite{Burbidge1958} and include the $VI$ photometry of \cite{Gim1998} as well as the variability surveys of \cite{Jahn1995}, \cite{MK1999}, and \cite{Zhang2003}. A proper-motion study of \cite{McNamara1981} focused on evolved stars in NGC 7789. Here, we adopt the astrometry of \textit{Gaia} Data Release 2 (\gaia; \citealt{GAIA2016, GaiaCollab2018a, GaiaCollab2018b}).  The use of \gaia~allows for more detailed study of the cluster, including the confident identification of cluster members based on their proper motions and parallaxes (\citealt{CG2018, Gao2018}).

The bright giants in NGC 7789 have also been the target of many spectroscopic abundance studies including \cite{Pancino2010}, \cite{Jacobson2011}, and \cite{Overbeek2015}. These abundance studies as well as the RV surveys of cluster giants done by \cite{Gim1998a} and \cite{Casamiquela2016} have provided one to several RV measurements for most of the evolved stars in NGC 7789. With this paper we provide a complete RV time-series survey from the most luminous stars down to $\sim$1 mag below the main-sequence turnoff. We also present the first comprehensive collection of spectroscopic binary orbit solutions for NGC 7789, and use these solutions for a first look at connections between binaries and the stellar population.
\vspace{0.5cm}

\section{Stellar Sample and Photometry}
\label{ngc7789rv:sec:sample}
The primary RV target sample for NGC 7789 that is the basis for this paper includes 1206 stars. The sample was initially built from the extensive $VI$ CCD photometry of \cite{Gim1998}. This photometry was obtained from the Dominion Astrophysical Observatory 1.8 m Plaskett telescope and covers $\sim$18$'$ radius from the center of NGC 7789. The full photometry set has over 15,000 stars, is complete from $V$ $\sim$ 10 to $V$ $\sim$ 21, and has a $V$ standard error under $\sim$0.01 mag down to our primary sample cutoff at $V$ $=$ 15.0. As in \cite{Milliman2014}, we applied cuts in $V$ and in $V - I$, such that $V < 15.0$ and that the stars above this cutoff had a measured $V-I$ in \cite{Gim1998}. We applied these criteria in order to select the upper main sequence, red clump, giant branch, and blue stragglers as well as other alternative stellar evolution products in the cluster.

\begin{figure*}[p!]
\begin{center}
\includegraphics[width=\linewidth]{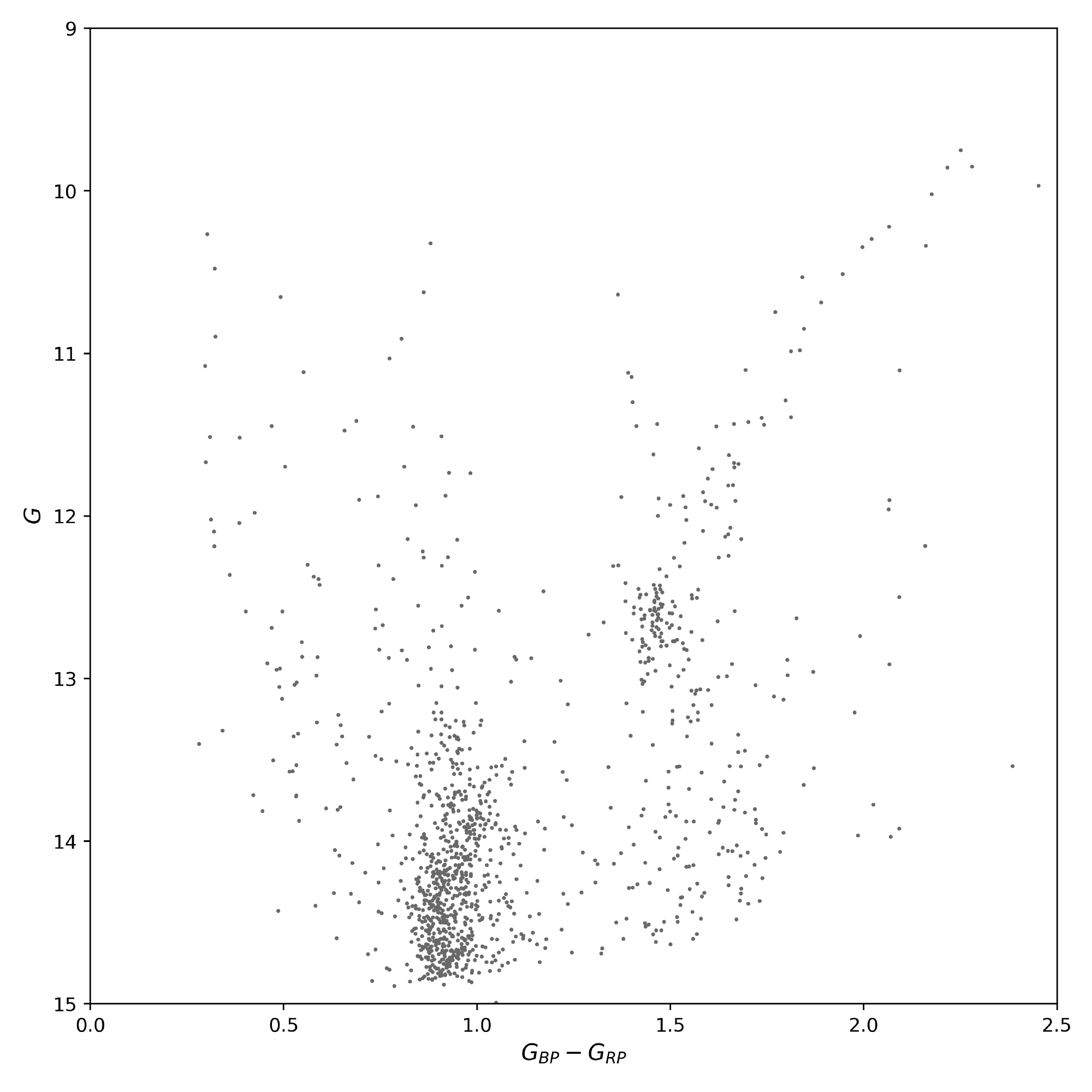}
\caption{CMD of the NGC 7789 stars in our primary RV target sample based on the photometry of \cite{GaiaCollab2018b}. The cluster's giant branch, red clump, and upper main sequence are clearly present, as well as a large number of blue straggler candidates.}
\label{ngc7789rv:fig:cmd.sample}
\end{center}
\end{figure*}

We then cross-matched this sample of stars with \gaia \, using TOPCAT (\citealt{Taylor2005}). From our sample of 1206 stars, we were able to retrieve matches for 1204 stars using a tolerance in position of 1". Of these, 1187 had full photometric and five-dimensional astrometric information. We use the proper motions to determine cluster memberships for our RV sample, discussed in detail in section \ref{ngc7789rv:sec:ppm}. The remaining 17 stars had insufficient proper-motion information for complete membership determination (Section \ref{ngc7789rv:sec:class}). Out of these 17 stars, 6 are RV members (Section \ref{ngc7789rv:sec:Prv}), and are classified as cluster members. We adopt the photometry of \gaia \, for the remainder of this paper. As shown in Figure~\ref{ngc7789rv:fig:cmd.sample}, 
our sample extends from $G$ $\sim$ 10 to $G$ $\sim$ 15, one mag below the main-sequence turnoff, and includes giant stars, a well-populated red clump, and a large number of BSS candidates.

\section{WIYN Observations and Data Reduction}
\label{ngc7789rv:sec:obs}

Our observations of NGC 7789 began in 2005 January using the Hydra Multi-Object Spectrograph (MOS; \citealt{Barden1994}) on the WIYN\footnote{The WIYN 3.5m Observatory is a joint facility of the University of Wisconsin-Madison, Indiana University, and the National Optical Astronomy Observatory.} 3.5m telescope. Our spectra are typically taken in a wavelength range of $\sim$500 \r{A} centered on 5125 \r{A}, which encompasses the Mg B triplet and a rich array of narrow metal lines. We conduct our observations at 11$^{\text{th}}$ order, which yields a spectral resolution of $\sim$15,000-20,000 with a dispersion of 0.13 \r{A}/pix. More details on our observational setup can be found in \cite{Geller2008}.

We follow the standard data acquisition and reduction procedures of the WOCS RV survey described in \cite{Geller2008}. Briefly, for each configuration our data include one 100 s dome flat and two 300 s thorium-argon emission lamp spectra, one each before and after three science integrations. We use standard IRAF routines to bias subtract, dispersion correct, and extract each spectrum. We then flat-field, throughput correct, and subtract the sky from these spectra. In 2014 August we incorporated the L.A. Cosmic routine (\citealt{Dokkum2001}) for improved cosmic ray rejection.

The goal of this study is to conduct time-series observations of evolved stars, stars in the upper main sequence and stars on alternative evolutionary tracks in order to detect binaries and determine their orbital properties. For each observing run we use a prioritization system described in \cite{Geller2008} that emphasizes high-amplitude velocity-variable, short-period binaries, followed by longer period binaries. For a complete orbital solution, we typically require a minimum of 12 observations of sufficient quality as defined in Section \ref{ngc7789rv:sec:precision}, and that the best fit orbital solution has an rms residual velocity comparable to the velocity measurement errors. As of 2019 July, we have over 9,000 spectra spanning nearly 15 years for 1,198 stars in the primary sample. The resulting RV measurements and first results from these data are presented in this paper.

\section{RV Measurement and Precision}
\label{ngc7789rv:sec:precision}
RVs for single narrow-lined stars (i.e., stars with \vsini \, $<$ 10 km s$^{-1}$) are derived from the centroid of a one-dimensional cross-correlation function (CCF) with an observed solar template at zero velocity, converted to a heliocentric velocity, and corrected for the individual fiber offsets of the Hydra MOS. RVs for double-lined stars are derived following the two-dimensional cross-correlation technique TODCOR (\citealt{Zucker1994}). Based on the analysis of \cite{Geller2008}, we apply a quality threshold requirement of CCF $\geq$ 0.4 for our RV measurements. We present all of our quality RV measurements for each star in Table~\ref{ngc7789rv:tab:rvs}, along with the Heliocentric Julian Date (HJD) of the observation and the height of the cross-correlation function. We also include the RV residual and the orbital phase of the observation for binary stars with completed orbital solutions. 

\centerwidetable
\begin{deluxetable*}{ccccccccc}
\tablewidth{0pt}
\tabletypesize{\footnotesize}
\tablecaption{Radial-Velocity Measurements\label{ngc7789rv:tab:rvs}}
\tablehead{ \colhead{ID} & \colhead{HJD-2,400,000} & \colhead{RV$_{1}$} & \colhead{Correlation Height$_{1}$} & \colhead{O$-$C$_{1}$} & \colhead{RV$_{2}$} & \colhead{Correlation Height$_{2}$} & \colhead{O$-$C$_{2}$} & \colhead{Phase} \\
\colhead{} & \colhead{(days)} & \colhead{(\kms)} & \colhead{} & \colhead{(\kms)} & \colhead{(\kms)} & \colhead{} & \colhead{(\kms)} & \colhead{}}
\startdata
7003 &  &  &  &  &  &  &  & \\
 & 53720.7009 & -95.464 & 0.41 & -1.64 & -12.457 & 0.29 & -2.26 & 0.132\\
 & 54275.9141 & -76.535 & 0.67 & -0.94 & -28.382 & 0.56 & 1.64 & 0.261\\
 & 54676.9439 & -58.680 & 0.85 & \nodata & \nodata & \nodata & \nodata & \nodata\\
 & 54683.9428 & -39.360 & 0.77 & -0.76 & -69.085 & 0.72 & 1.18 & 0.583\\
 & 54715.7860 & -36.947 & 0.67 & -2.05 & -74.254 & 0.62 & 0.04 & 0.622\\
 & 54723.8472 & -25.573 & 0.78 & -2.30 & -86.237 & 0.65 & 0.70 & 0.886\\
 & 54724.7236 & -26.386 & 0.76 & 2.20 & -79.336 & 0.63 & 1.82 & 0.915\\
 & 54747.8150 & -28.835 & 0.81 & 1.86 & -82.320 & 0.67 & -3.46 & 0.669\\
 & 54845.7016 & -20.716 & 0.52 & 0.62 & -91.026 & 0.34 & -1.99 & 0.865\\
 & 55018.7892 & -52.911 & 0.80 & \nodata & \nodata & \nodata & \nodata & \nodata\\
\enddata
\tablecomments{Table \ref{ngc7789rv:tab:rvs} is published in its entirety in machine-readable format. A portion is shown here for guidance regarding its form and content.}
\end{deluxetable*}

\begin{figure}[htbp]
\begin{center}
\includegraphics[width=0.75\linewidth]{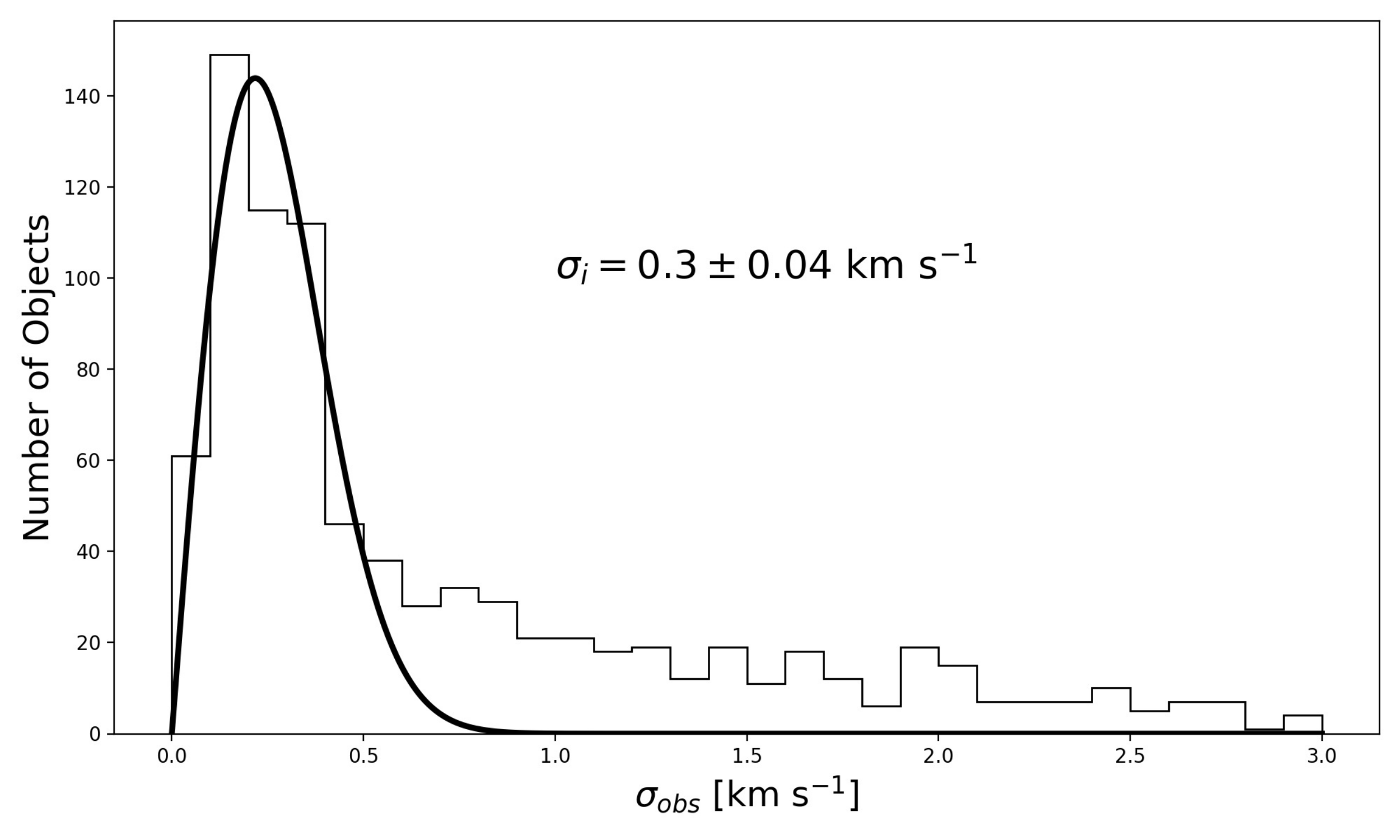}
\caption{Histogram of the RV standard deviations from the first three observations of the narrow-lined stars in our primary sample. We overplot the best-fit $\chi^{2}$ distribution function with a precision of 0.3 $\pm$ 0.04 \kms. The excesses above the theoretical distribution beginning at 0.5 \kms~are velocity-variable stars.}
\label{ngc7789rv:fig:precision}
\end{center}
\end{figure}

Following the procedure of \cite{Geller2008} and \cite{Tofflemire2014} of fitting a $\chi^{2}$ distribution to our observed distribution of RV standard deviations, we find our precision, $\sigma_{i}$, for the NGC 7789 narrow-lined stars to be 0.3 $\pm$ 0.04 \kms~(Figure~\ref{ngc7789rv:fig:precision}). This value is consistent with the value of $\sigma_i = 0.4$ \kms\: used in previous WOCS papers, which we adopt here as a conservative estimate. 

\begin{figure*}
    \gridline{\fig{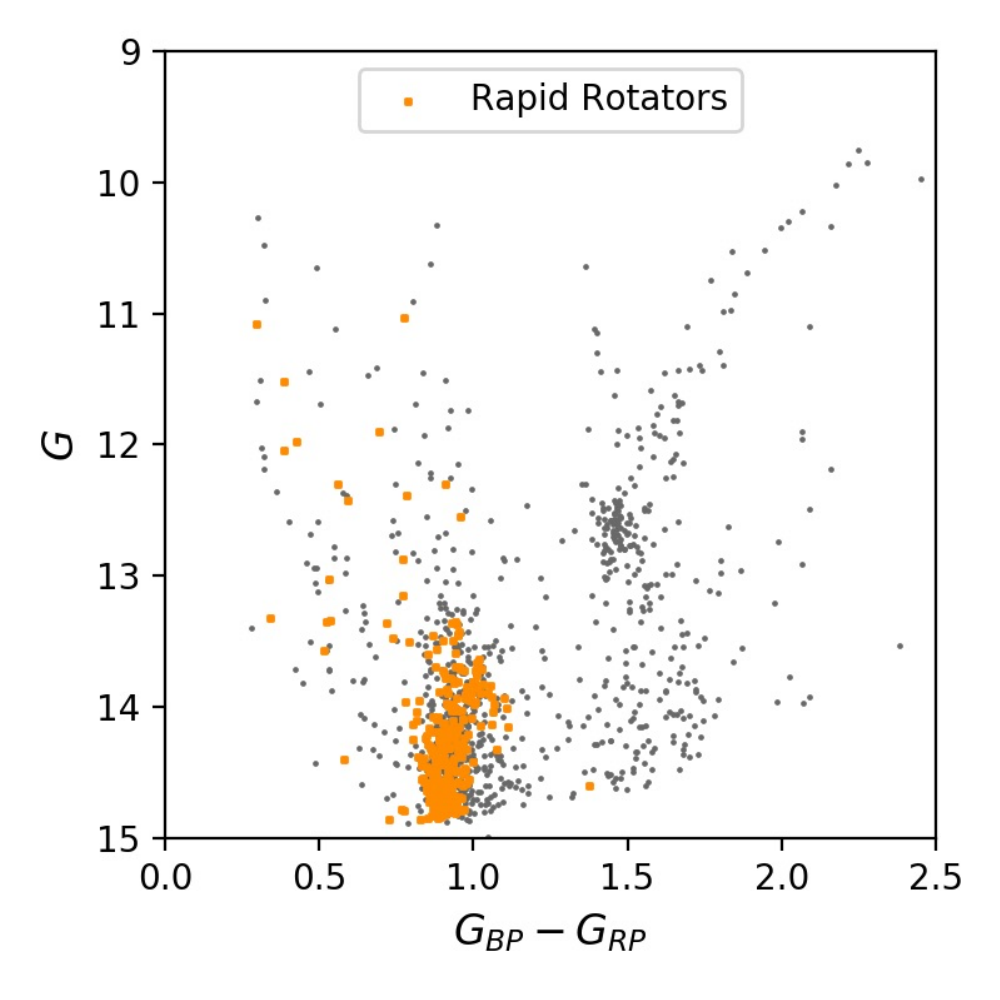}{0.5\linewidth}{}
            \fig{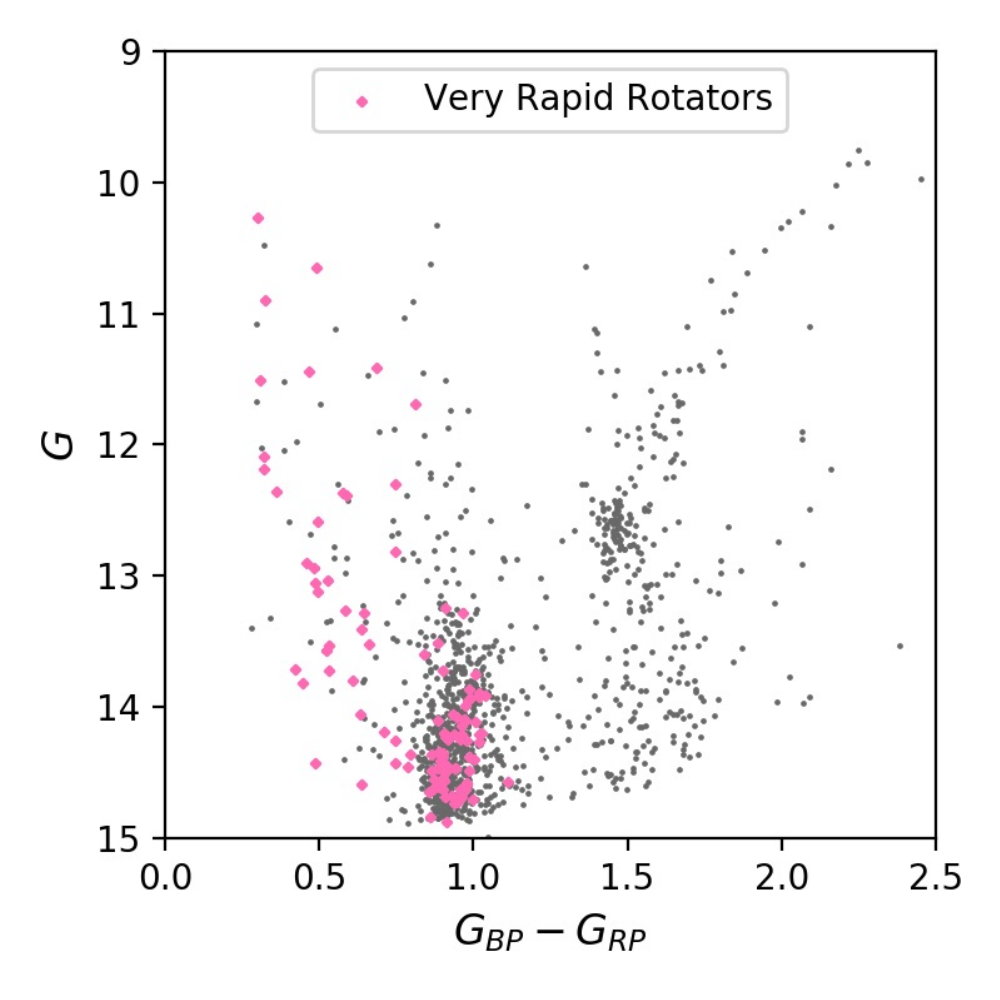}{0.5\linewidth}{}}
    \caption{NGC 7789 CMD of the primary sample presented in this paper, constructed with \gaia~ photometry. Left panel: Rapid rotators with 10 \kms~$<$~\vsini~$<$ 120 \kms~are shown by orange squares. Right panel: Very rapid rotators with \vsini~$>$ 120 \kms, for which we cannot derive accurate RVs, are shown with pink diamonds.}
    \label{ngc7789rv:fig:rrcmd}
\end{figure*}

For stars in our sample that are rapidly rotating (\vsini \, $>$ 10 km s$^{-1}$), using an observed solar template  results in a decrease in precision in our radial-velocity measurements due to their broadened CCFs. To achieve better RVs, we cross-correlate these stars against a grid of synthetic spectra\footnote{The library is based on ATLAS9 (http://kurucz.harvard.edu) and the companion program SYNTHE used to compute the synthetic spectrum from the model atmosphere and line list.} $\:$ (\citealt{Kurucz1993, Meibom2009}) with solar-metallicity, \logg~= 4.0, \teff~= 6500 K (unless otherwise noted) and spanning a range of projected rotation velocities. We selected these particular values for temperature and \logg~ based on the parameters near the turnoff of a 1.6 Gyr isochrone generated by the MESA Isochrones \& Stellar Tracks (MIST; \citealt{Dotter2016} and references therein), corresponding to the CMD position of the majority of the rapidly rotating stars (Figure~\ref{ngc7789rv:fig:rrcmd}). Adjusting the temperature of the synthetic spectra improved the RVs only for the hottest of stars in our sample. We note these stars and their best-matched temperatures (the highest cross-correlation peak height) in Table~\ref{ngc7789rv:tab:RVsum}. From the best-matched  template we derive the RVs presented in Table~\ref{ngc7789rv:tab:rvs} and note the \vsini~of the synthetic spectra in Table~\ref{ngc7789rv:tab:RVsum}. 

We assign a precision to these rapid rotators (RR) using the following relationship from \cite{Geller2010}:
\begin{equation}
\sigma_{i}= 0.38 + 0.012(v\,\textup{sin}\,i).
\label{eqn:rr_prec}
\end{equation}
The above procedure does not work consistently for stars with \vsini~greater than $\sim$120 \kms~ because of their very broad CCFs. We label stars in our sample that are rotating faster than this as very rapid rotators (VRR), and we are unable to get accurate RV measurements  or membership information for these stars.

In our sample, we identified 273 RR and 113 VRR stars. We plot the CMD location of 270 of the RR stars and all 113 VRR stars in Figure~\ref{ngc7789rv:fig:rrcmd}. As expected, most of these stars are hot and blue, with \gming\: $\leq$ 1.0, and are concentrated on the upper main sequence (MS). These stars represent approximately 50\% of the main-sequence stars in our primary sample.

\section{Completeness}
\label{ngc7789rv:sec:comp}
The primary target sample in this paper is comprised of 1206 stars with RV measurements starting in 2005. For our survey we constantly reevaluate the observing priority of each star. Generally we classify stars as single or velocity variable after three observations (\citealt{Mathieu1983}; \citealt{Geller2012}), and we then prioritize velocity-variable stars for continued observations until we determine an orbital solution. When we identify a VRR we move it to the bottom of our observation priority because we are unable to derive accurate RVs for such rapid rotators.

\begin{figure}[htbp]
    \gridline{\fig{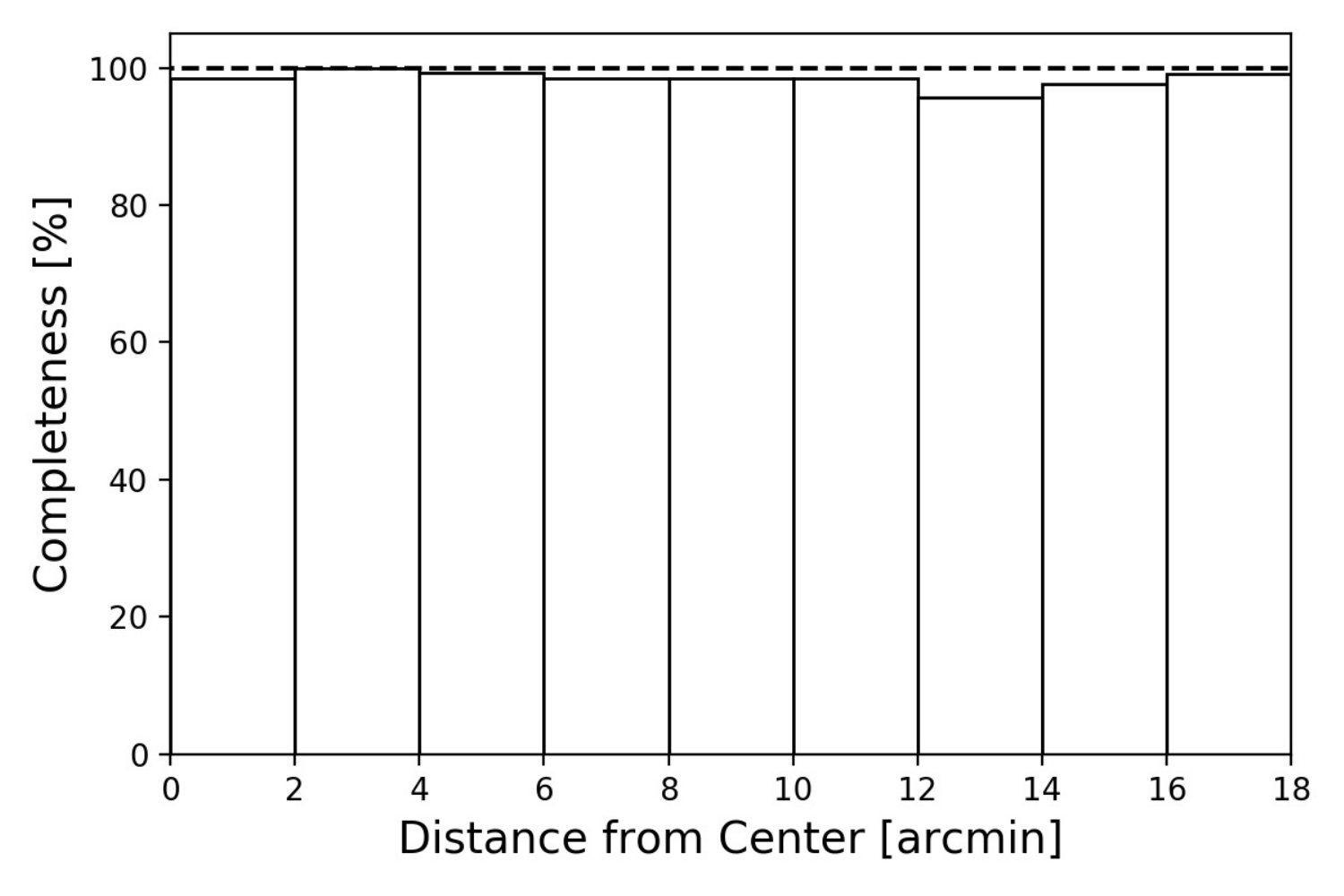}{0.5\linewidth}{}
    \fig{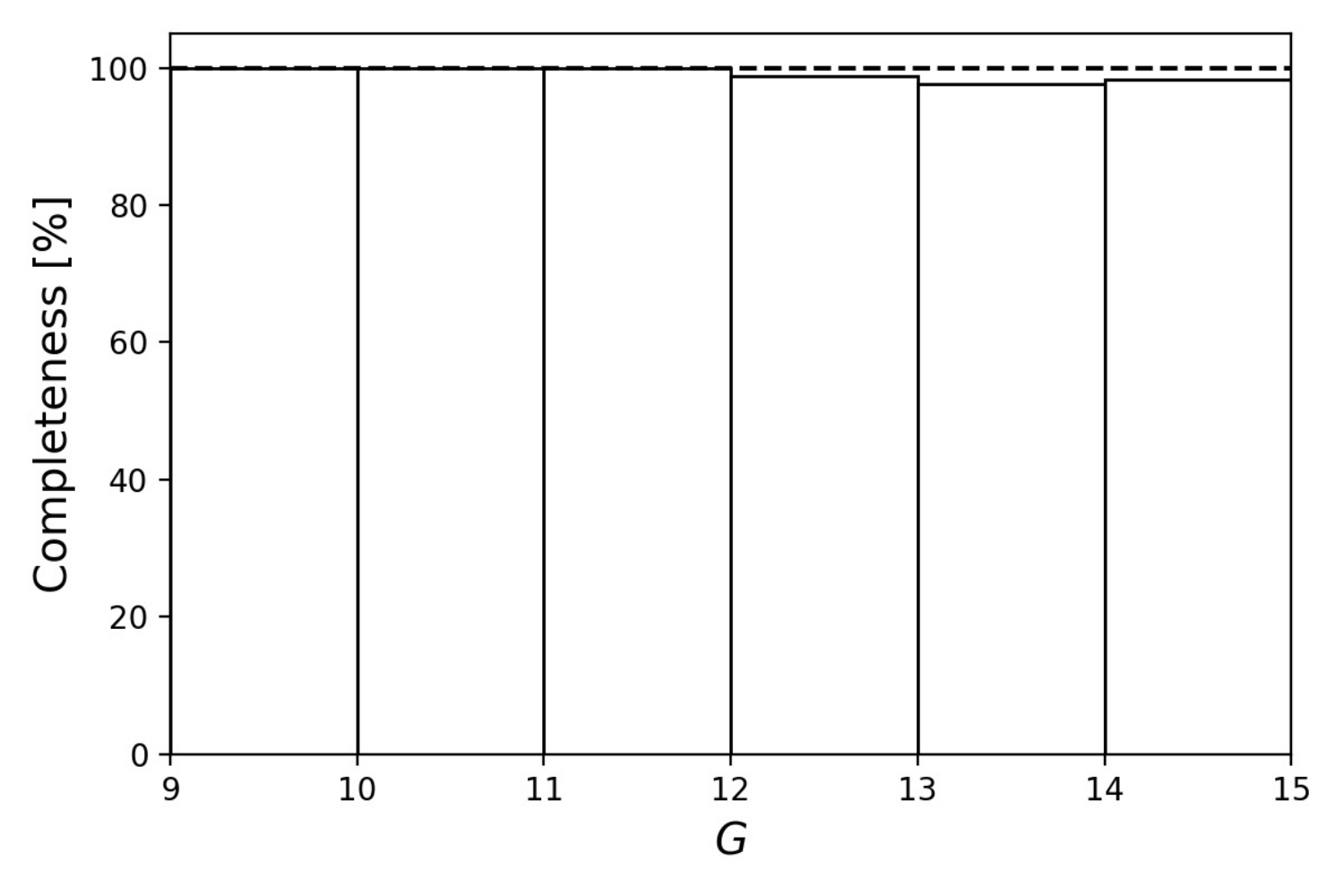}{0.5\linewidth}{}}
    \caption{Percentage of stars in our sample (excluding VRRs) that have three or more RV observations with respect to distance from the cluster center (left) and $G$ magnitude (right).}
    \label{ngc7789rv:fig:complete}
\end{figure}

The percentages of stars for which we have three or more observations (excluding VRRs) as a function of cluster radius and of $G$ magnitude are shown in Figure~\ref{ngc7789rv:fig:complete}. We are able to classify almost 99\% of stars in our primary stellar sample of NGC 7789 as single (by which we mean non-velocity variable), velocity-variable, or as VRRs. There are no significant biases in the classified stars with radius or magnitude.

\section{Results of RV Study}
\subsection{Velocity-Variable Stars}
We define velocity-variable stars as having RV standard deviations (the external error, $e$) greater than 4 times the precision (the internal error $\sigma_{i}$, or $i$, described in Section~\ref{ngc7789rv:sec:precision})\footnote{For consistency with previous WOCS papers and to prevent confusion with other uses of $\sigma$ in this paper, we will use $i$ to indicate internal precision for the rest of the paper.}, that is $e/i$ $>$ 4 (\citealt{Geller2008}). For each star with three or more observations we calculate the $\overline{\textup{RV}}$ and $e/i$, and classify it as single or velocity variable. Based on a Monte Carlo analysis for the NGC 7789 RV measurements, this procedure identifies 84\% of all binaries under 1,000 days and 60\% of all binaries with periods under 10,000 days (Section~\ref{ngc7789rv:sec:bin.detect}). 

We present the coordinates, \gaia \, photometry, number of observations, $\overline{\textup{RV}}$, $\overline{\textup{RV}}$ standard error, $e/i$, proper-motion membership probability (\PPM; Section~\ref{ngc7789rv:sec:ppm}), RV membership probability (\PRV), and membership classification for all of the stars we have observed in NGC 7789 as of 2019 July in Table~\ref{ngc7789rv:tab:RVsum}. For each binary star with a completed orbital solution, the center-of-mass velocity, $\gamma$, its error, and whether it is a single- or double-lined binary are also included in Table~\ref{ngc7789rv:tab:RVsum}.

\clearpage
\movetabledown=1.5in
\begin{rotatetable}
\begin{deluxetable*}{lcccccccccclccl}
\tabletypesize{\footnotesize}
\tablecaption{Radial-Velocity Summary Table\label{ngc7789rv:tab:RVsum}}
\tablehead{\colhead{ID} & \colhead{$\alpha$ (J2000)} & \colhead{$\delta$ (J2000)} & \colhead{$G$} & \colhead{$G_{BP}-G_{RP}$} & \colhead{$N$} & \colhead{$\overline{\mathrm{RV}}$} & \colhead{Std. Err.} & \colhead{$e/i$} & \colhead{\PPM} & \colhead{\PRV} & \colhead{Class} & \colhead{$\gamma_{RV}$} & \colhead{$\gamma_{RVe}$} & \colhead{Comment}\\
\colhead{} & \colhead{} & \colhead{} & \colhead{} & \colhead{} & \colhead{} & \colhead{(\kms)} & \colhead{(\kms)} & \colhead{} & \colhead{(\%)} & \colhead{(\%)} & \colhead{} & \colhead{(\kms)} & \colhead{(\kms)} & \colhead{}}
\tablewidth{0pt}
\startdata
1001 & 23 57 24.11 & 56 43 38.8 & 13.93 & 0.95 & 19 & -49.54 & 0.87 & 0.55 & 87 & 39 & SNM & \nodata & \nodata &              RR (100 \kms)\\
1002 & 23 57 18.72 & 56 43 50.8 & 13.73 & 0.94 & 3 & -36.48 & 0.22 & 0.55 & 0 & 0 & SNM & \nodata & \nodata & \nodata\\
1003 & 23 57 29.55 & 56 42 23.5 & 12.33 & 1.47 & 3 & -53.89 & 0.27 & 0.67 & 100 & 95 & SM & \nodata & \nodata & \nodata\\
1004 & 23 57 10.40 & 56 42 49.4 & 9.86 & 2.22 & 5 & -53.58 & 0.2 & 0.51 & 99 & 95 & SM & \nodata & \nodata & \nodata\\
1005 & 23 57 31.85 & 56 41 22.1 & 11.1 & 1.69 & 3 & -56.06 & 0.27 & 0.67 & \nodata & 84 & SM & \nodata & \nodata & \nodata\\
1006 & 23 57 10.06 & 56 40 56.7 & 10.74 & 1.77 & 16 & -53.08 & 2.09 & 5.24 & 89 & 93 & BM & -54.963 & 0.183 &      SB1\\
1007 & 23 57 34.30 & 56 46 2.9 & 11.69 & 0.5 & 3 & -4.05 & 0.1 & 0.26 & 0 & 0 & SNM & \nodata & \nodata & \nodata\\
1008 & 23 57 3.25 & 56 45 58.0 & 9.75 & 2.25 & 3 & -53.46 & 0.08 & 0.19 & 100 & 95 & SM & \nodata & \nodata & \nodata\\
1009 & 23 57 52.04 & 56 42 25.6 & 10.29 & 2.02 & 3 & -52.46 & 0.17 & 0.43 & 0 & 94 & SNM & \nodata & \nodata & \nodata\\
1010 & 23 57 43.94 & 56 39 42.8 & 11.51 & 0.31 & 11 & 18.93 & 3.79 & 9.48 & \nodata & 0 & VRR & \nodata & \nodata & VRR\\
\enddata
\tablecomments{Table \ref{ngc7789rv:tab:RVsum} is published in its entirety in machine-readable format. A portion is shown here for guidance regarding its form and content.}
\end{deluxetable*}
\end{rotatetable}

\clearpage

\subsection{RV Membership Probabilities}
\label{ngc7789rv:sec:Prv}
\begin{figure*}[htbp]
\begin{center}
\includegraphics[width=0.75\linewidth]{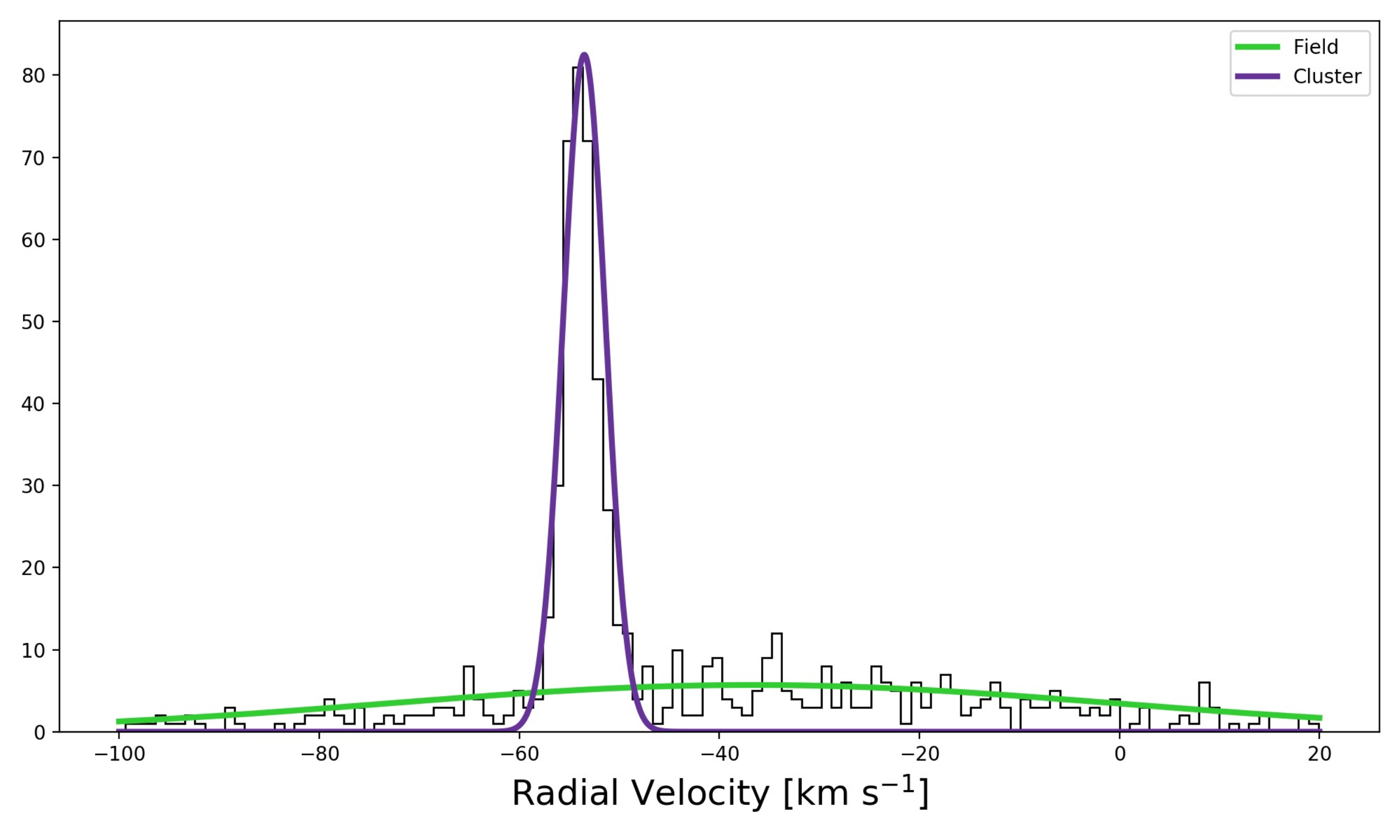}
\caption{Histogram of the RV distribution of single stars, $e/i$ $<$ 4, with three or more RV observations. Also plotted are the Gaussian distributions fit to the cluster (the large peak at a mean velocity of \ClusAvgRVCass~\kms;~purple line), and the field (green line).}
\label{ngc7789rv:fig:clusgauss}
\end{center}
\end{figure*}

We follow the standard WOCS procedure and calculate the RV membership probability of a given star, \PRV\,
in Table~\ref{ngc7789rv:tab:RVsum} using the equation:
 \begin{equation}
 P_{RV}(v)=\frac{F_{cluster}(v)}{F_{field}(v) + F_{cluster}(v)}.
 \end{equation}
$F_{cluster}$ and $F_{field}$ are  Gaussian functions simultaneously fit to the cluster and field-star populations using our sample of all single stars with three or more RV measurements. We plot these RVs and Gaussian functions in Figure~\ref{ngc7789rv:fig:clusgauss} and record the parameters for these Gaussian fits in Table~\ref{ngc7789rv:tab:gauss}. 

The RV survey of giant stars by \cite{Gim1998a} found an average RV of $-$54.9 $\pm$ 0.9 \kms~(s.d.) based on 50 stars and the abundance study of \cite{Jacobson2011} found an average RV of -54.7 $\pm$ 1.3 \kms~based on 26 evolved stars in NGC 7789. \cite{Overbeek2015} found an average RV of -54.6 $\pm$ 1.0 \kms~based on their study of 32 evolved stars. Our Gaussian fit to the cluster distribution is consistent with these studies, with a measured $\overline{\textup{RV}}$ of \ClusAvgRVCass~$\pm$ \ClusSigCass~\kms. 
Our result also matches that of \cite{Casamiquela2016} who found a median RV of -53.6 $\pm$ 0.6~\kms~for seven red clump stars in NGC 7789. \cite{Geller2008} and \cite{Geller2015} found that the WOCS RV system is on the same zero-point as the DAO (\citealt{Fletcher1982}) and CfA (\citealt{Latham1985}) systems.

\begin{figure}[htbp]
\begin{center}
\includegraphics[width=0.75\linewidth]{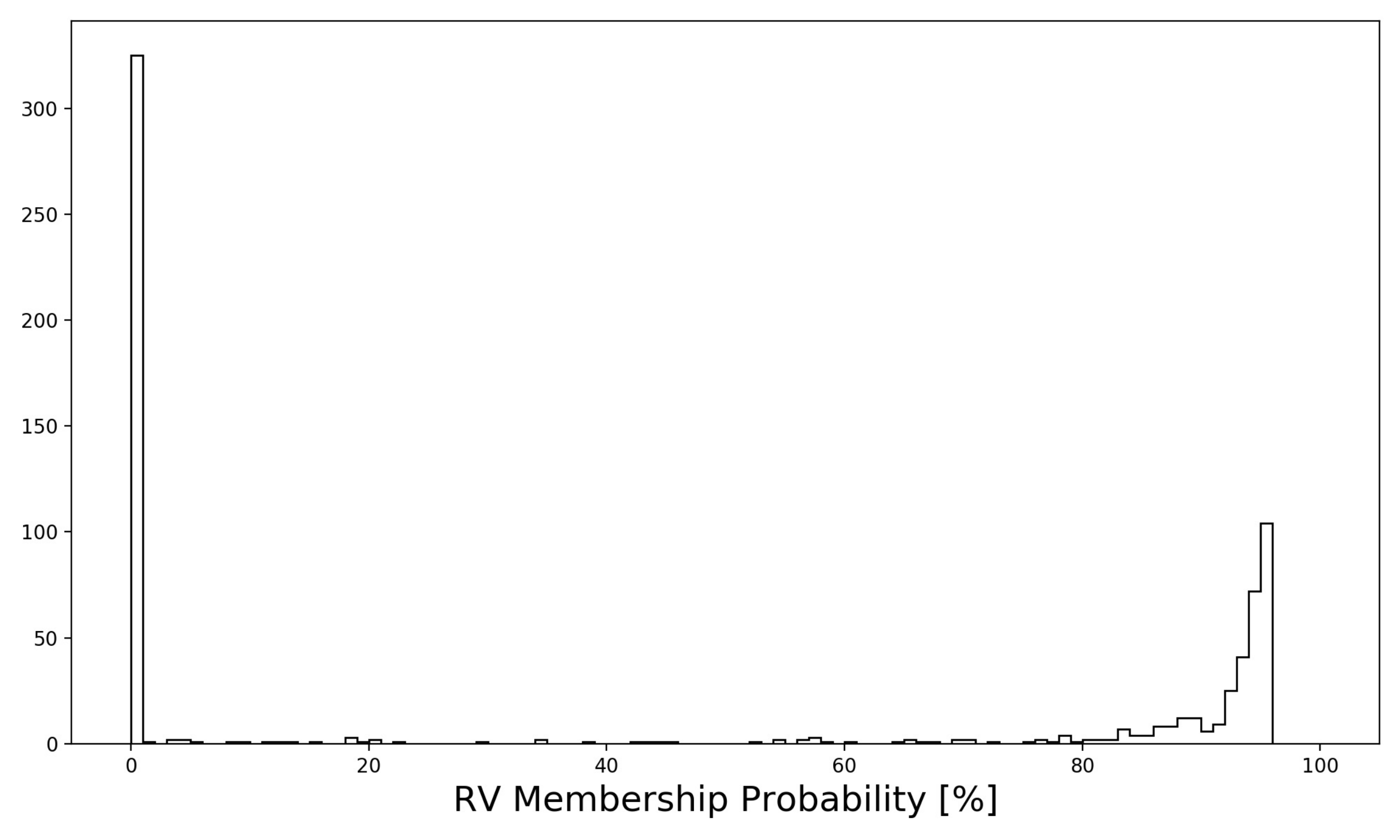}
\caption{Histogram distribution of the radial-velocity membership probabilities of single stars (those with $e/i$ $<$ 4) in NGC 7789.}
\label{ngc7789rv:fig:prvhist}
\end{center}
\end{figure}

We note a clear separation between the cluster and field populations based on \PRV~(Figure~\ref{ngc7789rv:fig:prvhist}) and we adopt the standard WOCS membership threshold of \PRV~$\geq$ 50\% for NGC 7789. We estimate from the cluster and field Gaussian functions a field star contamination of 5\% at this membership threshold. 

\begin{centering}
\begin{deluxetable}{ccc}
\tablewidth{0pc}
\tablecolumns{3}
\tablecaption{Gaussian Fit Parameters for Cluster and Field RV Distributions\label{ngc7789rv:tab:gauss}}
\tablehead{ \colhead{Parameter} & \colhead{Cluster} & \colhead{Field}}
\startdata
Ampl. [number] & \ClusAmpCass\ $\pm$ \ClusAmpErrCass & \FldAmpCass\ $\pm$ \FldAmpErrCass \\
$\overline{\textup{RV}}$ [\kms] & \ClusAvgRVCass\ $\pm$ \ClusAvgRVerrCass & \FldAvgRVCass\ $\pm$ \FldAvgRVerrCass\\
$\sigma$ [\kms] & \ClusSigCass\ $\pm$ \ClusSigErrCass & \FldSigCass\ $\pm$ \FldSigErrCass \\
\enddata
\end{deluxetable}
\end{centering}

\subsection{Proper-Motion and Radial-Velocity Memberships}
\label{ngc7789rv:sec:ppm}

We compare our RV membership results to the astrometric data retrieved from \gaia. We were able to obtain complete astrometric information for 1187 stars in our primary sample. For each of the sources returned, we check the associated astrometric excess noise ($\epsilon_i$) and the significance of the excess noise ($D$), which determine the disagreement between the observations of a source and the best-fitting astrometric model. For our derivation of \PPM~for this cluster, we reject those sources which have $\epsilon_i \geq 10^{-3}$, which corresponds to $D \gtrsim 10^{-2}$. By rejecting these sources we aim to minimize the effects of any systematic variations in the astrometric data. We rejected a total of 85 stars from our sample, leaving 1102 stars with complete and low-noise astrometry as determined by \gaia.

\begin{figure*}[p!]
    \centering
    \includegraphics[width=\linewidth]{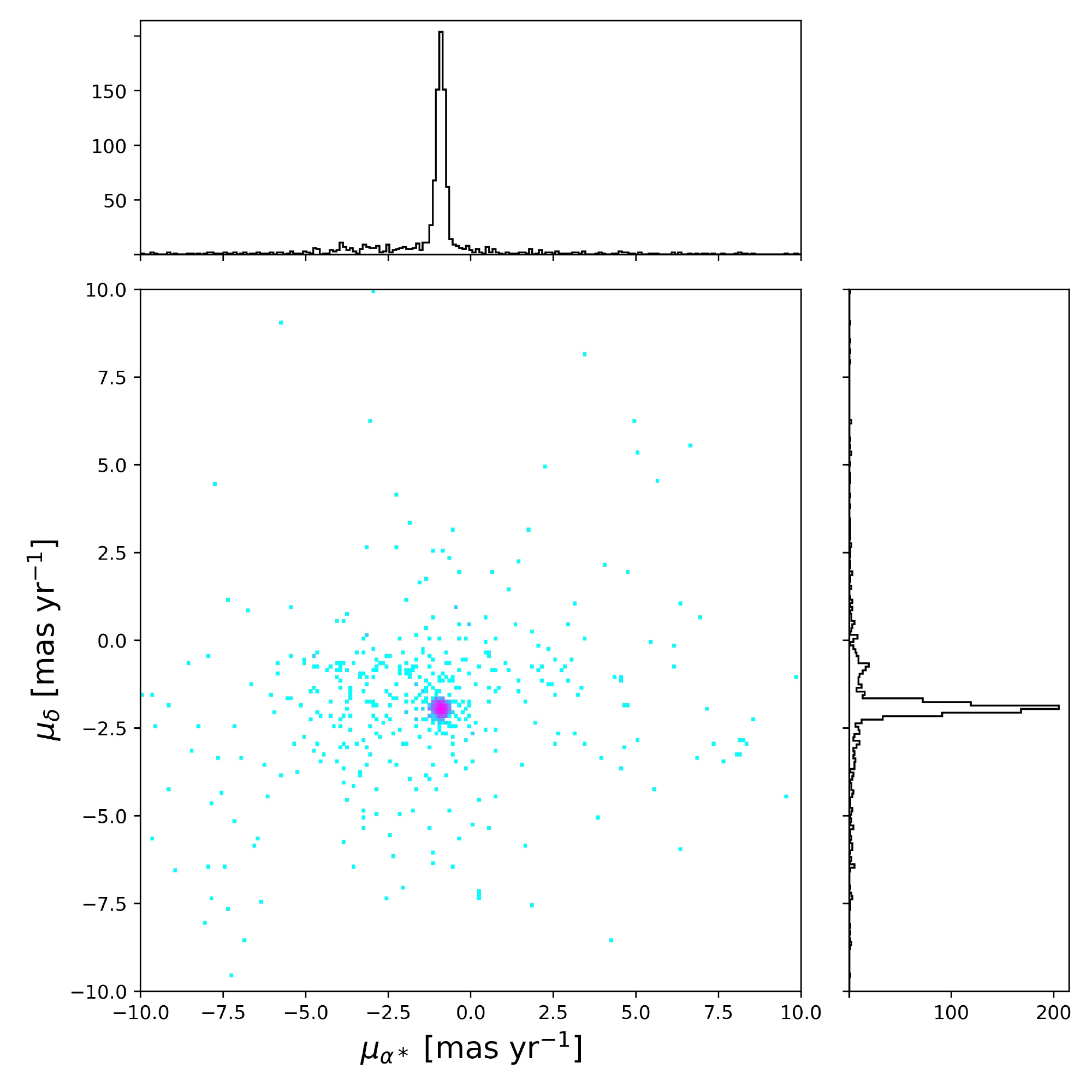}
    \caption{Two-dimensional histogram showing the \gaia \, proper motions of the stars in our primary sample in both right ascension and declination. On each axis we also plot the corresponding one-dimensional histogram in each direction. Cluster member stars appear as a clear overdensity in each plot.}
    \label{hist:twod}
\end{figure*}

\begin{figure*}[p!]
    \centering
    \includegraphics[width=\linewidth]{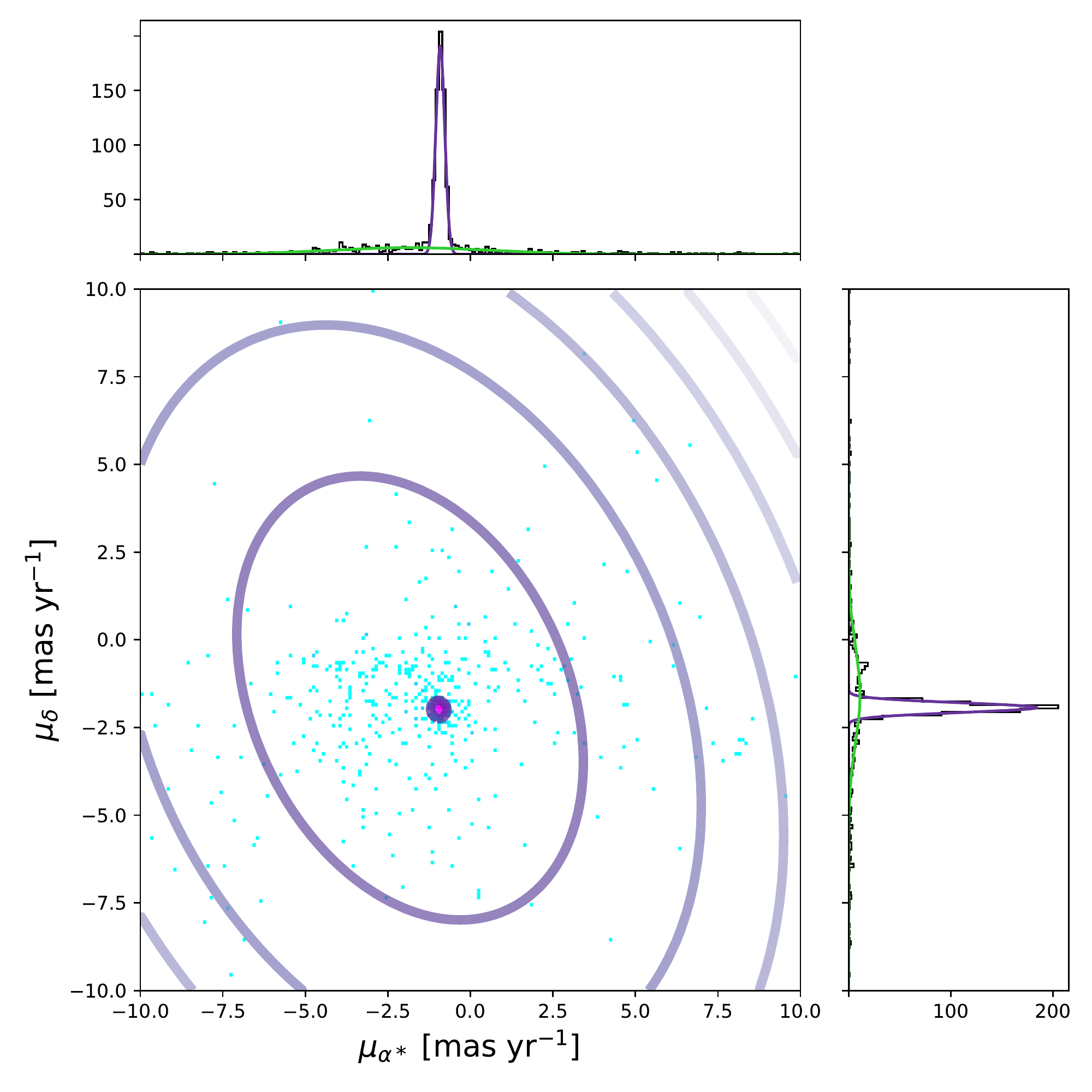}
    \caption{As in Figure 7, with the corresponding Gaussian fits overplotted. The fit to the cluster is tightly concentrated around the cluster proper motions in both the one-dimensional and two-dimensional cases, while the fit to the field is much more dispersed in proper-motion space. The contours in the central figure trace the  N-sigma confidence intervals for both of the fits; the confidence intervals to the cluster are very tightly concentrated.}
    \label{gauss:twod}
\end{figure*}

We use the astrometry from these 1102 stars to construct a two-dimensional Gaussian fit in proper-motion space in a similar manner as with the RVs. In this case, we fit two-dimensional Gaussian functions simultaneously to the proper motions of both cluster member and field stars.  A more general description of this process is laid out in \cite{Gao2018}. We show the two-dimensional histogram of the proper motions obtained from \gaia~ in Figure \ref{hist:twod}, and the Gaussian fits are shown in Figure \ref{gauss:twod}. We also show the parameters for these fits in Table \ref{ngc7789pm:tbl}.

\begin{figure}[htbp]
    \centering
    \includegraphics[width=0.75\linewidth]{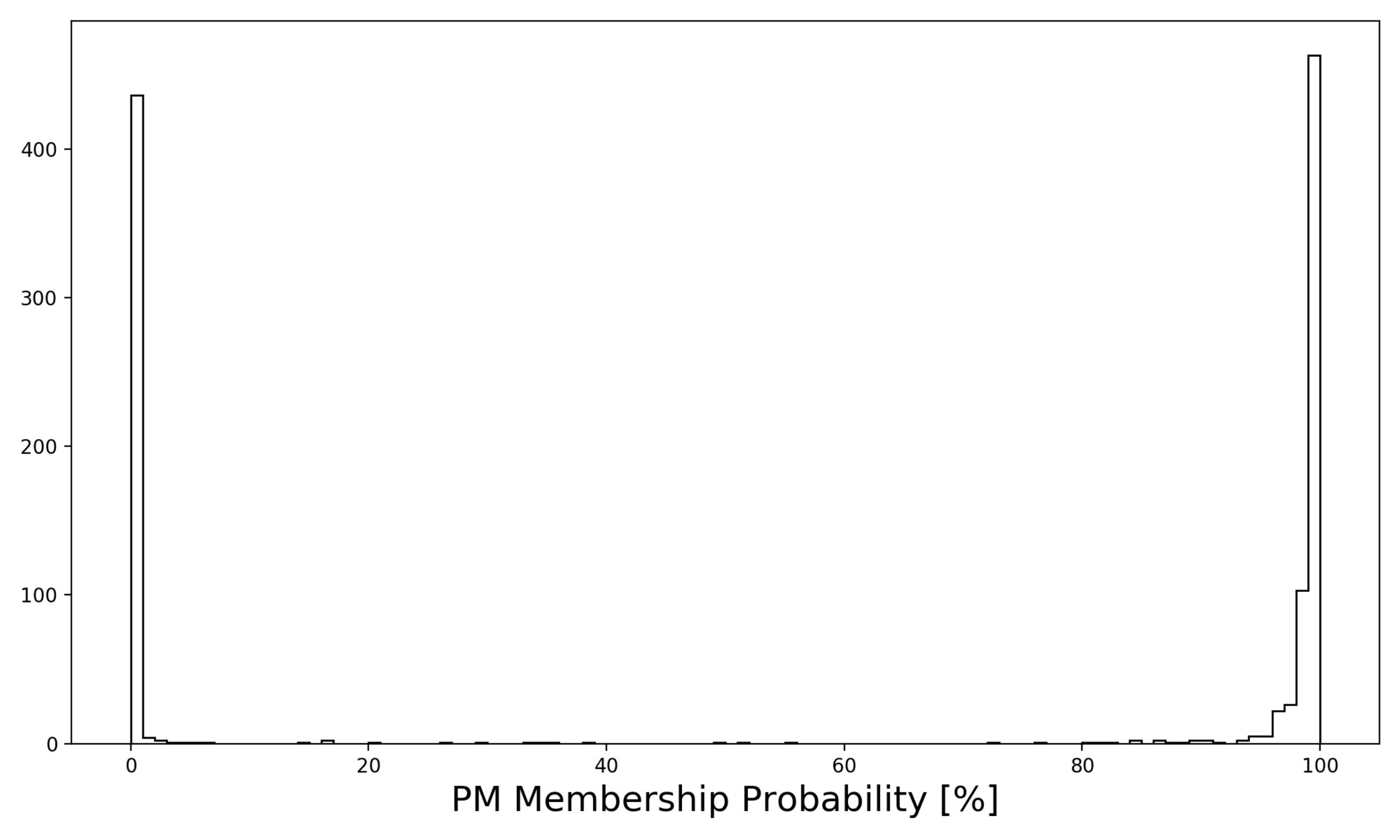}
    \caption{Histogram of the distribution of proper-motion membership probabilities for the stars with complete and low-noise astrometric data from \gaia.}
    \label{ngc7789:fig:ppm}
\end{figure}

As with the RVs, there is a clear separation in \PPM \, between cluster and field populations, as shown in Figure \ref{ngc7789:fig:ppm}. We again adopt the WOCS standard membership threshold of \PPM \, $\geq$ 50\%. We estimate from the amplitudes of the one-dimensional Gaussian functions fit to the cluster and field populations a field-star contamination of $\sim$4\% at this threshold.

\begin{figure}[htbp]
\begin{center}
\includegraphics[width=0.6\linewidth]{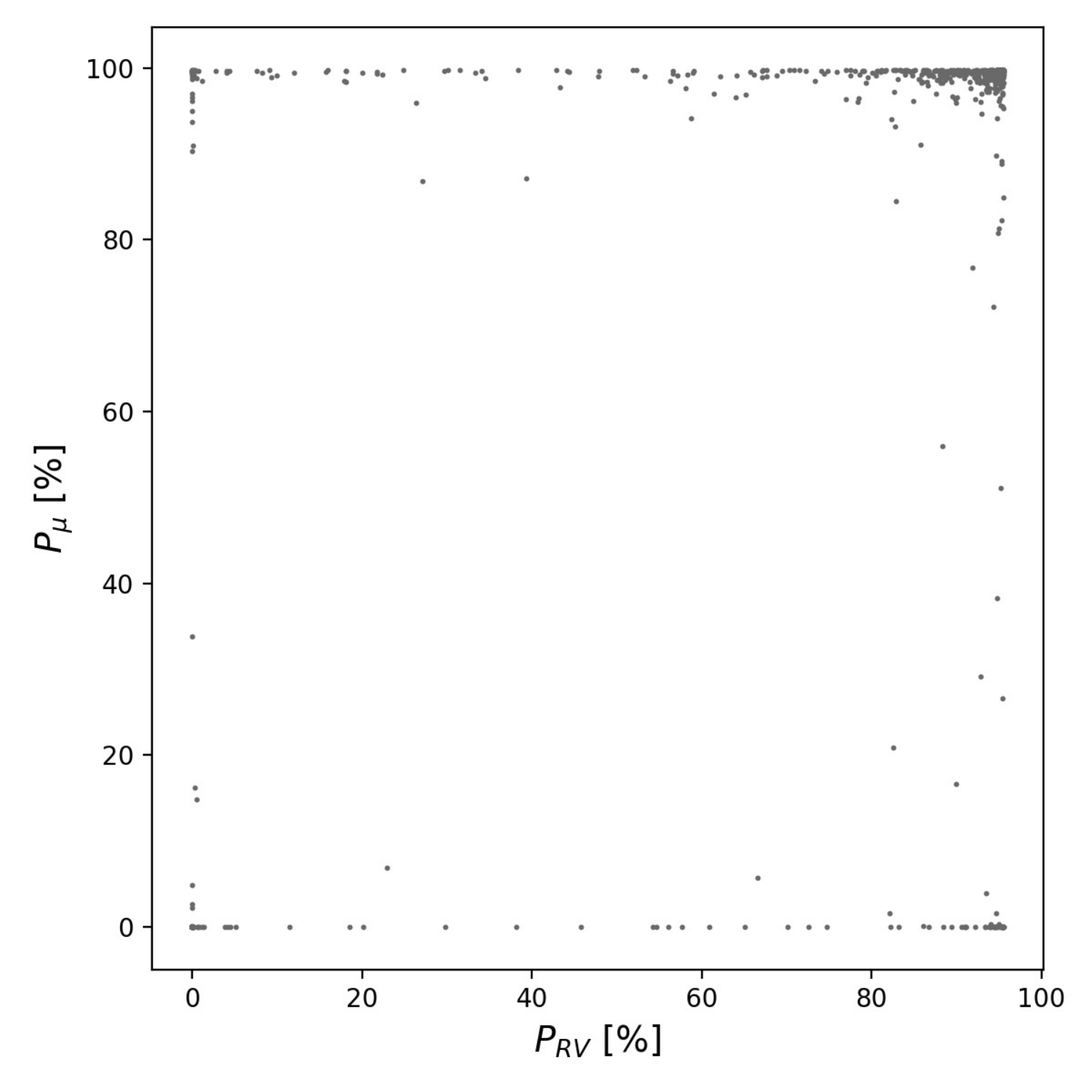}
\caption{Our radial-velocity membership probabilities compared with our proper-motion membership probabilities derived from \gaia \, for the stars in our sample that have been observed at least three times and have complete and low-noise astrometry available, excluding VRRs ($n$ = 979).}
\label{ngc7789rv:fig:prv.ppm}
\end{center}
\end{figure}

Where available we list the \PPM~in Table~\ref{ngc7789rv:tab:RVsum}. In Figure~\ref{ngc7789rv:fig:prv.ppm} we plot the membership percentages from our RV study and \gaia \, proper motions. We find the majority of stars that overlap between the data sets have membership classifications that agree. As noted above, we expect a field star contamination of 5\% within our sample of stars with \PRV~$\geq$ 50\% based on the Gaussian fits to the cluster and field star distribution. Comparing the \PPM~and \PRV~values we indeed find $\sim$7\% of the stars with \PRV~$\geq$ 50\% have proper motions that identify them as field stars, as can be seen in Figure \ref{ngc7789rv:fig:prv.ppm}.  

We note that in Figure \ref{ngc7789rv:fig:prv.ppm} there also appear a substantial number of stars for which \PRV~$<$ 50\% and \PPM~$>$ 50\%. Most of these stars are highly velocity variable, such that their mean RVs do not coincide with the mean RV of the cluster, but do include the mean within the range of their minimum and maximum measured RVs. There are 29 stars, however, that are either not velocity variable, rapidly rotating, or whose velocity variability does not include the mean RV of the cluster. Under the conservative assumption that these stars are all cluster nonmembers, they comprise $\sim$3\% of our sample, consistent with our estimate.

\begin{deluxetable}{ccc}
\tabletypesize{\footnotesize}
\tablewidth{0pt}
\tablecaption{\label{ngc7789pm:tbl}Two-Dimensional Gaussian Fit Parameters for Cluster and Field Proper Motions}
\tablehead{ \colhead{Parameter} & \colhead{Cluster} & \colhead{Field}}
\startdata
Ampl. [number] & \ClusAmpPM\ $\pm$ \ClusAmpPMerr & \FldAmpPM\ $\pm$ \FldAmpPMerr \\
$\overline{\mu_{\alpha*}}$ [mas yr$^{-1}$] & \ClusAvgPMRA\ $\pm$ \ClusAvgPMRAerr & \FldAvgPMRA\ $\pm$ \FldAvgPMRAerr\\
$\overline{\mu_{\delta}}$ [mas yr$^{-1}$] & \ClusAvgPMDEC\ $\pm$ \ClusAvgPMDECerr & \FldAvgPMDEC\ $\pm$ \FldAvgPMDECerr\\
$\sigma_{\alpha*}$ [mas yr$^{-1}$] & \ClusPMRASig\ $\pm$ \ClusPMRASigerr & \FldPMRASig\ $\pm$ \FldPMRASigerr \\
$\sigma_{\delta}$ [mas yr$^{-1}$] & \ClusPMDECSig\ $\pm$ \ClusPMDECSigerr & \FldPMDECSig\ $\pm$ \FldPMDECSigerr \\
\enddata
\end{deluxetable}

\pagebreak

\subsection{Membership Classification of Stars}

\label{ngc7789rv:sec:class}
Following previous WOCS procedure (\citealt{Geller2015}), we give in Table 2 membership classifications for stars in NGC 7789. The classifications are defined  below for ease of reference here. Table~\ref{ngc7789rv:t:cat} lists the number of stars in each membership classification. 

\textit{Single Member (SM)}: Stars that have $e/i$ $<$ 4, \PRV\ $\geq$ 50\%, and \PPM\ $\geq$ 50\%.

\textit{Single Non-member (SNM)}: Stars that have $e/i$ $<$ 4 and either \PRV\ $<$ 50\% or \PPM\ $<$ 50\%.

\textit{Binary Member (BM)}: Velocity-variable stars that have a completed orbital solution, \PRV\ $\geq$ 50\%, and \PPM\ $\geq$ 50\%.

\textit{Binary Non-member (BNM)}: Velocity-variable stars that have a completed orbital solution and either \PRV\ $<$ 50\% or  \PPM\ $<$ 50\%.

\textit{Binary Likely Member (BLM)}: Velocity-variable stars that do not have a completed orbital solution, \PRV\ $\geq$ 50\%, and \PPM\ $\geq$ 50\%.

\textit{Binary Likely Non-member (BLN)}: Velocity-variable stars that do not have a completed orbital solution. Either \PRV\ or \PPM\ is $<$ 50\% and the range of RVs does not include the cluster mean, making it unlikely that the orbital solution will place the star within the cluster distribution.

\textit{Binary Unknown (BU)}: Velocity-variable stars that do not have a completed orbital solution. \PRV\ is $<$ 50\%, but \PPM\ is $\geq$ 50\%, and the range of individual RVs includes the cluster mean, making it possible that the binary could be a member.

\textit{Very Rapid Rotator (VRR)}: Stars that are too rapidly rotating (\vsini~$>$ 120 \kms) for accurate RV measurements. 

For those 85 stars discussed in Section \ref{ngc7789rv:sec:ppm} and the 6 discussed in Section \ref{ngc7789rv:sec:sample} which do not have PM information available, we assign membership based solely on their RV information, using similar criteria as listed above.

\begin{deluxetable}{cc}
\tablecolumns{2}
\tablecaption{Number of Stars Within Each Classification
\label{ngc7789rv:t:cat}}
\tablehead{ \colhead{Classification} & \colhead{N Stars}}
\startdata
SM & \numcatSMCass \\
SNM & \numcatSNCass \\
BM & \numcatBMCass \\
BNM & \numcatBNCass \\
BLM & \numcatBLMCass \\
BLN & \numcatBLNCass \\
BU & \numcatBUCass \\
VRR & \numcatVRRCass \\
\enddata
\end{deluxetable}

\subsection{Color-Magnitude Diagram}
\label{ngc7789rv:sec:cmd}
\begin{figure*}[p!]
\includegraphics[width=1.0\linewidth]{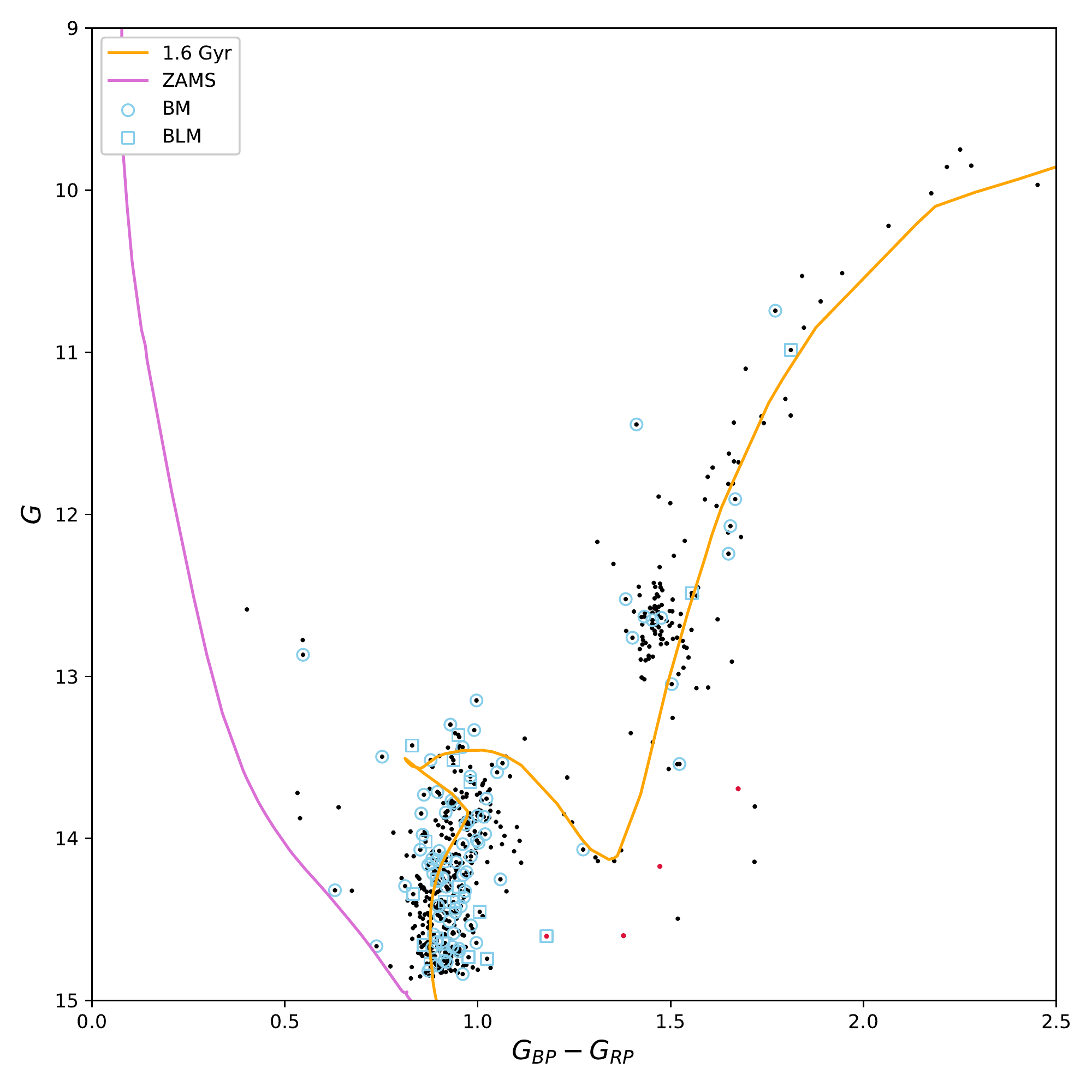}
\caption{NGC 7789 CMD of all three-dimensional members (SM, BLM, and BM) and a few members determined only by RV (Section \ref{ngc7789rv:sec:class}). Velocity-variable stars are indicated in blue, with circles for BMs with completed orbital solutions and squares for BLMs without completed orbital solutions. We overplot a 1.6 Gyr MIST (\citealt{Dotter2016}) isochrone in orange with a distance of 1.8 kpc, A$_V$ = 0.85, and [Fe/H] = 0.023, as well as a ZAMS in lavender with the same parameters. The SSGs and RSS discussed in Section \ref{ngc7789rv:sec:ssg_rss} are marked with red points.} 
\label{ngc7789rv:fig:cmd.member}
\end{figure*}

Using the \gaia\ photometry and our three-dimensional membership information we present a cleaned CMD of NGC 7789 members in Figure~\ref{ngc7789rv:fig:cmd.member}. This CMD includes all three-dimensional BM, BLM, and SM stars. The cleaned CMD clearly shows the main sequence, crowded with velocity-variable stars, a very well-defined red giant branch (RGB), and a populous red clump at G $\sim$ 13. A number of BSS remain, which we discuss in more detail in Section~\ref{ngc7789rv:sec:bs}.

We also show a 1.6 Gyr MIST (\citealt{Dotter2016}) isochrone with a distance of 1.8 kpc, A$_V$ = 0.85, and [Fe/H] = 0.023. These figures are roughly consistent with the heliocentric distance of 2074.7 $\pm$ 4.4 pc as measured by \citealt{CG2018}, $E(B-V)$ = 0.28 $\pm$ 0.02 as measured by \cite{Wu2007}, and [Fe/H] = 0.02 $\pm$ 0.04 as measured by \cite{Jacobson2011}. The parameters used in our isochrone fitting are not meant to supersede those already published. Our isochrone is meant rather to guide the eye and serve as a reference point to distinguish between MS stars and BSS candidates.

\subsection{Sub-Subgiants and Red Stragglers}
\label{ngc7789rv:sec:ssg_rss}

We note two three-dimensional-member stars are sub-subgiant (SSG) candidates that fall significantly to the red and fainter than the subgiant branch and RGB with \gming\: $\sim$\: 1.2 to 1.4. (Additional background on SSGs can be found in \citealt{Geller2017a}.) The candidates are WOCS 20035 and WOCS 35033. WOCS 20035 has \gming\: $=$ 1.18, \PRV\: $=$ 95\%, and \PPM\: $=$ 99\%. WOCS 35033 in turn has \gming\: $=$ 1.38, \PRV\: $=$ 94\%, and \PPM\: $=$ 93\%. WOCS 20035 is velocity variable with a relatively low amplitude of $\sim$7 \kms, but does not yet have a complete orbital solution. WOCS 35033 is a rapid rotator with a best-fit \vsini~of 120 \kms; at the consequent poorer measurement precision of $\sim$2 \kms, we have not detected velocity variability (see Equation \ref{eqn:rr_prec}). They are both targets of ongoing study.

We also note two three-dimensional member stars that lie to the red of the RGB: WOCS 19016 ($G=13.7$, \gming~ $=1.68$) and WOCS 22023 ($G=14.2$, \gming~ $=1.47$). These stars lie in the region of the CMD consistent with red stragglers (RSS; more background also can be found in \citealt{Geller2017a}). \cite{Geller2017a} note that out of the seven RSS candidates they identify, four of them were found to be X-ray sources and three were found to be photometric or velocity variables.  The three observations of each of these stars conducted to date, however, do not meet our velocity-variability threshold.

\cite{Geller2017b} suggest that SSGs and RSS may be linked through evolution; that is, SSGs may evolve into RSS. We might therefore expect these RSS candidates to be velocity variables, as are most SSGs, if the larger red stragglers can fit within the SSG orbits. We have obtained effective temperatures and luminosities of both WOCS 19016 and WOCS 22023 from \gaia: WOCS 19016 has \teff~ $= 4224 \text{K}$ and $L = 73 L_{\odot}$, while WOCS 22023 has \teff~ $= 4531 \text{K}$ and $L = 9.5 L_{\odot}$. From these values we derive estimated stellar radii of 16 \Rsolar\, and 5 \Rsolar\,, respectively. Following the formalism of \cite{Paczy1971}, we derive the minimum orbital periods of these two stars with the relation 

\begin{equation}
    \frac{R_{L1}}{a} = 0.38+0.2\log_{10} \left(\frac{M_1}{M_2}\right),
\end{equation}

\noindent where $R_{L1}$ is the Roche lobe radius of the primary star. For each star we set $R_{L1}$ equal to the derived radius of the star. Assuming each star is near the cluster turnoff with a mass of $\sim$1.8 \Msolar~and each has a 1 \Msolar~companion orbiting at the Roche radius, we find limiting orbital periods of $\sim$17\,d for WOCS 19016 and $\sim$3\,d for WOCS 22023. These periods are typical of SSGs (albeit perhaps not WOCS 20035 here), and so do not evidently explain the lack of velocity variability of these RSS candidates.

We also examine the respective critical separations for stability against tidal mergers, using the relation from \cite{Rasio1995}:

\begin{equation}
    \frac{a_{\text{inst}}}{R_1} = \left[\frac{3(1+q)}{q}\right]^{1/2}k_1,
\end{equation}

\noindent where $q$ is defined to be $M_2$/$M_1$, $R_1$ is the radius of the primary star, and $k_1$ is the dimensionless gyroradius of the primary star. If we assume $k_1^2 \sim 0.1$, as does \cite{Webbink1976}, the critical separation for each star corresponds to a period of $\sim$120\,d for WOCS 19016 and $\sim$22\,d for WOCS 22023, respectively, both substantially longer than the minimum periods at their Roche radii. We note that these results depend somewhat sensitively on the choice of $k_1$. We also note that the relation of \cite{Rasio1995} is a good approximation only for a mostly radiative main-sequence star. Stars with more substantial convective envelopes are more compressible than primarily radiative stars,  which tends to shrink $a_{\text{inst}}$ for a fixed primary radius. Presuming sub-subgiants and their descendants do have substantial convective envelopes, our derived $a_{\text{inst}}$ could be too large by as much as $\sim$30\%. It is therefore possible that these two stars are the result of merged sub-subgiants in close binaries, once the primary star began evolving along its RGB. Neither of these two stars, however, have been observed to be rapidly rotating, challenging the merger hypothesis. These two stars remain targets of ongoing study.

There are three other stars in the sub-subgiant and red straggler regions of the CMD: WOCS 24035, WOCS 29015, and WOCS 35029. All of these stars have high-noise astrometry, and as such we do not have reliable \PPM\, for them. These stars are placed on the CMD based solely on their \PRV, and none of them show velocity variability or unusually large radii. While these stars are also the subject of continued study, we do not label them as RSS or SSGs at this time because we cannot say with the same level of certainty that they are cluster members.

\clearpage
\clearpage

\section{Spectroscopic Binary Orbits}
\label{ngc7789rv:sec:Specs}

\subsection{Single-Lined Orbital Solutions}
\label{ngc7789rv:sec:SB1}
\begin{figure*}
\gridline{\fig{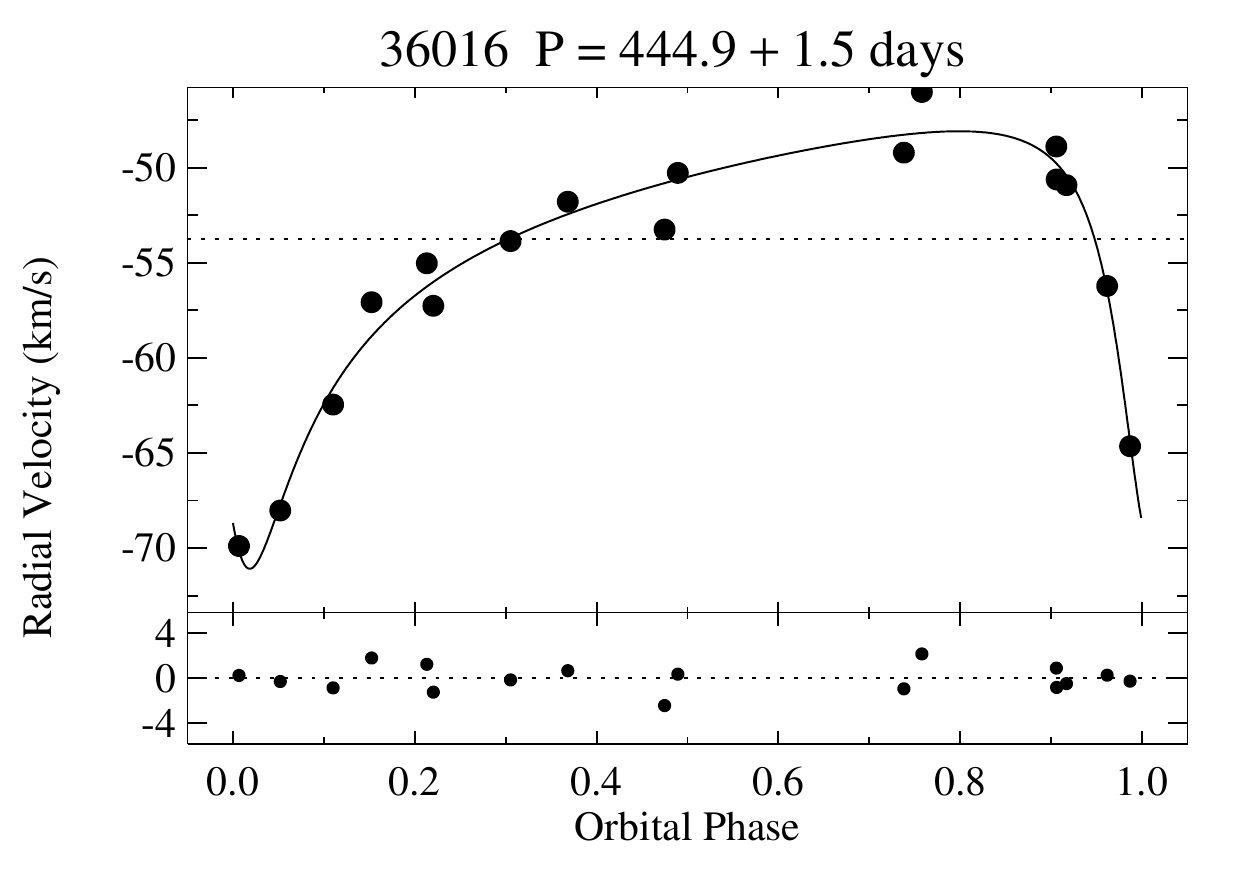}{0.3\linewidth}{}
  \fig{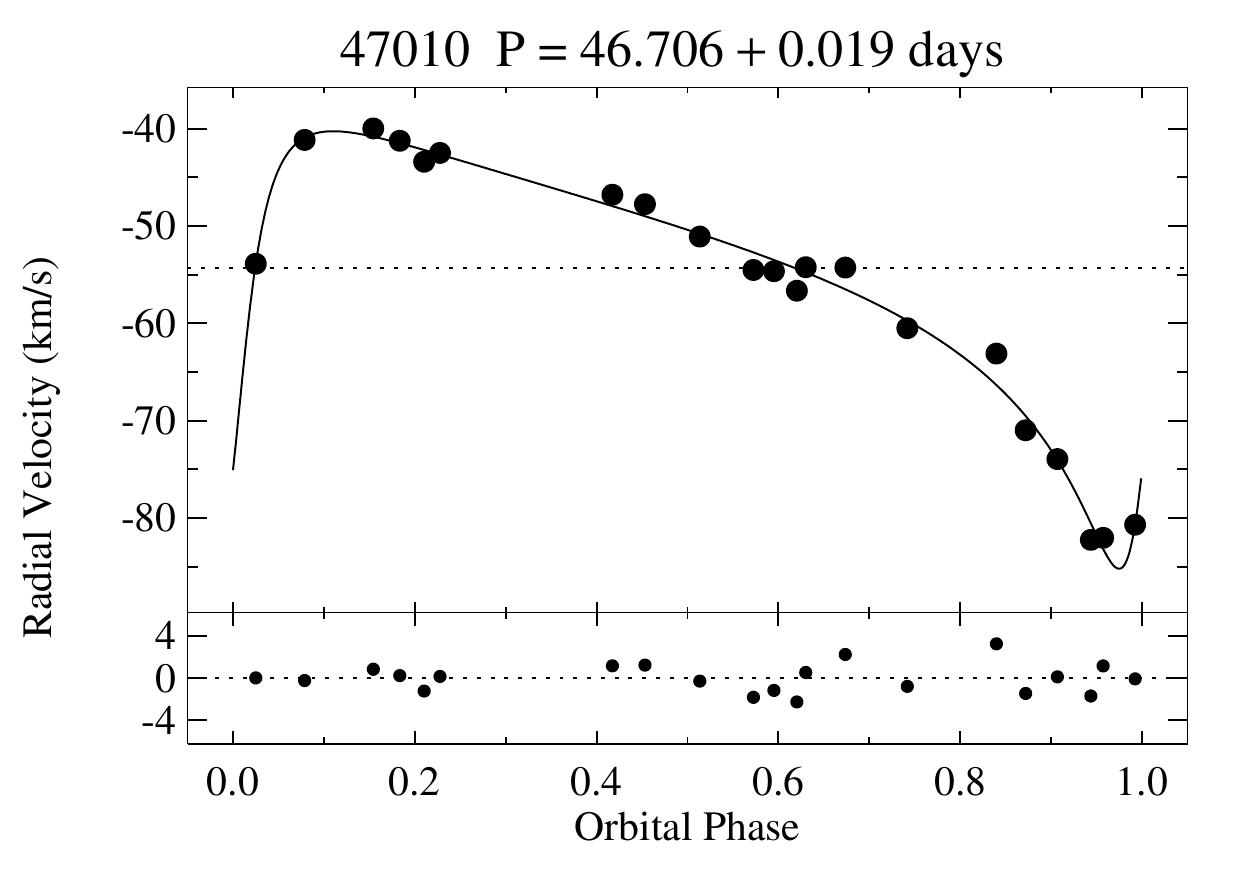}{0.3\linewidth}{}}
\caption{\footnotesize NGC 7789 SB1 orbit plots.  We plot RV against orbital phase for each binary with the WOCS ID and orbital period above each plot. Filled circles indicate the data points, the solid line is the orbital fit to the data, and the dotted line marks the $\gamma$-velocity. We show the residuals from the fit below each binary plot.\normalsize}
\label{fig:Sb1plots}
\end{figure*}

Using the data given in Table~\ref{ngc7789rv:tab:rvs} we determine orbital solutions for 62
single-lined spectroscopic binaries (SB1). We show these orbit solutions in Figure~\ref{fig:Sb1plots}. 
For each, the orbit solution is plotted in the top panel and the observed-minus-computed-residuals (O$-$C) are plotted in the bottom panel. The orbital elements and the 1$\sigma$ errors for each binary are listed in Table~\ref{SB1tab}. The first row contains the binary ID, the orbital period ($P$), the number of orbital cycles observed, the center-of-mass RV ($\gamma$), the orbital amplitude ($K$), the eccentricity ($e$), the longitude of periastron ($\omega$), a Julian Date of periastron passage ($T_{\circ}$), the projected semi-major axis ($a\,$sin$\,i$), the mass function ($f(m)$), the rms residual velocity from the orbital solution ($\sigma$), and the number of RV measurements ($N$). The second row contains the respective errors.

\clearpage
\startlongtable
\centerwidetable
\begin{deluxetable*}{l r c r r r r r r r c c}
\tabletypesize{\footnotesize}
\tablewidth{0pt}
\centering
\tablecaption{Orbital Parameters for NGC 7789 Single-Lined Binaries}\label{SB1tab}
\tablehead{\colhead{ID} & \colhead{P} & \colhead{Orbital} & \colhead{$\gamma$} & \colhead{K} & \colhead{e} & \colhead{$\omega$} & \colhead{T$_\circ$} & \colhead{a$\sin$ i} & \colhead{f(m)} & \colhead{$\sigma$} & \colhead{N} \\
\colhead{} & \colhead{(days)} & \colhead{Cycles} & \colhead{(\kms)} & \colhead{(\kms)} & \colhead{} & \colhead{(deg)} & \colhead{(HJD-2400000 d)} & \colhead{(10$^6$ km)} & \colhead{(\Msolar)} & \colhead{(\kms)} & \colhead{}}
\startdata
    1006 &           783.9 &    2.3 &          -54.96 &            14.3 &           0.413 &              54 &           55421 &             141 &         1.81e-1 &  0.63 &   16 \\
         &       $\pm$ 2.3 &       &      $\pm$ 0.18 &       $\pm$ 0.3 &     $\pm$ 0.014 &         $\pm$ 3 &         $\pm$ 4 &         $\pm$ 3 &    $\pm$ 1.2e-2 &       &      \\
    2001 &          58.581 &   55.7 &           -52.5 &            26.5 &           0.151 &              72 &         55049.4 &            21.1 &         1.10e-1 &  1.75 &   36 \\
         &     $\pm$ 0.017 &       &       $\pm$ 0.4 &       $\pm$ 0.6 &     $\pm$ 0.022 &         $\pm$ 6 &       $\pm$ 1.0 &       $\pm$ 0.4 &    $\pm$ 0.7e-2 &       &      \\
    4004 &            1546 &    1.3 &          -55.09 &              20 &            0.90 &             286 &           55144 &             190 &          1.2e-1 &  0.39 &   26 \\
         &         $\pm$ 3 &       &      $\pm$ 0.19 &         $\pm$ 5 &      $\pm$ 0.05 &         $\pm$ 6 &         $\pm$ 4 &        $\pm$ 60 &    $\pm$ 1.0e-1 &       &      \\
    4035 &          155.79 &   13.4 &          -55.02 &           26.77 &           0.442 &            50.8 &         54693.2 &            51.4 &         2.23e-1 &  0.35 &   17 \\
         &      $\pm$ 0.04 &       &      $\pm$ 0.10 &      $\pm$ 0.15 &     $\pm$ 0.004 &       $\pm$ 0.8 &       $\pm$ 0.3 &       $\pm$ 0.3 &    $\pm$ 0.4e-2 &       &      \\
    5008 &           217.3 &    9.4 &          -54.77 &            4.97 &            0.06 &             330 &           55042 &            14.8 &          2.8e-3 &  0.51 &   15 \\
         &       $\pm$ 0.5 &       &      $\pm$ 0.14 &      $\pm$ 0.20 &      $\pm$ 0.05 &        $\pm$ 40 &        $\pm$ 22 &       $\pm$ 0.6 &    $\pm$ 0.3e-3 &       &      \\
    5009 &          202.20 &   24.9 &          -54.50 &           21.48 &           0.011 &             140 &           54500 &            59.7 &         2.07e-1 &  0.39 &   21 \\
         &      $\pm$ 0.03 &       &      $\pm$ 0.09 &      $\pm$ 0.14 &     $\pm$ 0.006 &        $\pm$ 40 &        $\pm$ 21 &       $\pm$ 0.4 &    $\pm$ 0.4e-2 &       &      \\
    5010 &           451.0 &    4.9 &          -54.51 &           17.03 &           0.262 &            71.8 &         55146.4 &           101.9 &         2.07e-1 &  0.20 &   13 \\
         &       $\pm$ 0.4 &       &      $\pm$ 0.07 &      $\pm$ 0.10 &     $\pm$ 0.005 &       $\pm$ 1.4 &       $\pm$ 1.6 &       $\pm$ 0.6 &    $\pm$ 0.4e-2 &       &      \\
    5011 &            2710 &    1.4 &          -53.78 &            4.15 &            0.32 &               6 &           55980 &             147 &         1.71e-2 &  0.67 &   34 \\
         &        $\pm$ 60 &       &      $\pm$ 0.13 &      $\pm$ 0.20 &      $\pm$ 0.04 &         $\pm$ 9 &        $\pm$ 60 &         $\pm$ 7 &    $\pm$ 2.5e-3 &       &      \\
    7014 &          359.16 &   14.0 &          -55.15 &           12.30 &           0.484 &             262 &         55093.9 &            53.1 &         4.64e-2 &  0.54 &   22 \\
         &      $\pm$ 0.23 &       &      $\pm$ 0.14 &      $\pm$ 0.19 &     $\pm$ 0.011 &         $\pm$ 3 &       $\pm$ 1.9 &       $\pm$ 0.9 &    $\pm$ 2.2e-3 &       &      \\
    8007 &          55.563 &   37.7 &          -54.28 &           22.12 &           0.168 &              33 &         55027.5 &           16.66 &         5.97e-2 &  0.56 &   15 \\
         &     $\pm$ 0.018 &       &      $\pm$ 0.23 &      $\pm$ 0.20 &     $\pm$ 0.011 &         $\pm$ 4 &       $\pm$ 0.6 &      $\pm$ 0.15 &    $\pm$ 1.6e-3 &       &      \\
    8029 &          267.72 &    3.3 &          -56.17 &           14.07 &           0.269 &           289.4 &         56962.3 &            49.9 &         6.90e-2 &  0.17 &   13 \\
         &      $\pm$ 0.22 &       &      $\pm$ 0.05 &      $\pm$ 0.08 &     $\pm$ 0.005 &       $\pm$ 1.3 &       $\pm$ 0.9 &       $\pm$ 0.3 &    $\pm$ 1.2e-3 &       &      \\
    9011 &             680 &    3.5 &           -54.4 &             6.9 &            0.05 &             240 &           54950 &              64 &          2.3e-2 &  0.76 &   19 \\
         &         $\pm$ 4 &       &       $\pm$ 0.3 &       $\pm$ 0.4 &      $\pm$ 0.05 &        $\pm$ 60 &       $\pm$ 120 &         $\pm$ 4 &    $\pm$ 0.4e-2 &       &      \\
   10011 &             517 &    5.5 &           -53.7 &             6.5 &            0.37 &              79 &           56773 &              43 &         1.19e-2 &  1.14 &   22 \\
         &         $\pm$ 6 &       &       $\pm$ 0.3 &       $\pm$ 0.4 &      $\pm$ 0.07 &        $\pm$ 10 &        $\pm$ 11 &         $\pm$ 3 &    $\pm$ 2.3e-3 &       &      \\
   11017 &         13.7827 &  160.5 &          -53.21 &            9.50 &           0.053 &             102 &         55167.6 &            1.80 &         1.22e-3 &  0.46 &   17 \\
         &    $\pm$ 0.0009 &       &      $\pm$ 0.13 &      $\pm$ 0.15 &     $\pm$ 0.024 &        $\pm$ 19 &       $\pm$ 0.8 &      $\pm$ 0.03 &    $\pm$ 0.6e-4 &       &      \\
   12002 &         13.1721 &  217.2 &           -54.1 &            46.7 &           0.015 &             130 &         56546.5 &            8.46 &         1.39e-1 &  1.55 &   15 \\
         &    $\pm$ 0.0008 &       &       $\pm$ 0.5 &       $\pm$ 0.6 &     $\pm$ 0.018 &        $\pm$ 60 &       $\pm$ 2.0 &      $\pm$ 0.12 &    $\pm$ 0.6e-2 &       &      \\
   12004 &          9.4119 &  192.7 &           -54.5 &            27.2 &           0.044 &             175 &         54234.7 &            3.52 &         1.96e-2 &  1.06 &   17 \\
         &    $\pm$ 0.0005 &       &       $\pm$ 0.3 &       $\pm$ 0.4 &     $\pm$ 0.015 &        $\pm$ 18 &       $\pm$ 0.5 &      $\pm$ 0.05 &    $\pm$ 0.8e-3 &       &      \\
   12006 &          114.56 &   18.2 &          -54.49 &           18.34 &           0.151 &              73 &         54636.8 &            28.6 &         7.08e-2 &  0.74 &   29 \\
         &      $\pm$ 0.03 &       &      $\pm$ 0.16 &      $\pm$ 0.21 &     $\pm$ 0.013 &         $\pm$ 5 &       $\pm$ 1.5 &       $\pm$ 0.3 &    $\pm$ 2.4e-3 &       &      \\
   12007 &          85.509 &   58.8 &           -54.8 &            20.9 &           0.064 &             127 &           54961 &            24.6 &          8.1e-2 &  1.34 &   27 \\
         &     $\pm$ 0.018 &       &       $\pm$ 0.3 &       $\pm$ 0.4 &     $\pm$ 0.019 &        $\pm$ 18 &         $\pm$ 4 &       $\pm$ 0.5 &    $\pm$ 0.5e-2 &       &      \\
   13007 &           532.2 &    9.4 &          -54.57 &             8.2 &            0.30 &              68 &           54913 &            57.0 &          2.6e-2 &  0.74 &   26 \\
         &       $\pm$ 1.4 &       &      $\pm$ 0.16 &       $\pm$ 0.3 &      $\pm$ 0.03 &         $\pm$ 5 &         $\pm$ 7 &       $\pm$ 1.9 &    $\pm$ 0.3e-2 &       &      \\
   13017 &          45.047 &  104.3 &          -53.42 &            11.3 &            0.20 &             340 &         55229.3 &            6.84 &          6.3e-3 &  0.77 &   15 \\
         &     $\pm$ 0.009 &       &      $\pm$ 0.22 &       $\pm$ 0.4 &      $\pm$ 0.03 &         $\pm$ 9 &       $\pm$ 1.1 &      $\pm$ 0.24 &    $\pm$ 0.6e-3 &       &      \\
   13021 &           205.8 &   17.8 &           -52.1 &            10.0 &            0.47 &              22 &           55522 &            25.1 &          1.5e-2 &  1.13 &   16 \\
         &       $\pm$ 0.3 &       &       $\pm$ 0.3 &       $\pm$ 0.6 &      $\pm$ 0.04 &        $\pm$ 11 &         $\pm$ 7 &       $\pm$ 1.7 &    $\pm$ 0.3e-2 &       &      \\
   14008 &           225.4 &    9.6 &          -55.28 &            7.79 &            0.02 &             310 &           54590 &            24.1 &         1.10e-2 &  0.68 &   17 \\
         &       $\pm$ 0.4 &       &      $\pm$ 0.19 &      $\pm$ 0.23 &      $\pm$ 0.04 &       $\pm$ 120 &        $\pm$ 80 &       $\pm$ 0.7 &    $\pm$ 1.0e-3 &       &      \\
   14011 &            1420 &    1.8 &          -52.08 &             6.1 &            0.73 &             143 &           55459 &              81 &          1.0e-2 &  0.80 &   18 \\
         &        $\pm$ 60 &       &      $\pm$ 0.20 &       $\pm$ 0.5 &      $\pm$ 0.03 &         $\pm$ 8 &        $\pm$ 15 &         $\pm$ 7 &    $\pm$ 0.3e-2 &       &      \\
   15012 &          56.360 &   83.3 &           -52.4 &            19.8 &           0.188 &             274 &         55761.2 &            15.1 &         4.30e-2 &  0.90 &   15 \\
         &     $\pm$ 0.014 &       &       $\pm$ 0.3 &       $\pm$ 0.3 &     $\pm$ 0.020 &         $\pm$ 6 &       $\pm$ 0.9 &       $\pm$ 0.3 &    $\pm$ 2.3e-3 &       &      \\
   15016 &            1840 &    1.4 &          -53.25 &             4.9 &            0.43 &             141 &           54180 &             111 &          1.6e-2 &  0.71 &   17 \\
         &        $\pm$ 60 &       &      $\pm$ 0.23 &       $\pm$ 0.5 &      $\pm$ 0.07 &         $\pm$ 8 &        $\pm$ 70 &        $\pm$ 11 &    $\pm$ 0.5e-2 &       &      \\
   16013 &           277.7 &    9.3 &          -54.11 &             4.8 &            0.33 &              54 &           56935 &            17.3 &          2.7e-3 &  0.47 &   14 \\
         &       $\pm$ 0.8 &       &      $\pm$ 0.21 &       $\pm$ 0.3 &      $\pm$ 0.06 &        $\pm$ 12 &         $\pm$ 6 &       $\pm$ 1.0 &    $\pm$ 0.4e-3 &       &      \\
   18003 &         3.09148 &  925.2 &          -54.66 &            27.6 &           0.148 &             299 &        56395.40 &           1.162 &          6.6e-3 &  0.74 &   15 \\
         &   $\pm$ 0.00005 &       &      $\pm$ 0.25 &       $\pm$ 0.5 &     $\pm$ 0.012 &         $\pm$ 5 &      $\pm$ 0.04 &     $\pm$ 0.020 &    $\pm$ 0.3e-3 &       &      \\
   18009 &          130.05 &   19.9 &           -54.8 &            13.3 &            0.27 &             313 &         56733.9 &            22.9 &          2.8e-2 &  0.91 &   14 \\
         &      $\pm$ 0.07 &       &       $\pm$ 0.3 &       $\pm$ 0.5 &      $\pm$ 0.03 &         $\pm$ 6 &       $\pm$ 2.1 &       $\pm$ 0.8 &    $\pm$ 0.3e-2 &       &      \\
   18015 &          57.105 &   45.4 &          -54.33 &            24.1 &           0.386 &            90.2 &         56778.1 &           17.48 &          6.5e-2 &  0.63 &   13 \\
         &     $\pm$ 0.017 &       &      $\pm$ 0.18 &       $\pm$ 0.3 &     $\pm$ 0.011 &       $\pm$ 2.3 &       $\pm$ 0.3 &      $\pm$ 0.24 &    $\pm$ 0.3e-2 &       &      \\
   18028 &          32.551 &   22.7 &          -51.81 &           21.07 &           0.289 &           234.8 &        56913.66 &            9.03 &         2.77e-2 &  0.14 &   14 \\
         &     $\pm$ 0.003 &       &      $\pm$ 0.06 &      $\pm$ 0.07 &     $\pm$ 0.003 &       $\pm$ 0.7 &      $\pm$ 0.05 &      $\pm$ 0.03 &    $\pm$ 0.3e-3 &       &      \\
   19008 &         2.62664 &  576.0 &           -53.8 &            30.6 &           0.006 &             270 &         54877.9 &           1.107 &          7.8e-3 &  1.30 &   22 \\
         &   $\pm$ 0.00003 &       &       $\pm$ 0.3 &       $\pm$ 0.4 &     $\pm$ 0.016 &       $\pm$ 110 &       $\pm$ 0.8 &     $\pm$ 0.016 &    $\pm$ 0.3e-3 &       &      \\
   19011 &           206.1 &    8.9 &          -55.31 &             3.0 &            0.44 &              72 &           56277 &             7.6 &          4.1e-4 &  0.95 &   17 \\
         &       $\pm$ 1.6 &       &      $\pm$ 0.24 &       $\pm$ 0.4 &      $\pm$ 0.10 &        $\pm$ 20 &         $\pm$ 8 &       $\pm$ 1.1 &    $\pm$ 1.7e-4 &       &      \\
   19019 &          7.1108 &  159.9 &          -53.50 &           10.85 &           0.148 &             160 &        56087.61 &           1.049 &          9.1e-4 &  0.42 &   12 \\
         &    $\pm$ 0.0003 &       &      $\pm$ 0.14 &      $\pm$ 0.22 &     $\pm$ 0.018 &         $\pm$ 8 &      $\pm$ 0.15 &     $\pm$ 0.022 &    $\pm$ 0.6e-4 &       &      \\
   20006 &           360.5 &    5.5 &           -52.9 &            14.5 &            0.50 &              25 &           55066 &              62 &          7.3e-2 &  1.83 &   16 \\
         &       $\pm$ 1.4 &       &       $\pm$ 0.5 &       $\pm$ 0.9 &      $\pm$ 0.04 &         $\pm$ 7 &         $\pm$ 4 &         $\pm$ 4 &    $\pm$ 1.4e-2 &       &      \\
   20009 &            4190 &    0.9 &          -56.21 &             4.5 &            0.27 &              11 &           58390 &             253 &          3.6e-2 &  1.19 &   44 \\
         &       $\pm$ 230 &       &      $\pm$ 0.21 &       $\pm$ 0.4 &      $\pm$ 0.09 &        $\pm$ 17 &       $\pm$ 160 &        $\pm$ 23 &    $\pm$ 1.0e-2 &       &      \\
   20014 &         16.5000 &  304.8 &           -54.5 &            10.4 &            0.05 &              20 &         55194.7 &            2.36 &         1.92e-3 &  0.98 &   19 \\
         &    $\pm$ 0.0015 &       &       $\pm$ 0.3 &       $\pm$ 0.3 &      $\pm$ 0.03 &        $\pm$ 50 &       $\pm$ 2.3 &      $\pm$ 0.07 &    $\pm$ 1.7e-4 &       &      \\
   21006 &           549.1 &    4.6 &           -56.8 &            21.1 &           0.657 &              38 &           54898 &             120 &          2.3e-1 &  1.20 &   17 \\
         &       $\pm$ 1.8 &       &       $\pm$ 0.3 &       $\pm$ 0.9 &     $\pm$ 0.019 &         $\pm$ 3 &         $\pm$ 3 &         $\pm$ 6 &    $\pm$ 0.3e-1 &       &      \\
   21010 &         24.1219 &  208.5 &          -54.04 &           21.46 &           0.017 &              80 &           55326 &            7.12 &         2.47e-2 &  0.78 &   21 \\
         &    $\pm$ 0.0015 &       &      $\pm$ 0.19 &      $\pm$ 0.25 &     $\pm$ 0.013 &        $\pm$ 40 &         $\pm$ 3 &      $\pm$ 0.08 &    $\pm$ 0.9e-3 &       &      \\
   22008 &           413.7 &    9.4 &           -52.2 &               7 &            0.79 &             231 &           55598 &              25 &          4.0e-3 &  1.19 &   24 \\
         &       $\pm$ 0.6 &       &       $\pm$ 0.4 &         $\pm$ 4 &      $\pm$ 0.16 &        $\pm$ 17 &         $\pm$ 3 &        $\pm$ 17 &    $\pm$ 0.7e-2 &       &      \\
   22010 &         9.05801 &  315.9 &          -54.58 &           20.33 &           0.019 &              72 &         56572.3 &           2.532 &         7.88e-3 &  0.39 &   14 \\
         &   $\pm$ 0.00022 &       &      $\pm$ 0.13 &      $\pm$ 0.19 &     $\pm$ 0.010 &        $\pm$ 22 &       $\pm$ 0.5 &     $\pm$ 0.024 &    $\pm$ 2.2e-4 &       &      \\
   22018 &        1.883543 &  691.1 &           -51.4 &            34.8 &           0.038 &               6 &        56442.86 &           0.900 &          8.2e-3 &  0.97 &   12 \\
         &  $\pm$ 0.000019 &       &       $\pm$ 0.3 &       $\pm$ 0.5 &     $\pm$ 0.019 &        $\pm$ 17 &      $\pm$ 0.09 &     $\pm$ 0.013 &    $\pm$ 0.3e-3 &       &      \\
   24005 &            1054 &    4.8 &          -53.09 &             4.3 &            0.32 &              78 &           54920 &              59 &          7.2e-3 &  0.72 &   19 \\
         &        $\pm$ 15 &       &      $\pm$ 0.18 &       $\pm$ 0.3 &      $\pm$ 0.06 &        $\pm$ 13 &        $\pm$ 40 &         $\pm$ 4 &    $\pm$ 1.4e-3 &       &      \\
   24025 &           208.7 &   10.6 &           -54.2 &            10.4 &            0.27 &             307 &           55136 &            28.7 &          2.2e-2 &  0.98 &   17 \\
         &       $\pm$ 0.3 &       &       $\pm$ 0.3 &       $\pm$ 0.4 &      $\pm$ 0.04 &         $\pm$ 8 &         $\pm$ 4 &       $\pm$ 1.3 &    $\pm$ 0.3e-2 &       &      \\
   25015 &           50.58 &   25.5 &           -54.1 &            21.9 &            0.10 &             179 &           56335 &            15.1 &          5.4e-2 &  1.22 &   12 \\
         &      $\pm$ 0.03 &       &       $\pm$ 0.5 &       $\pm$ 0.6 &      $\pm$ 0.03 &        $\pm$ 20 &         $\pm$ 3 &       $\pm$ 0.4 &    $\pm$ 0.5e-2 &       &      \\
   26019 &             885 &    3.8 &          -53.71 &             5.9 &            0.34 &             339 &           57460 &              67 &          1.5e-2 &  1.08 &   31 \\
         &         $\pm$ 7 &       &      $\pm$ 0.23 &       $\pm$ 0.6 &      $\pm$ 0.09 &         $\pm$ 9 &        $\pm$ 19 &         $\pm$ 7 &    $\pm$ 0.5e-2 &       &      \\
   27008 &             735 &    2.3 &          -54.18 &            10.1 &            0.37 &             186 &           55893 &              95 &          6.2e-2 &  0.47 &   12 \\
         &         $\pm$ 3 &       &      $\pm$ 0.16 &       $\pm$ 0.5 &      $\pm$ 0.04 &         $\pm$ 5 &         $\pm$ 8 &         $\pm$ 5 &    $\pm$ 0.9e-2 &       &      \\
   27012 &          143.55 &   25.6 &           -53.3 &            13.9 &            0.31 &              53 &         55905.7 &            26.2 &          3.5e-2 &  0.91 &   14 \\
         &      $\pm$ 0.13 &       &       $\pm$ 0.3 &       $\pm$ 0.4 &      $\pm$ 0.03 &         $\pm$ 6 &       $\pm$ 2.2 &       $\pm$ 0.8 &    $\pm$ 0.3e-2 &       &      \\
   28010 &           277.1 &   14.9 &           -55.4 &             7.9 &            0.51 &              14 &           55402 &            26.0 &          9.1e-3 &  1.01 &   16 \\
         &       $\pm$ 0.4 &       &       $\pm$ 0.3 &       $\pm$ 0.5 &      $\pm$ 0.04 &         $\pm$ 7 &         $\pm$ 3 &       $\pm$ 1.7 &    $\pm$ 1.7e-3 &       &      \\
   28032 &          9.2830 &  113.5 &           -54.5 &             9.1 &            0.18 &             348 &         56465.5 &            1.14 &          6.9e-4 &  0.82 &   12 \\
         &    $\pm$ 0.0022 &       &       $\pm$ 0.3 &       $\pm$ 0.8 &      $\pm$ 0.06 &        $\pm$ 18 &       $\pm$ 0.5 &      $\pm$ 0.10 &    $\pm$ 1.9e-4 &       &      \\
   28035 &        3.825608 &  543.0 &           -55.2 &            58.4 &           0.006 &             190 &         54712.0 &           3.073 &         7.90e-2 &  1.54 &   29 \\
         &  $\pm$ 0.000025 &       &       $\pm$ 0.3 &       $\pm$ 0.4 &     $\pm$ 0.007 &        $\pm$ 80 &       $\pm$ 0.9 &     $\pm$ 0.021 &    $\pm$ 1.6e-3 &       &      \\
   29011 &           80.35 &   62.5 &          -53.93 &            10.1 &            0.19 &              42 &         55087.6 &            11.0 &          8.1e-3 &  0.67 &   16 \\
         &      $\pm$ 0.04 &       &      $\pm$ 0.17 &       $\pm$ 0.3 &      $\pm$ 0.03 &         $\pm$ 8 &       $\pm$ 1.6 &       $\pm$ 0.3 &    $\pm$ 0.6e-3 &       &      \\
   29022 &         13.8619 &  362.7 &           -54.5 &            33.3 &           0.007 &             210 &           54936 &            6.34 &          5.3e-2 &  1.72 &   31 \\
         &    $\pm$ 0.0003 &       &       $\pm$ 0.4 &       $\pm$ 0.7 &     $\pm$ 0.017 &       $\pm$ 110 &         $\pm$ 4 &      $\pm$ 0.13 &    $\pm$ 0.3e-2 &       &      \\
   31008 &         26.6262 &   82.7 &          -54.56 &            16.2 &           0.103 &             343 &         56334.2 &            5.90 &         1.15e-2 &  0.36 &   16 \\
         &    $\pm$ 0.0019 &       &      $\pm$ 0.12 &       $\pm$ 0.3 &     $\pm$ 0.018 &         $\pm$ 6 &       $\pm$ 0.4 &      $\pm$ 0.13 &    $\pm$ 0.7e-3 &       &      \\
   31025 &         2.57599 & 1823.0 &          -53.59 &           10.28 &            0.05 &              10 &        55158.94 &           0.364 &         2.89e-4 &  0.75 &   17 \\
         &   $\pm$ 0.00003 &       &      $\pm$ 0.21 &      $\pm$ 0.23 &      $\pm$ 0.03 &        $\pm$ 30 &      $\pm$ 0.21 &     $\pm$ 0.008 &    $\pm$ 2.0e-5 &       &      \\
   32009 &           46.76 &   88.5 &          -54.09 &            13.5 &           0.196 &             154 &         55585.5 &            8.49 &         1.11e-2 &  0.66 &   16 \\
         &      $\pm$ 0.03 &       &      $\pm$ 0.21 &       $\pm$ 0.3 &     $\pm$ 0.023 &         $\pm$ 8 &       $\pm$ 1.2 &      $\pm$ 0.21 &    $\pm$ 0.8e-3 &       &      \\
   33021 &           250.1 &    7.3 &          -54.02 &             7.3 &            0.46 &               6 &           55875 &            22.3 &          7.0e-3 &  0.66 &   15 \\
         &       $\pm$ 0.9 &       &      $\pm$ 0.20 &       $\pm$ 0.5 &      $\pm$ 0.03 &         $\pm$ 9 &         $\pm$ 7 &       $\pm$ 1.4 &    $\pm$ 1.3e-3 &       &      \\
   34028 &          48.414 &   97.0 &           -55.2 &            33.3 &           0.393 &              52 &         55374.1 &            20.4 &         1.44e-1 &  1.99 &   38 \\
         &     $\pm$ 0.006 &       &       $\pm$ 0.4 &       $\pm$ 0.8 &     $\pm$ 0.018 &         $\pm$ 3 &       $\pm$ 0.3 &       $\pm$ 0.5 &    $\pm$ 1.1e-2 &       &      \\
   35011 &        3.194552 & 1574.6 &           -52.2 &            37.5 &           0.058 &              13 &        55348.59 &           1.643 &         1.73e-2 &  1.12 &   25 \\
         &  $\pm$ 0.000010 &       &       $\pm$ 0.3 &       $\pm$ 0.4 &     $\pm$ 0.008 &        $\pm$ 11 &      $\pm$ 0.10 &     $\pm$ 0.017 &    $\pm$ 0.5e-3 &       &      \\
   35028 &         11.3078 &  415.2 &          -53.66 &             8.3 &            0.08 &              69 &         55289.3 &            1.29 &          6.7e-4 &  0.73 &   20 \\
         &    $\pm$ 0.0006 &       &      $\pm$ 0.19 &       $\pm$ 0.3 &      $\pm$ 0.04 &        $\pm$ 22 &       $\pm$ 0.7 &      $\pm$ 0.04 &    $\pm$ 0.6e-4 &       &      \\
   36011 &           217.6 &   17.9 &          -52.49 &             2.5 &            0.74 &             186 &           56029 &             5.0 &          1.1e-4 &  0.54 &   13 \\
         &       $\pm$ 0.4 &       &      $\pm$ 0.17 &       $\pm$ 0.3 &      $\pm$ 0.09 &        $\pm$ 14 &         $\pm$ 6 &       $\pm$ 1.0 &    $\pm$ 0.5e-4 &       &      \\
   36016 &           444.9 &    7.5 &           -53.7 &            11.5 &            0.64 &             142 &           55864 &              54 &          3.1e-2 &  1.39 &   17 \\
         &       $\pm$ 1.5 &       &       $\pm$ 0.4 &       $\pm$ 0.7 &      $\pm$ 0.03 &         $\pm$ 5 &         $\pm$ 3 &         $\pm$ 4 &    $\pm$ 0.6e-2 &       &      \\
   47010 &          46.706 &  103.6 &           -54.3 &            22.5 &            0.68 &             237 &        55374.74 &            10.6 &          2.2e-2 &  1.58 &   21 \\
         &     $\pm$ 0.019 &       &       $\pm$ 0.4 &       $\pm$ 1.1 &      $\pm$ 0.03 &         $\pm$ 3 &      $\pm$ 0.21 &       $\pm$ 0.6 &    $\pm$ 0.4e-2 &       &      \\
\enddata
\tablecomments{Table \ref{SB1tab} is also published in machine-readable format.}
\end{deluxetable*}

\clearpage
\begin{figure*}
\gridline{\fig{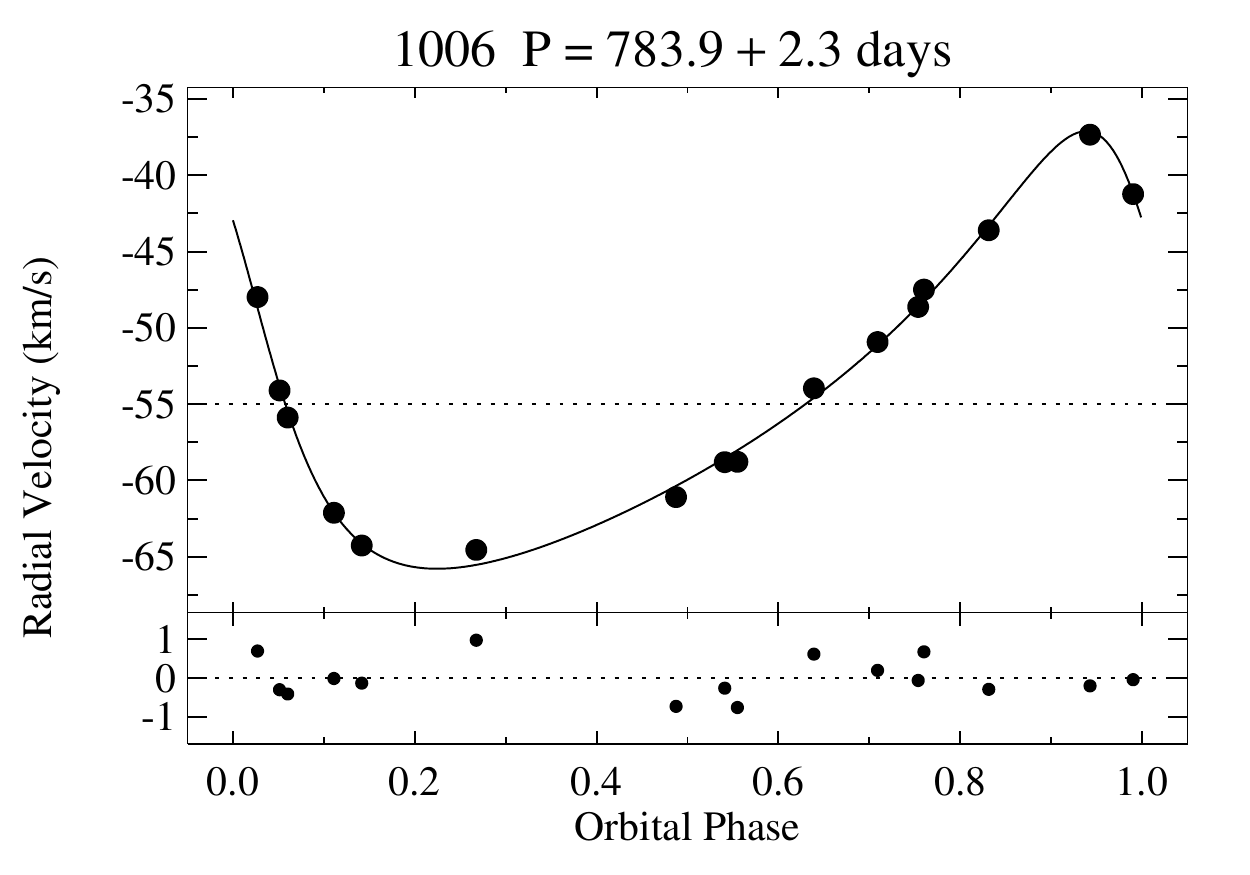}{0.3\linewidth}{}
  \fig{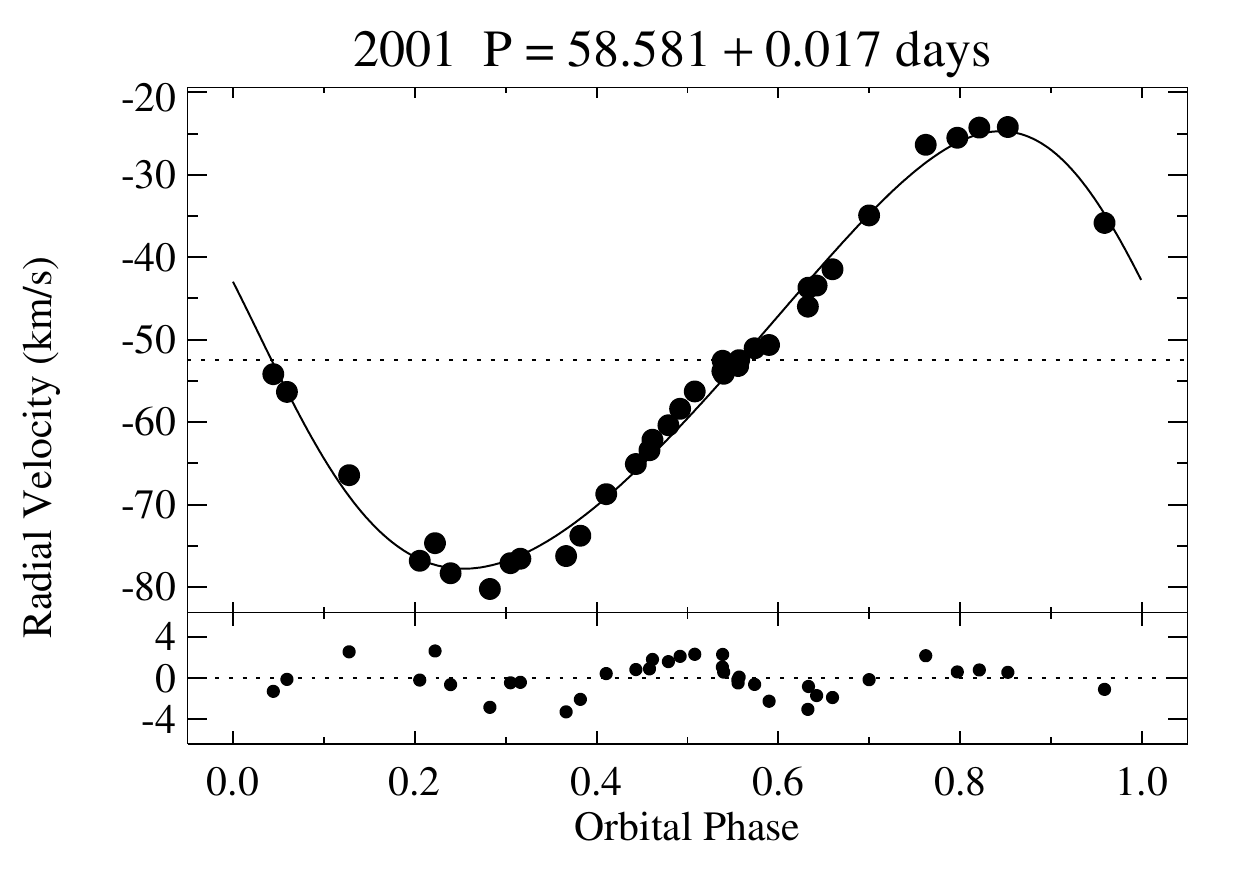}{0.3\linewidth}{}
  \fig{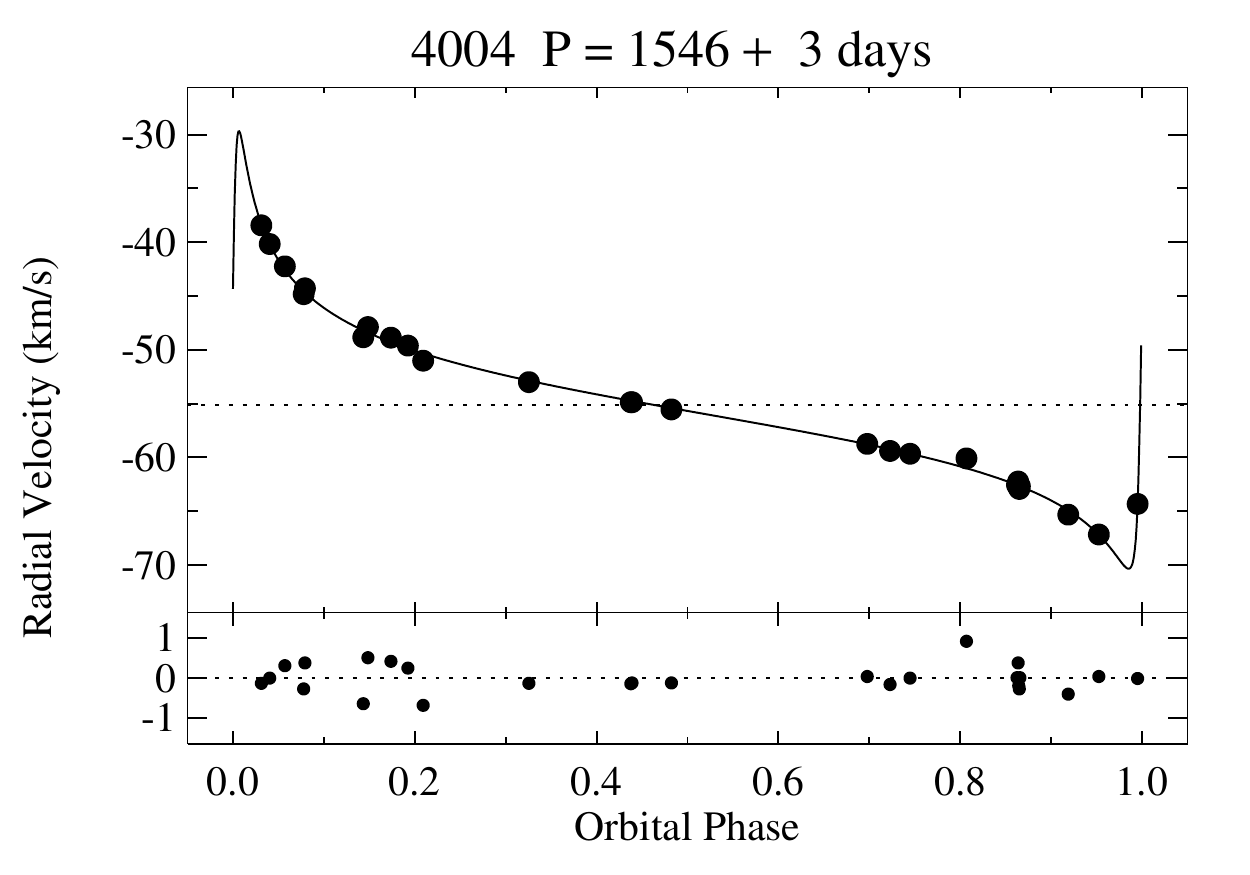}{0.3\linewidth}{}}
\gridline{\fig{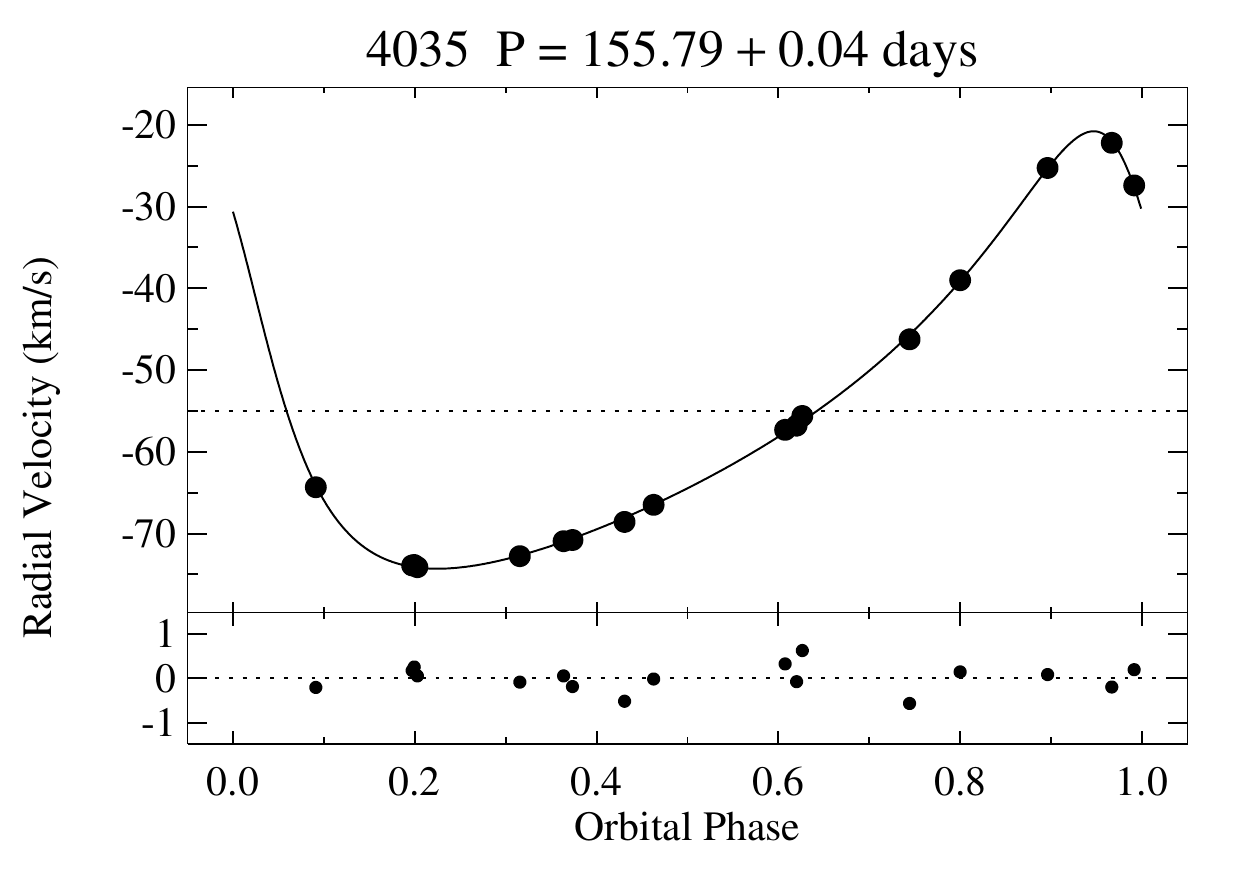}{0.3\linewidth}{}
  \fig{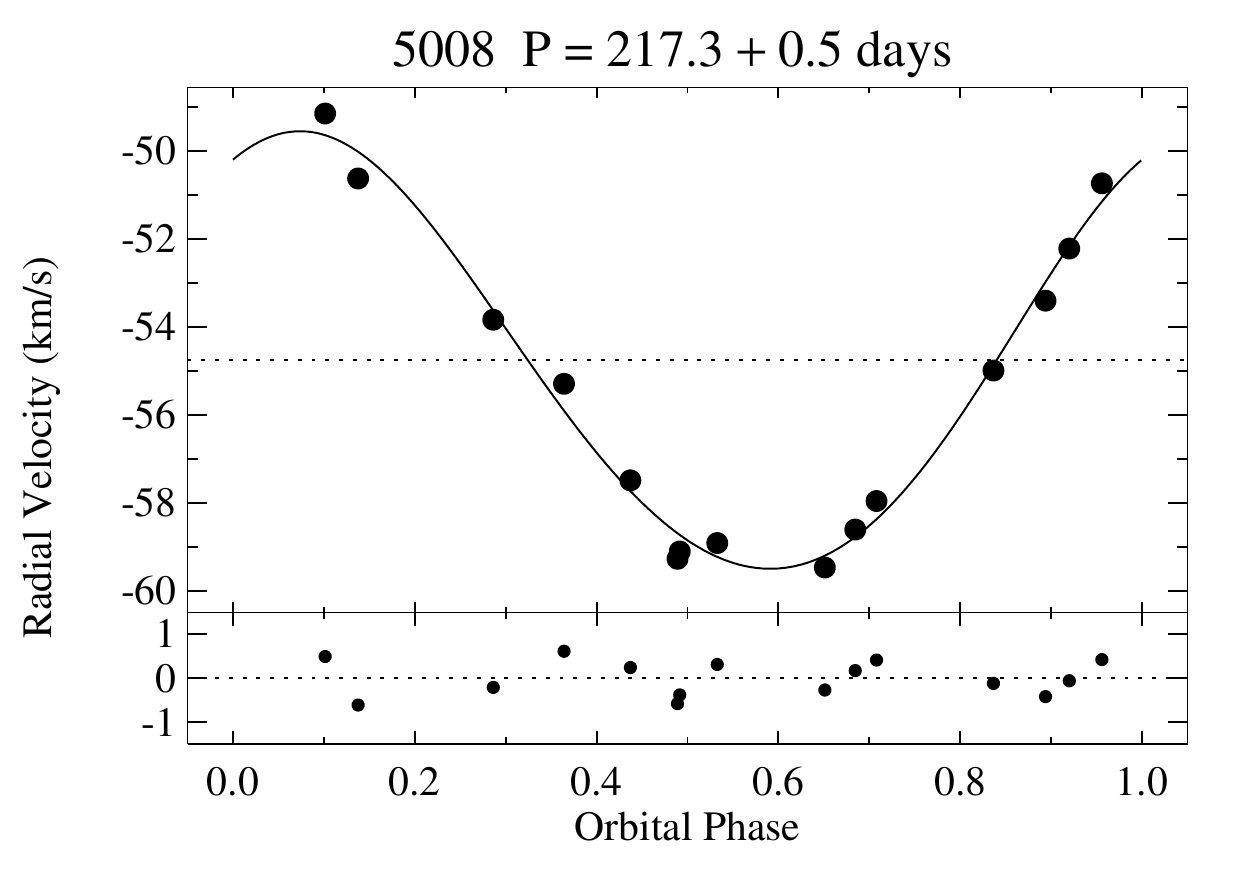}{0.3\linewidth}{}
  \fig{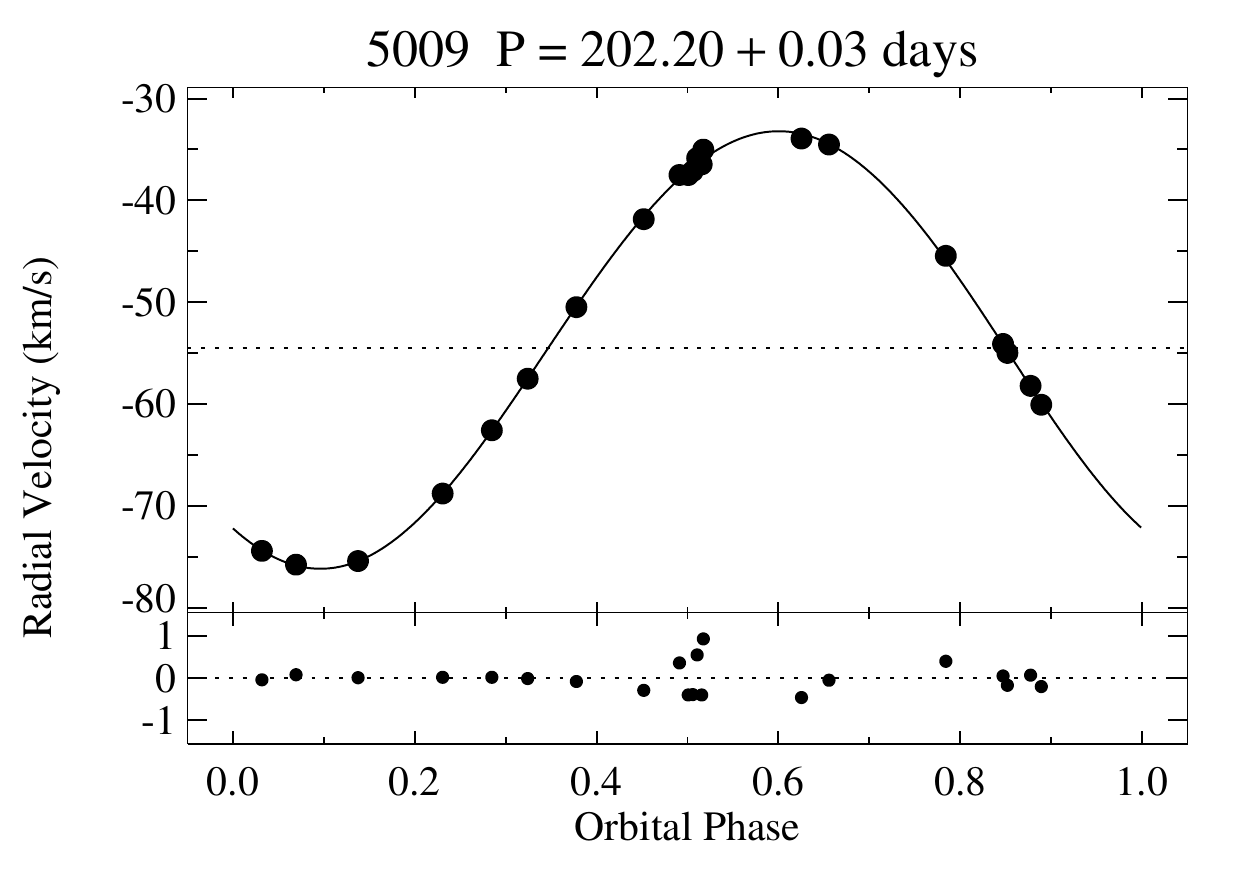}{0.3\linewidth}{}}
\gridline{\fig{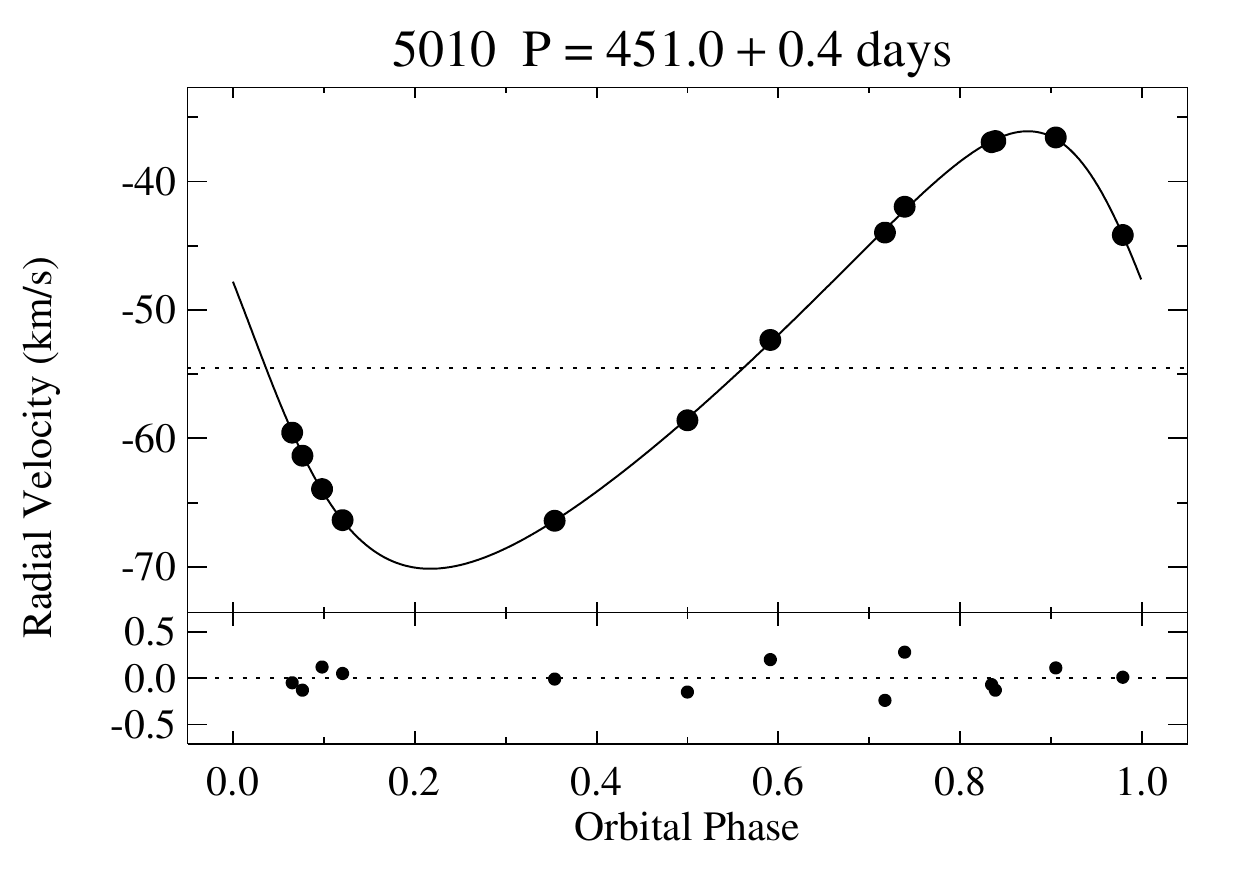}{0.3\linewidth}{}
  \fig{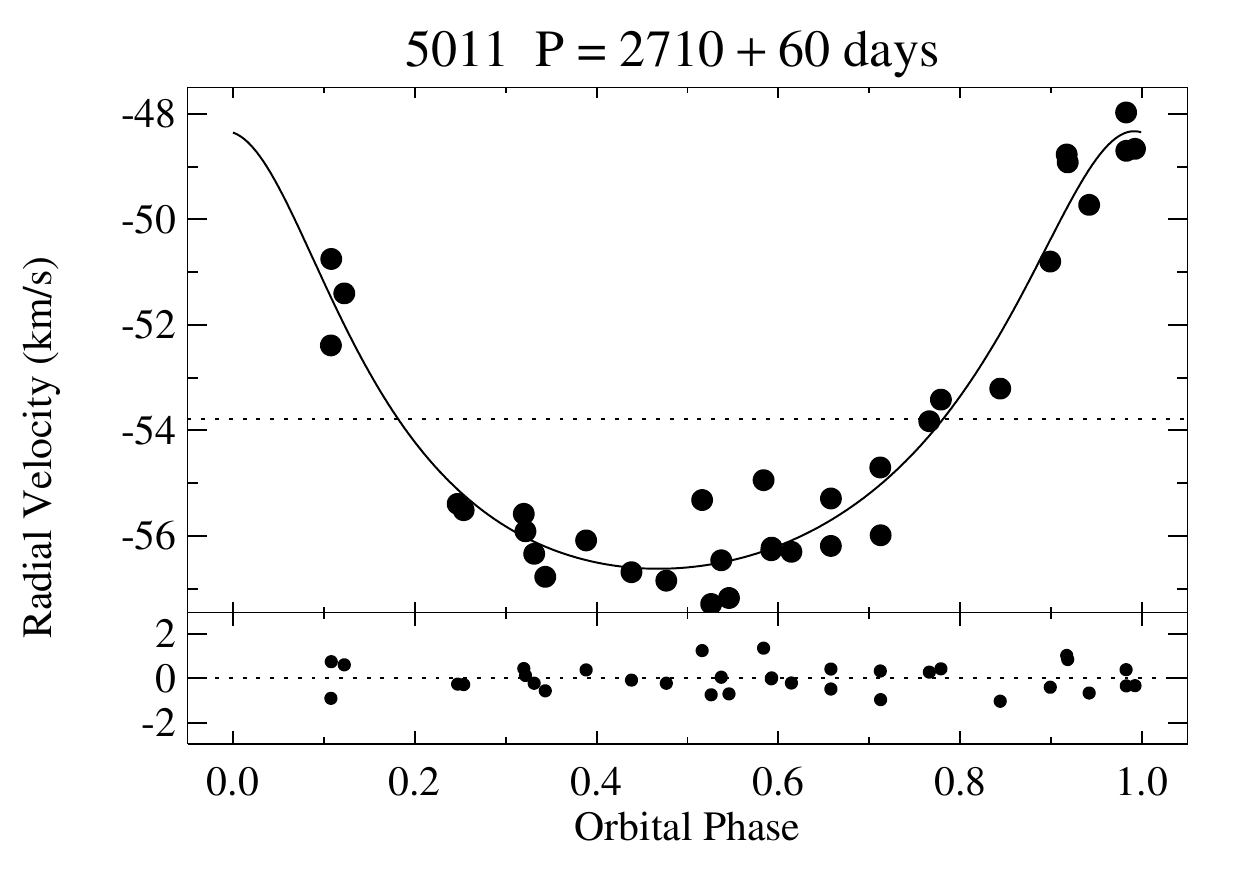}{0.3\linewidth}{}
  \fig{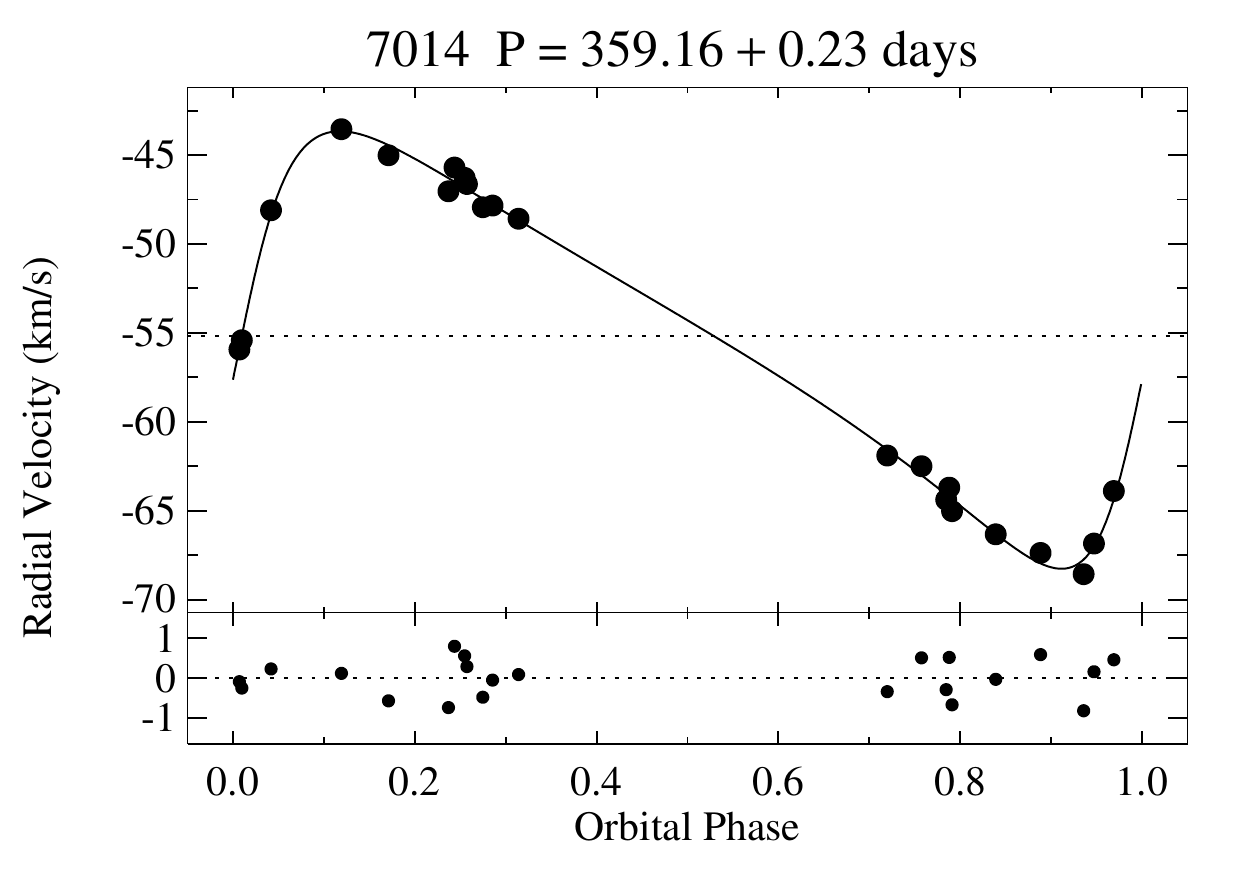}{0.3\linewidth}{}}
\gridline{\fig{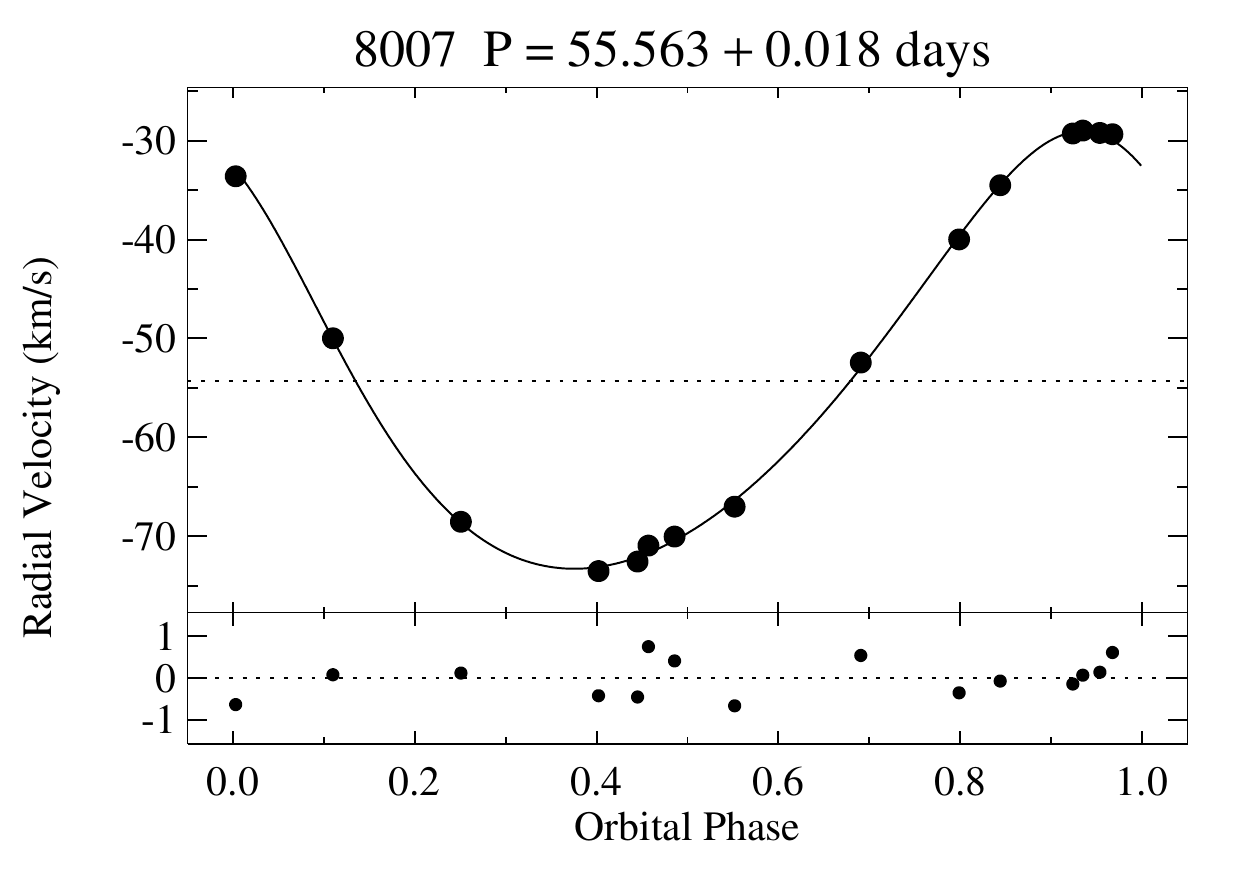}{0.3\linewidth}{}
  \fig{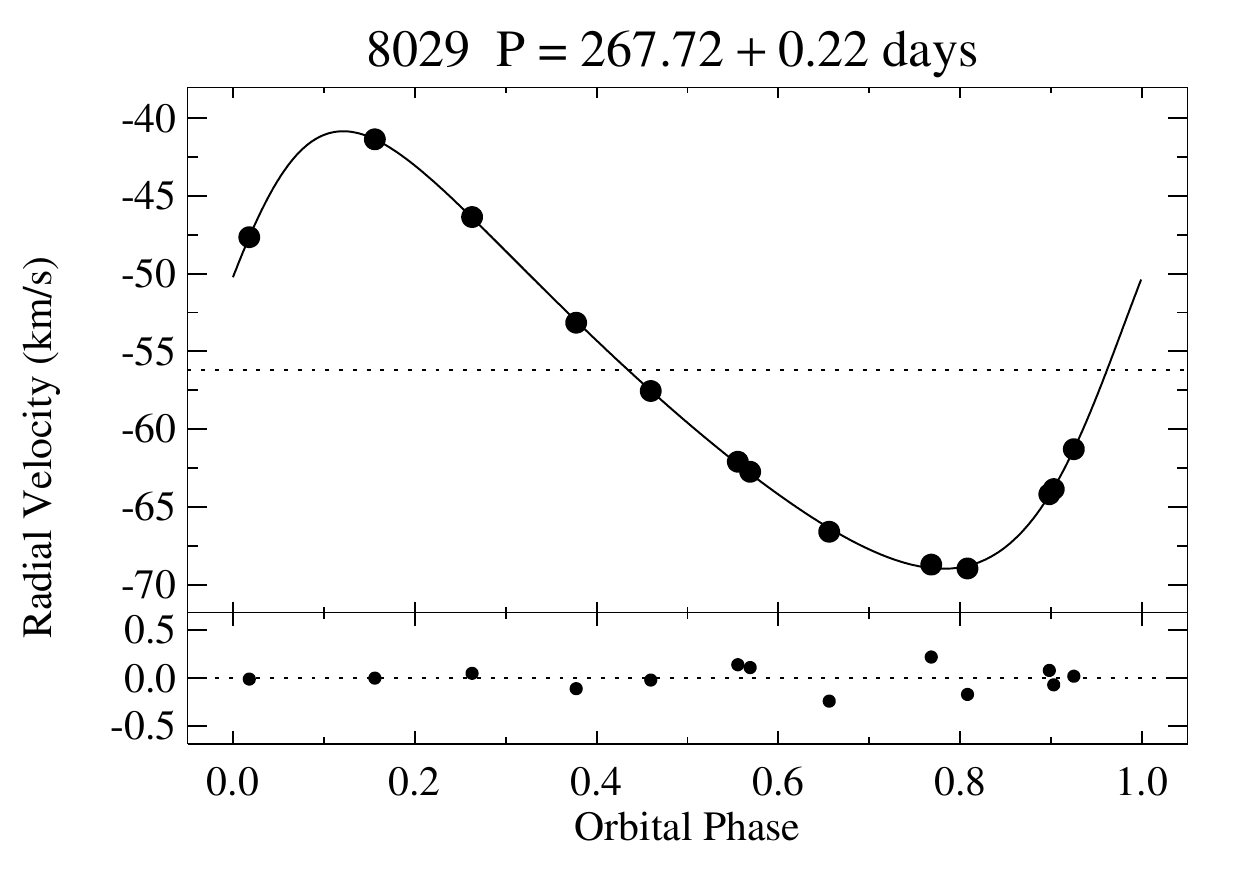}{0.3\linewidth}{}
  \fig{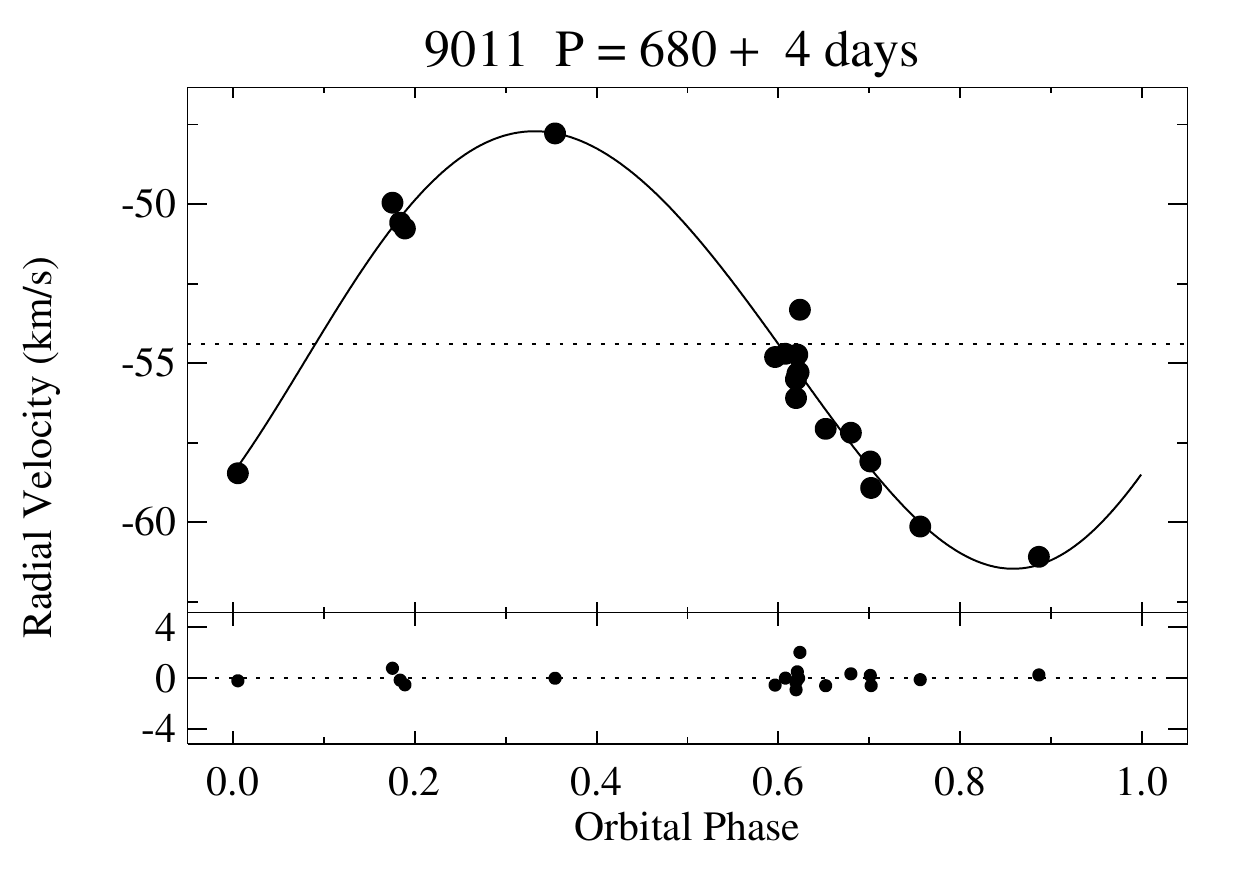}{0.3\linewidth}{}}
\end{figure*}
\begin{figure*}
\gridline{\fig{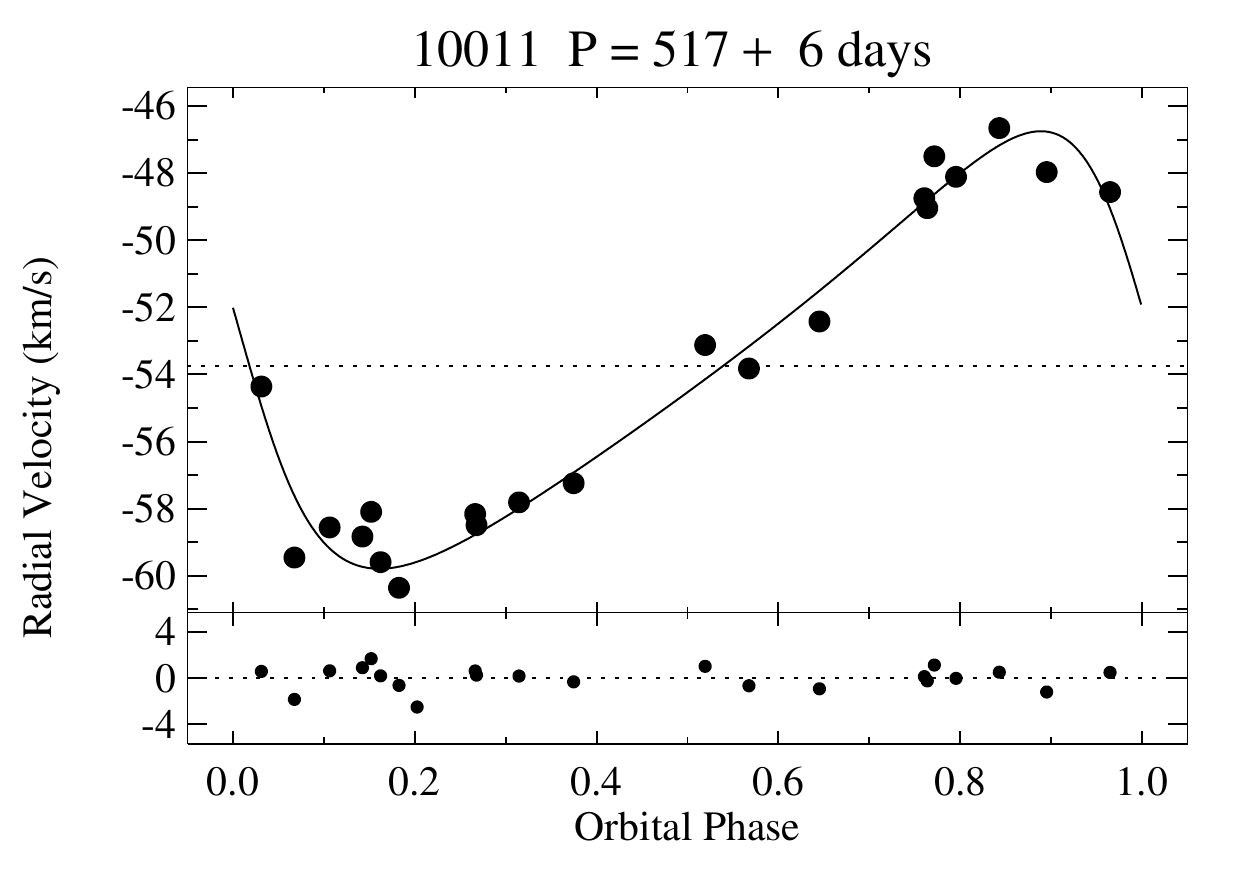}{0.3\linewidth}{}
  \fig{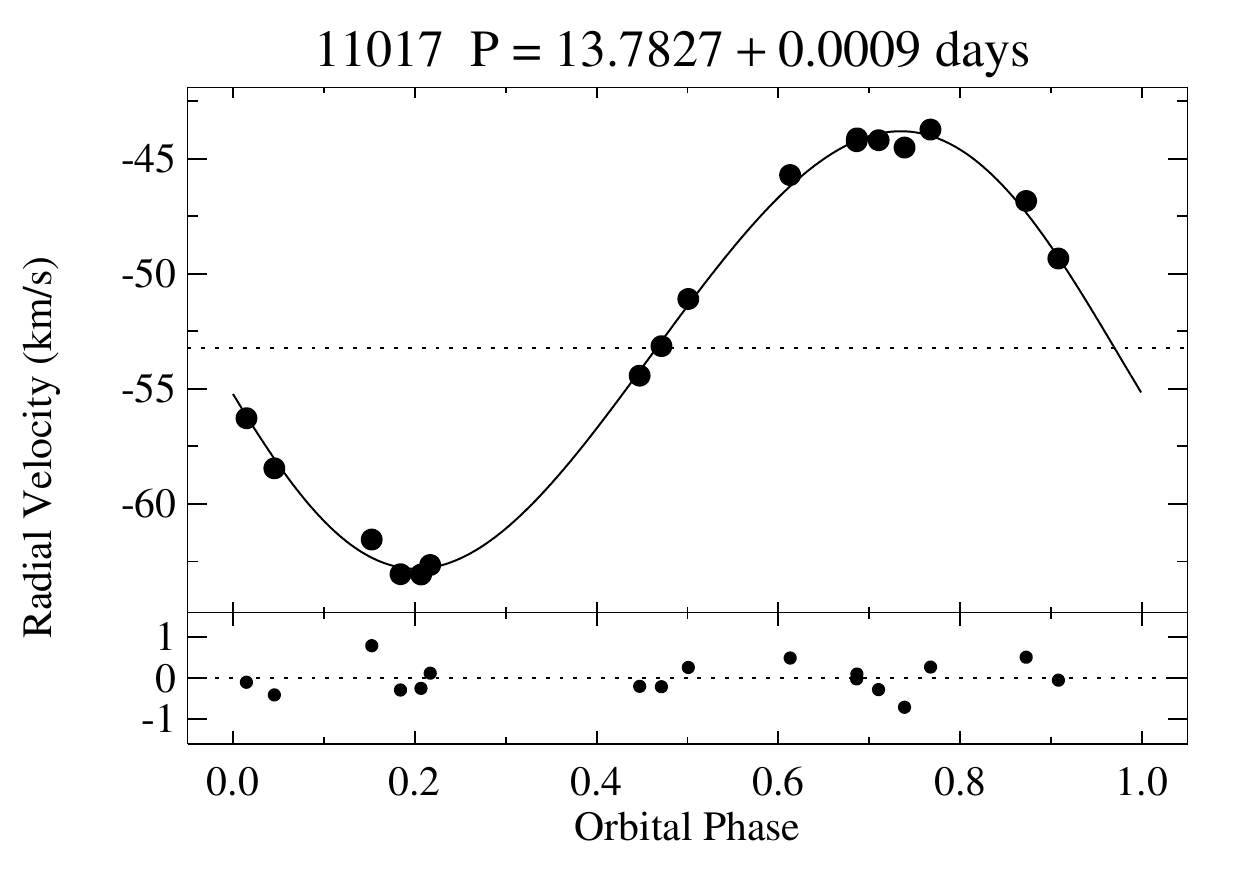}{0.3\linewidth}{}
  \fig{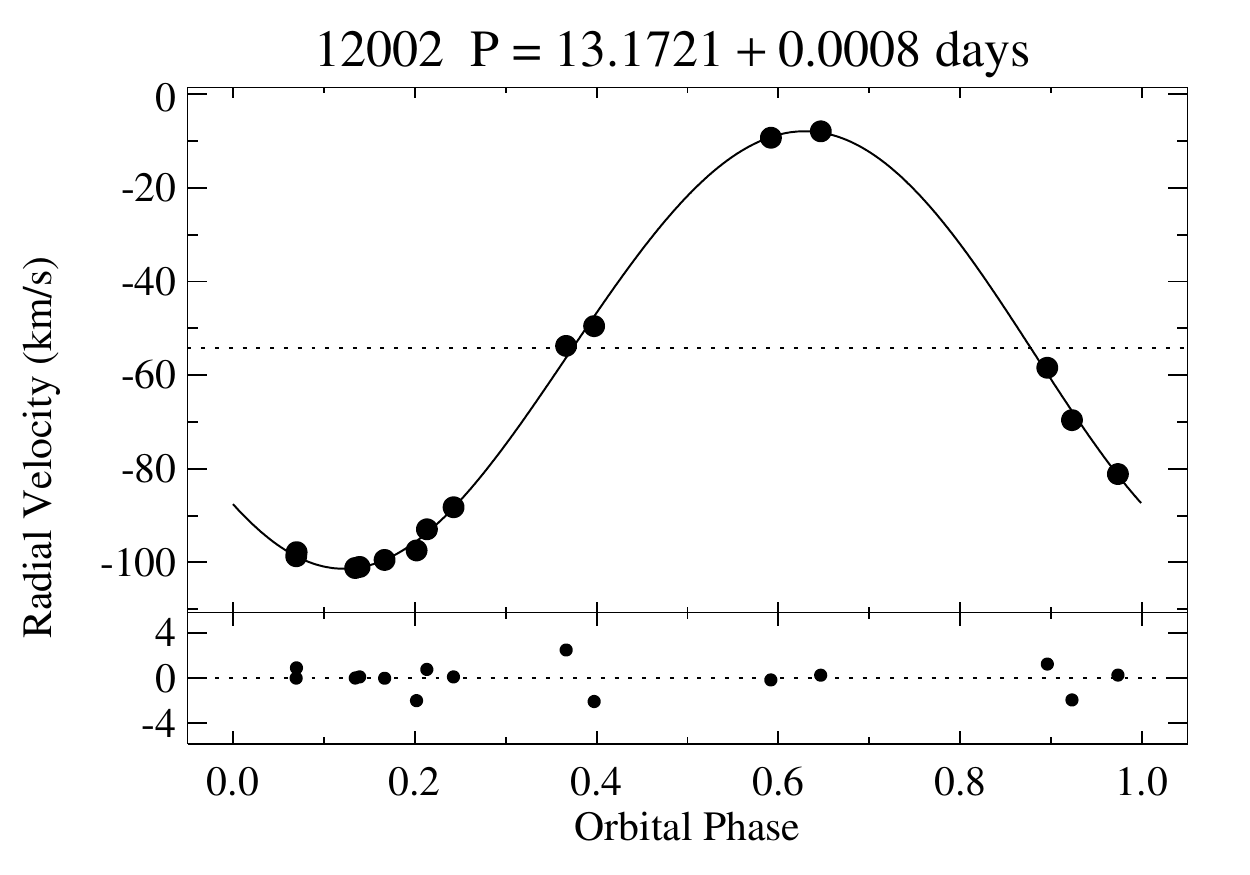}{0.3\linewidth}{}}
\gridline{\fig{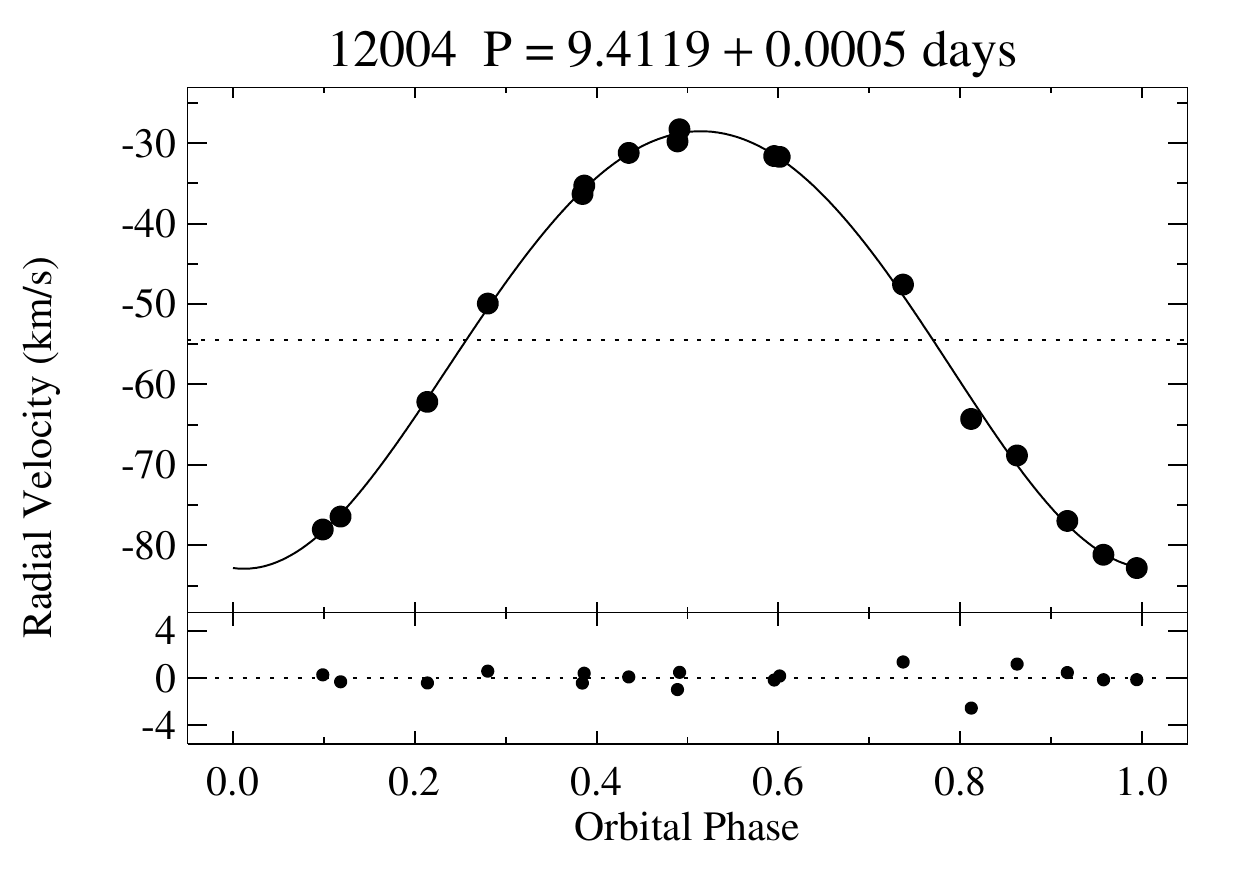}{0.3\linewidth}{}
  \fig{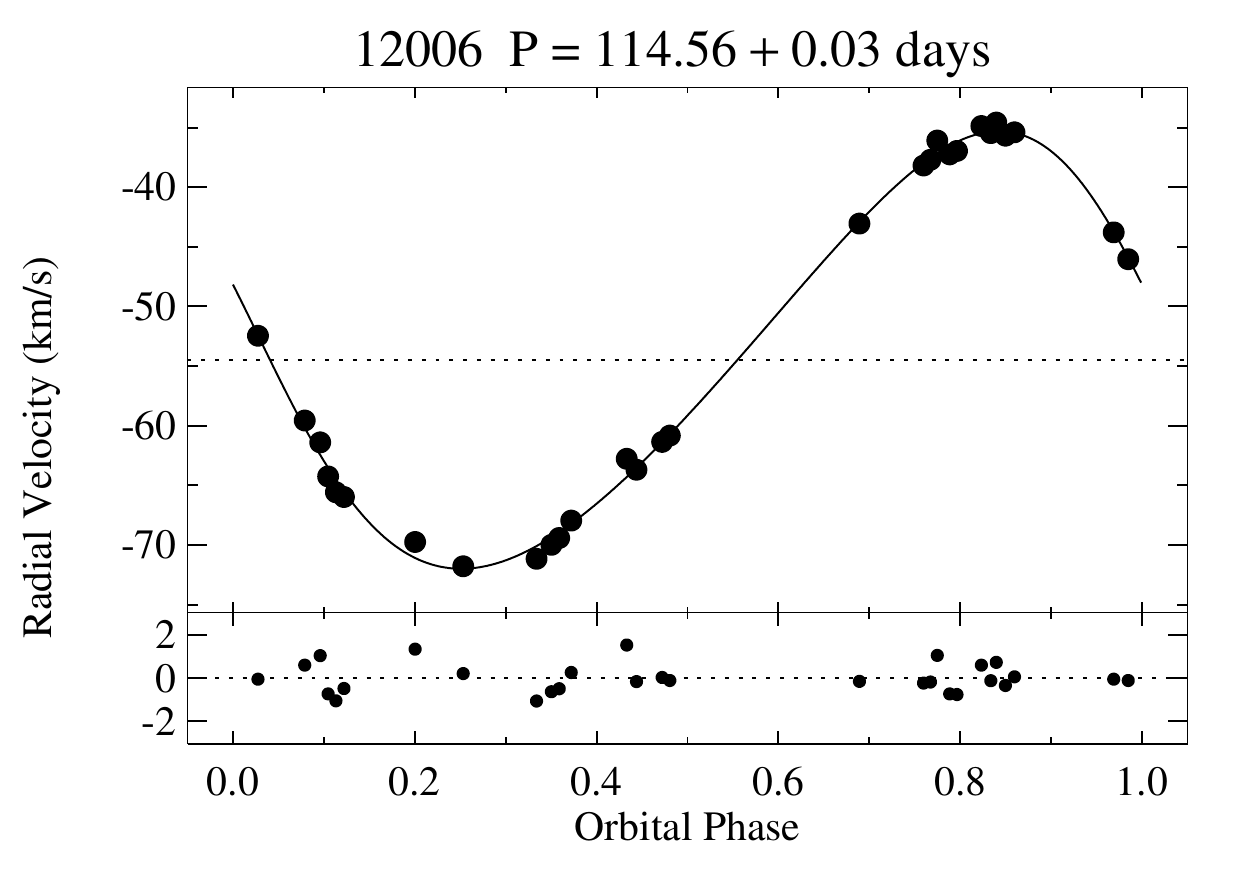}{0.3\linewidth}{}
  \fig{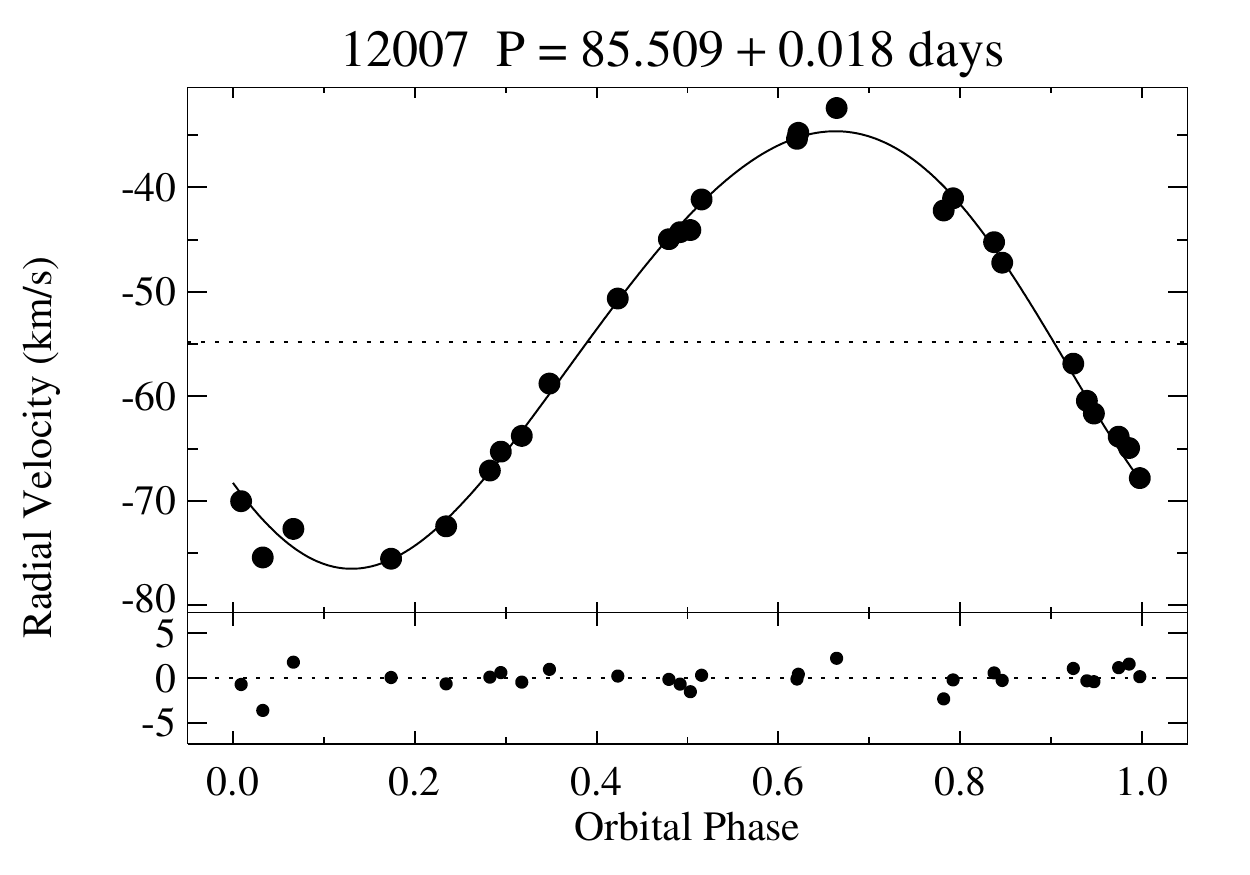}{0.3\linewidth}{}}
\gridline{\fig{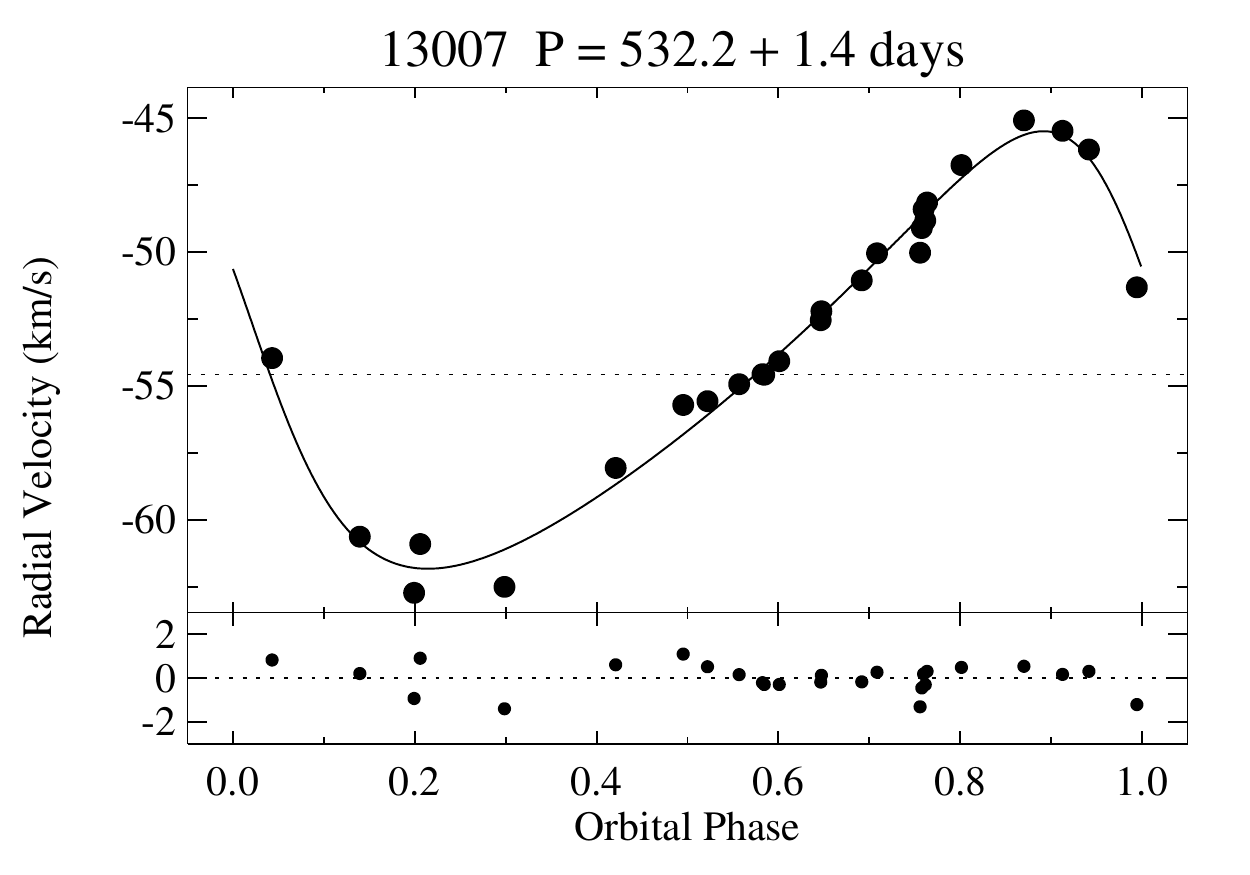}{0.3\linewidth}{}
  \fig{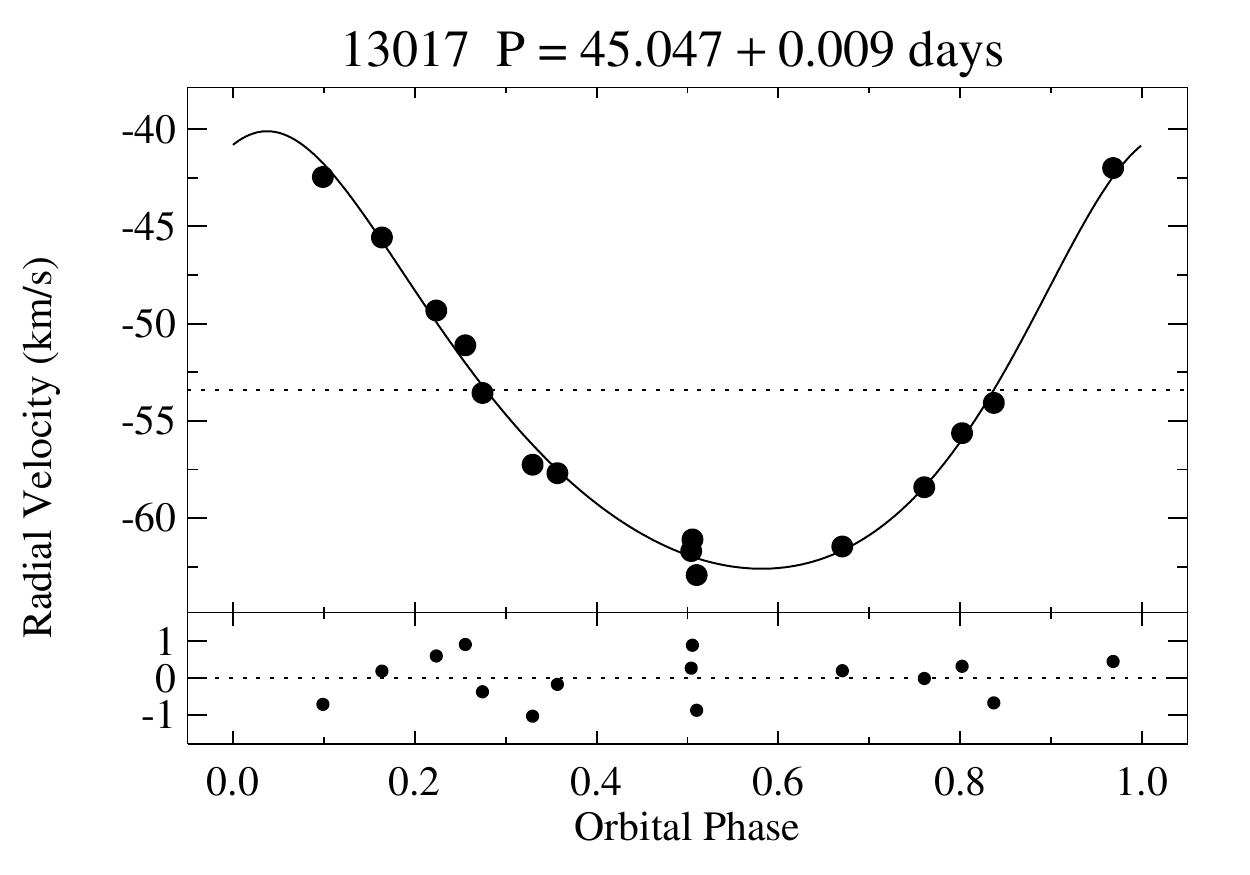}{0.3\linewidth}{}
  \fig{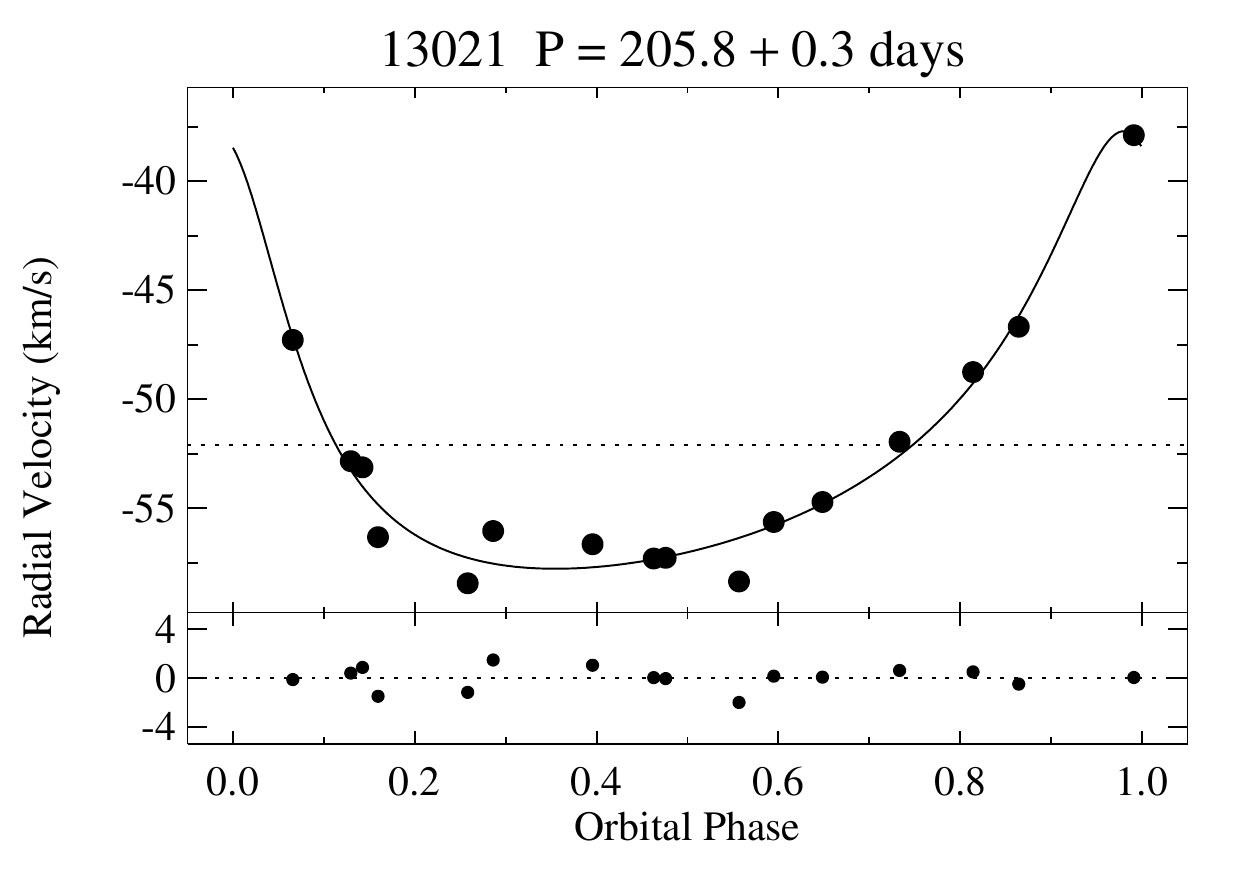}{0.3\linewidth}{}}
\gridline{\fig{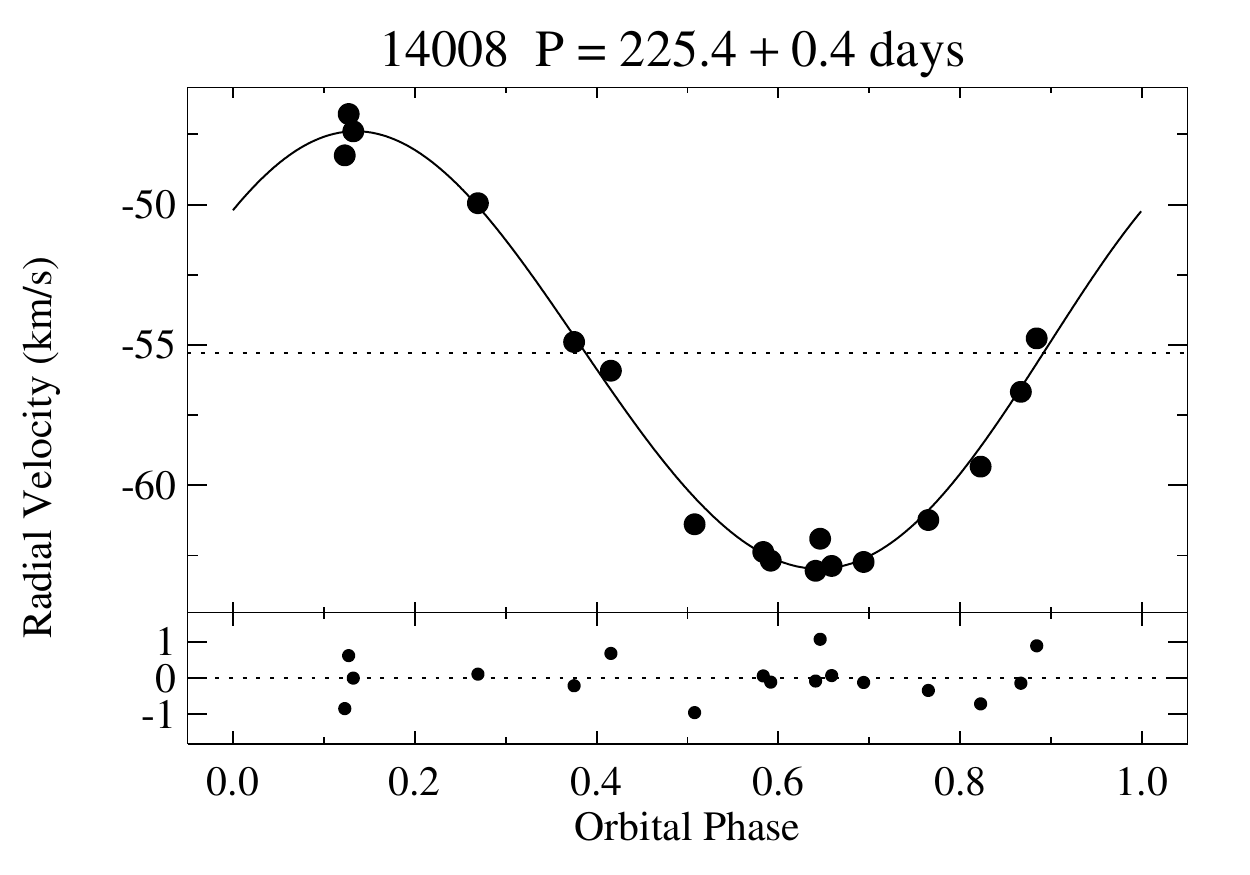}{0.3\linewidth}{}
  \fig{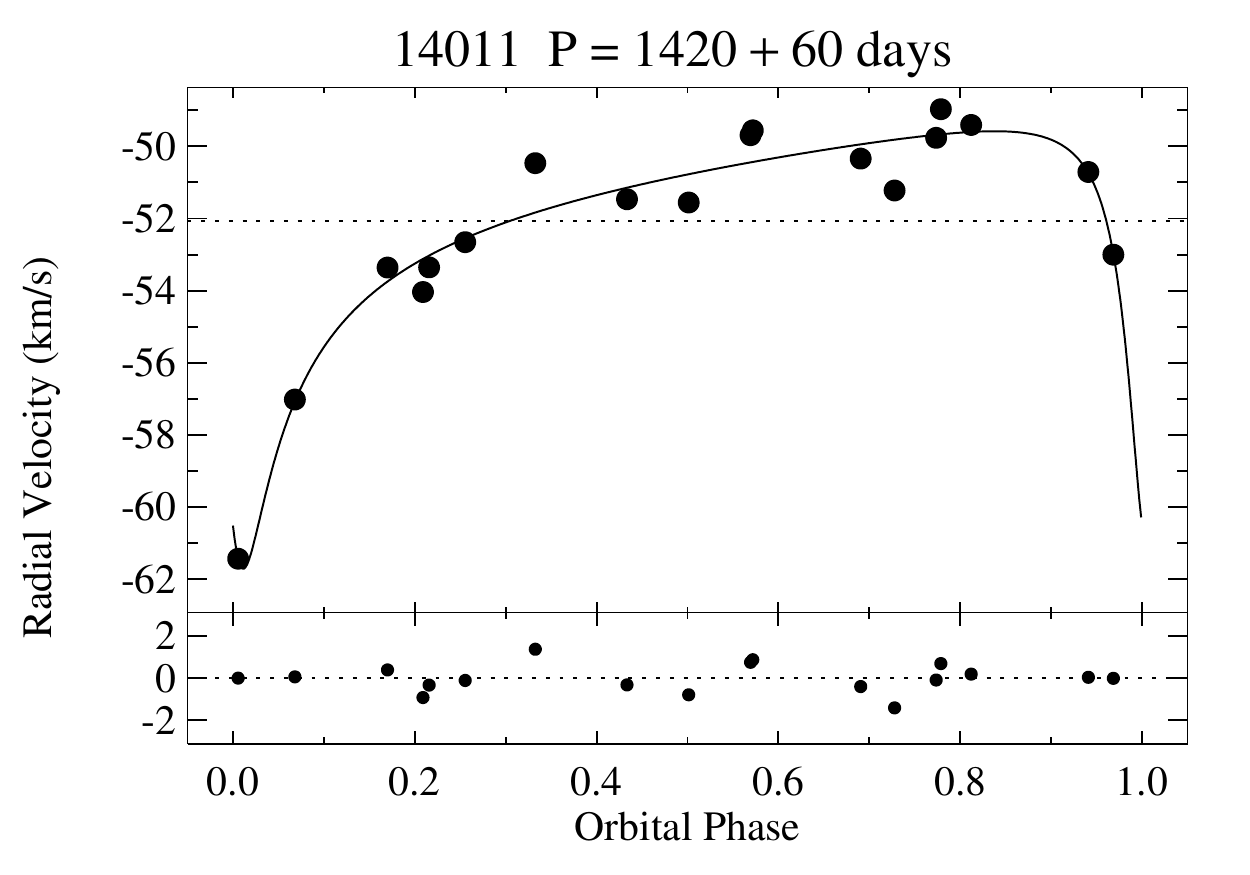}{0.3\linewidth}{}
  \fig{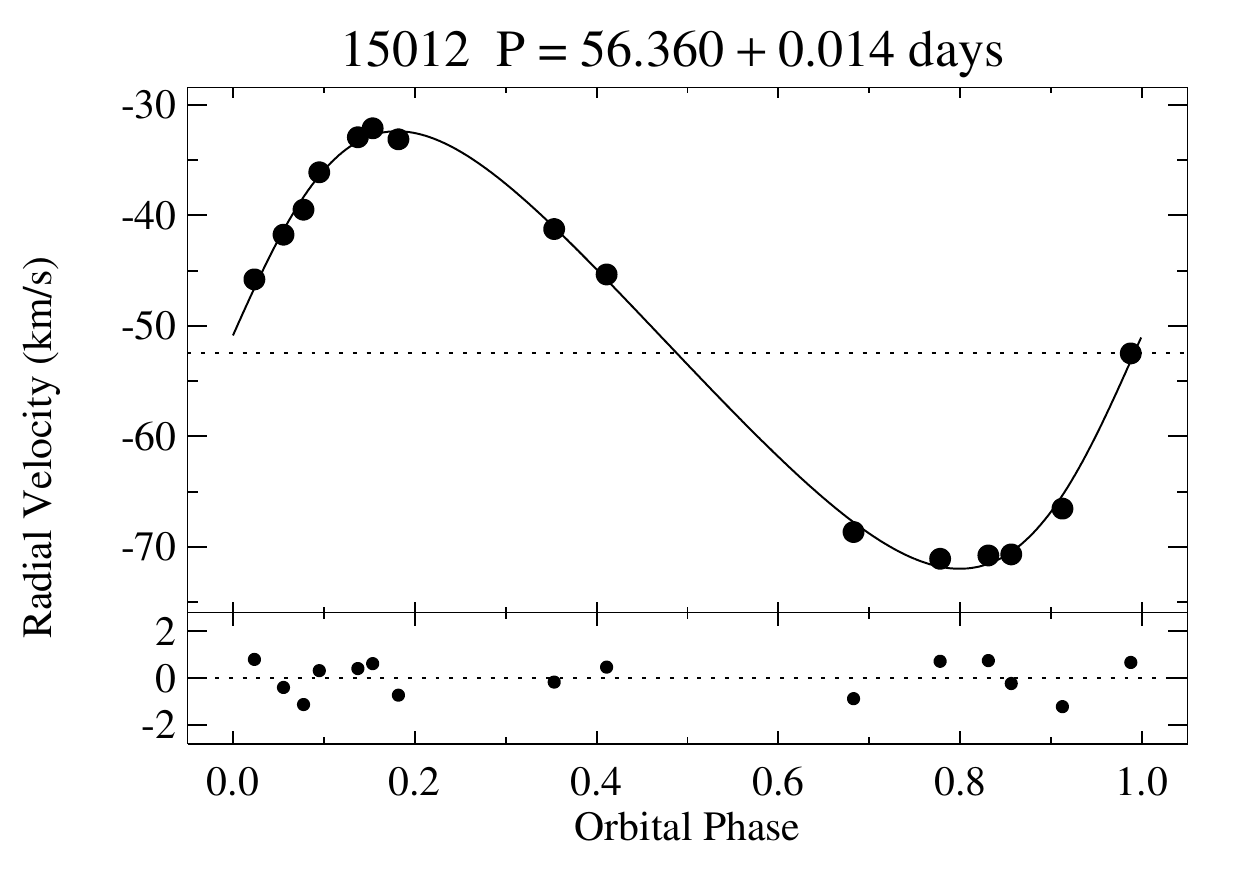}{0.3\linewidth}{}}
\end{figure*}
\begin{figure*}
\gridline{\fig{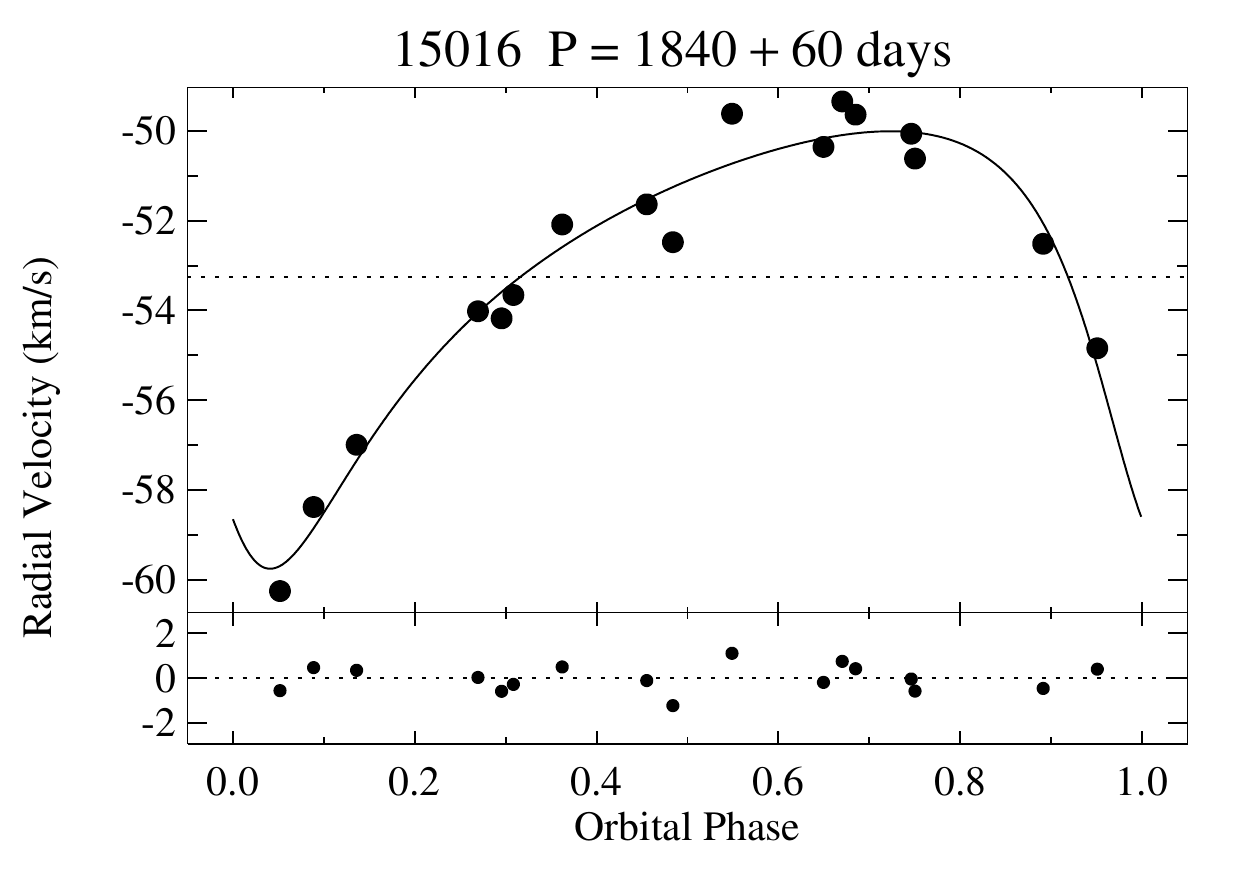}{0.3\linewidth}{}
  \fig{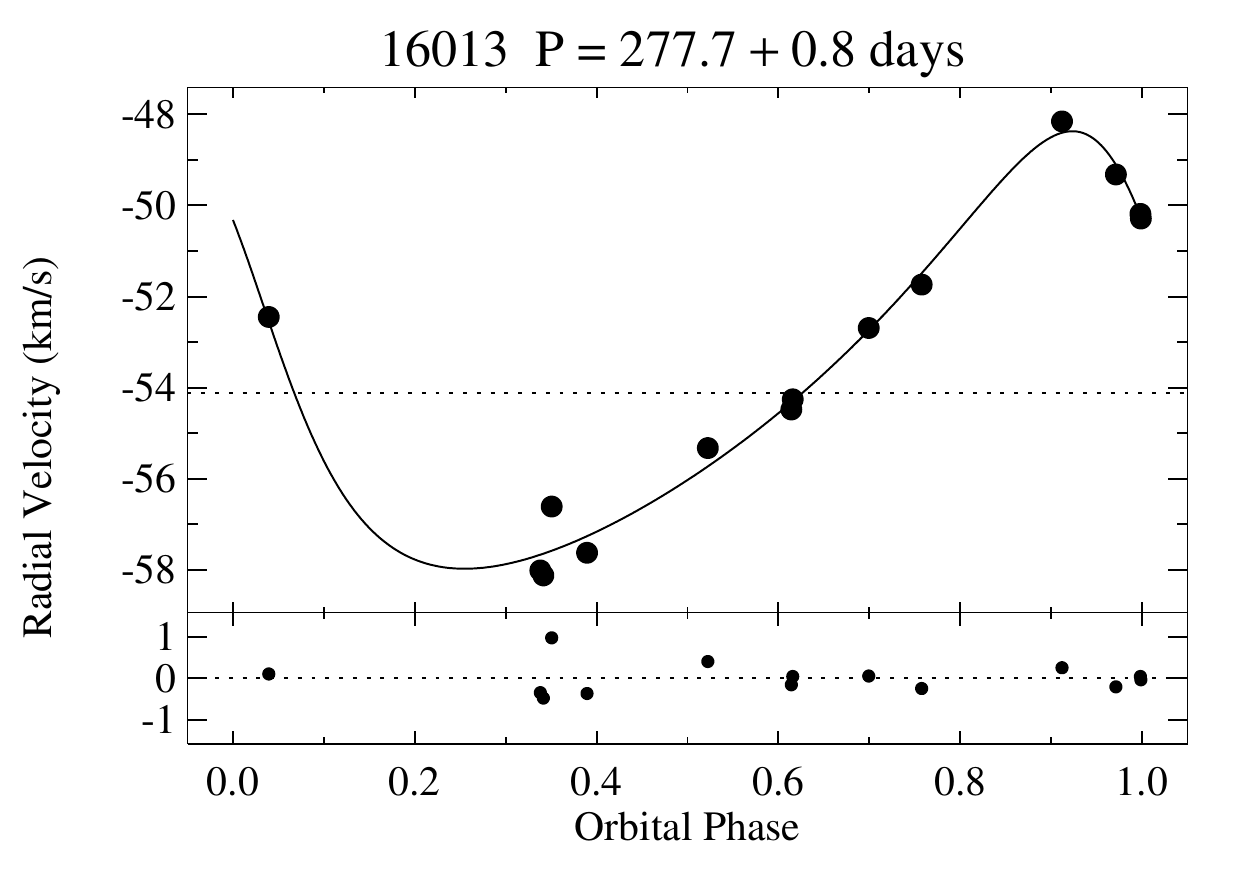}{0.3\linewidth}{}
  \fig{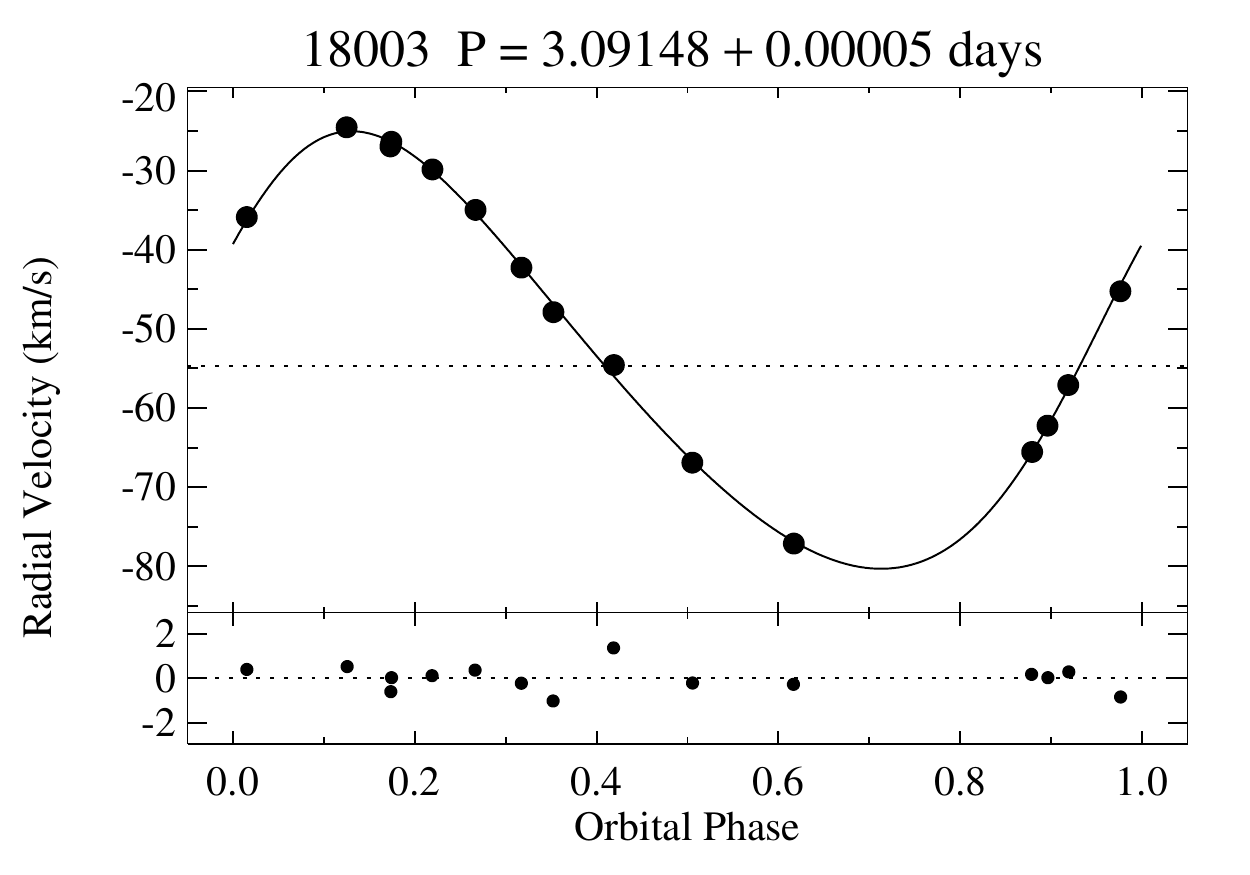}{0.3\linewidth}{}}
\gridline{\fig{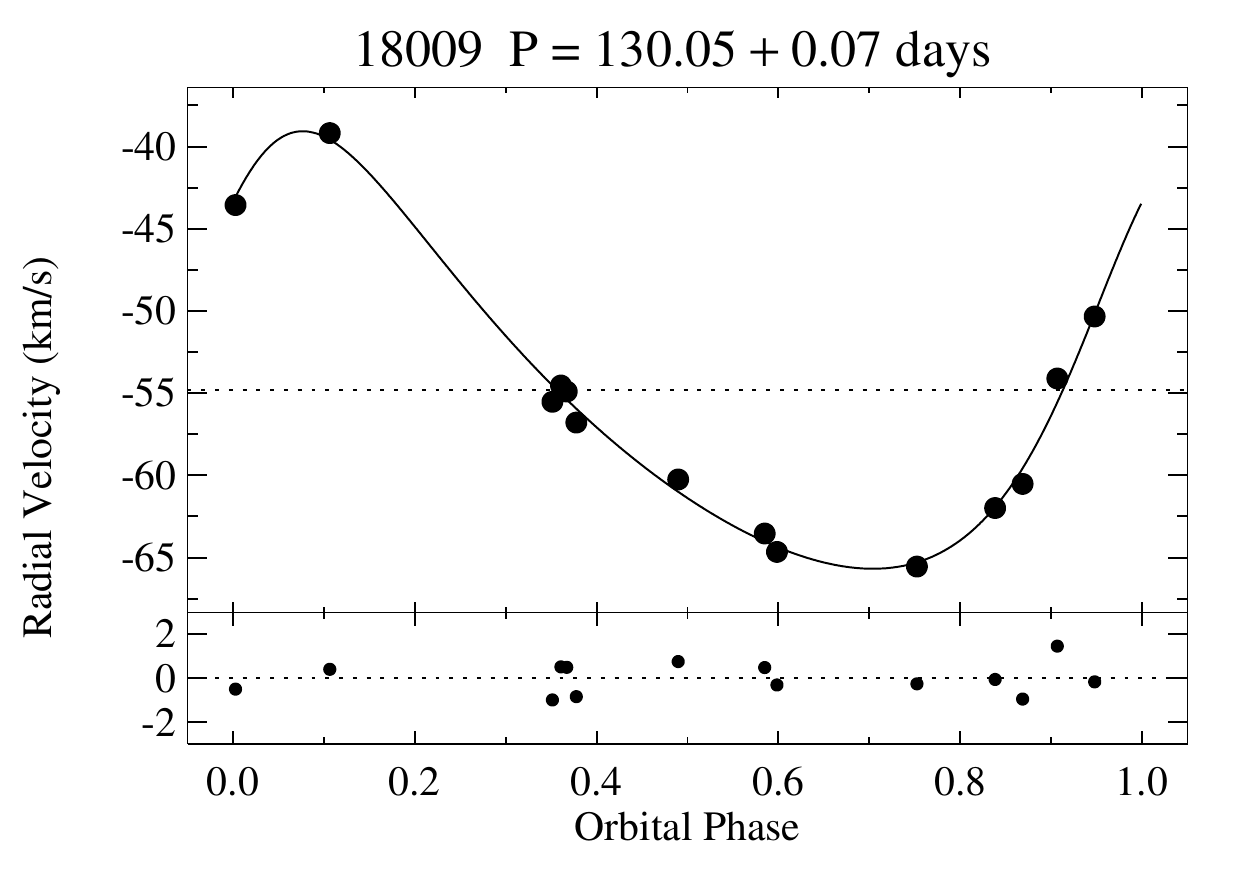}{0.3\linewidth}{}
  \fig{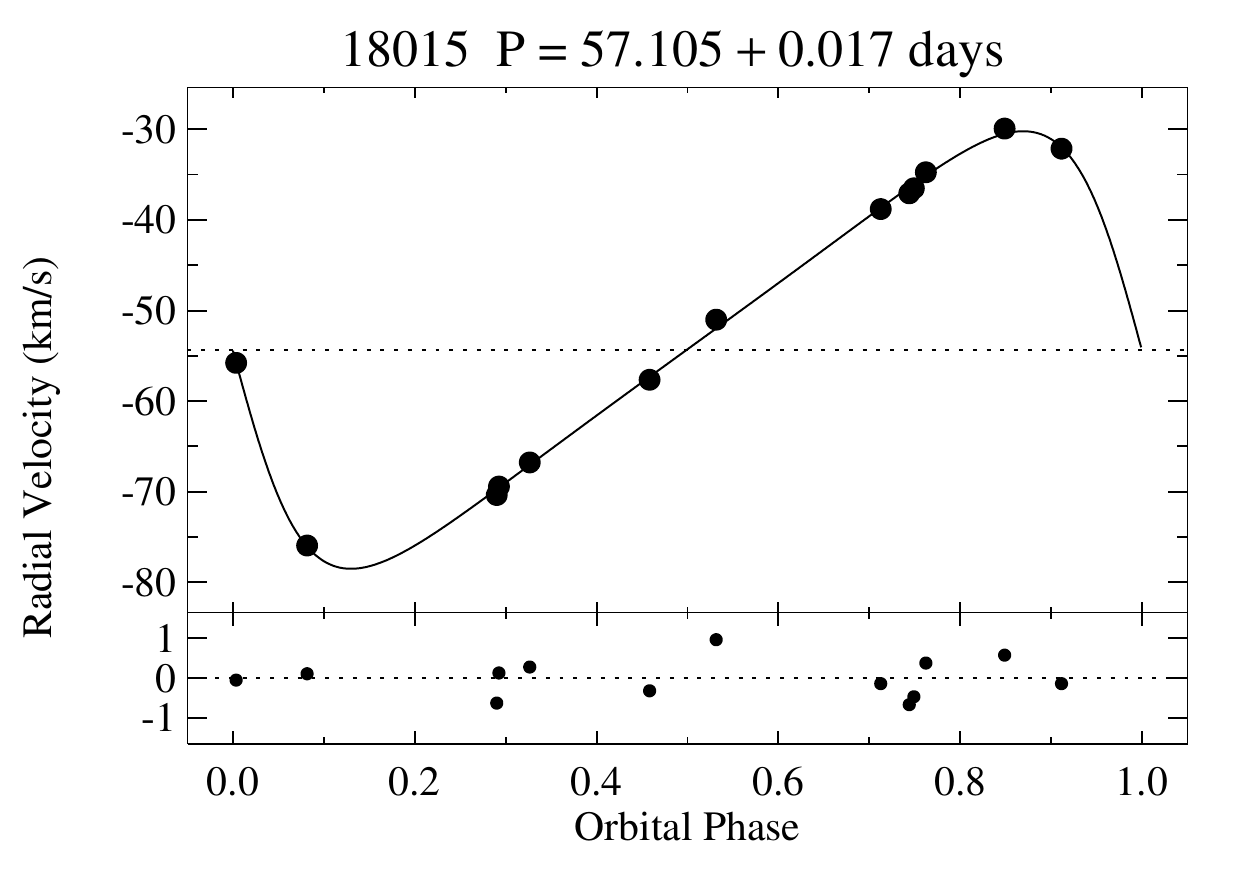}{0.3\linewidth}{}
  \fig{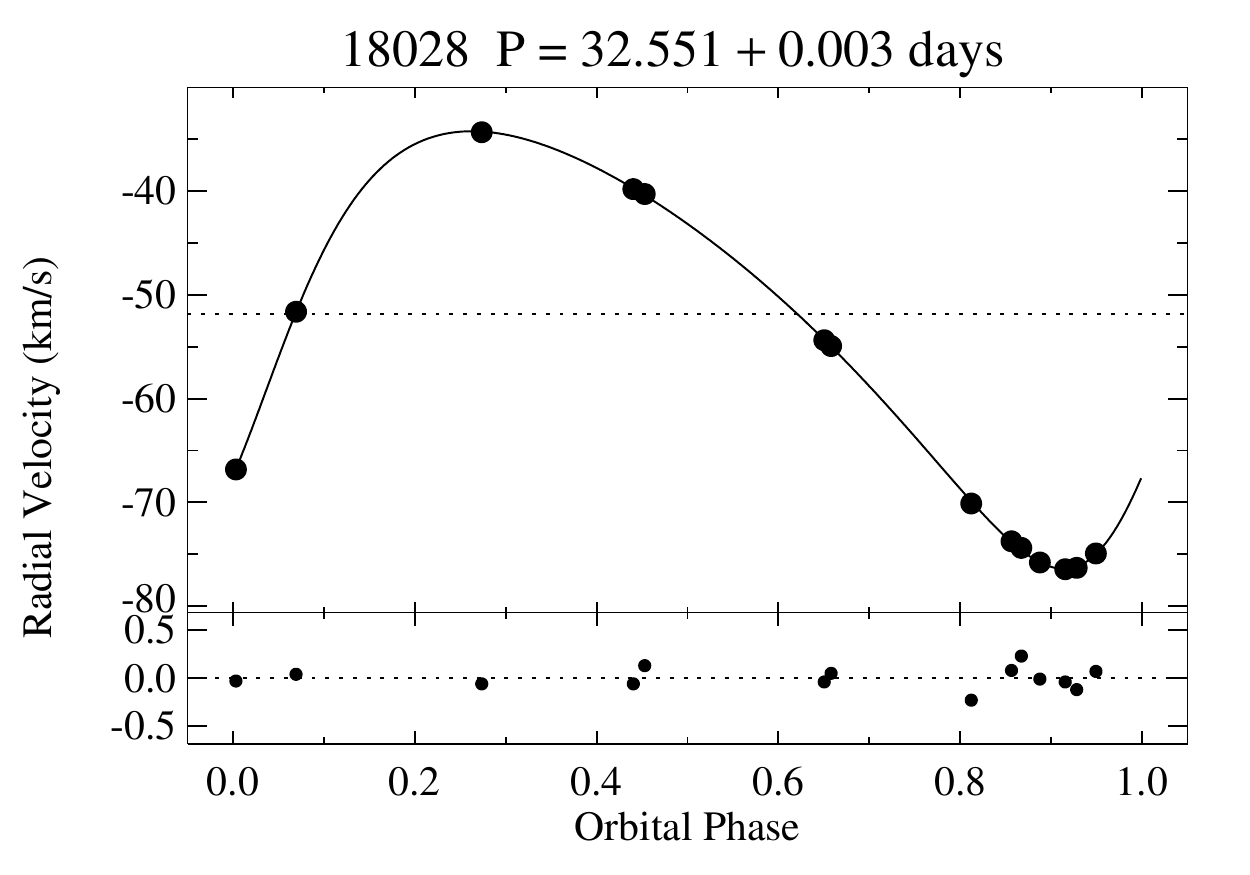}{0.3\linewidth}{}}
\gridline{\fig{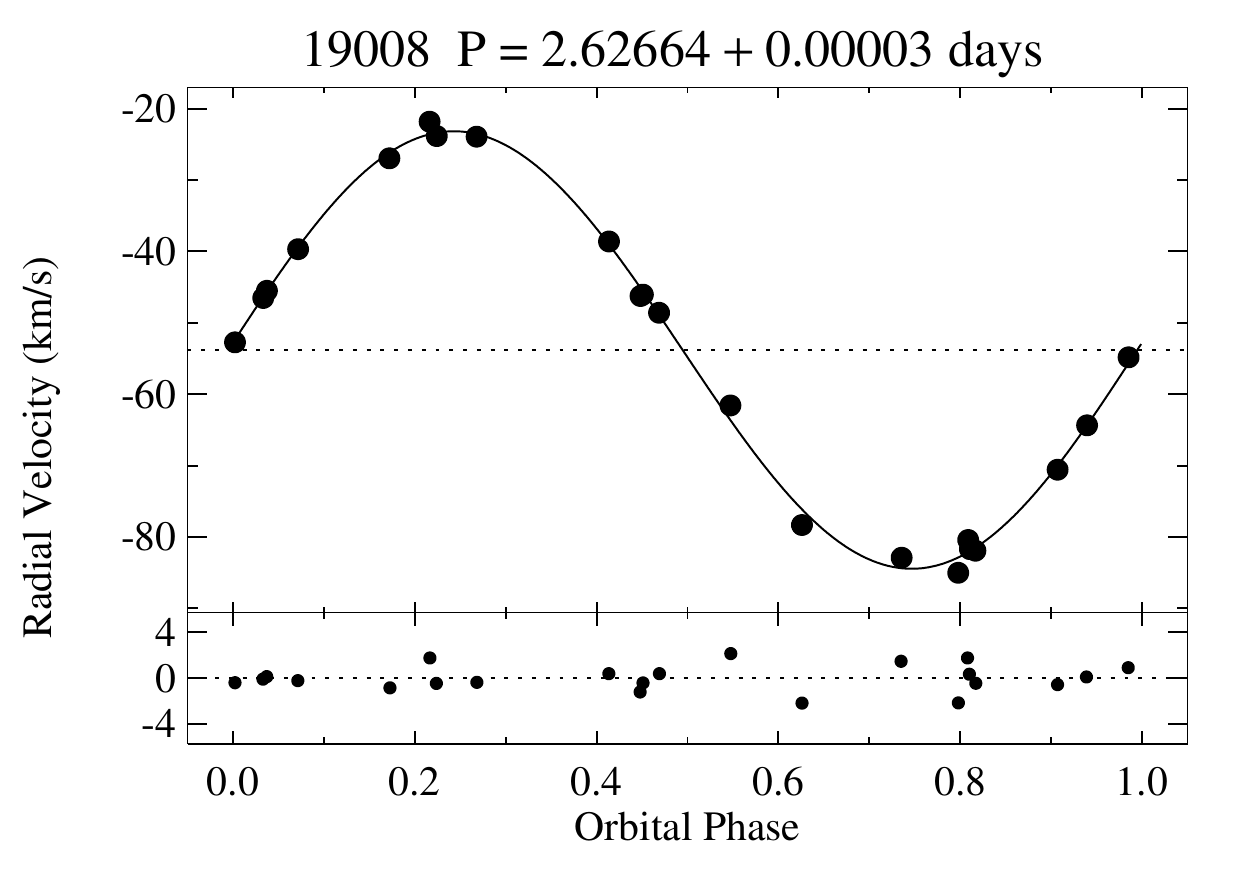}{0.3\linewidth}{}
  \fig{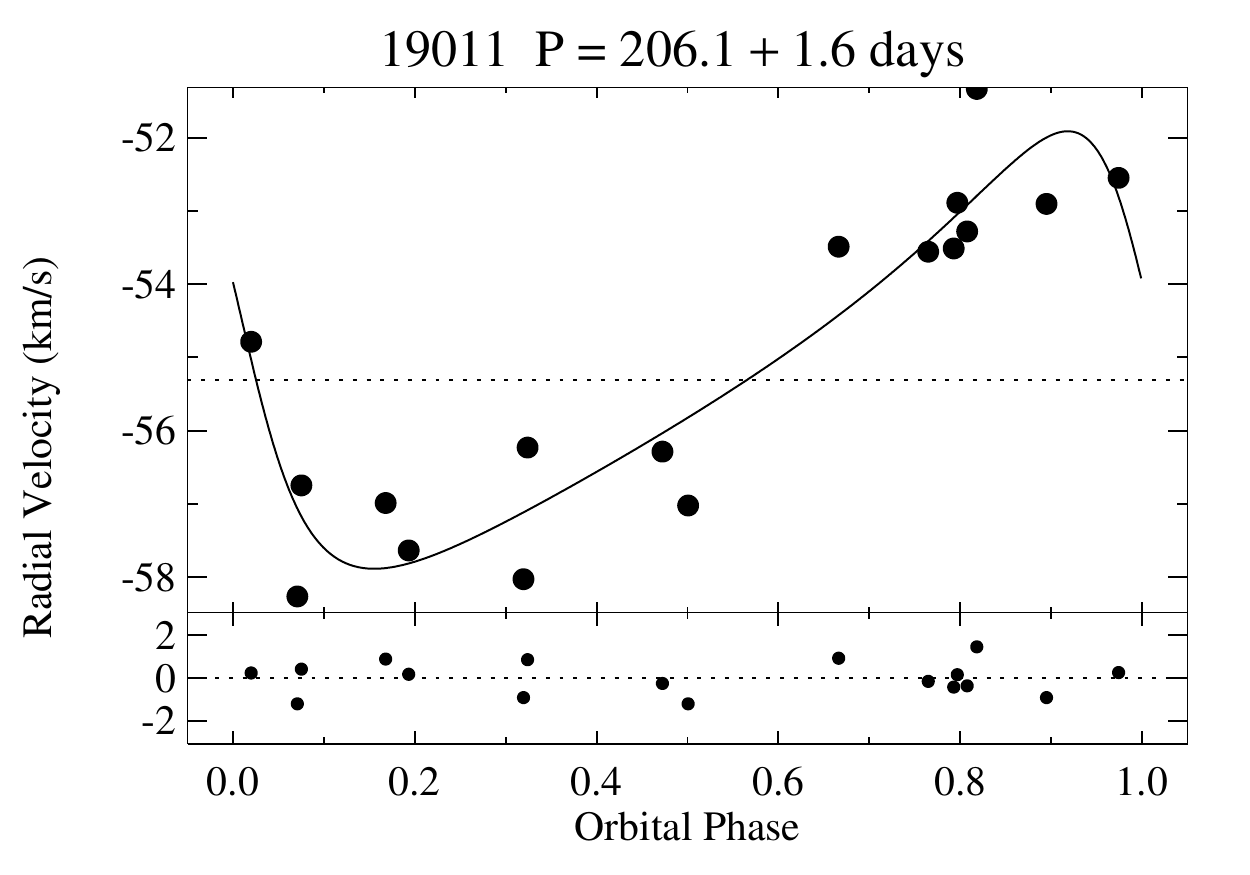}{0.3\linewidth}{}
  \fig{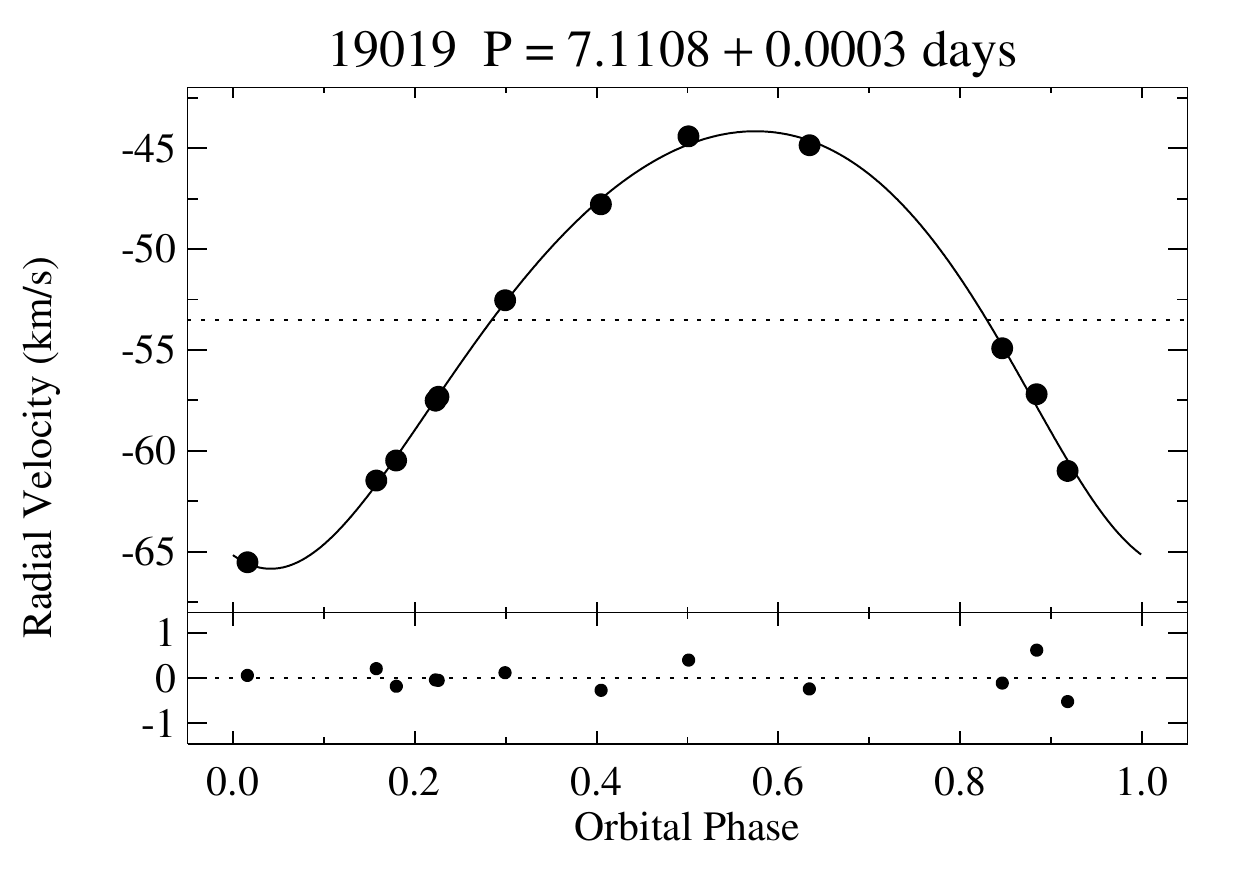}{0.3\linewidth}{}}
\gridline{\fig{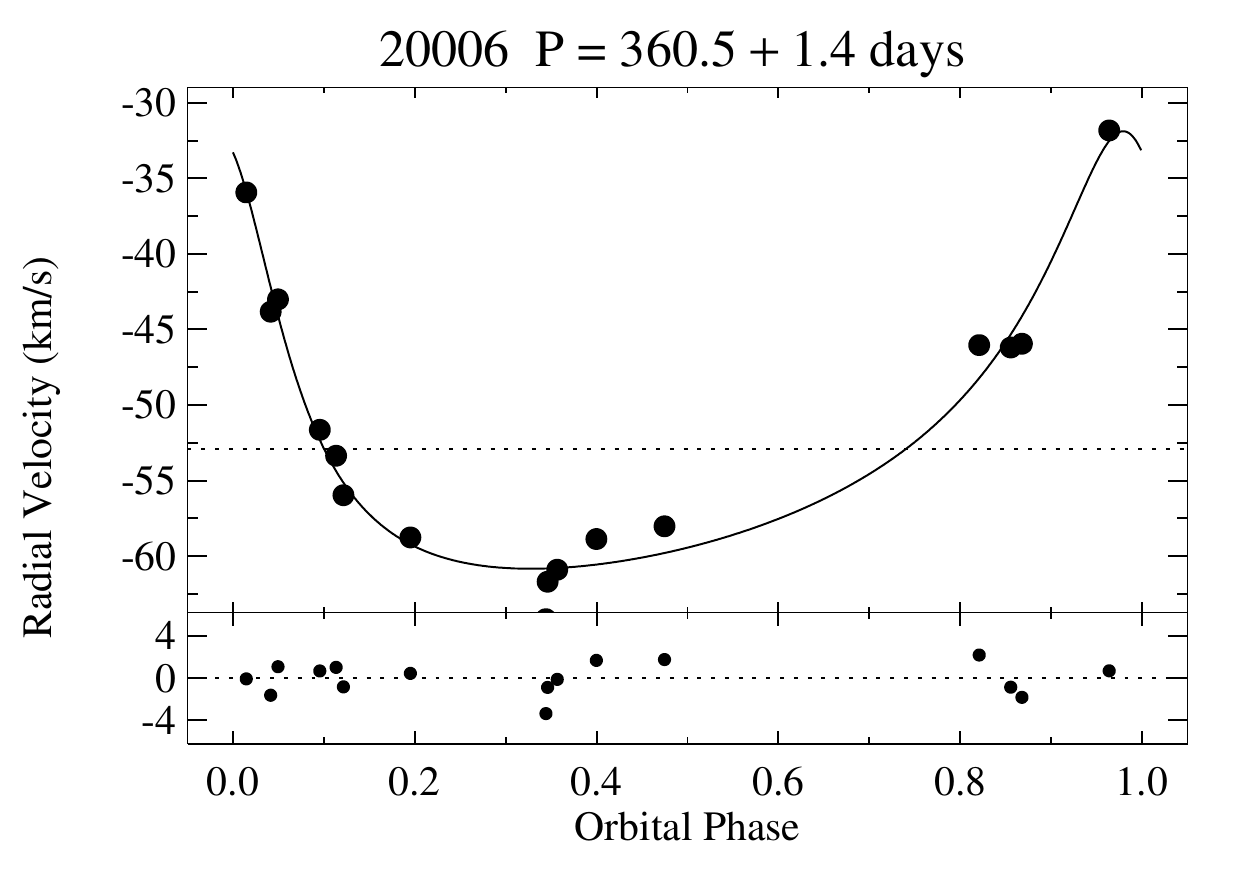}{0.3\linewidth}{}
  \fig{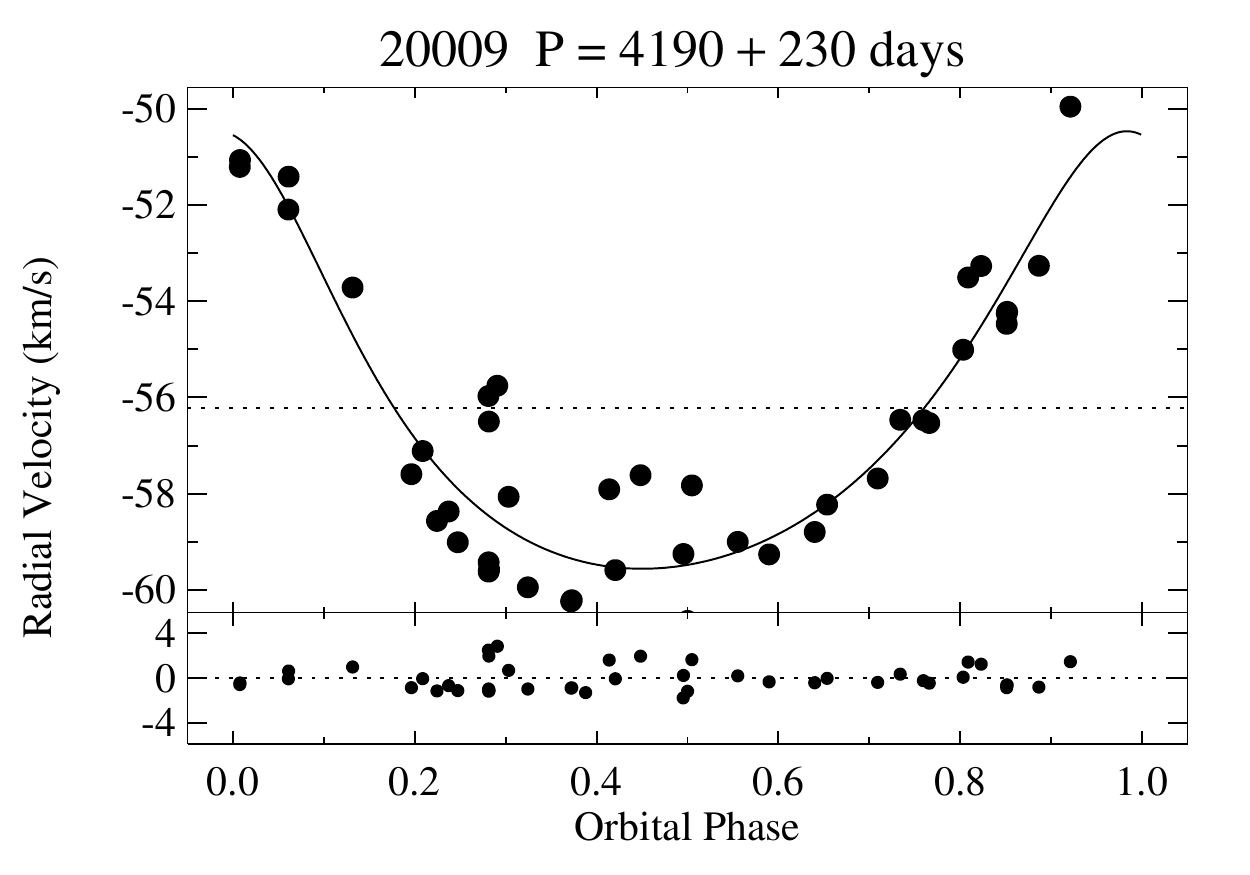}{0.3\linewidth}{}
  \fig{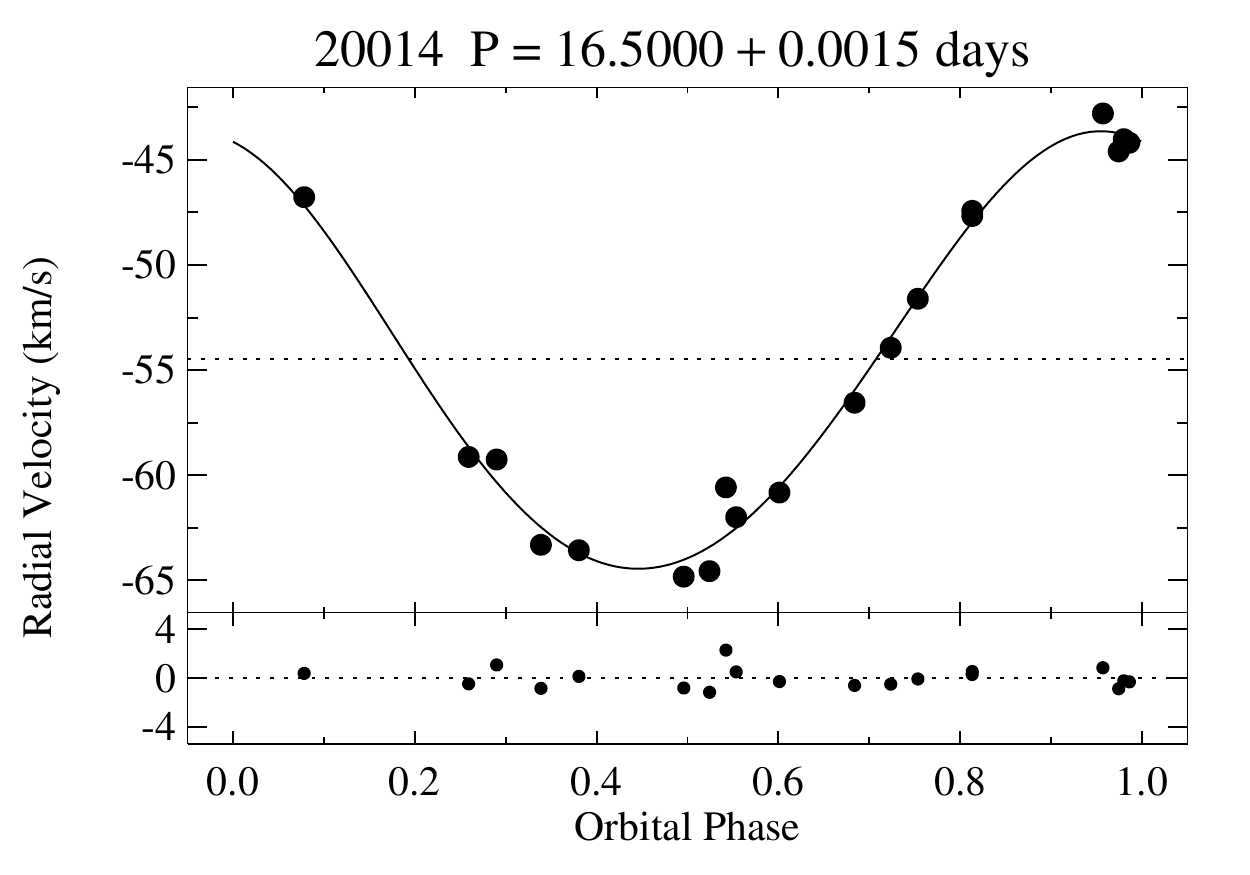}{0.3\linewidth}{}}
\end{figure*}
\begin{figure*}
\gridline{\fig{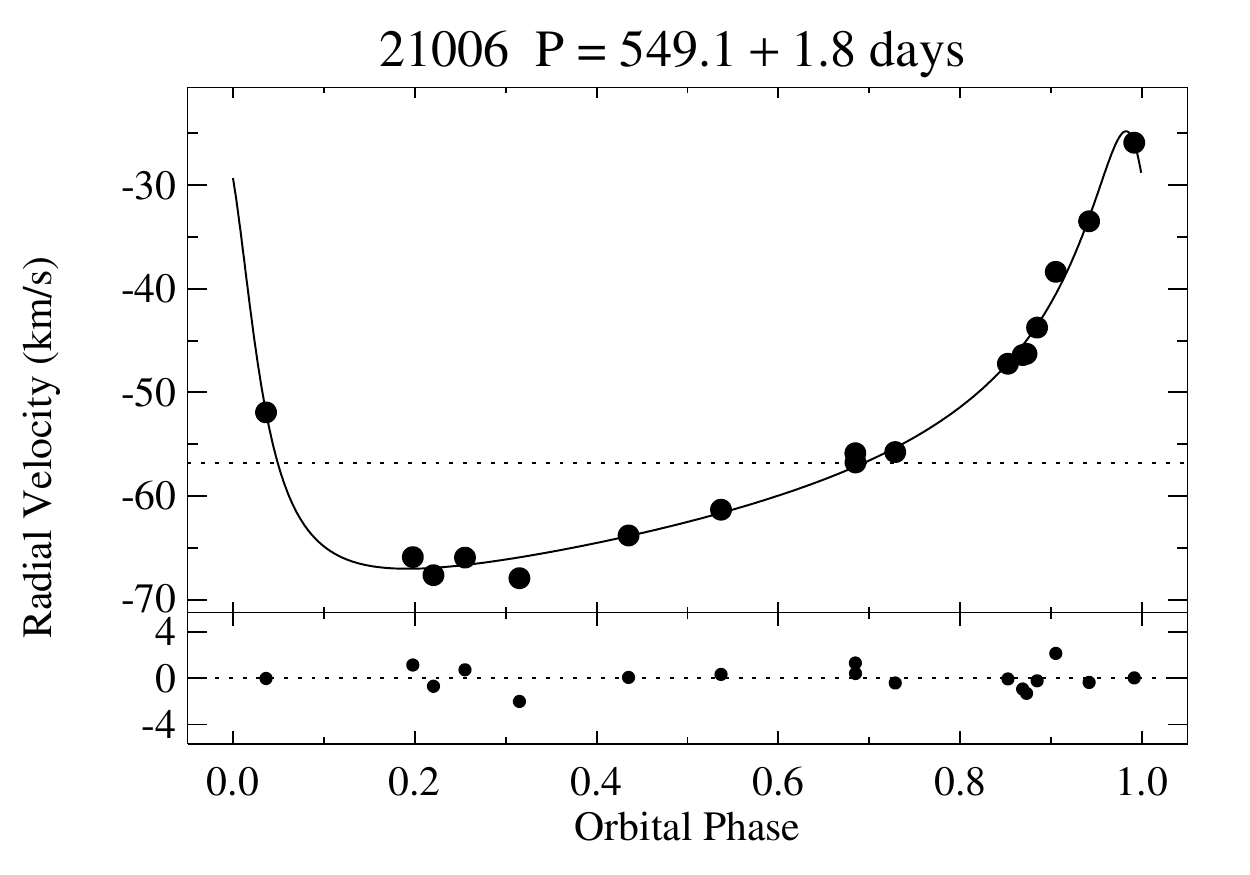}{0.3\linewidth}{}
  \fig{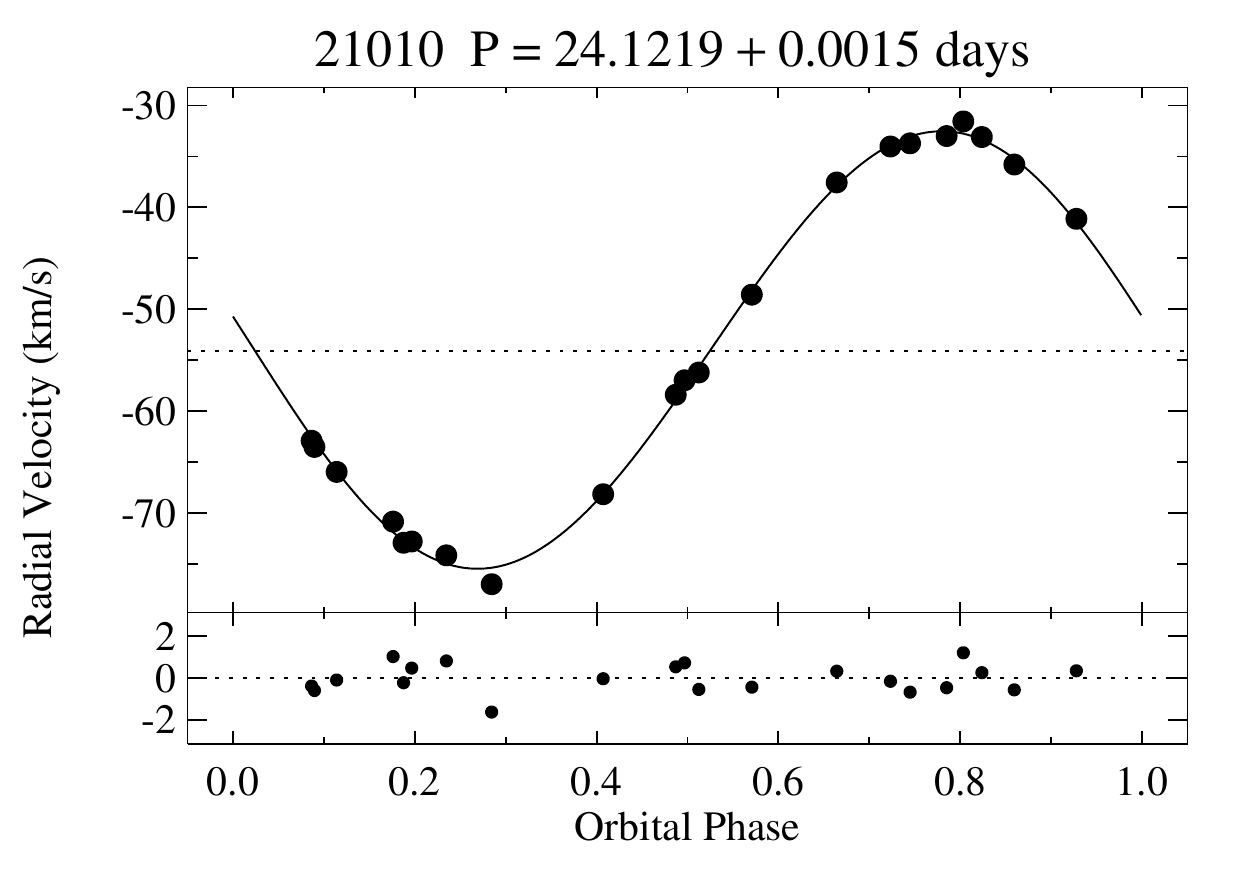}{0.3\linewidth}{}
  \fig{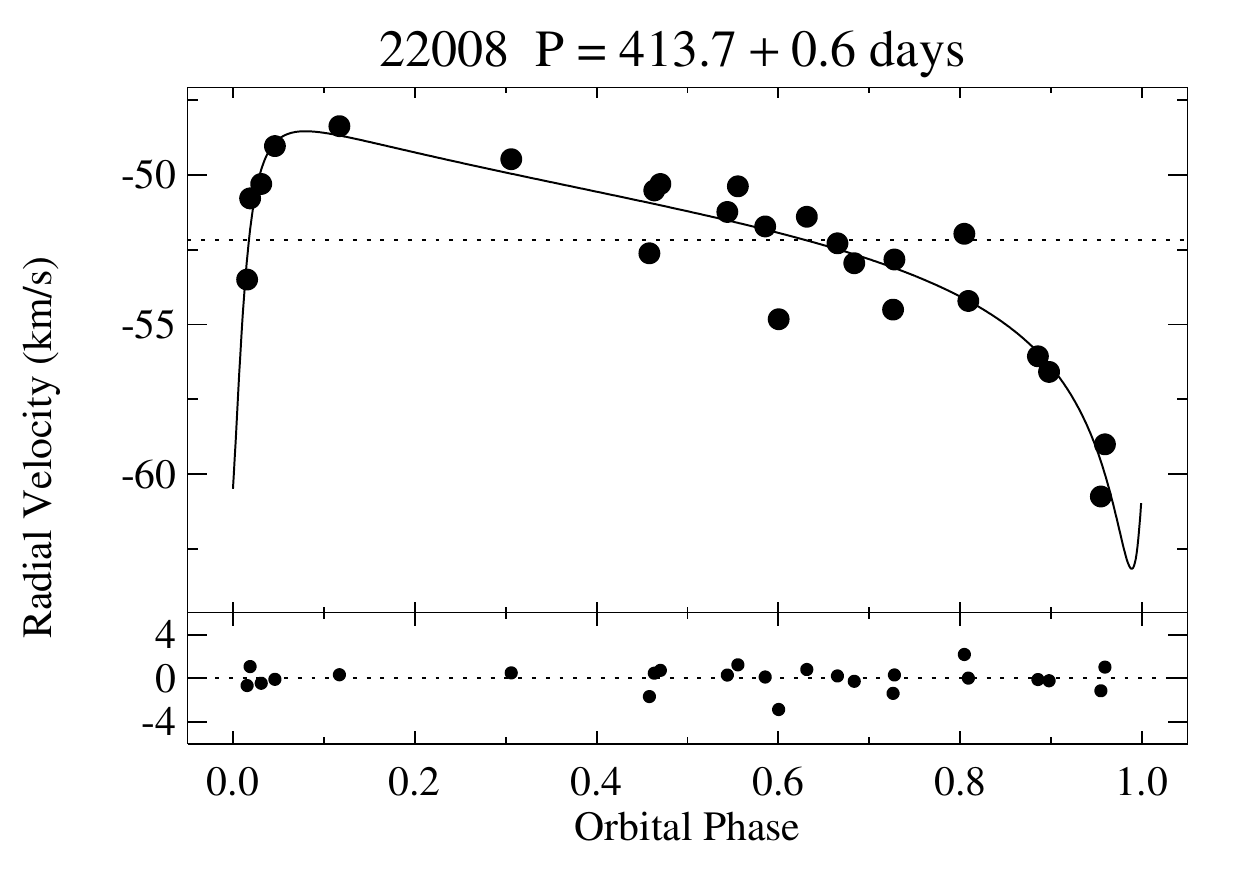}{0.3\linewidth}{}}
\gridline{\fig{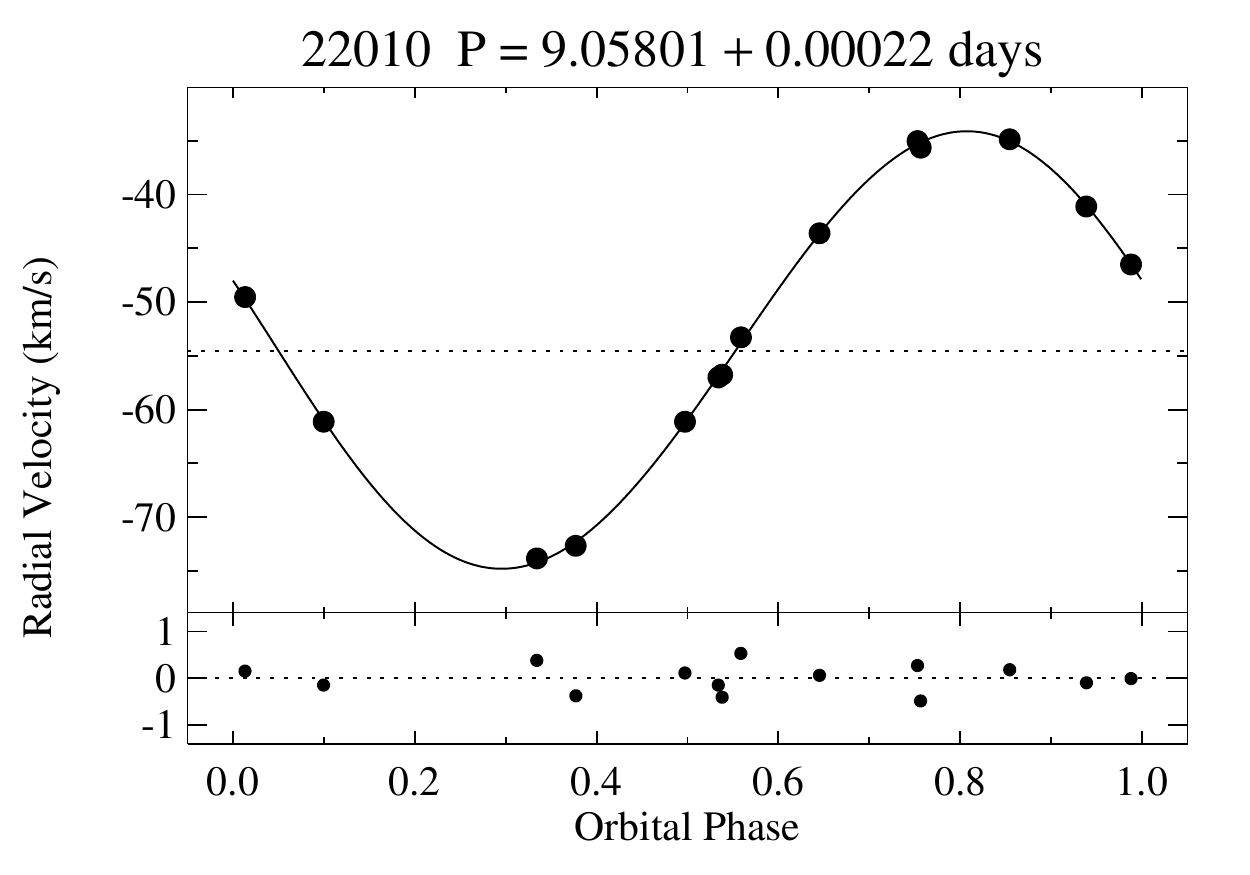}{0.3\linewidth}{}
  \fig{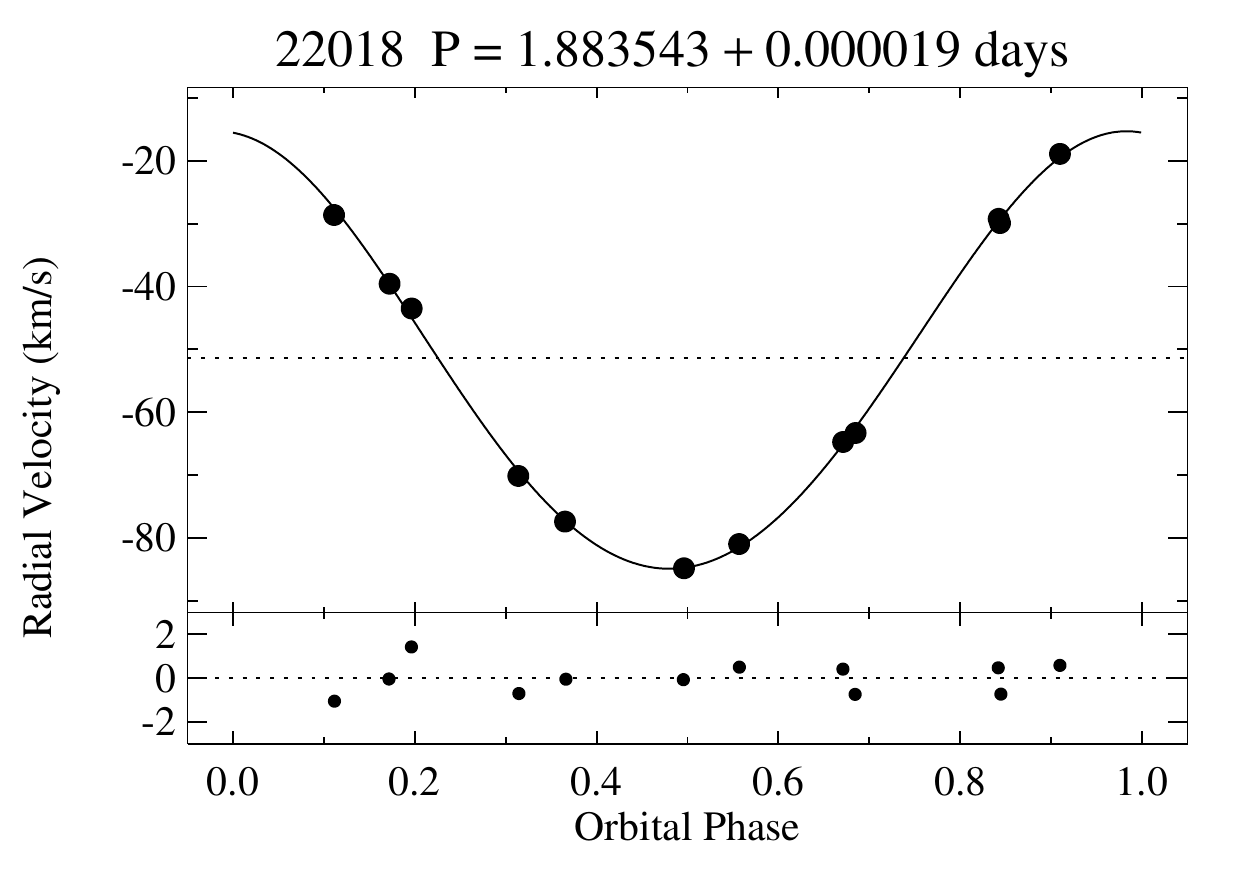}{0.3\linewidth}{}
  \fig{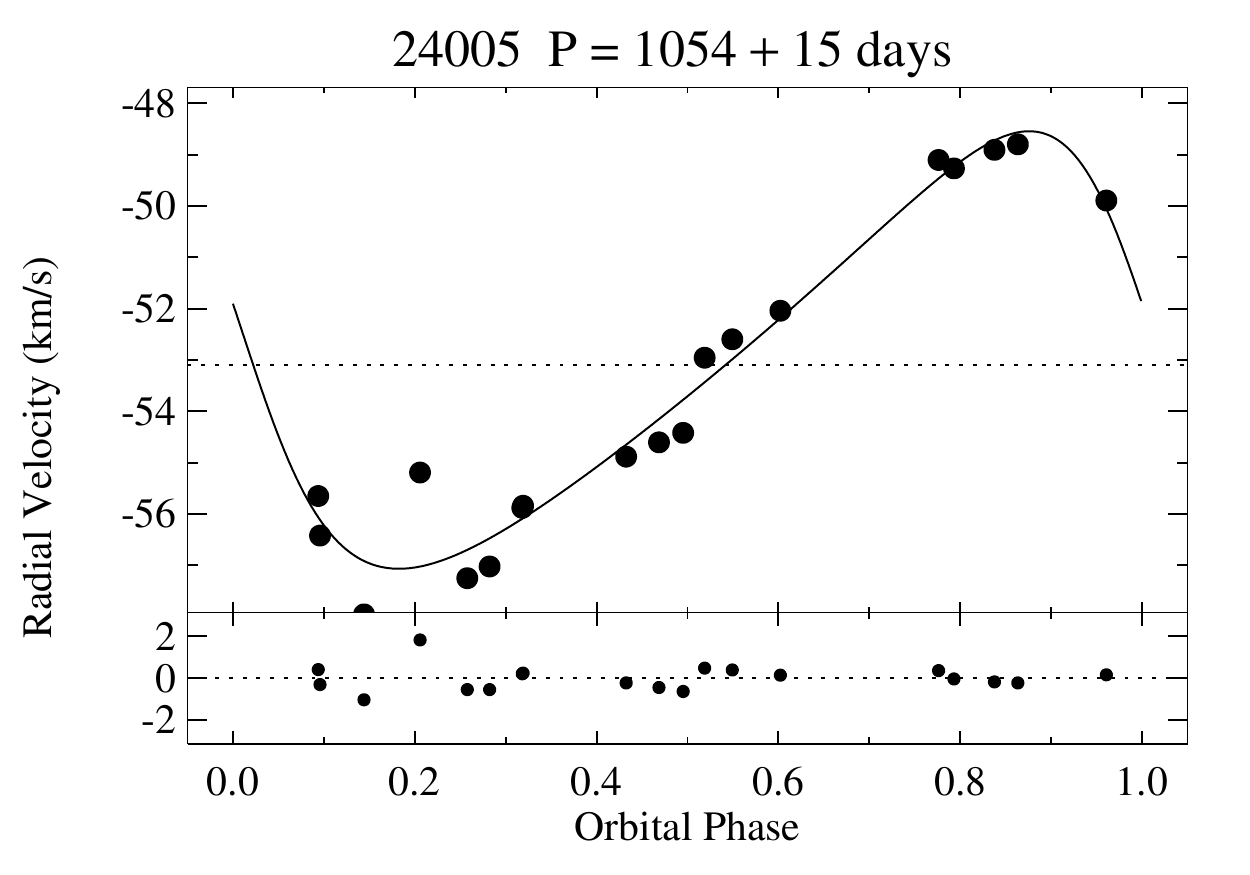}{0.3\linewidth}{}}
\gridline{\fig{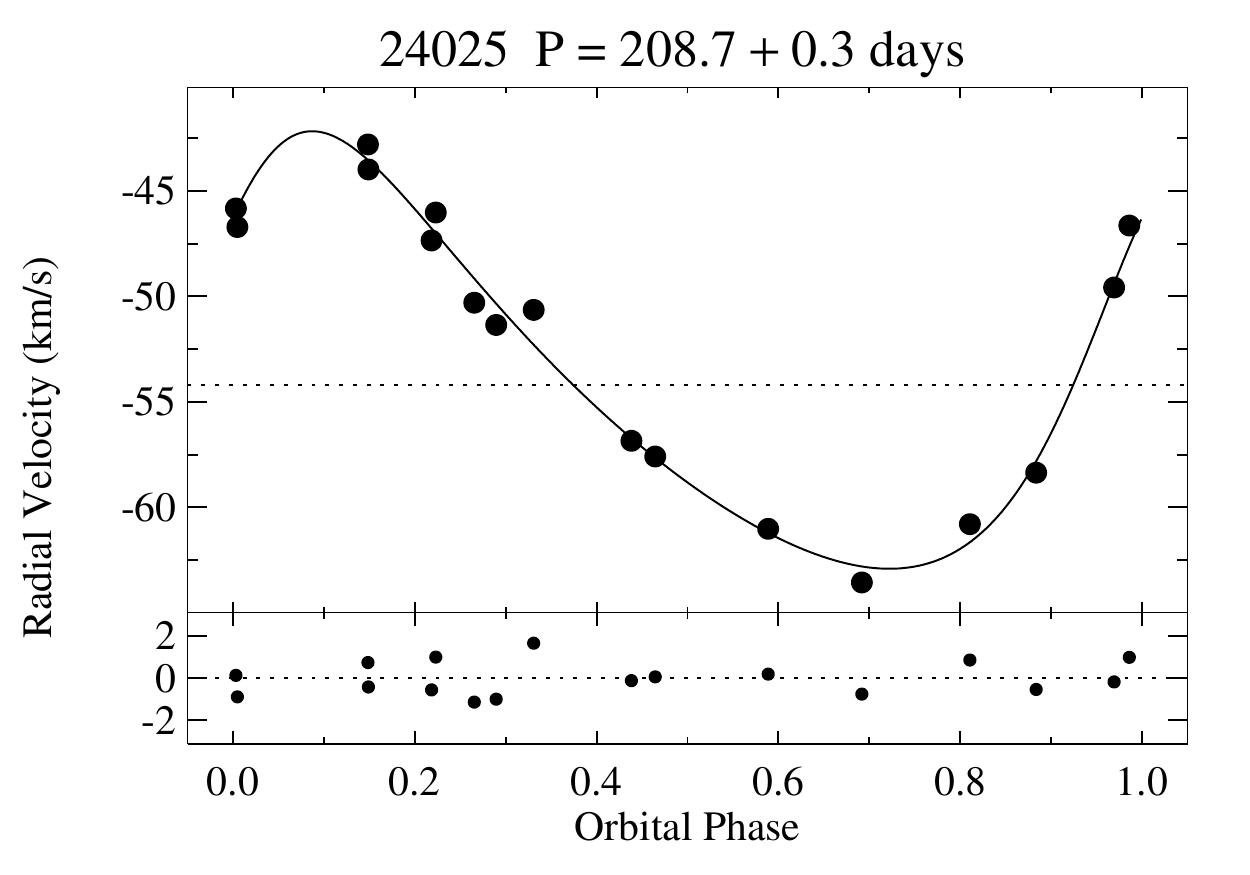}{0.3\linewidth}{}
  \fig{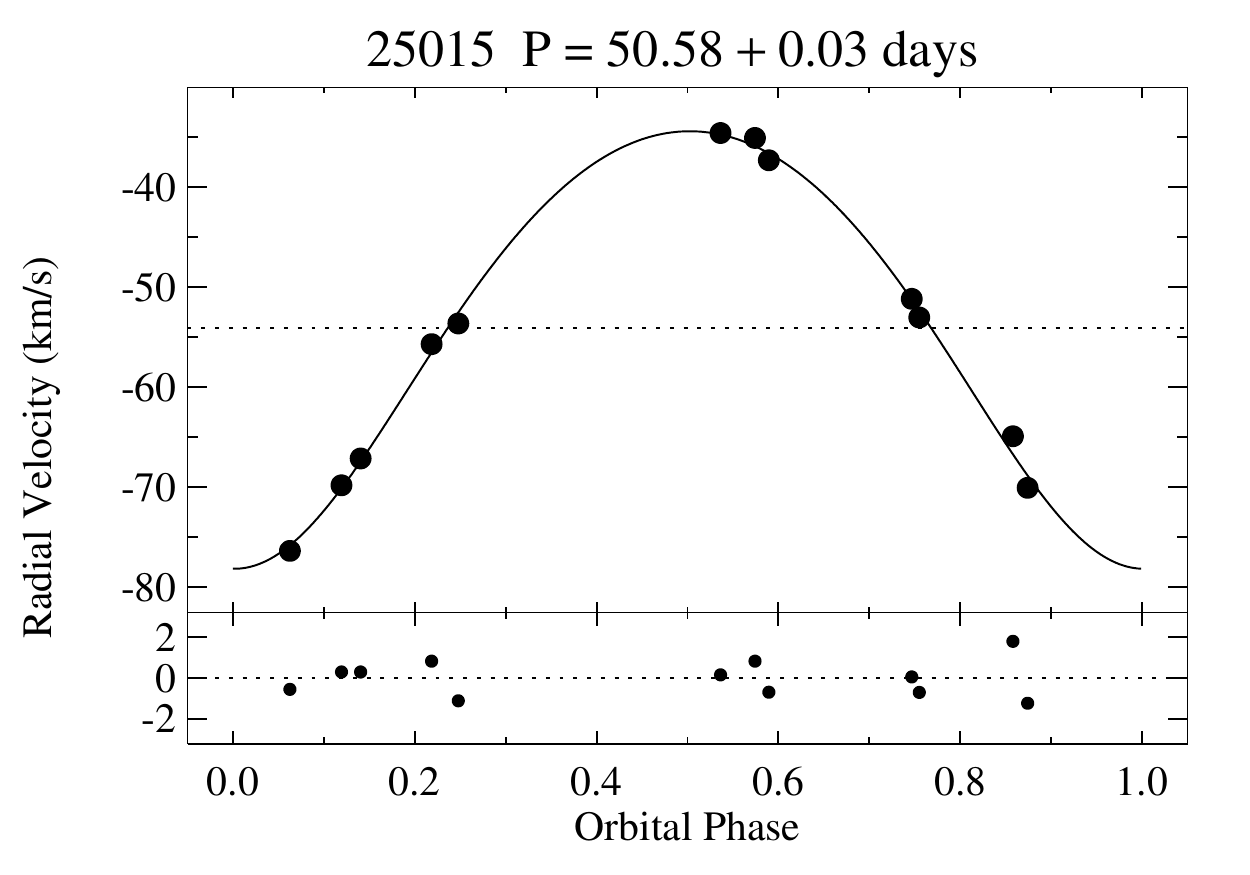}{0.3\linewidth}{}
  \fig{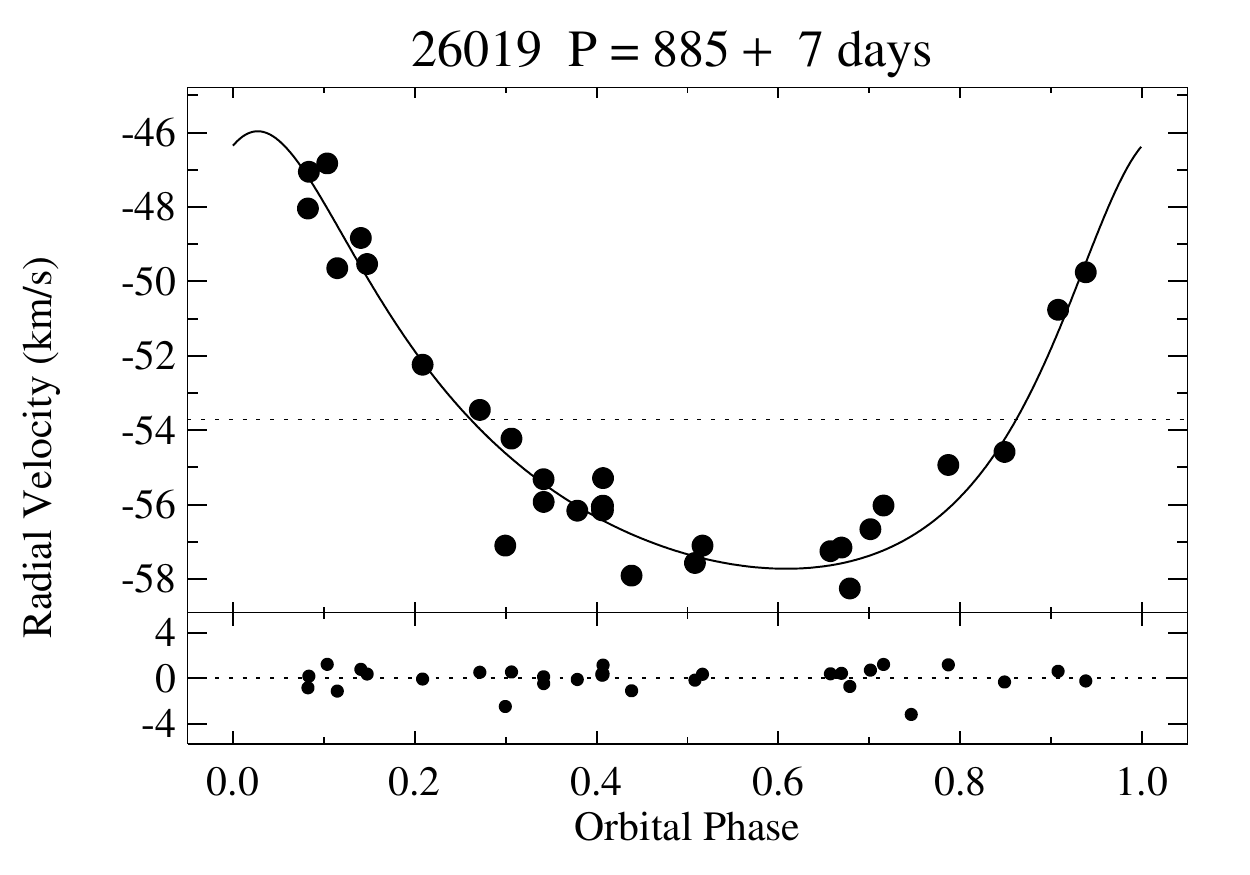}{0.3\linewidth}{}}
\gridline{\fig{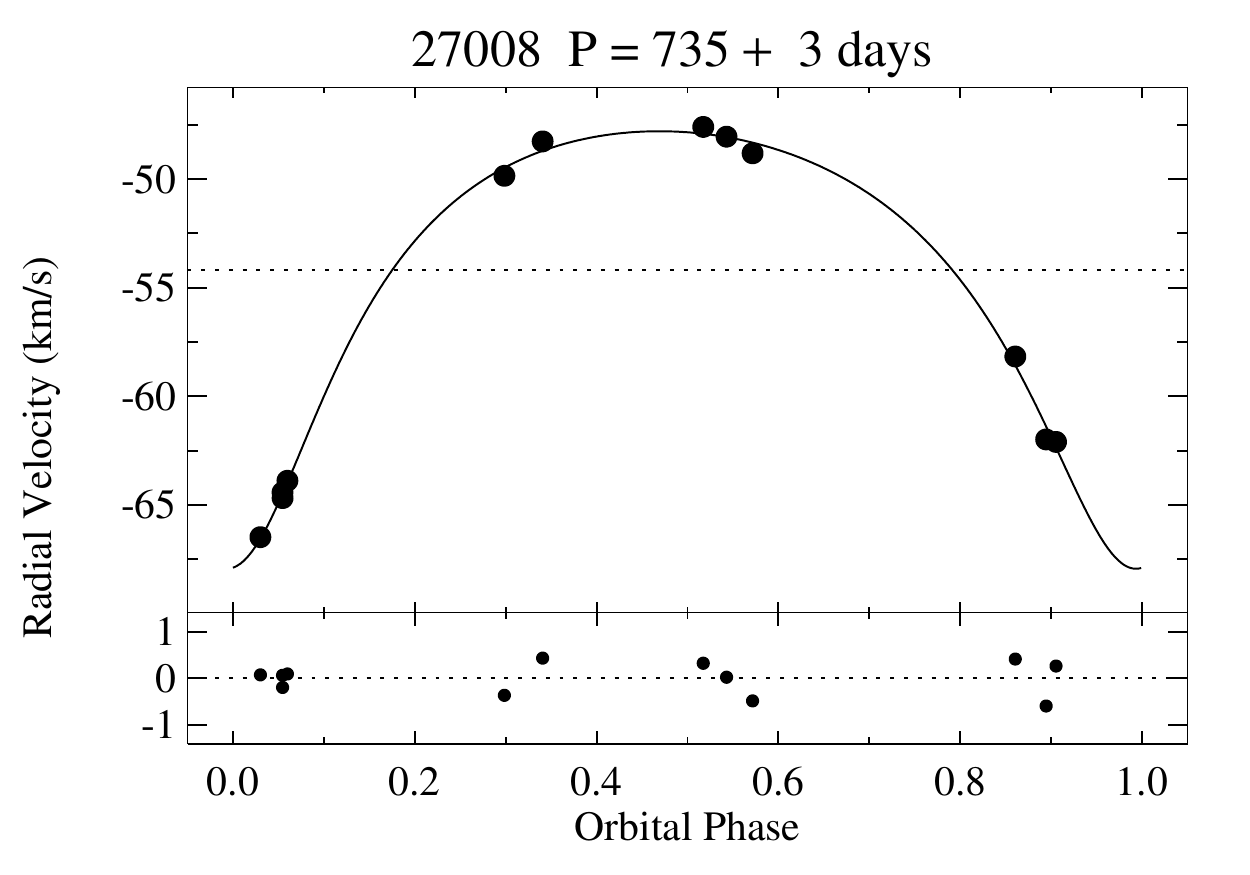}{0.3\linewidth}{}
  \fig{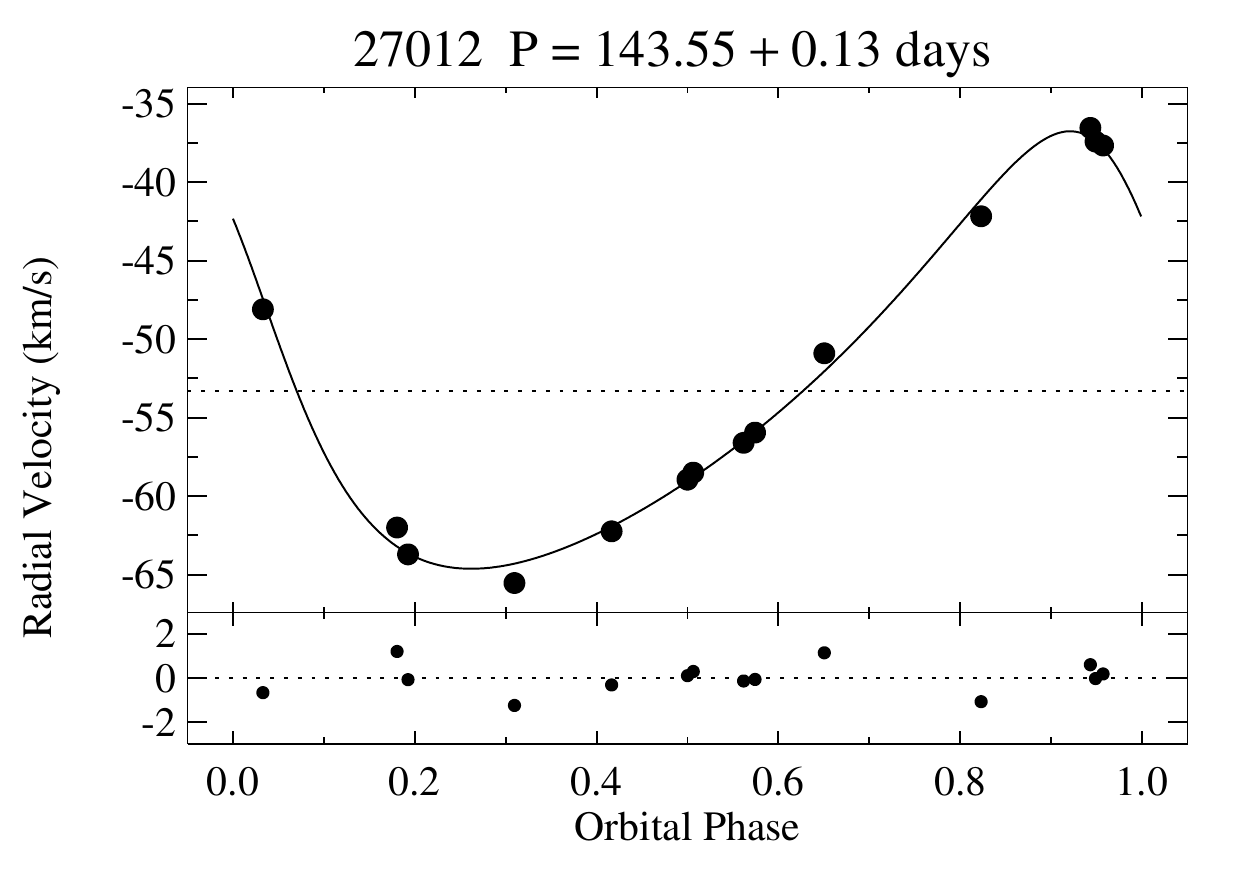}{0.3\linewidth}{}
  \fig{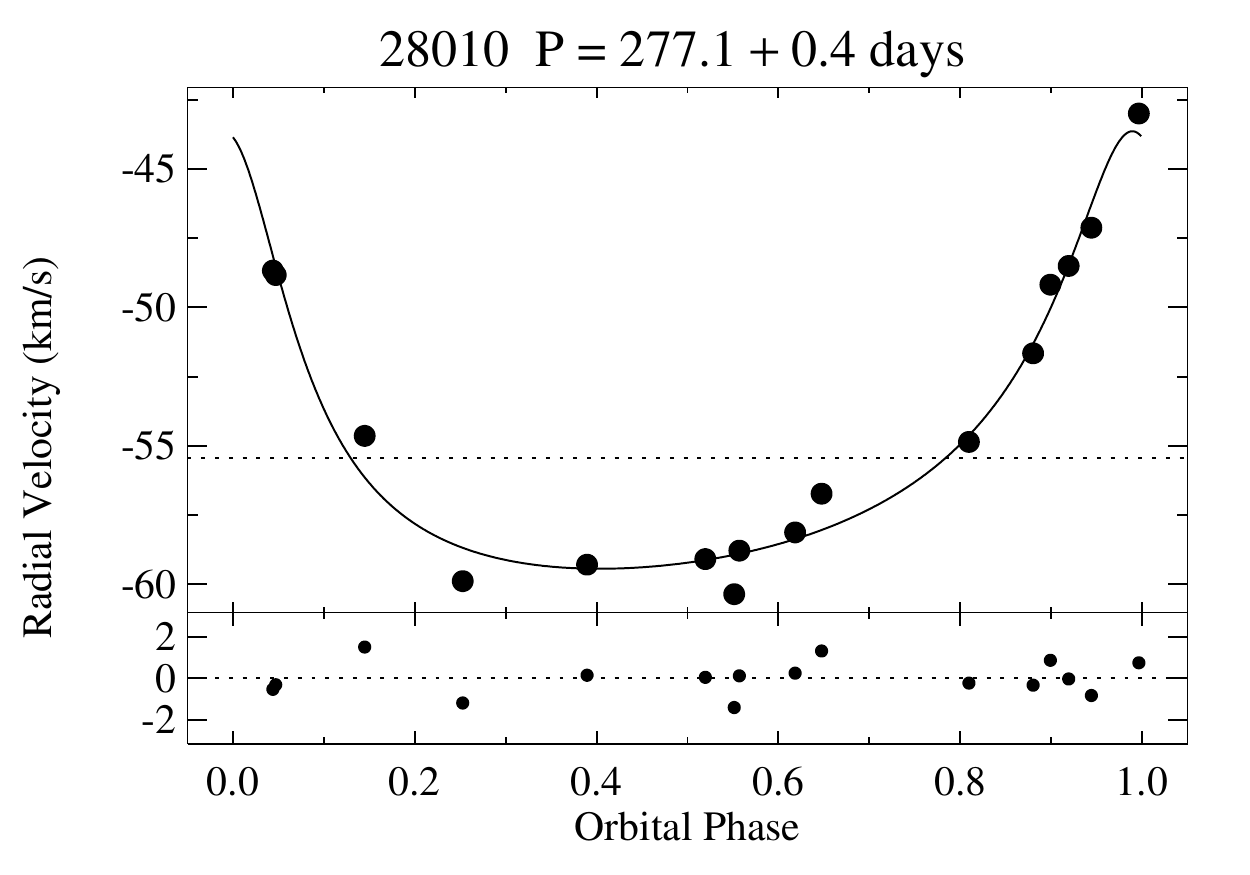}{0.3\linewidth}{}}
\end{figure*}
\begin{figure*}
\gridline{\fig{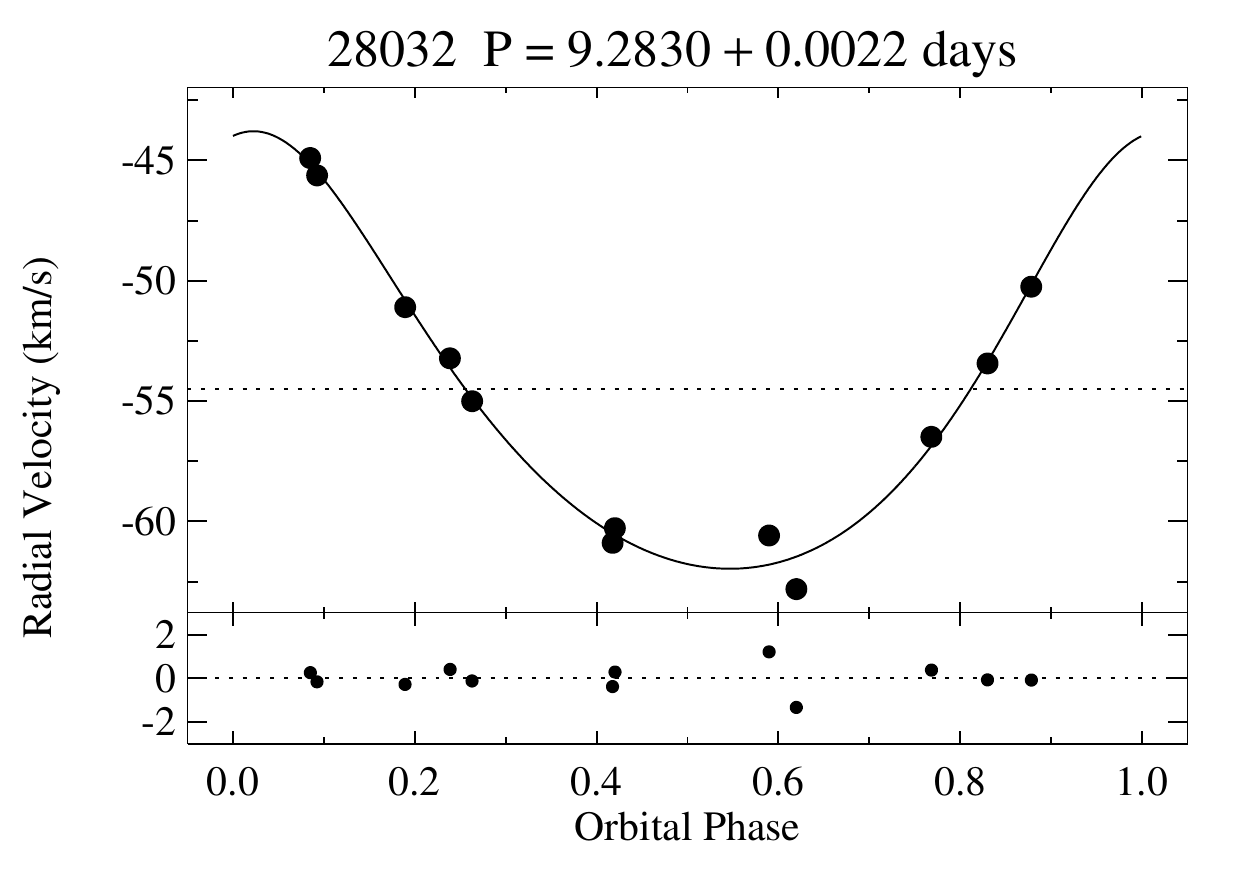}{0.3\linewidth}{}
  \fig{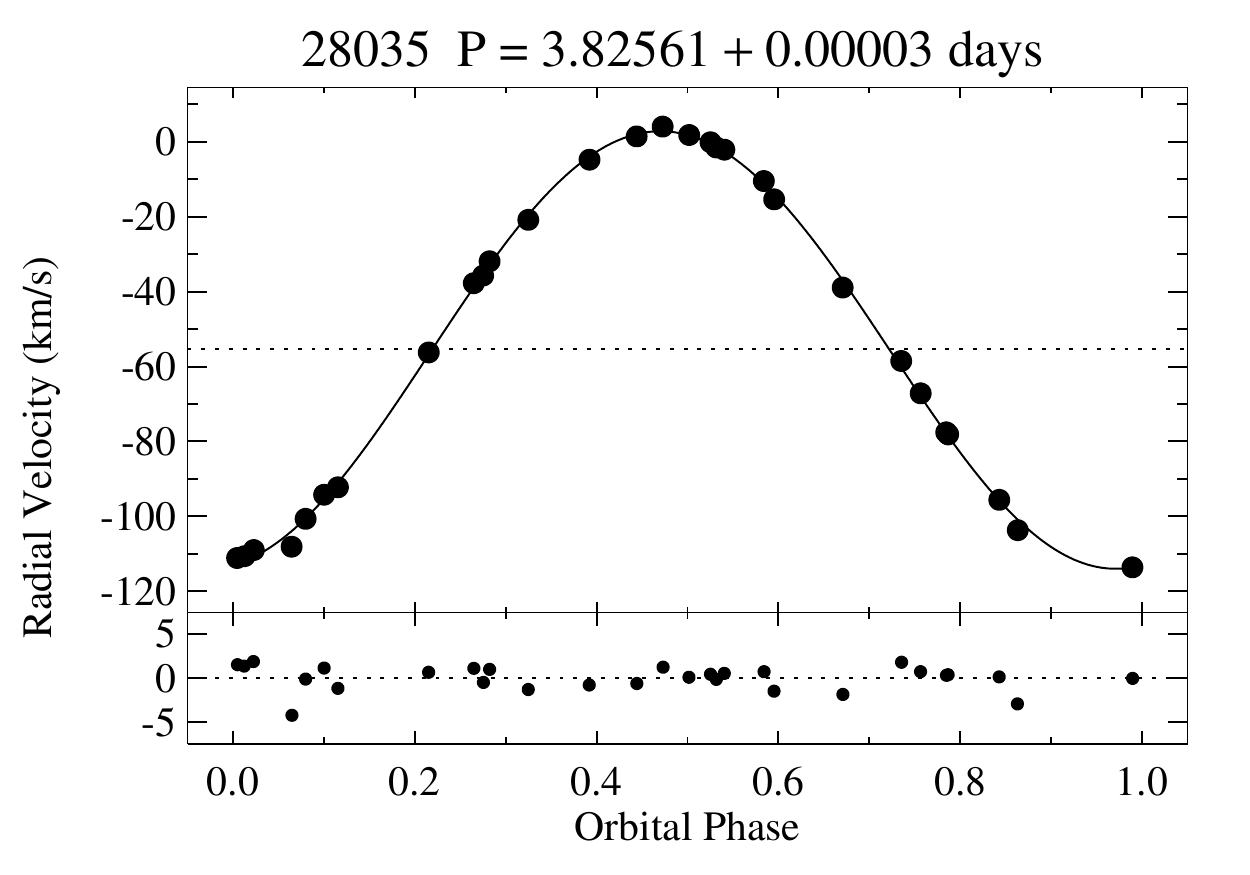}{0.3\linewidth}{}
  \fig{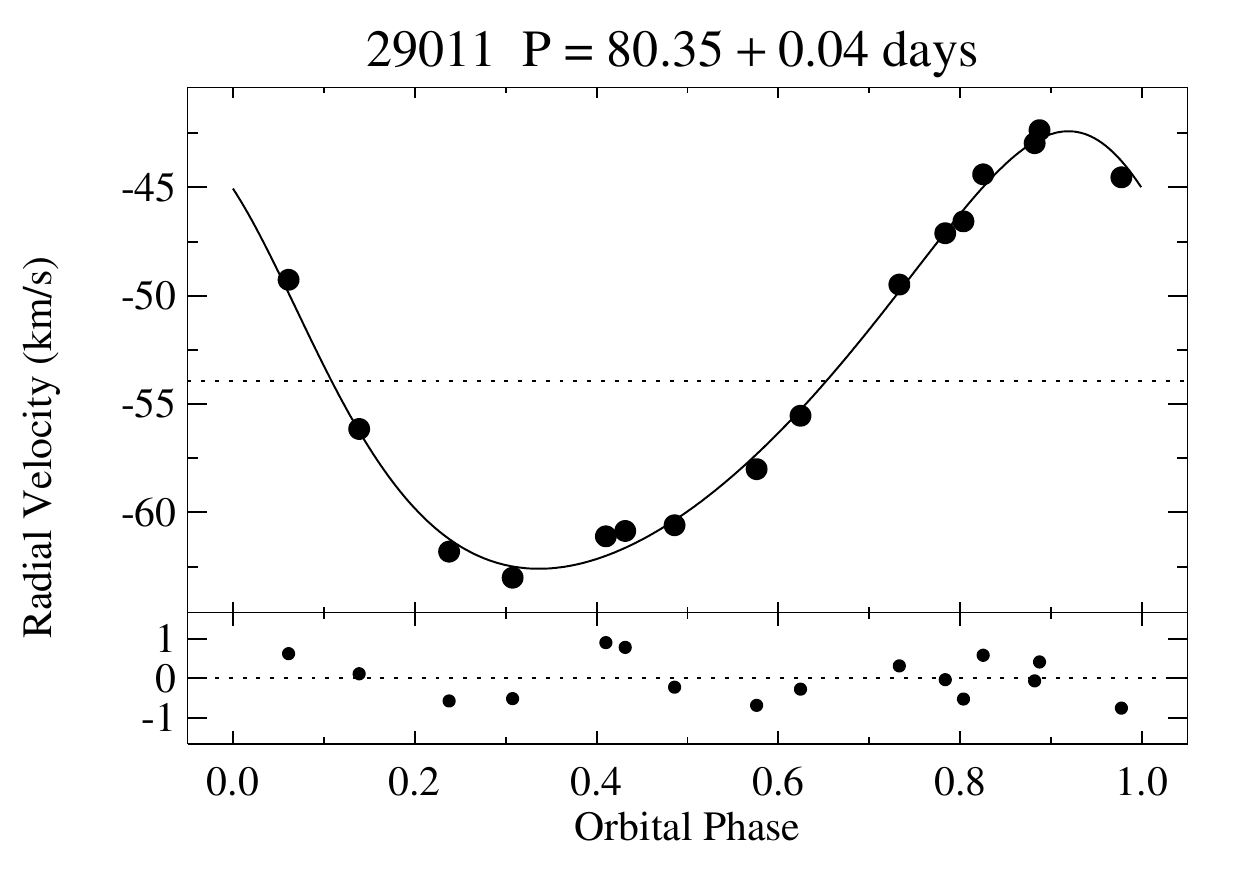}{0.3\linewidth}{}}
\gridline{\fig{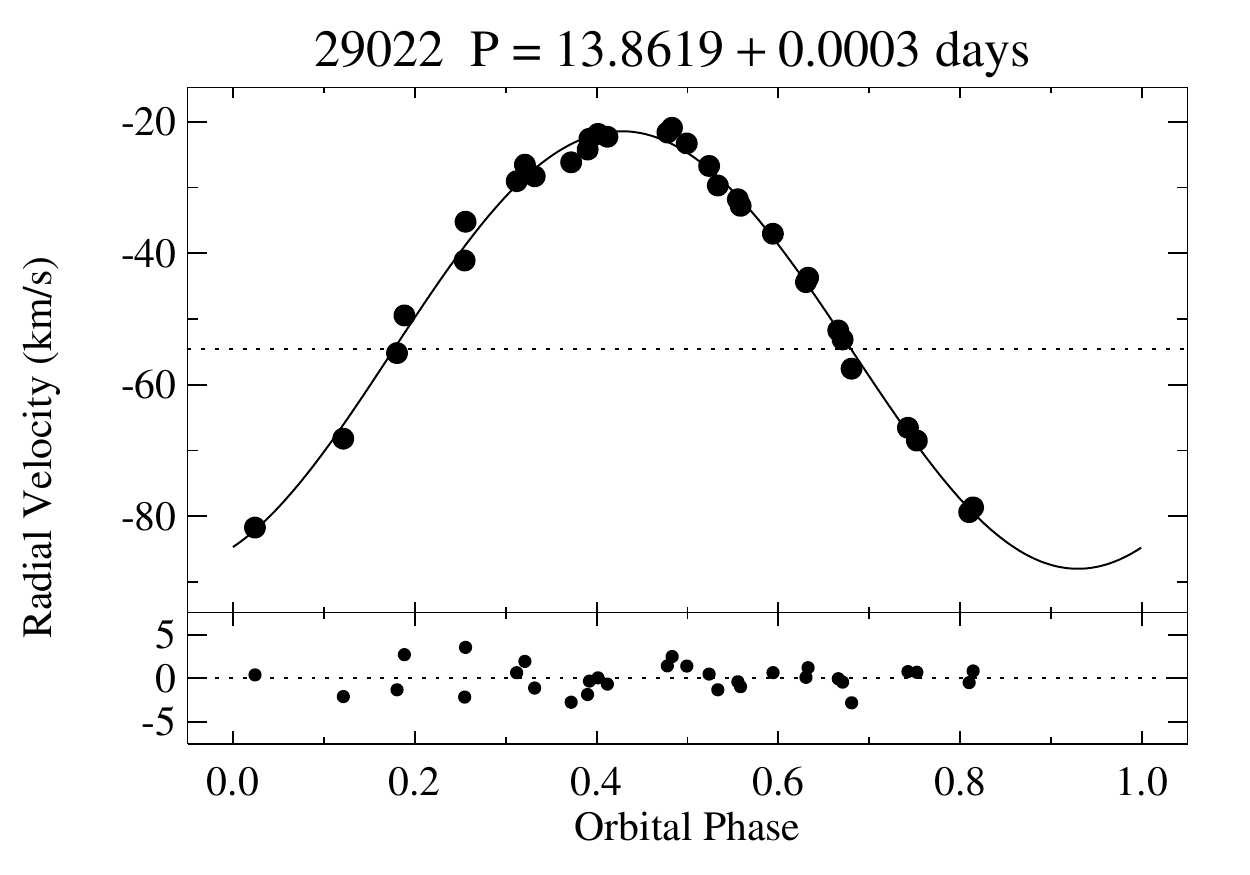}{0.3\linewidth}{}
  \fig{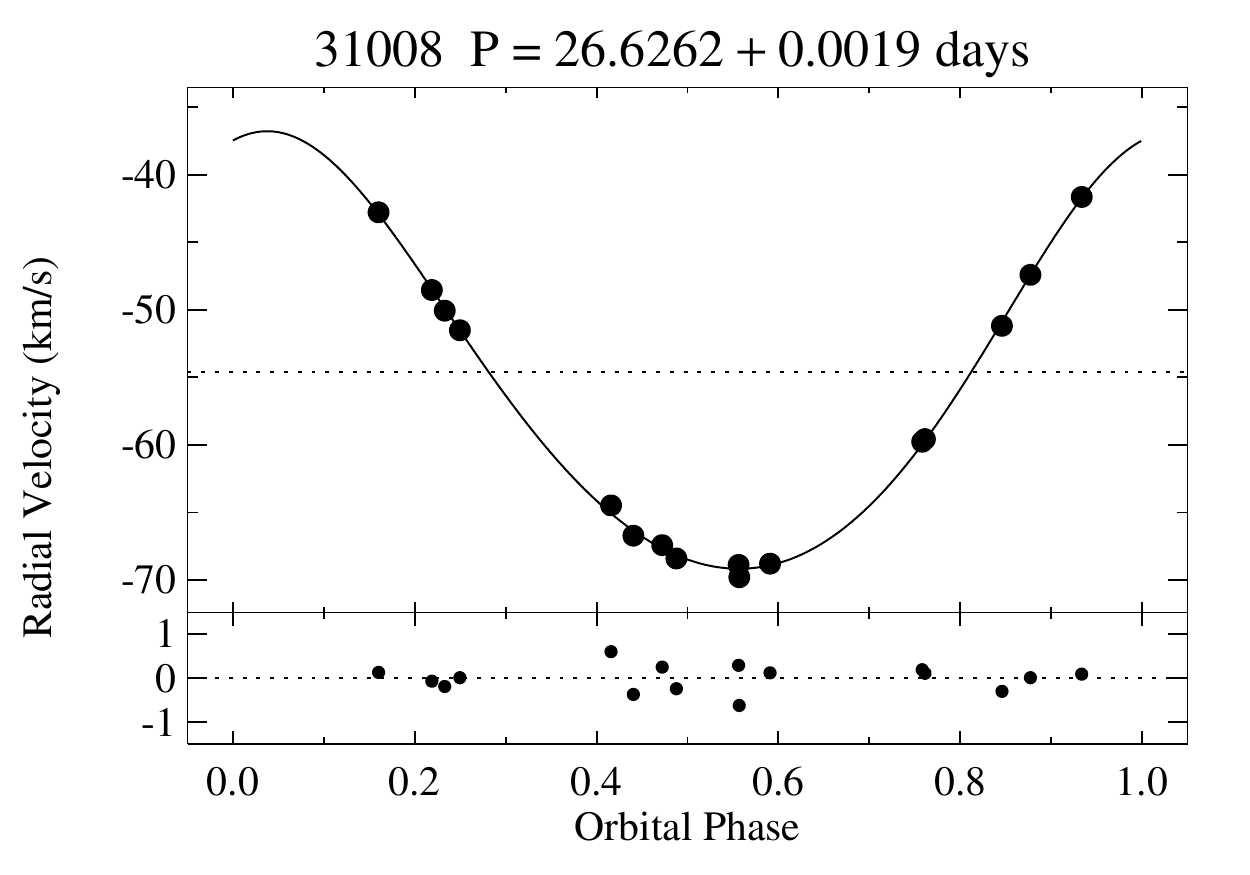}{0.3\linewidth}{}
  \fig{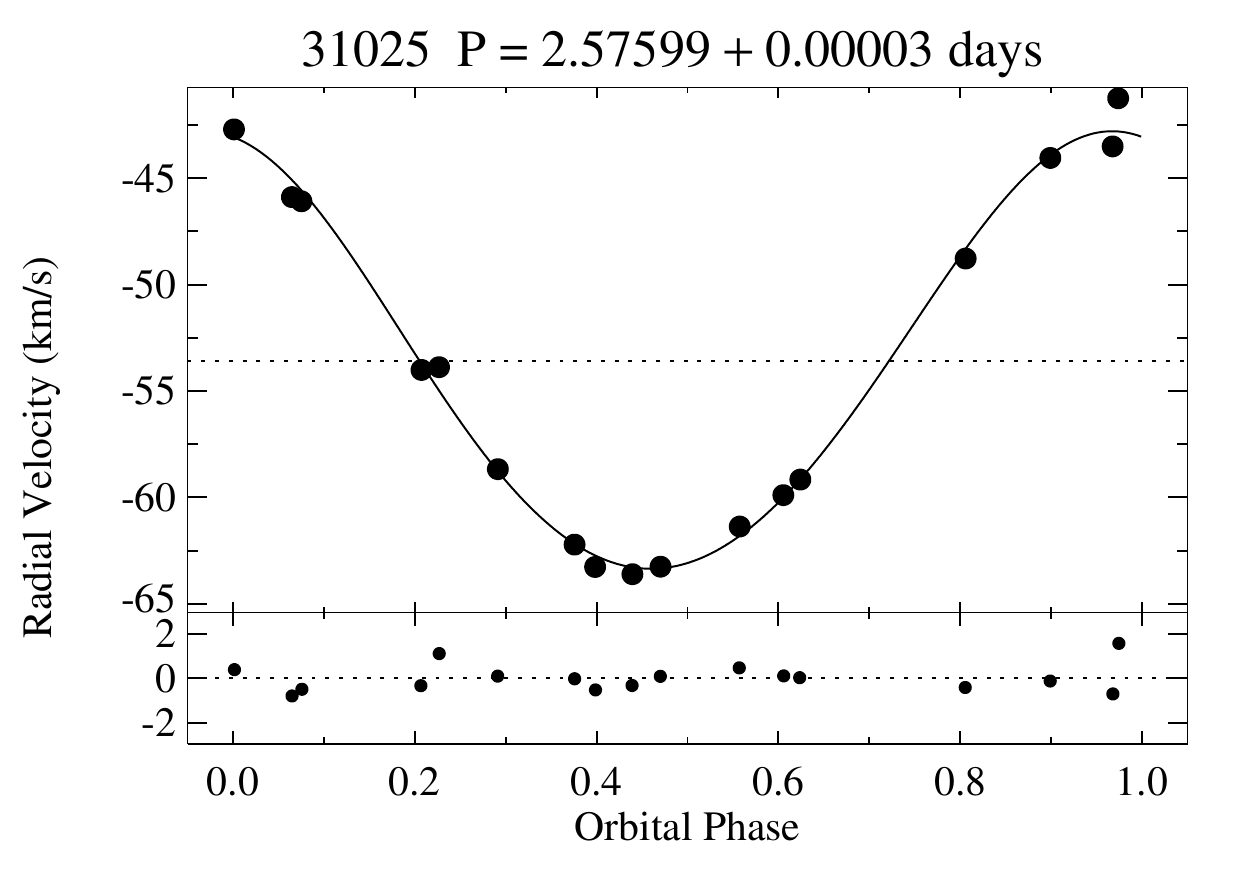}{0.3\linewidth}{}}
\gridline{\fig{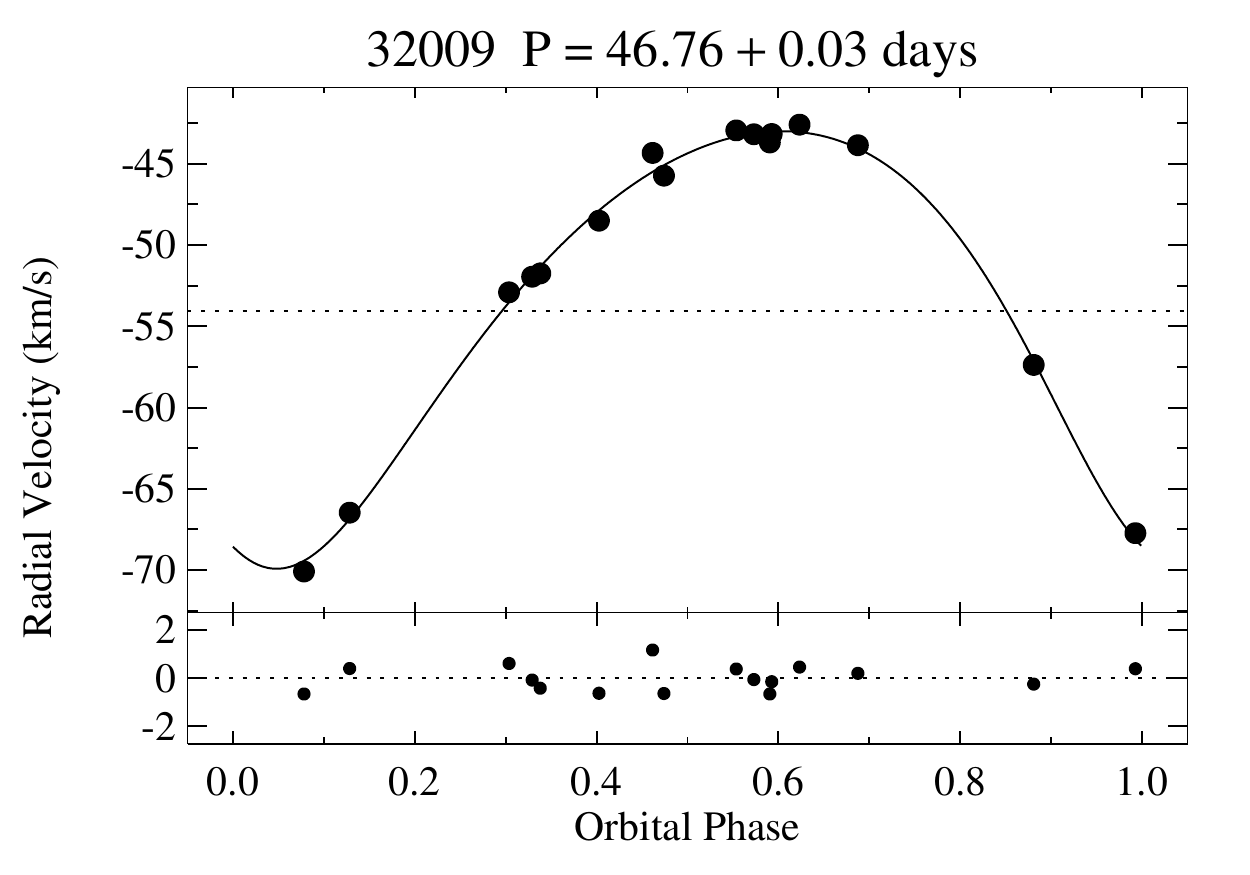}{0.3\linewidth}{}
  \fig{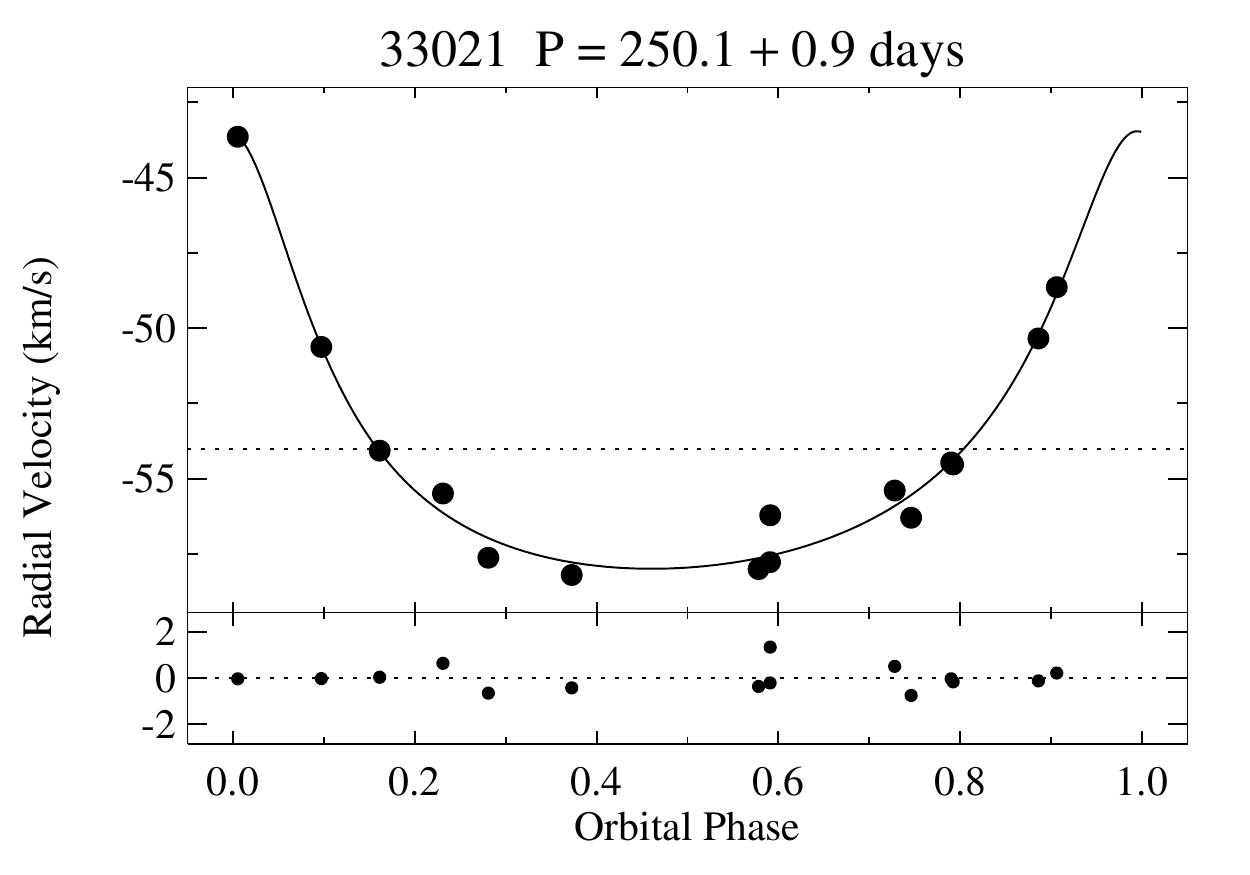}{0.3\linewidth}{}
  \fig{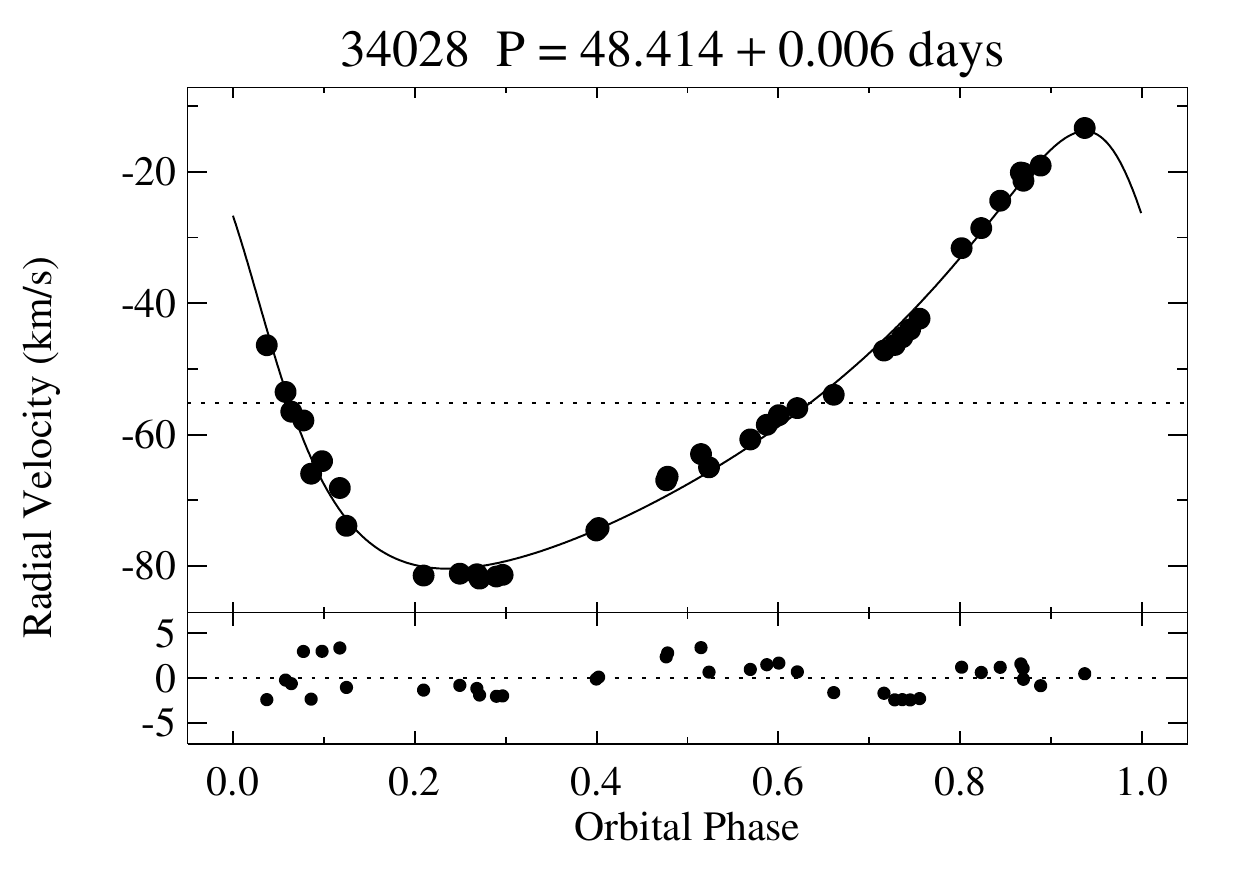}{0.3\linewidth}{}}
\gridline{\fig{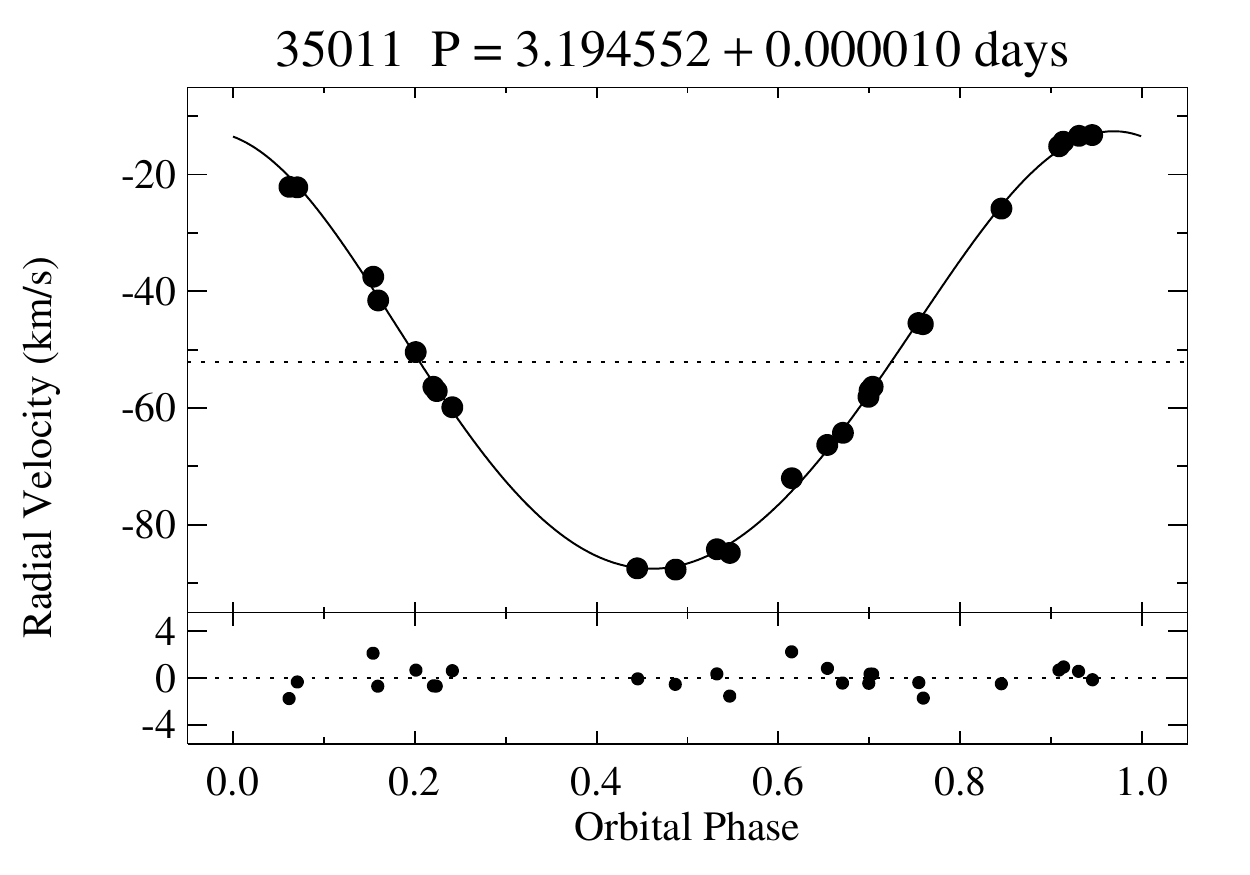}{0.3\linewidth}{}
  \fig{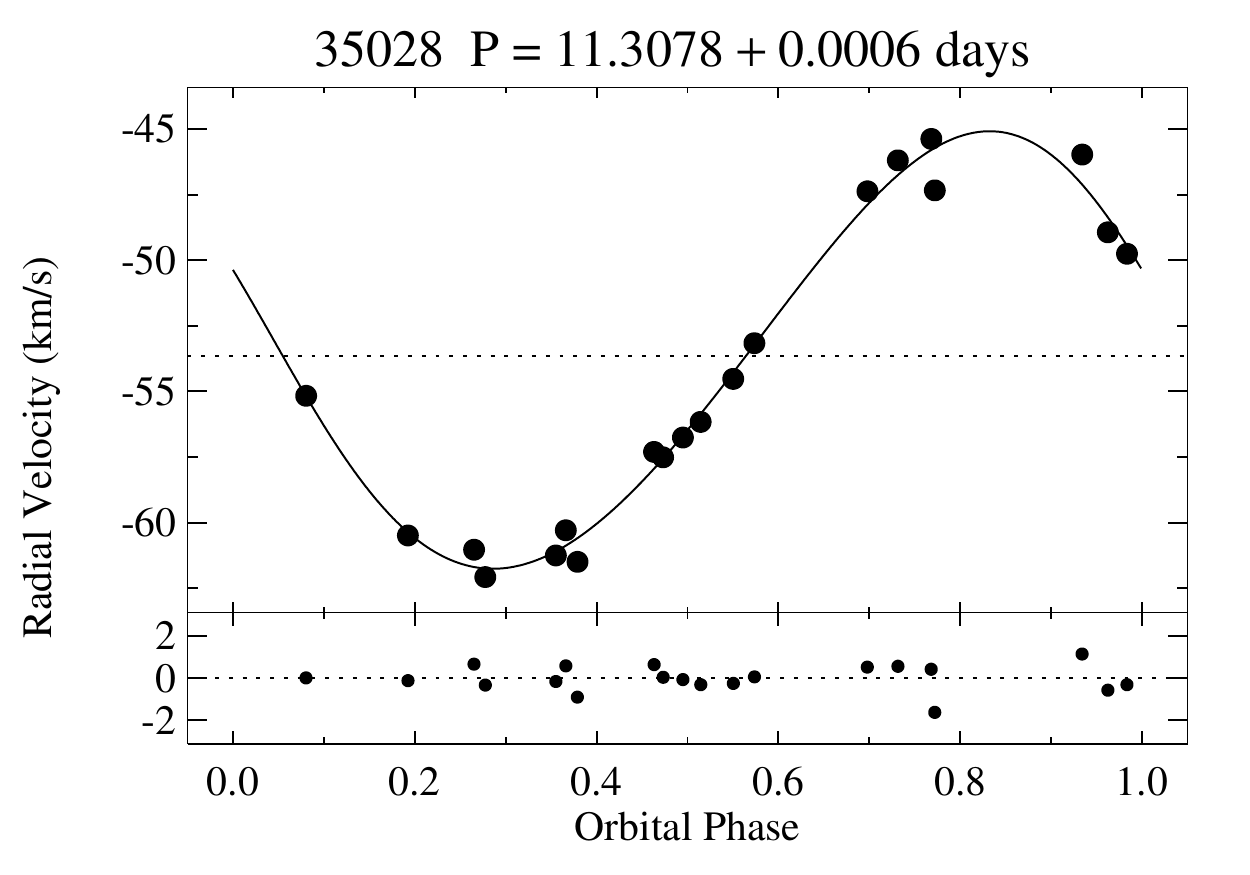}{0.3\linewidth}{}
  \fig{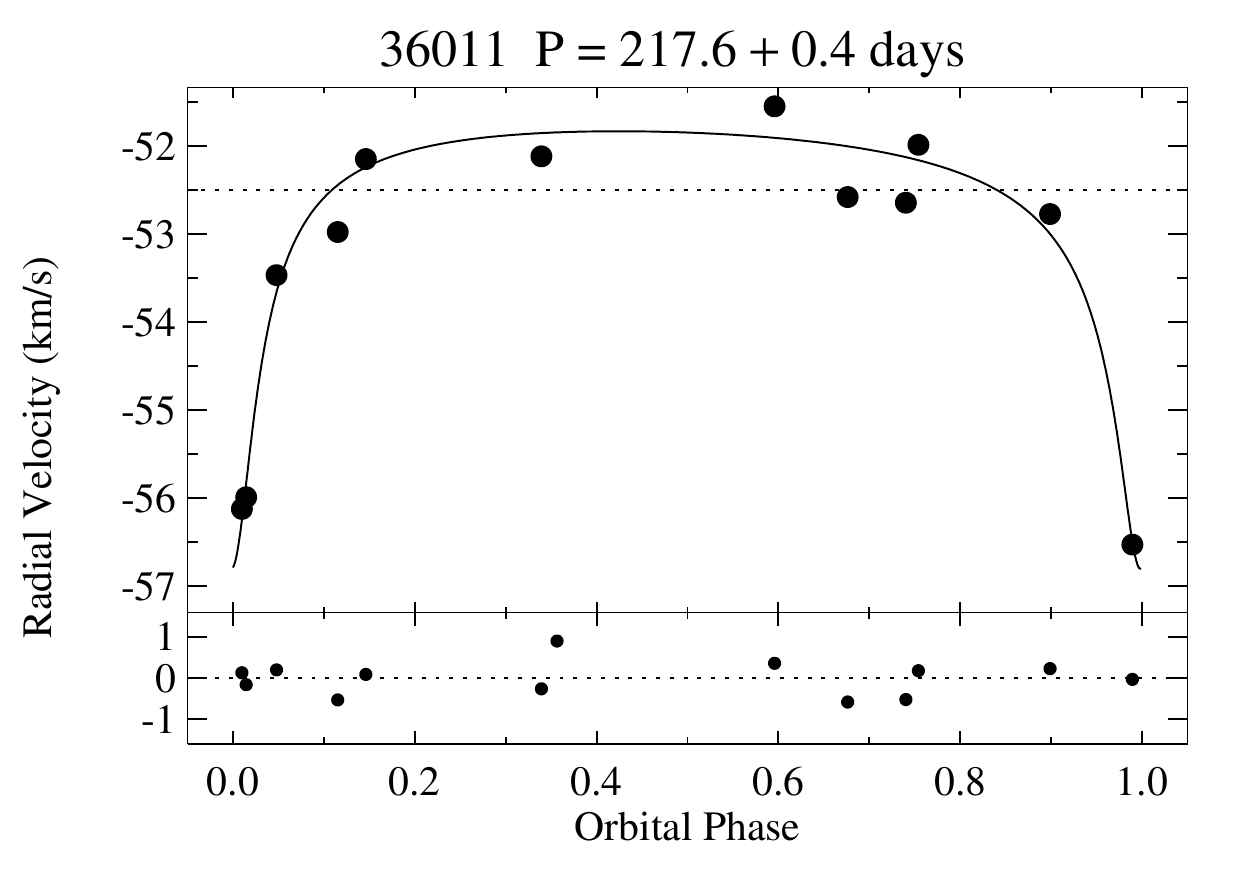}{0.3\linewidth}{}}
\end{figure*}

\clearpage

\subsection{Double-Lined Orbital Solutions}
\label{ngc7789rv:sec:SB2}
For double-lined spectroscopic binaries (SB2) we use TODCOR (\citealt{Zucker1994}) which utilizes two template spectra to simultaneously derive RVs for the primary and secondary stars. For these binaries we used an observed solar spectrum for both templates.

\begin{figure*}
\gridline{\fig{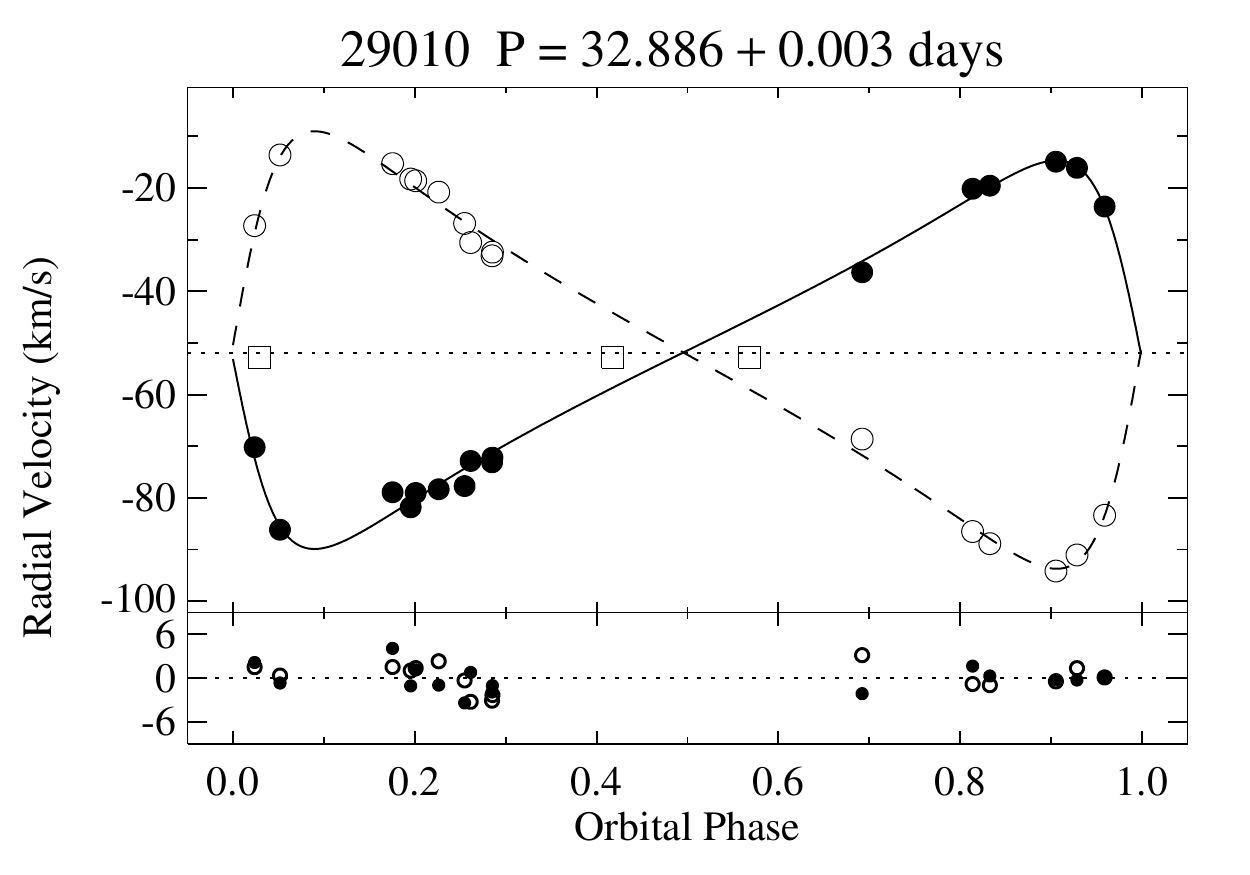}{0.3\linewidth}{}
  \fig{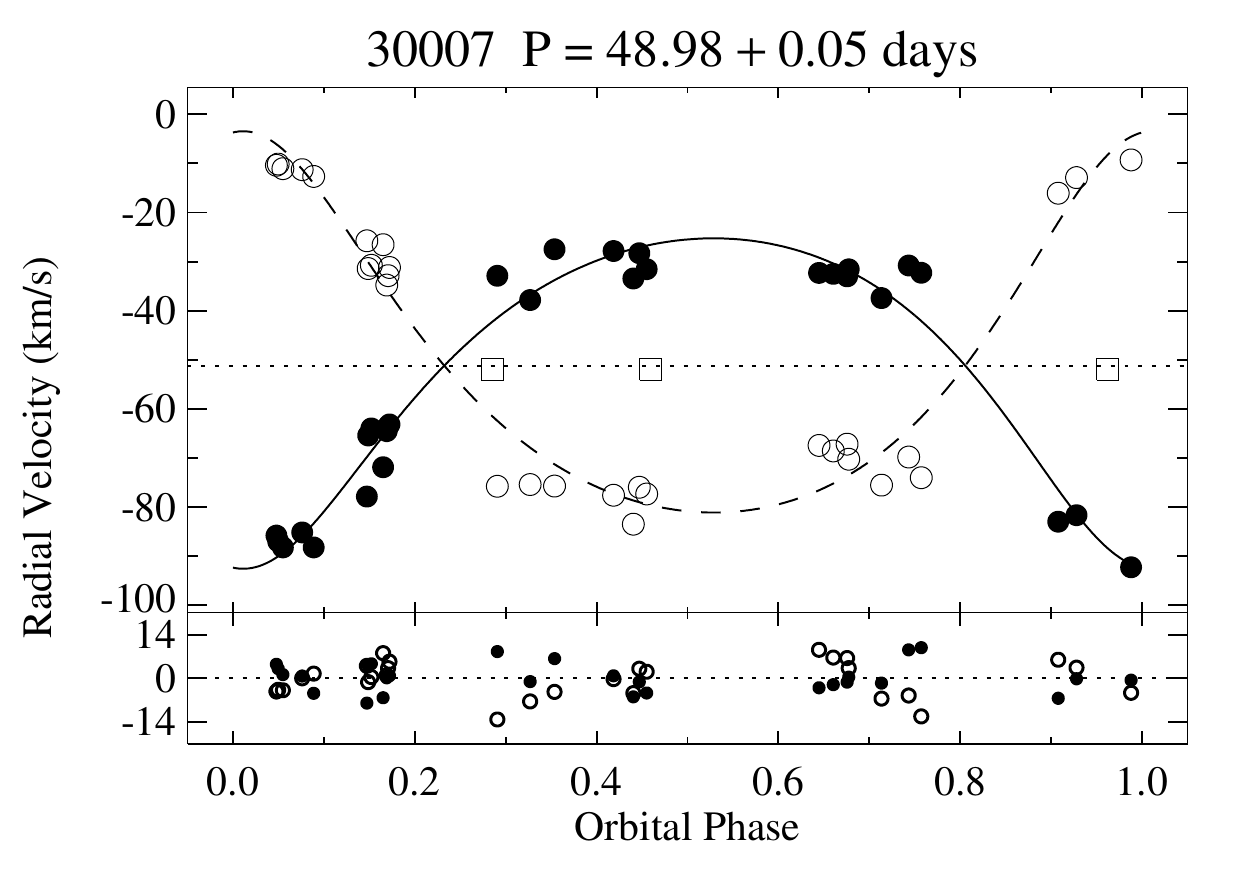}{0.3\linewidth}{}
  \fig{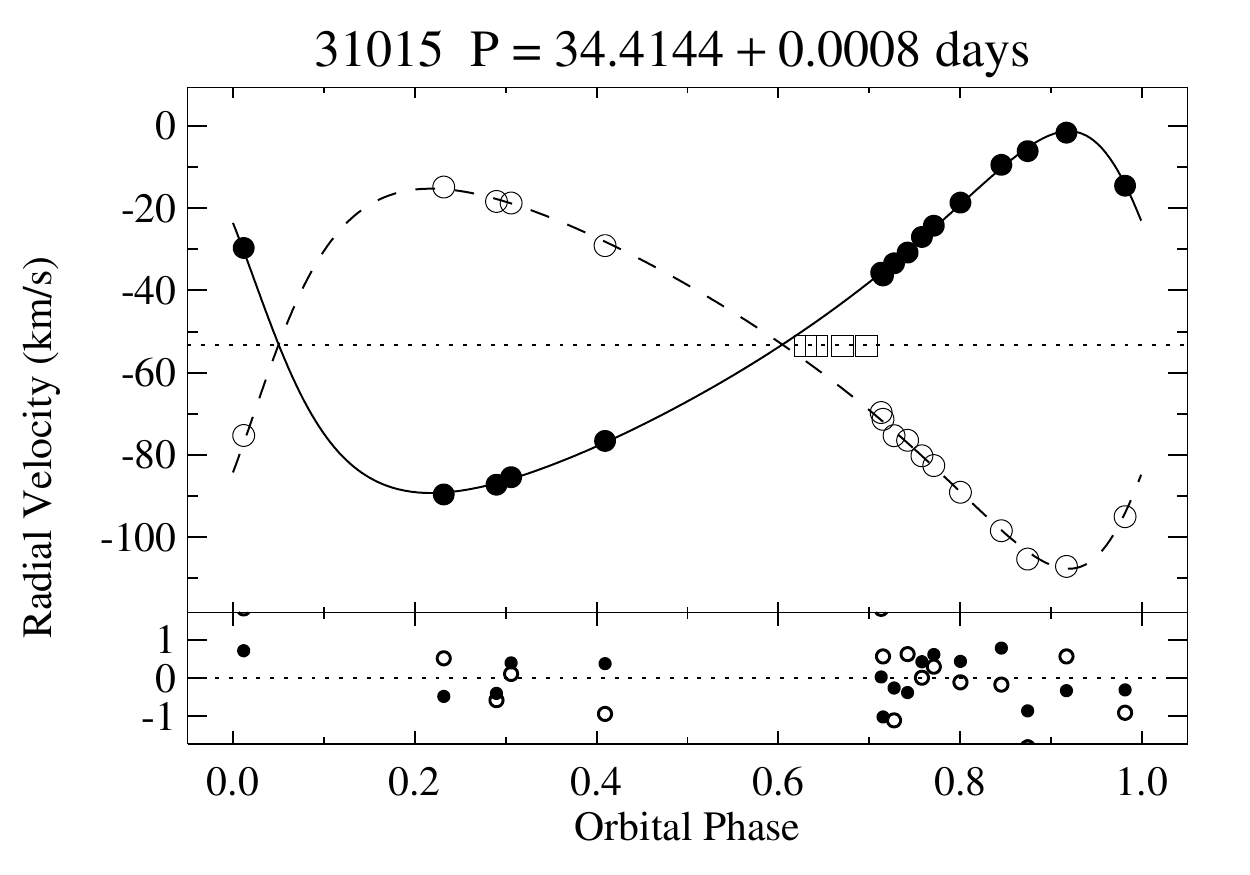}{0.3\linewidth}{}}
\gridline{\fig{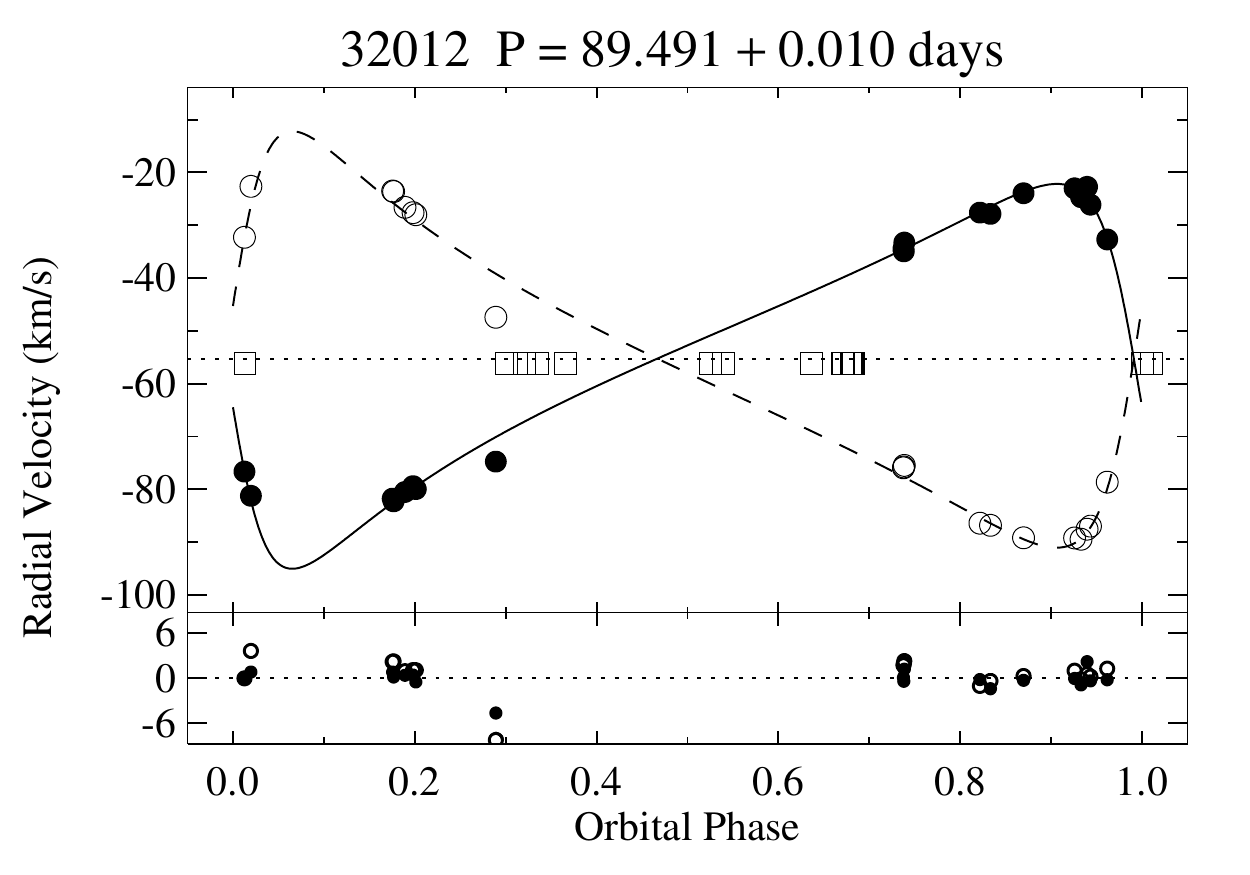}{0.3\linewidth}{}
  \fig{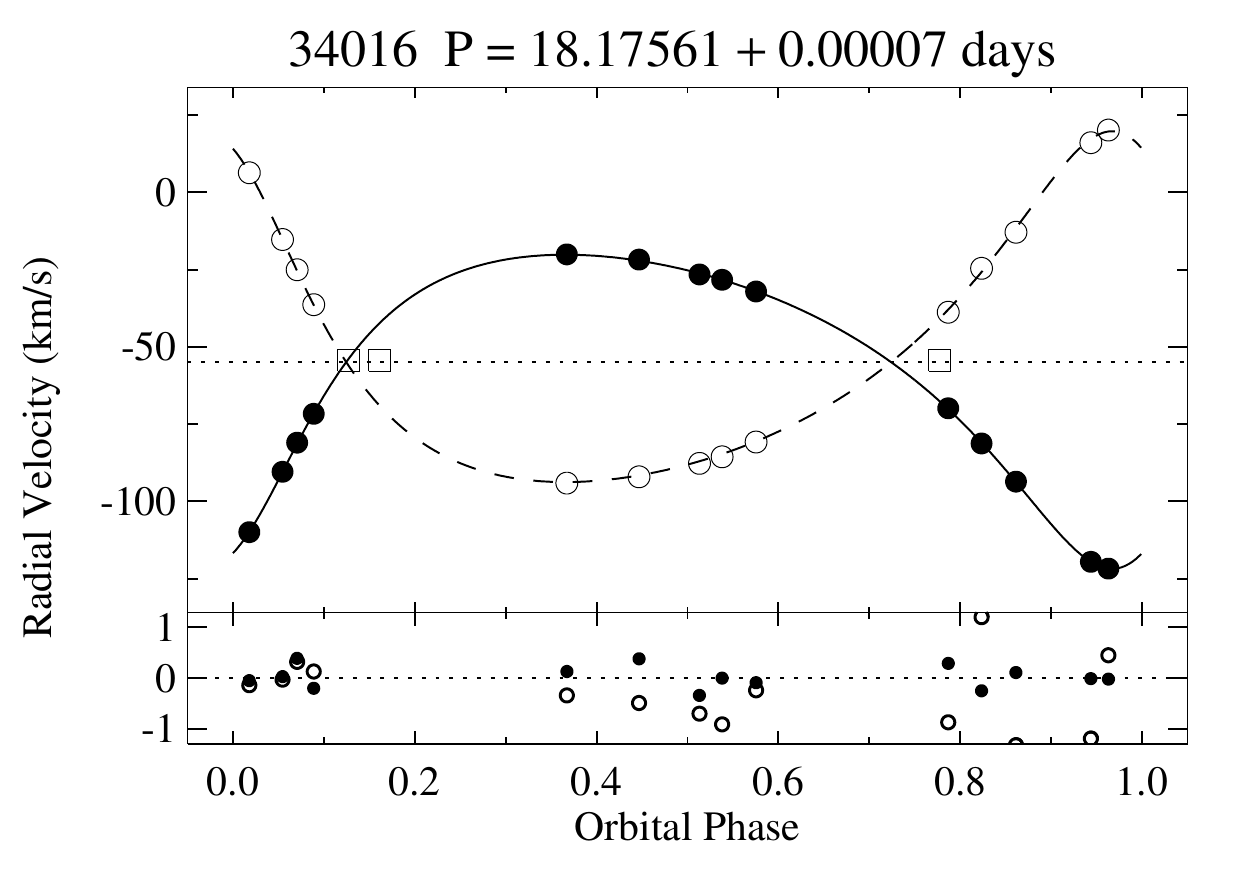}{0.3\linewidth}{}
  \fig{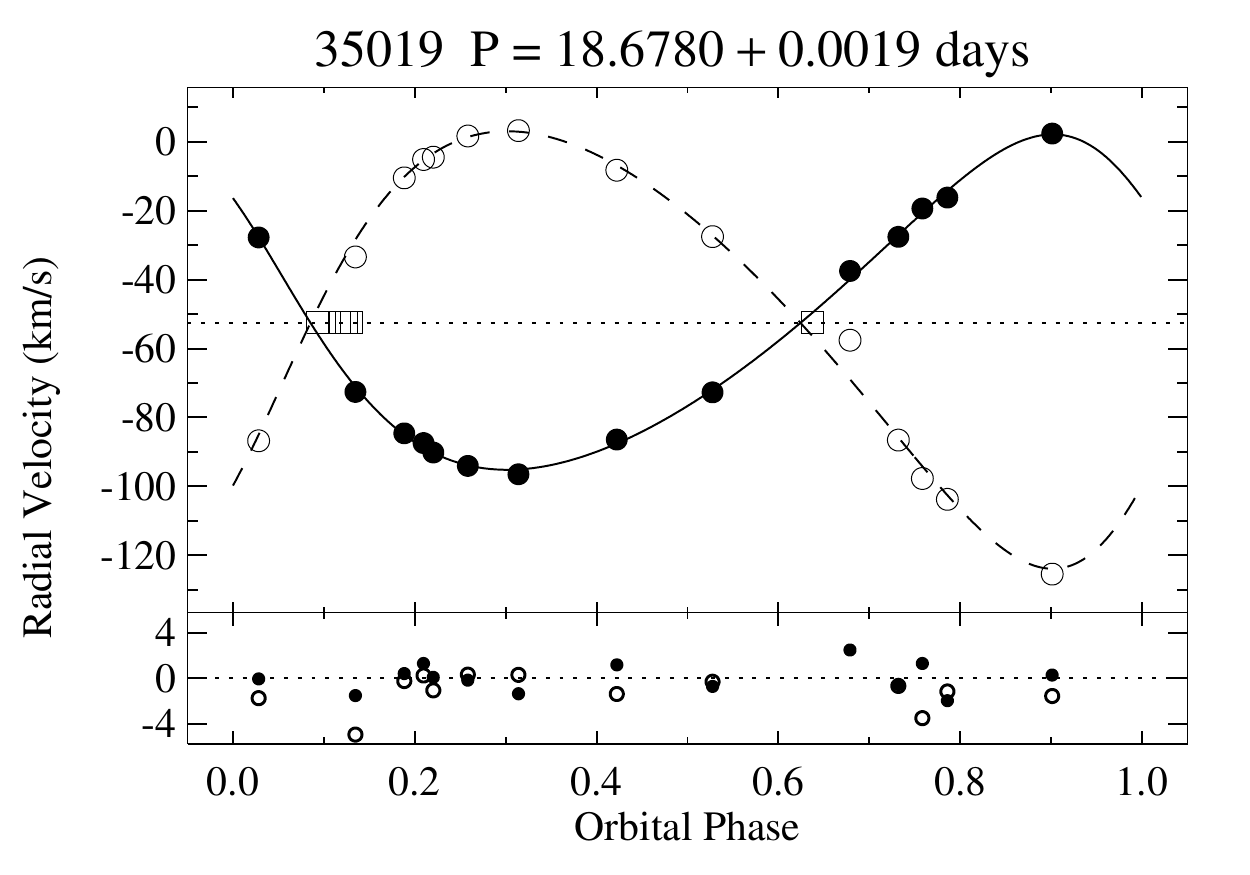}{0.3\linewidth}{}}
\gridline{\fig{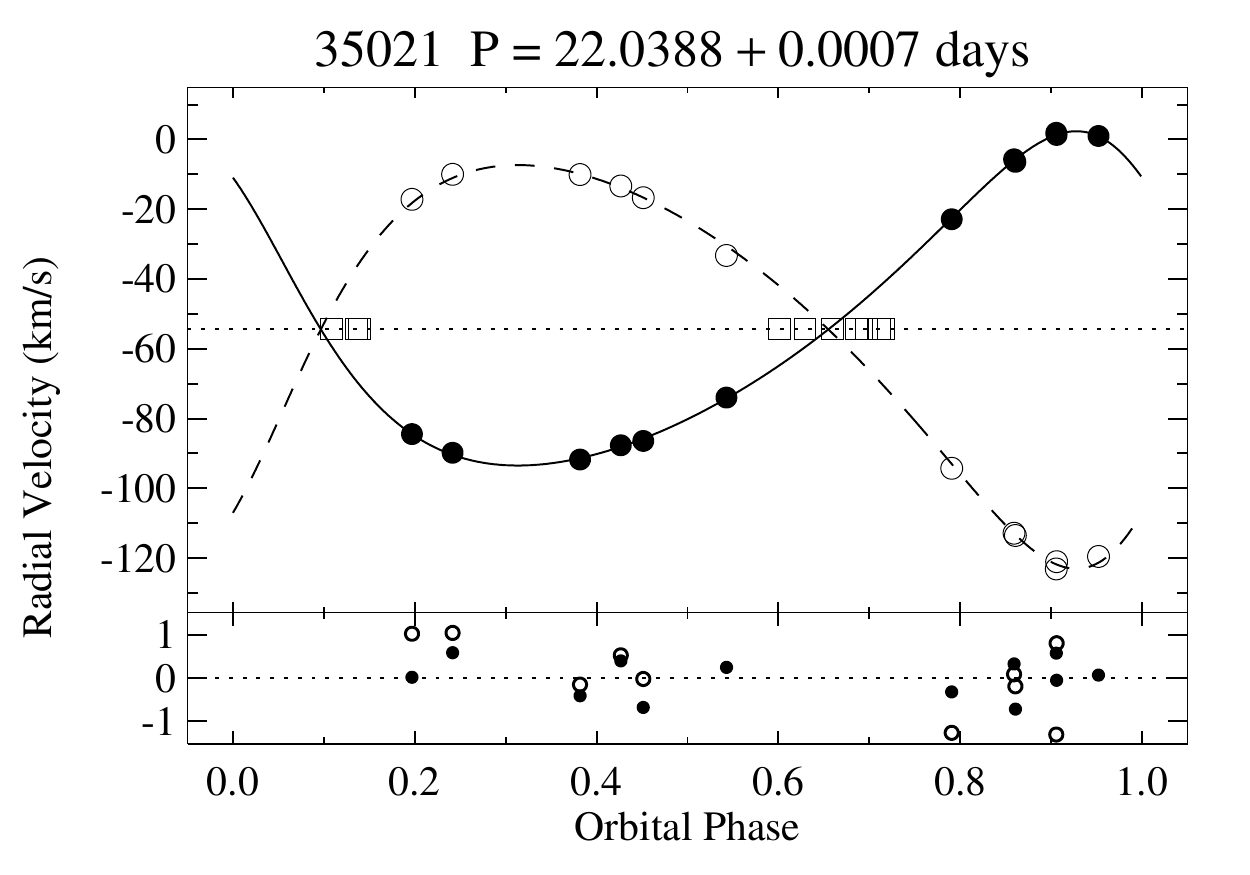}{0.3\linewidth}{}
  \fig{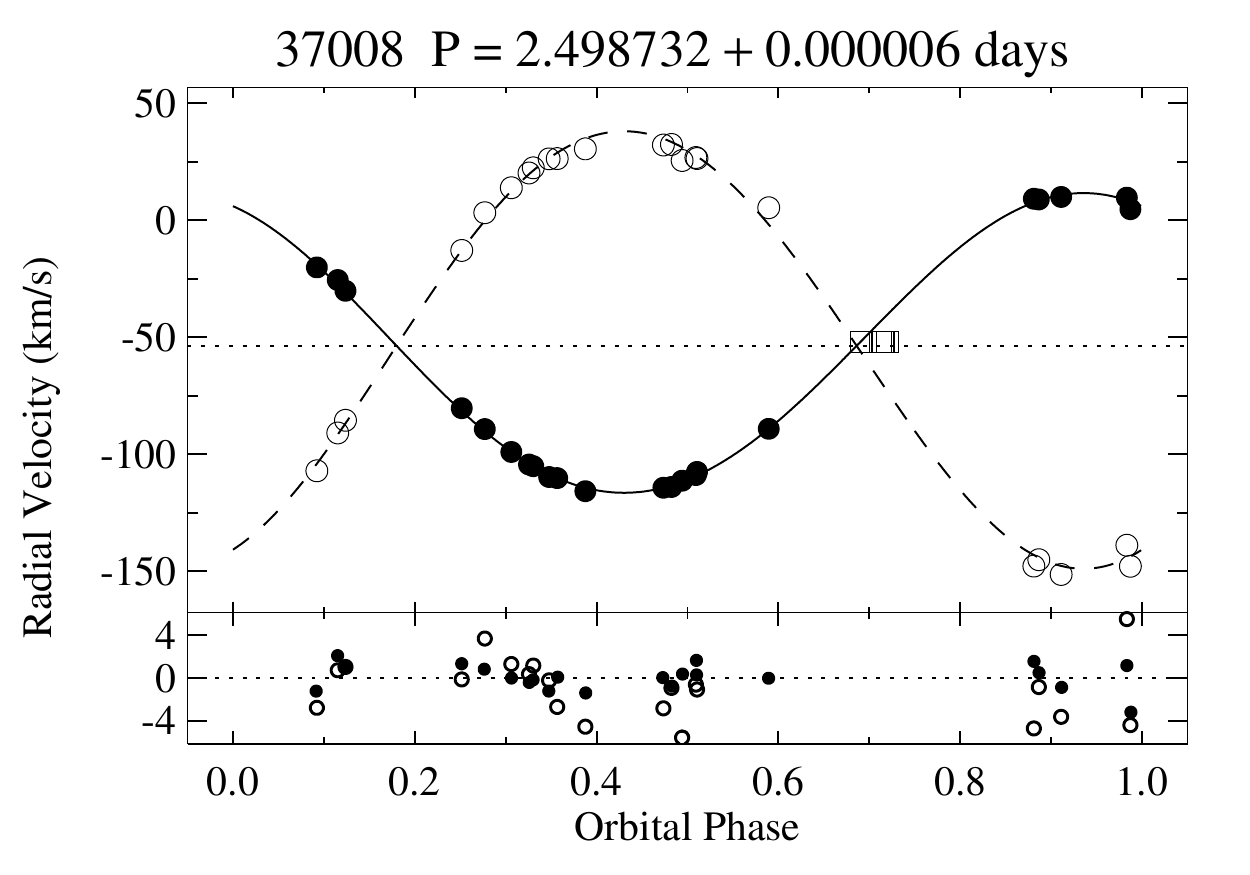}{0.3\linewidth}{}
  \fig{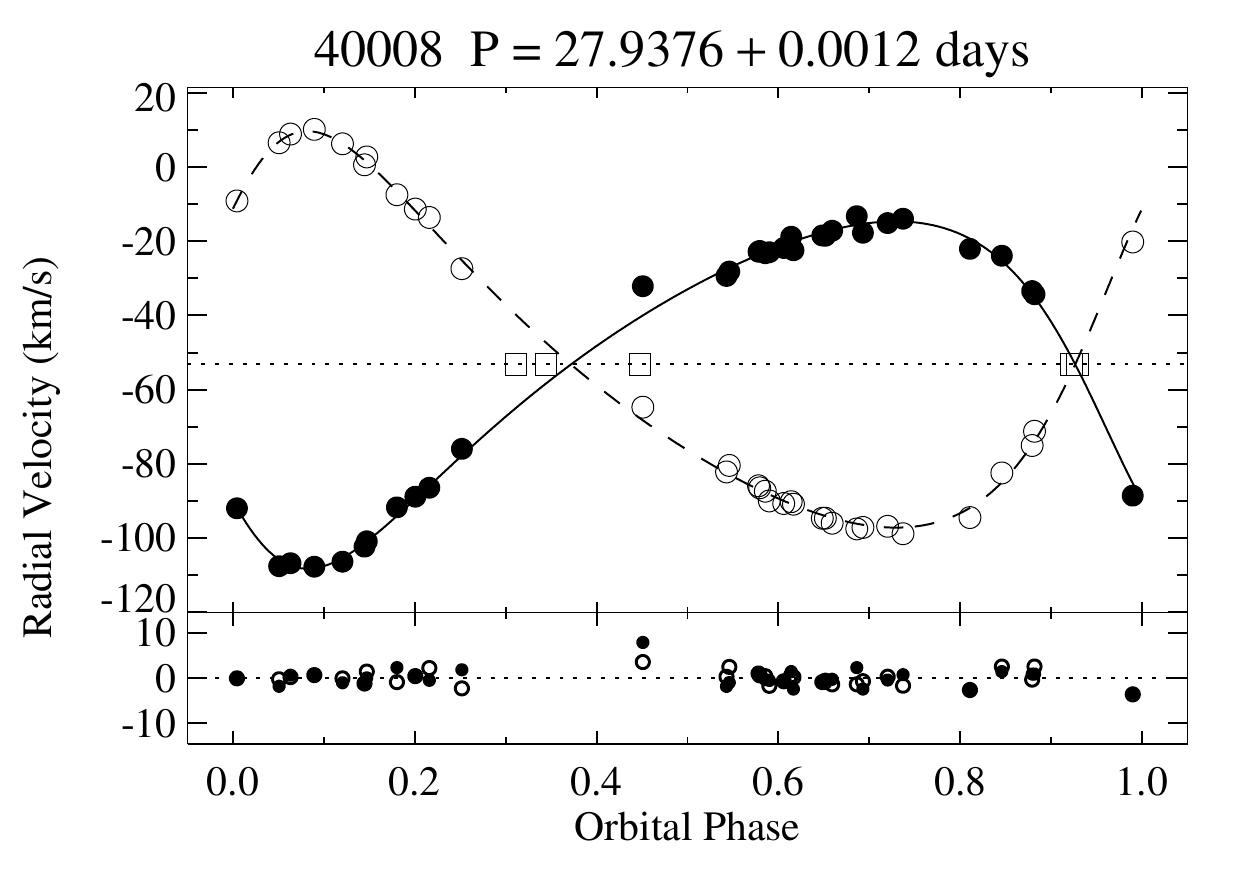}{0.3\linewidth}{}}
\caption{\footnotesize NGC 7789 SB2 orbit plots. We plot RV against orbital phase for each binary with the WOCS ID and orbital period above each plot. The data points and orbital fits for the primary star are represented by filled circles and a solid line. For the secondary star open circles and a dashed line represent the data points and the orbital fit, respectively. We mark the $\gamma$-velocity with a dotted line. We show the residuals from the fit below each binary plot, using the same symbols.}
\label{fig:Sb2plots}
\end{figure*}

We plot the SB2 orbital solutions in Figure~\ref{fig:Sb2plots}. The top panel shows the primary RVs (filled circles) and fitted solution (solid line), the secondary RVs (open circles) and fitted solution (dashed line), as well as the $\gamma$ velocity (dotted line). Square symbols ($\square$) represent data where the difference in velocities were below our spectral resolution of 20 \kms, and were therefore not included in our orbital solutions. The bottom panel shows the O$-$C residuals for the primary (filled circles) and secondary (open circles). Table~\ref{SB2tab} lists the orbital elements and the 1$\sigma$ errors for each SB2. In the first row is the binary ID, the orbital period ($P$), the number of orbital cycles observed, the center-of-mass RV ($\gamma$), the orbital amplitude of the primary ($K$), the eccentricity ($e$), the longitude of periastron ($\omega$), a Julian Date of periastron passage ($T_{\circ}$), the projected primary semi-major axis ($a\,$sin$\,i$), $m\,$sin$^{3}\,i$, the mass ratio ($q$), the rms residual velocity for the primary from the orbital solution ($\sigma$), and the number of RV measurements used in the orbital solutions after excluding measurements where the two velocities could not be resolved ($N$). The second row contains the respective errors on each of these values. The third and fourth rows are the amplitude ($K$), projected semi-major axis ($a\,$sin$\,i$), $m\,$sin$^{3}\,i$, the rms residual velocity from the orbital solution ($\sigma$), and the number of resolved RV measurements ($N$) for the secondary star and the respective errors.  

\clearpage
\startlongtable
\begin{longrotatetable}
\begin{deluxetable}{l r c r r r r r r r r c c}
\tabletypesize{\footnotesize}
\tablewidth{0pt}
\tablecaption{Orbital Parameters For NGC 7789 Double-Lined Binaries\label{SB2tab}}
\tablehead{\colhead{ID} & \colhead{$P$} & \colhead{Orbital} & \colhead{$\gamma$} & \colhead{$K$} & \colhead{$e$} & \colhead{$\omega$} & \colhead{$T_\circ$} & \colhead{a$\,\sin\,$i} & \colhead{m$\,\sin^3\,$i} & \colhead{q} & \colhead{$\sigma$} & \colhead{$N$} \\
\colhead{} & \colhead{(days)} & \colhead{Cycles} & \colhead{(\kms)} & \colhead{(\kms)} & \colhead{} & \colhead{(deg)} & \colhead{(HJD-2400000 d)} & \colhead{(10$^6$ km)} & \colhead{(\Msolar)} & \colhead{} & \colhead{(\kms)} & \colhead{}}
\startdata
7003 & 30.627 & 66.5 & -53.8 & 38.1 & 0.415 & 108 & 55064.24 & 14.59 & 0.626 & 0.919 & 1.68 & 21\\
         &     $\pm$0.005  &     \nodata &       $\pm$ 0.3 &       $\pm$0.6 &     $\pm$ 0.012 &       $\pm$ 2 &      $\pm$ 0.14 &       $\pm$ 0.24 &      $\pm$ 0.025 &     $\pm$ 0.020 &     \nodata &     \nodata \\
         &         \nodata &     \nodata &         \nodata &            41.4 &         \nodata &         \nodata &         \nodata &            15.9 &            0.575 &         \nodata &        1.91 &          21 \\
         &         \nodata &     \nodata &         \nodata &       $\pm$ 0.7&         \nodata &         \nodata &         \nodata &       $\pm$ 0.3 &      $\pm$ 0.022 &         \nodata &     \nodata &     \nodata\\
8003 & 11.7576 & 399.2 & -54.78 & 51.1 & 0.011 & 80 & 55045 & 8.26 & 0.677 & 0.981 & 1.43 & 26\\
         &     $\pm$0.0003  &     \nodata &       $\pm$ 0.24 &       $\pm$0.4 &     $\pm$ 0.006 &       $\pm$ 40 &      $\pm$ 1 &       $\pm$ 0.07 &      $\pm$ 0.016 &     $\pm$ 0.013 &     \nodata &     \nodata \\
         &         \nodata &     \nodata &         \nodata &            52.1 &         \nodata &         \nodata &         \nodata &            8.43 &            0.663 &         \nodata &        1.99 &          26 \\
         &         \nodata &     \nodata &         \nodata &       $\pm$ 0.5&         \nodata &         \nodata &         \nodata &       $\pm$ 0.09 &      $\pm$ 0.013 &         \nodata &     \nodata &     \nodata\\
8016 & 30.9783 & 151.6 & -52.4 & 52.3 & 0.393 & 308.7 & 55313.130 & 20.52 & 1.73 & 0.913 & 0.34 & 15\\
         &     $\pm$0.0005  &     \nodata &       $\pm$ 0.1 &       $\pm$0.1 &     $\pm$ 0.002 &       $\pm$ 0.5 &      $\pm$ 0.025 &       $\pm$ 0.07 &      $\pm$ 0.06 &     $\pm$ 0.015 &     \nodata &     \nodata \\
         &         \nodata &     \nodata &         \nodata &            57.4 &         \nodata &         \nodata &         \nodata &            22.5 &            1.58 &         \nodata &        2.55 &          15 \\
         &         \nodata &     \nodata &         \nodata &       $\pm$ 0.8 &         \nodata &         \nodata &         \nodata &       $\pm$ 0.4 &      $\pm$ 0.03 &         \nodata &     \nodata &     \nodata\\
9020 & 5.50869 & 852.8 & -54.59 & 79.1 & 0.137 & 213 & 54878.02 & 5.94 & 1.26 & 0.936 & 1.23 & 29\\
         &     $\pm$0.00005  &     \nodata &       $\pm$ 0.24 &       $\pm$0.4 &     $\pm$ 0.004 &       $\pm$ 2 &      $\pm$ 0.03 &       $\pm$ 0.03 &      $\pm$ 0.04 &     $\pm$ 0.014 &     \nodata &     \nodata \\
         &         \nodata &     \nodata &         \nodata &            84 &         \nodata &         \nodata &         \nodata &            6.34 &            1.175 &         \nodata &        3.96 &          29 \\
         &         \nodata &     \nodata &         \nodata &       $\pm$ 1&         \nodata &         \nodata &         \nodata &       $\pm$ 0.09 &      $\pm$ 0.022 &         \nodata &     \nodata &     \nodata\\
14007 & 23.167 & 217.2 & -53.0 & 38 & 0.219 & 227 & 54571.61 & 11.9 & 0.61 & 0.905 & 1.25 & 17\\
         &     $\pm$0.001  &     \nodata &       $\pm$ 0.3 &       $\pm$1 &     $\pm$ 0.022 &       $\pm$ 4 &      $\pm$ 0.23 &       $\pm$ 0.3 &      $\pm$ 0.05 &     $\pm$ 0.023 &     \nodata &     \nodata \\
         &         \nodata &     \nodata &         \nodata &            42 &         \nodata &         \nodata &         \nodata &            13.1 &            0.55 &         \nodata &        2.33 &          17 \\
         &         \nodata &     \nodata &         \nodata &       $\pm$ 1&         \nodata &         \nodata &         \nodata &       $\pm$ 0.4 &      $\pm$ 0.04 &         \nodata &     \nodata &     \nodata\\
14014 & 2.36601 & 1748.0 & -51.6 & 107.5 & 0.020 & 94 & 55371.60 & 3.50 & 1.58 & 0.880 & 2.87 & 19\\
         &     $\pm$0.00001  &     \nodata &       $\pm$ 0.7 &       $\pm$0.9 &     $\pm$ 0.009 &       $\pm$ 26 &      $\pm$ 0.17 &       $\pm$ 0.03 &      $\pm$ 0.07 &     $\pm$ 0.019 &     \nodata &     \nodata \\
         &         \nodata &     \nodata &         \nodata &            122.2 &         \nodata &         \nodata &         \nodata &            3.97 &            1.39 &         \nodata &        7.68 &          19 \\
         &         \nodata &     \nodata &         \nodata &       $\pm$ 2.2 &         \nodata &         \nodata &         \nodata &       $\pm$ 0.08 &      $\pm$ 0.04 &         \nodata &     \nodata &     \nodata\\
14022 & 11.5494 & 180.1 & -53.5 & 33.9 & 0.015 & 40 & 54550 & 5.38 & 0.201 & 0.960 & 1.13 & 22\\
         &     $\pm$0.0003  &     \nodata &       $\pm$ 0.2 &       $\pm$0.3 &     $\pm$ 0.009 &       $\pm$ 30 &      $\pm$ 1 &       $\pm$ 0.06 &      $\pm$ 0.005 &     $\pm$ 0.014 &     \nodata &     \nodata \\
         &         \nodata &     \nodata &         \nodata &            35.2 &         \nodata &         \nodata &         \nodata &            5.60 &            0.193 &         \nodata &        1.23 &          22 \\
         &         \nodata &     \nodata &         \nodata &       $\pm$ 0.4 &         \nodata &         \nodata &         \nodata &       $\pm$ 0.06 &      $\pm$ 0.005 &         \nodata &     \nodata &     \nodata\\
17008 & 2.76075 & 1701.8 & -53.8 & 78.6 & 0.007 & 40 & 55011.4 & 2.984 & 1.070 & 0.730 & 1.85 & 23\\
         &     $\pm$0.00001  &     \nodata &       $\pm$ 0.4 &       $\pm$0.5 &     $\pm$ 0.006 &       $\pm$ 50 &      $\pm$ 0.4 &       $\pm$ 0.021 &      $\pm$ 0.021 &     $\pm$ 0.008 &     \nodata &     \nodata \\
         &         \nodata &     \nodata &         \nodata &            107.7 &         \nodata &         \nodata &         \nodata &            4.09 &            0.781 &         \nodata &        3.13 &          23 \\
         &         \nodata &     \nodata &         \nodata &       $\pm$ 0.8&         \nodata &         \nodata &         \nodata &       $\pm$ 0.04 &      $\pm$ 0.013 &         \nodata &     \nodata &     \nodata\\
17023 & 191.64 & 8.6 & -52.6 & 27.1 & 0.401 & 176 & 57035 & 65.5 & 1.18 & 1.02 & 1.73 & 14\\
         &     $\pm$0.23  &     \nodata &       $\pm$ 0.3 &       $\pm$0.5 &     $\pm$ 0.012 &       $\pm$ 6 &      $\pm$ 2 &       $\pm$ 1.7 &      $\pm$ 0.06 &     $\pm$ 0.03 &     \nodata &     \nodata \\
         &         \nodata &     \nodata &         \nodata &            26.7 &         \nodata &         \nodata &         \nodata &            64.4 &            1.20 &         \nodata &        1.18 &          14 \\
         &         \nodata &     \nodata &         \nodata &       $\pm$ 0.4&         \nodata &         \nodata &         \nodata &       $\pm$ 1.2 &      $\pm$ 0.07 &         \nodata &     \nodata &     \nodata\\
17028 & 3.553964 & 1415.9 & -56.5 & 86.9 & 0.119 & 166 & 54711.71 & 4.22 & 1.89 & 0.717 & 3.86 & 24\\
         &     $\pm$0.000014  &     \nodata &       $\pm$ 0.6 &       $\pm$1.5 &     $\pm$ 0.014 &       $\pm$ 5 &      $\pm$ 0.04 &       $\pm$ 0.07 &      $\pm$ 0.08 &     $\pm$ 0.014 &     \nodata &     \nodata \\
         &         \nodata &     \nodata &         \nodata &            121.2 &         \nodata &         \nodata &         \nodata &            5.88 &            1.39 &         \nodata &        5.28 &          24 \\
         &         \nodata &     \nodata &         \nodata &       $\pm$ 2.0 &         \nodata &         \nodata &         \nodata &       $\pm$ 0.10 &      $\pm$ 0.06 &         \nodata &     \nodata &     \nodata\\
21025 & 3.555133 & 1227.0 & -52.8 & 74.9 & 0.106 & 37 & 56430.56 & 3.64 & 0.880 & 0.84 & 2.53 & 35\\
         &     $\pm$0.000011  &     \nodata &       $\pm$ 0.3 &       $\pm$0.6 &     $\pm$ 0.006 &       $\pm$ 3 &      $\pm$ 0.03 &       $\pm$ 0.03 &      $\pm$ 0.017 &     $\pm$ 0.01 &     \nodata &     \nodata \\
         &         \nodata &     \nodata &         \nodata &            89.7 &         \nodata &         \nodata &         \nodata &            4.36 &            0.735 &         \nodata &        3.18 &          35 \\
         &         \nodata &     \nodata &         \nodata &       $\pm$ 0.7&         \nodata &         \nodata &         \nodata &       $\pm$ 0.04 &      $\pm$ 0.013 &         \nodata &     \nodata &     \nodata\\
23012 & 8.06720 & 354.4 & -53.39 & 63.89 & 0.009 & 330 & 56378.9 & 7.09 & 1.367 & 0.803 & 1.50 & 12\\
         &     $\pm$0.00013  &     \nodata &       $\pm$ 0.21 &       $\pm$0.52 &     $\pm$ 0.004 &       $\pm$ 20 &      $\pm$ 0.4 &       $\pm$ 0.07 &      $\pm$ 0.016 &     $\pm$ 0.008 &     \nodata &     \nodata \\
         &         \nodata &     \nodata &         \nodata &            79.53 &         \nodata &         \nodata &         \nodata &            8.82 &            1.099 &         \nodata &        0.64 &          12 \\
         &         \nodata &     \nodata &         \nodata &       $\pm$ 0.24&         \nodata &         \nodata &         \nodata &       $\pm$ 0.03 &      $\pm$ 0.021 &         \nodata &     \nodata &     \nodata\\
29010 & 32.886 & 56.8 & -51.9 & 37.6 & 0.522 & 91 & 56904.30 & 14.50 & 0.57 & 0.888 & 1.98 & 16\\
         &     $\pm$0.004  &     \nodata &       $\pm$ 0.4 &       $\pm$0.9 &     $\pm$ 0.012 &       $\pm$ 2 &      $\pm$ 0.14 &       $\pm$ 0.34 &      $\pm$ 0.03 &     $\pm$ 0.025 &     \nodata &     \nodata \\
         &         \nodata &     \nodata &         \nodata &            42.34 &         \nodata &         \nodata &         \nodata &            16.3 &            0.51 &         \nodata &        2.04 &          16 \\
         &         \nodata &     \nodata &         \nodata &       $\pm$ 0.98&         \nodata &         \nodata &         \nodata &       $\pm$ 0.4 &      $\pm$ 0.03 &         \nodata &     \nodata &     \nodata\\
30007 & 48.98 & 53.6 & -51.2 & 33.6 & 0.23 & 173 & 57123 & 22.0 & 0.952 & 0.87 & 4.91 & 29\\
         &     $\pm$0.05  &     \nodata &       $\pm$ 0.7 &       $\pm$1.6 &     $\pm$ 0.04 &       $\pm$ 9 &      $\pm$ 1 &       $\pm$ 1.0 &      $\pm$ 0.114 &     $\pm$ 0.05 &     \nodata &     \nodata \\
         &         \nodata &     \nodata &         \nodata &            38.8 &         \nodata &         \nodata &         \nodata &            25.4 &            0.83 &         \nodata &        6.07 &          29 \\
         &         \nodata &     \nodata &         \nodata &       $\pm$ 1.9&         \nodata &         \nodata &         \nodata &       $\pm$ 1.2 &      $\pm$ 0.09 &         \nodata &     \nodata &     \nodata\\
31015 & 34.4144 & 95.3 & -53.16 & 44.0 & 0.364 & 60 & 55558.35 & 19.38 & 1.081 & 0.952 & 0.62 & 16\\
         &     $\pm$0.0008  &     \nodata &       $\pm$ 0.16 &       $\pm$0.4 &     $\pm$ 0.009 &       $\pm$ 1 &      $\pm$ 0.07 &       $\pm$ 0.13 &      $\pm$ 0.023 &     $\pm$ 0.011 &     \nodata &     \nodata \\
         &         \nodata &     \nodata &         \nodata &            46.2 &         \nodata &         \nodata &         \nodata &            20.36 &            1.029 &         \nodata &        1.09 &          16 \\
         &         \nodata &     \nodata &         \nodata &       $\pm$ 0.5&         \nodata &         \nodata &         \nodata &       $\pm$ 0.20 &      $\pm$ 0.018 &         \nodata &     \nodata &     \nodata\\
32012 & 89.49 & 30.3 & -55.3 & 36.4 & 0.573 & 99 & 57120.5 & 36.8 & 1.16 & 0.92 & 1.46 & 19\\
         &     $\pm$0.01  &     \nodata &       $\pm$ 0.3 &       $\pm$0.8 &     $\pm$ 0.010 &       $\pm$ 3 &      $\pm$ 0.4 &       $\pm$ 0.7 &      $\pm$ 0.08 &     $\pm$ 0.03 &     \nodata &     \nodata \\
         &         \nodata &     \nodata &         \nodata &            39.4 &         \nodata &         \nodata &         \nodata &            39.8 &            1.07 &         \nodata &        2.66 &          19 \\
         &         \nodata &     \nodata &         \nodata &       $\pm$ 1.1 &         \nodata &         \nodata &         \nodata &       $\pm$ 1.1 &      $\pm$ 0.06 &         \nodata &     \nodata &     \nodata\\
34016 & 18.17561 & 276.7 & -54.92 & 50.77 & 0.351 & 205.8 & 55639.781 & 11.88 & 1.013 & 0.895 & 0.25 & 14\\
         &     $\pm$0.00007  &     \nodata &       $\pm$ 0.08 &       $\pm$0.11 &     $\pm$ 0.002 &       $\pm$ 0.5 &      $\pm$ 0.018 &       $\pm$ 0.03 &      $\pm$ 0.013 &     $\pm$ 0.006 &     \nodata &     \nodata \\
         &         \nodata &     \nodata &         \nodata &            56.72 &         \nodata &         \nodata &         \nodata &            13.27 &            0.907 &         \nodata &        0.84 &          14 \\
         &         \nodata &     \nodata &         \nodata &       $\pm$ 0.30 &         \nodata &         \nodata &         \nodata &       $\pm$ 0.08 &      $\pm$ 0.007 &         \nodata &     \nodata &     \nodata\\
35019 & 18.6780 & 125.8 & -52.6 & 48.6 & 0.200 & 52 & 56098.16 & 12.24 & 1.5 & 0.767 & 1.39 & 14\\
         &     $\pm$0.0019  &     \nodata &       $\pm$ 0.4 &       $\pm$0.6 &     $\pm$ 0.011 &       $\pm$ 3 &      $\pm$ 0.16 &       $\pm$ 0.17 &      $\pm$ 0.1 &     $\pm$ 0.024 &     \nodata &     \nodata \\
         &         \nodata &     \nodata &         \nodata &            63.5 &         \nodata &         \nodata &         \nodata &            16.0 &            1.11 &         \nodata &        4.12 &          14 \\
         &         \nodata &     \nodata &         \nodata &       $\pm$ 1.6 &         \nodata &         \nodata &         \nodata &       $\pm$ 0.5 &      $\pm$ 0.05 &         \nodata &     \nodata &     \nodata\\
35021 & 22.03877 & 96.5 & -54.4 & 47.90 & 0.257 & 44 & 56601.18 & 14.03 & 1.33 & 0.828 & 0.52 & 12\\
         &     $\pm$0.0007  &     \nodata &       $\pm$ 0.2 &       $\pm$0.21 &     $\pm$ 0.005 &       $\pm$ 1 &      $\pm$ 0.08 &       $\pm$ 0.07 &      $\pm$ 0.03 &     $\pm$ 0.009 &     \nodata &     \nodata \\
         &         \nodata &     \nodata &         \nodata &            57.8 &         \nodata &         \nodata &         \nodata &            16.94 &            1.104 &         \nodata &        1.41 &          12 \\
         &         \nodata &     \nodata &         \nodata &       $\pm$ 0.5 &         \nodata &         \nodata &         \nodata &       $\pm$ 0.18 &      $\pm$ 0.018 &         \nodata &     \nodata &     \nodata\\
37008 & 2.498732 & 1751.3 & -53.64 & 64.00 & 0.021 & 24 & 54975.3 & 2.199 & 0.599 & 0.685 & 1.30 & 22\\
         &     $\pm$0.000006  &     \nodata &       $\pm$ 0.32 &       $\pm$0.36 &     $\pm$ 0.008 &       $\pm$ 15 &      $\pm$ 0.1 &       $\pm$ 0.014 &      $\pm$ 0.015 &     $\pm$ 0.009 &     \nodata &     \nodata \\
         &         \nodata &     \nodata &         \nodata &            93.43 &         \nodata &         \nodata &         \nodata &            3.21 &            0.410 &         \nodata &        3.53 &          22 \\
         &         \nodata &     \nodata &         \nodata &       $\pm$ 0.95 &         \nodata &         \nodata &         \nodata &       $\pm$ 0.04 &      $\pm$ 0.007 &         \nodata &     \nodata &     \nodata\\
40008 & 27.9376 & 179.9 & -53.2 & 46.8 & 0.290 & 128 & 55230.117 & 17.22 & 1.364 & 0.876 & 2.08 & 33\\
         &     $\pm$0.0012  &     \nodata &       $\pm$ 0.2 &       $\pm$0.5 &     $\pm$ 0.006 &       $\pm$ 2 &      $\pm$ 0.112 &       $\pm$ 0.18 &      $\pm$ 0.025 &     $\pm$ 0.011 &     \nodata &     \nodata \\
         &         \nodata &     \nodata &         \nodata &            53.4 &         \nodata &         \nodata &         \nodata &            19.65 &            1.20 &         \nodata &        1.65 &          33 \\
         &         \nodata &     \nodata &         \nodata &       $\pm$ 0.4 &         \nodata &         \nodata &         \nodata &       $\pm$ 0.15 &      $\pm$ 0.03 &         \nodata &     \nodata &     \nodata\\
\enddata
\tablecomments{Table \ref{SB2tab} is also published in machine-readable format.}
\end{deluxetable}
\end{longrotatetable}

\clearpage
\begin{figure*}
\gridline{\fig{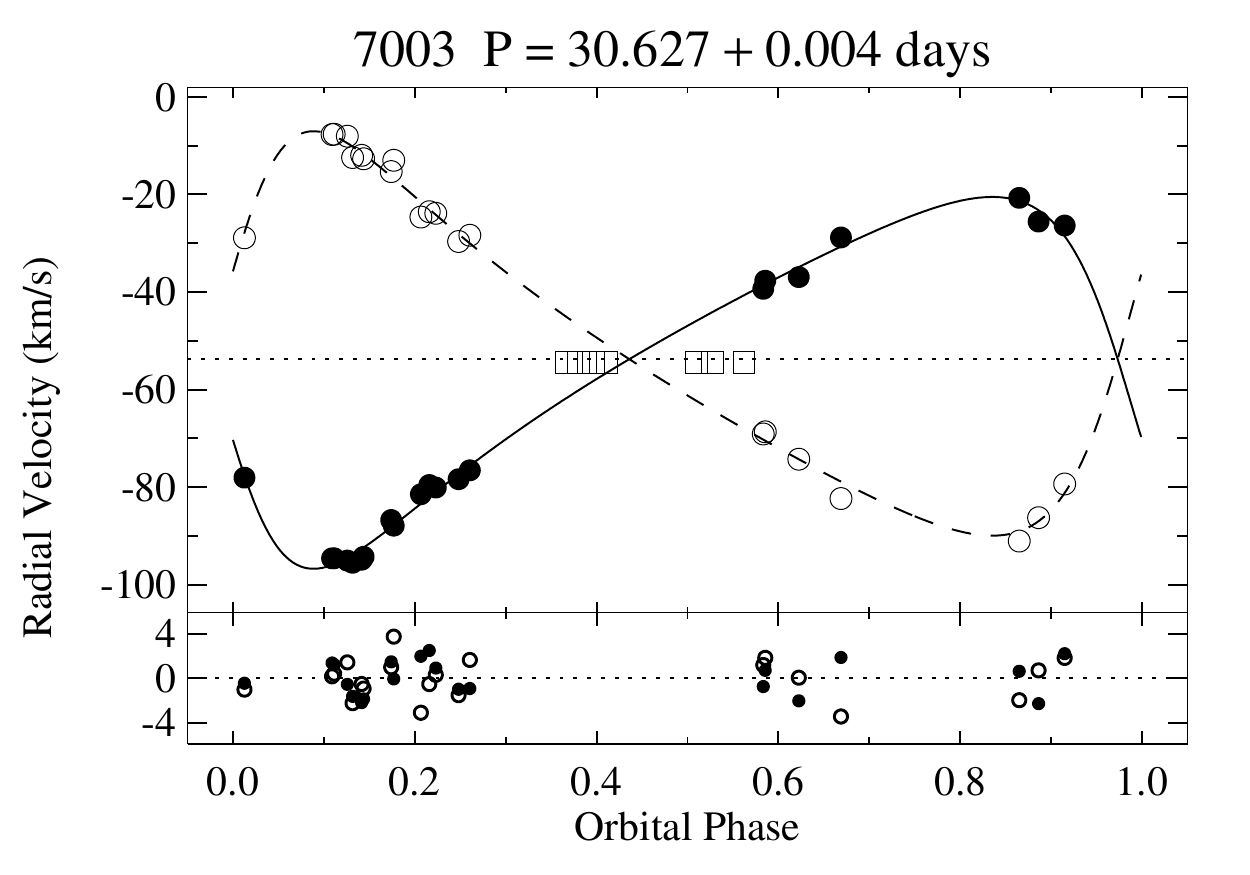}{0.3\linewidth}{}
  \fig{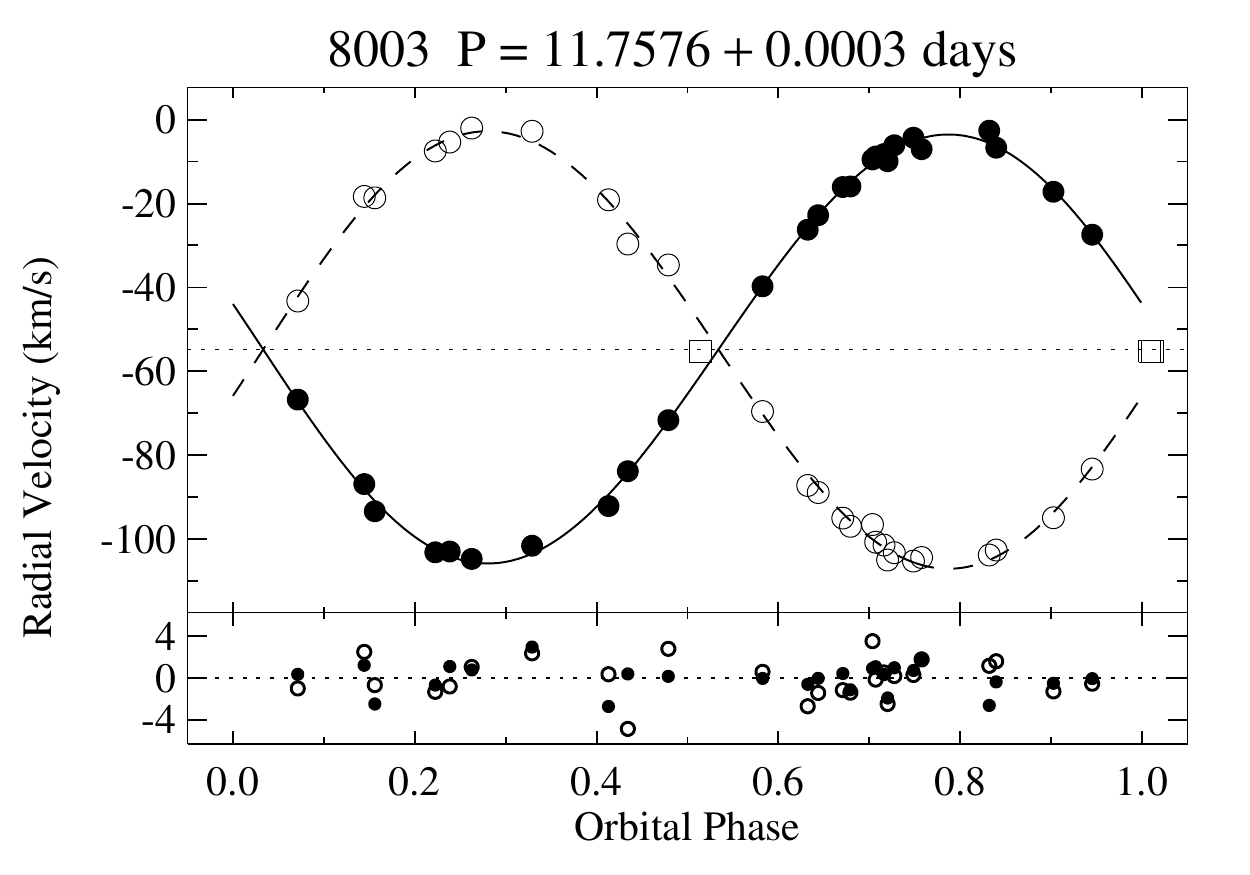}{0.3\linewidth}{}
  \fig{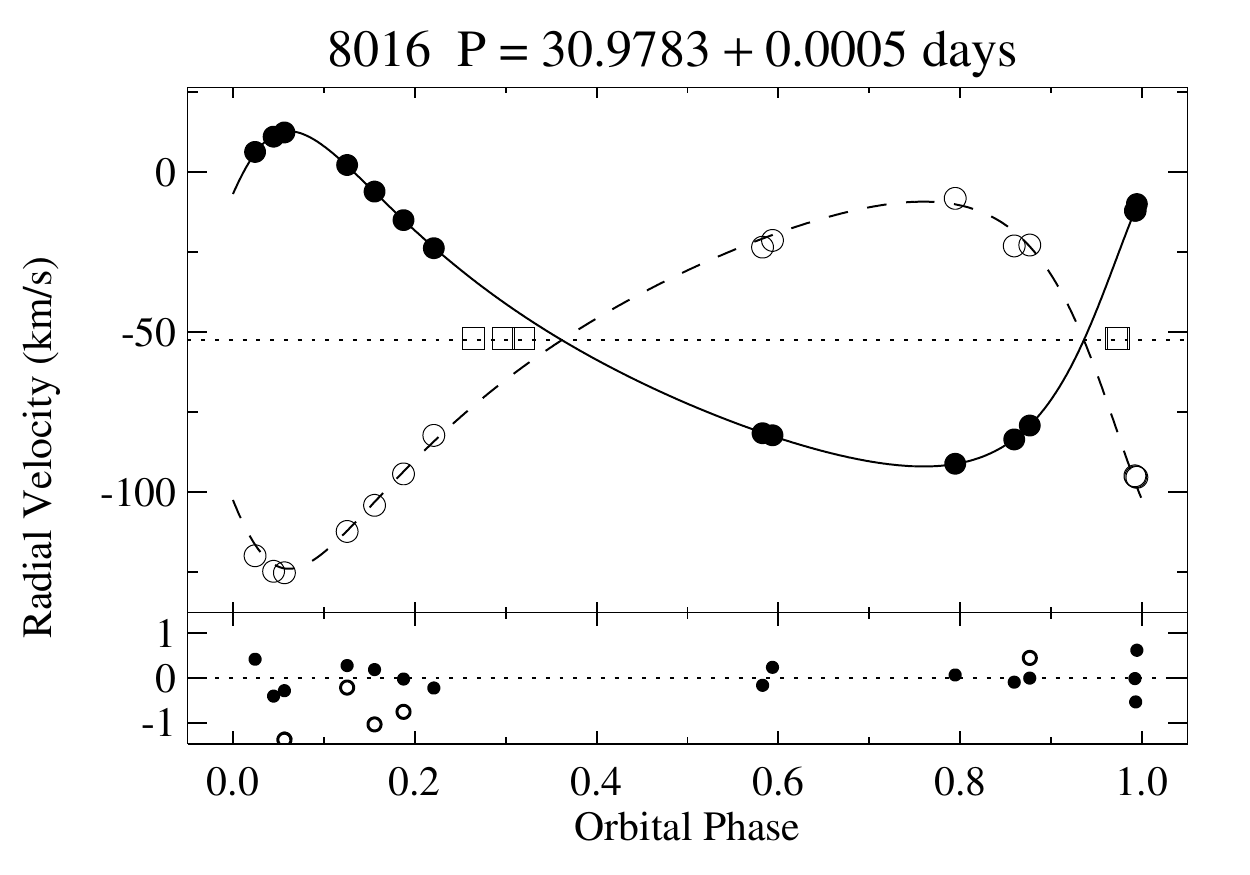}{0.3\linewidth}{}}
\gridline{\fig{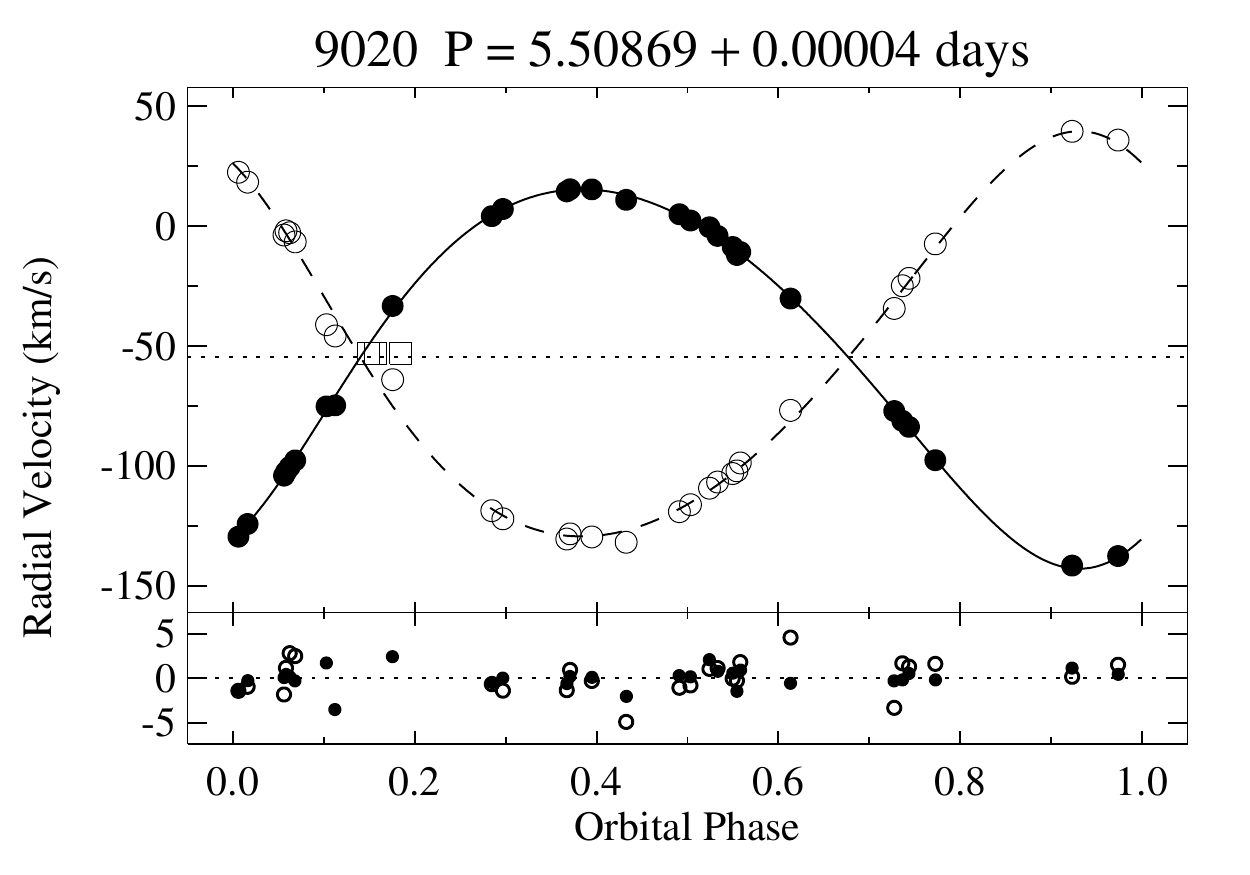}{0.3\linewidth}{}
  \fig{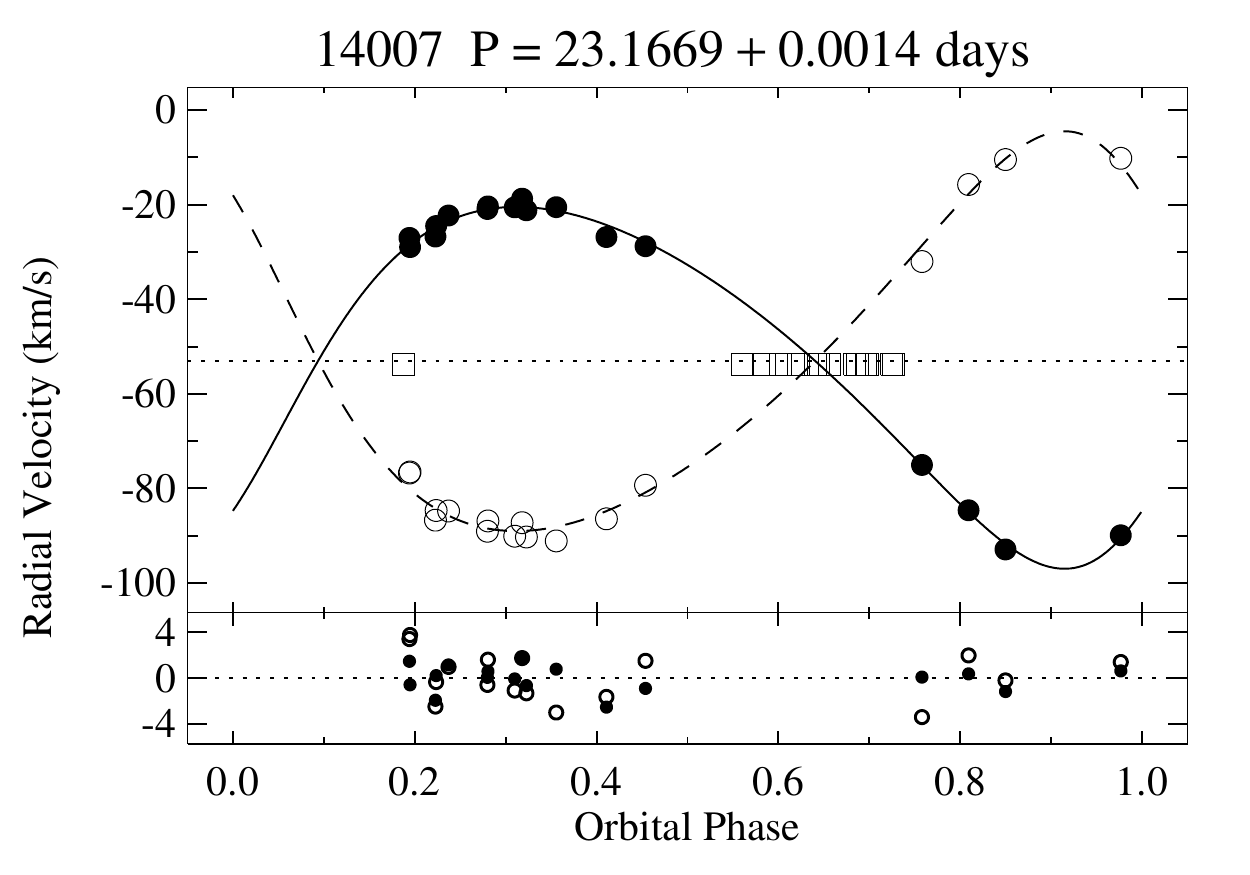}{0.3\linewidth}{}
  \fig{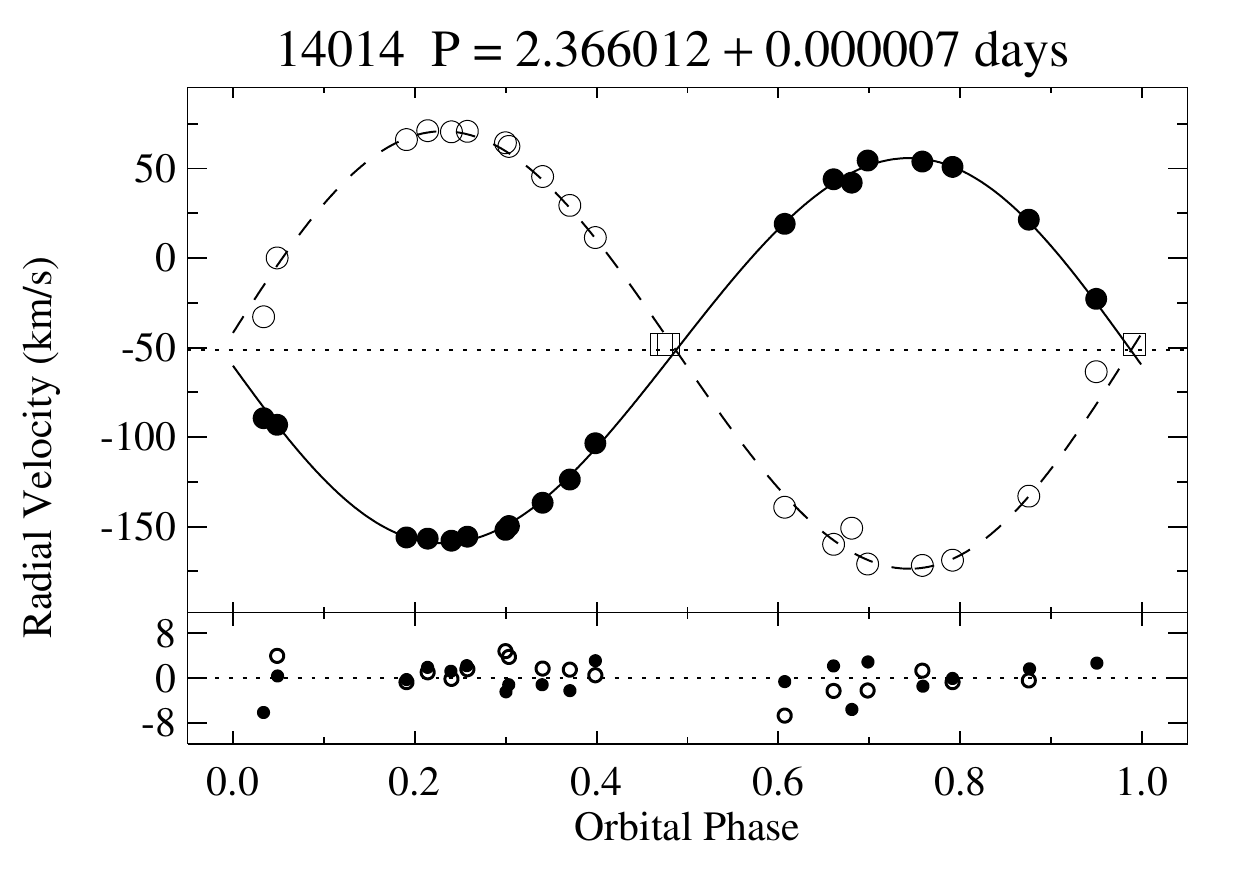}{0.3\linewidth}{}}
\gridline{\fig{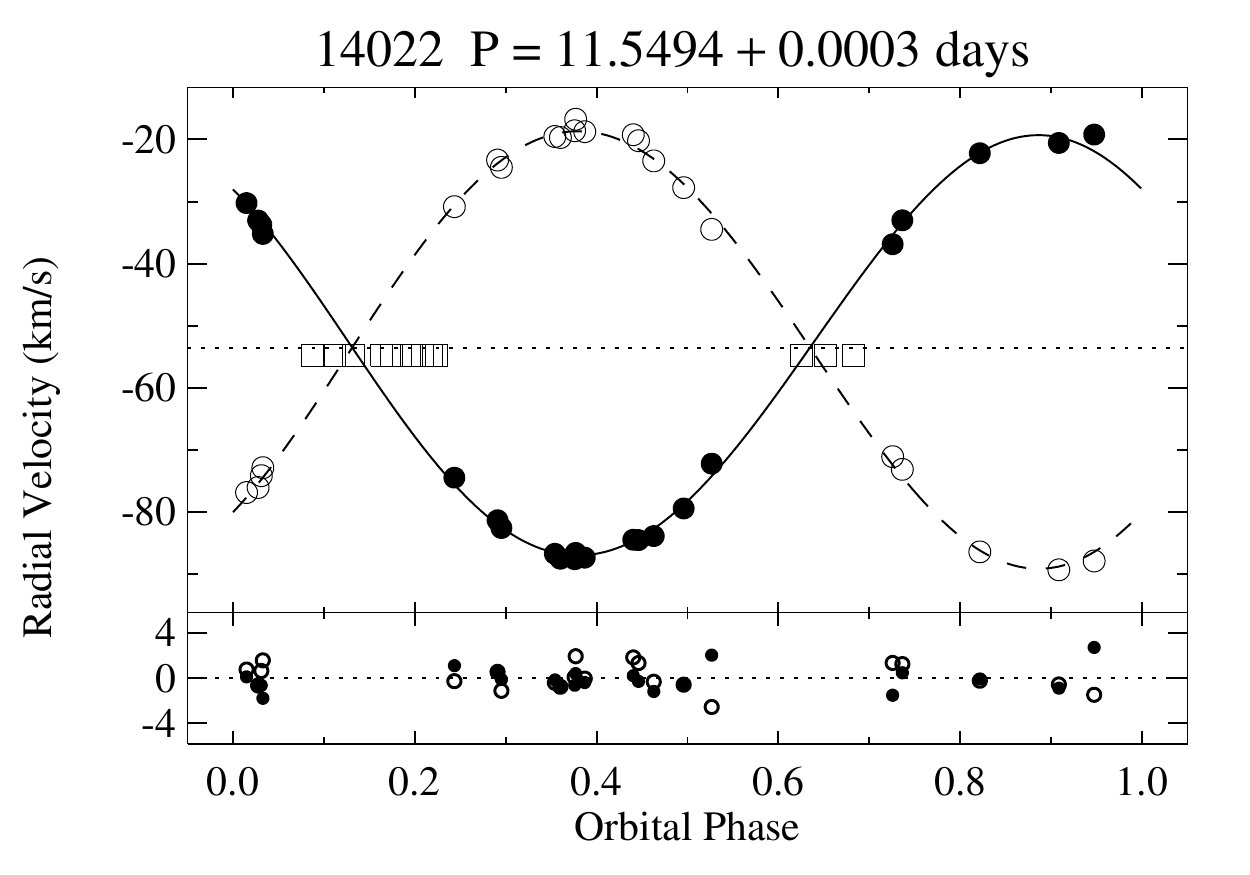}{0.3\linewidth}{}
  \fig{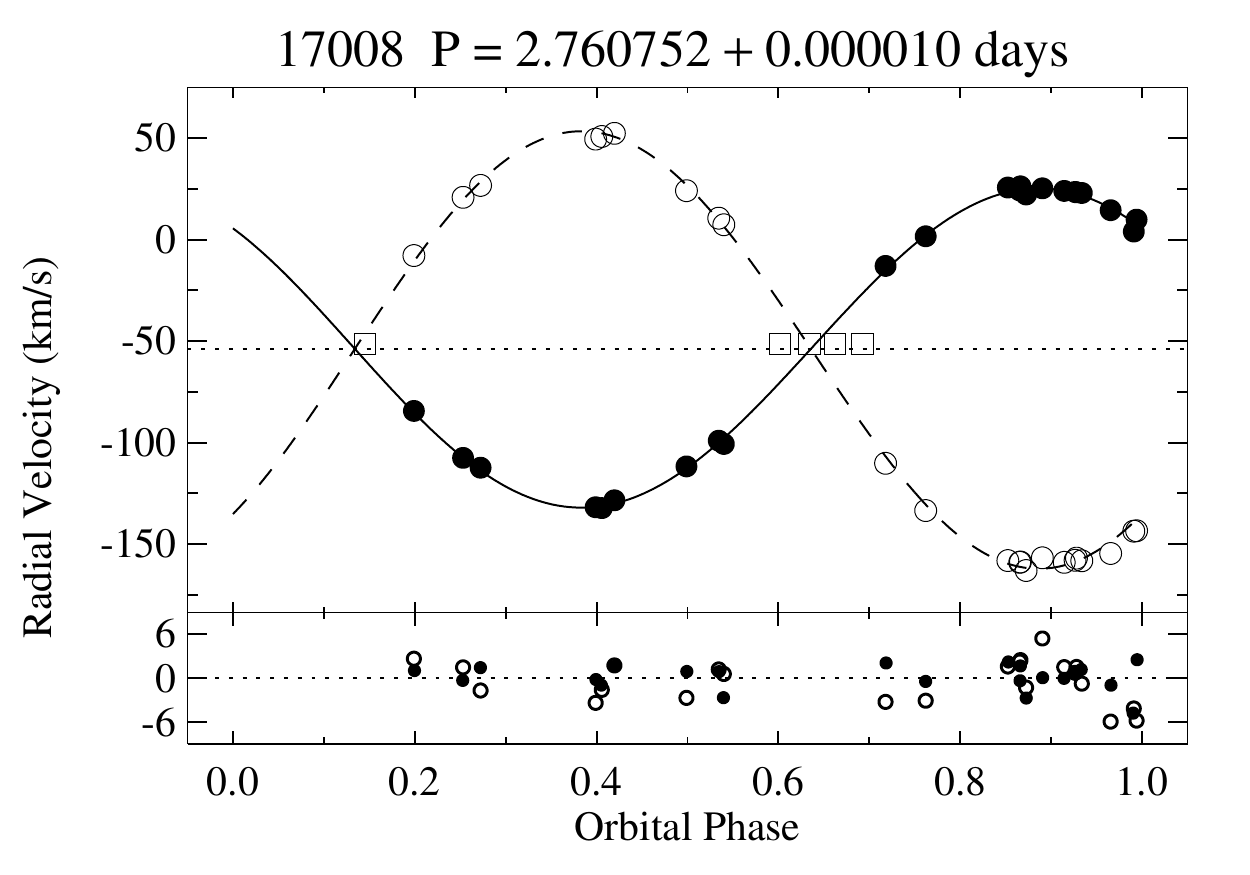}{0.3\linewidth}{}
  \fig{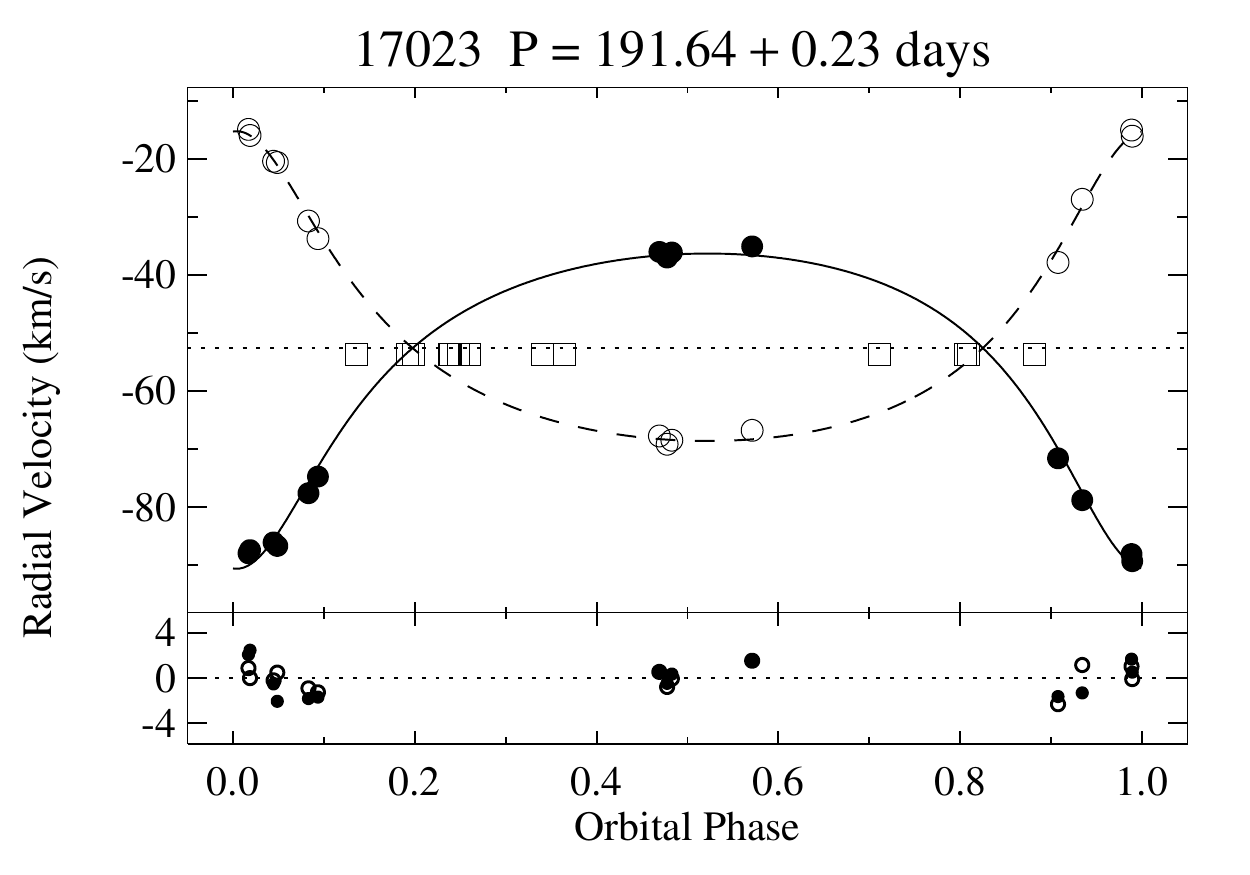}{0.3\linewidth}{}}
\gridline{\fig{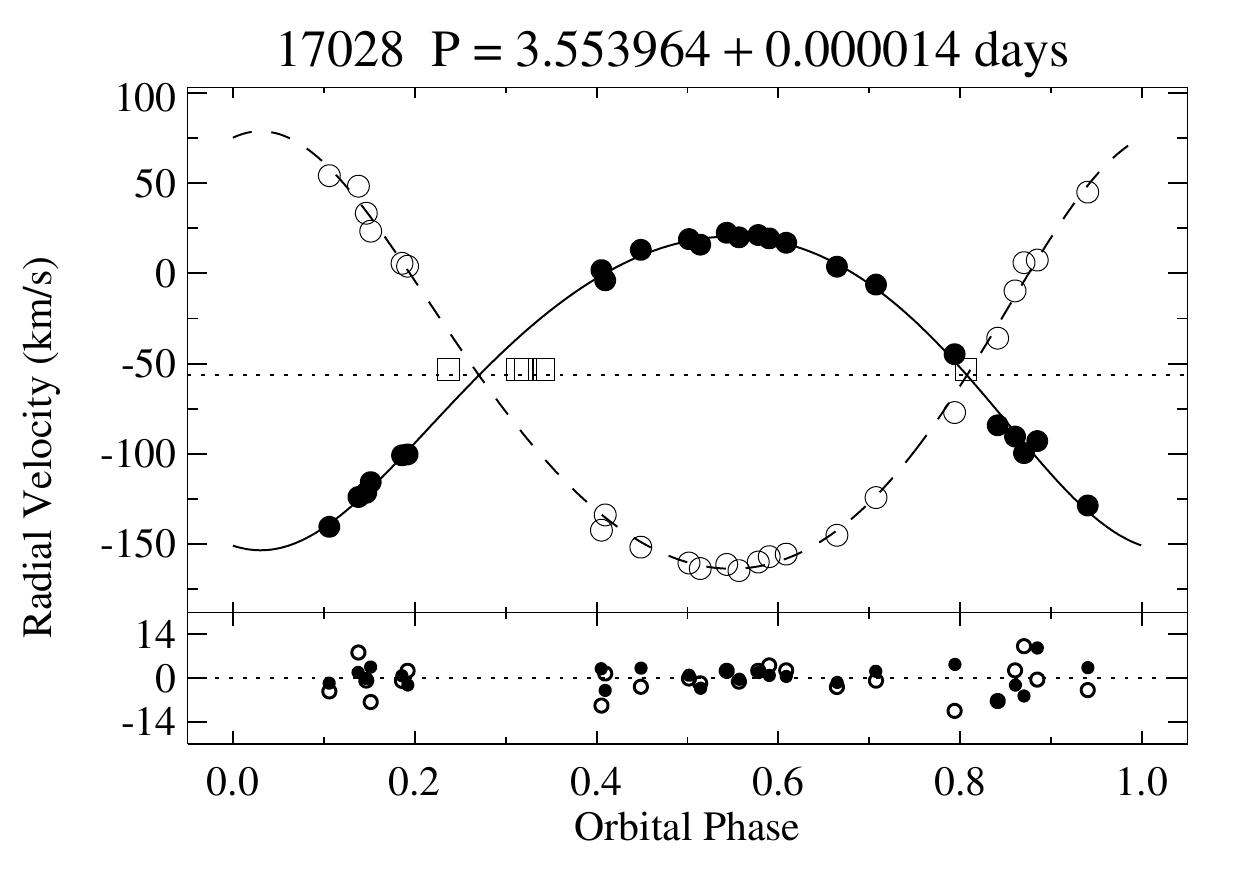}{0.3\linewidth}{}
  \fig{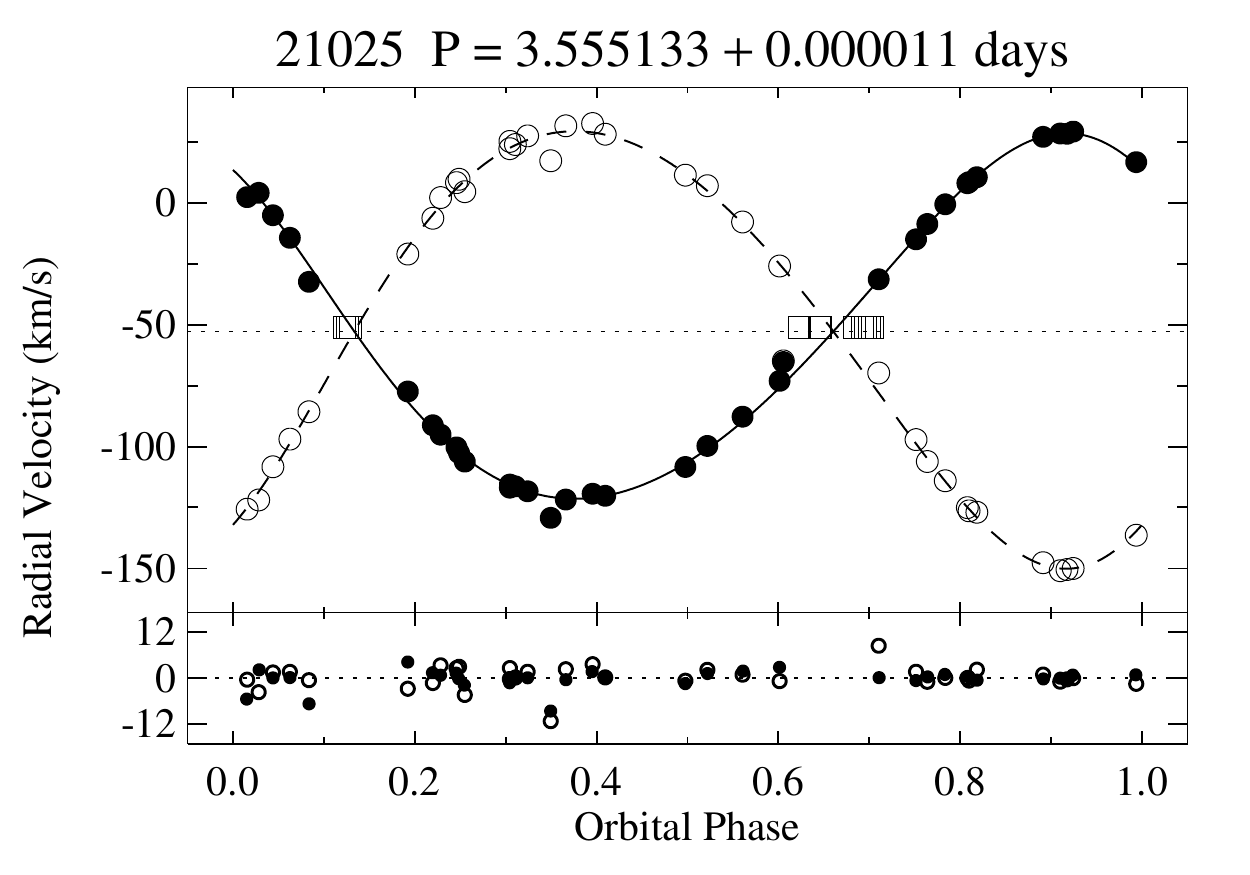}{0.3\linewidth}{}
  \fig{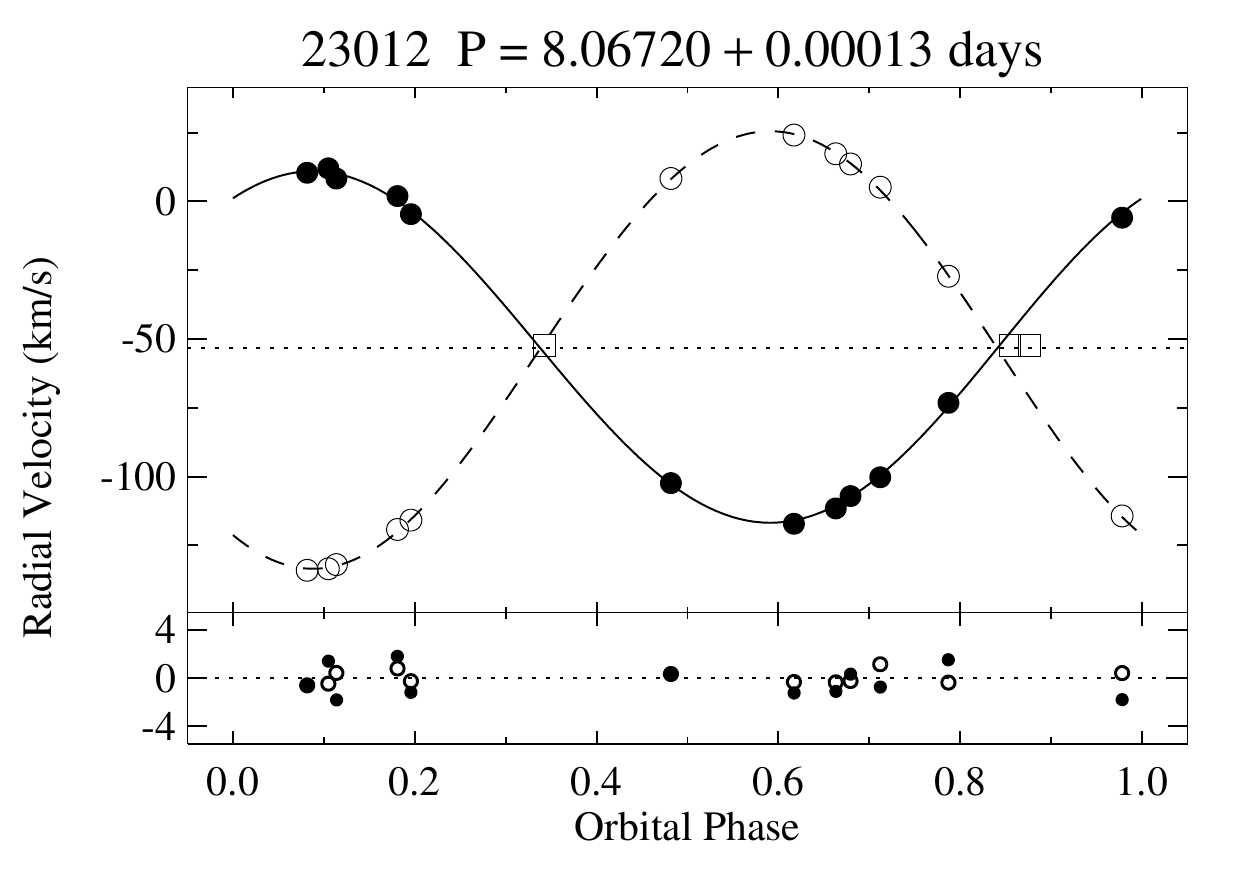}{0.3\linewidth}{}}
\end{figure*}

\clearpage

\section{Binary Frequency}
\label{ngc7789rv:sec:bin.detect}
\subsection{Completeness in Binary Detection}
We follow the Monte Carlo approach to our binary detection completeness described in \cite{Geller2012}. In brief, we generate observations of artificial binaries based on the Galactic field binary period and eccentricity distributions of \cite{R2010} and a random selection of mass ratio, orbital inclination, $\omega$, and phase. We incorporate the measured tidal circularization period of NGC 7789 determined in Section \ref{ngc7789rv:sec:circ}.
For each star we use the actual observation dates and precision value based on its \vsini. From the detectability of these synthetic binaries we calculated a detection percentage of 84\% for binaries with periods under 1,000 days, 75\% for periods under 3,000 days, and 60\% for periods under 10,000 days. These numbers are slightly less than the completeness percentages for NGC 188 (\citealt{Geller2012}) and NGC 6819 (\citealt{Milliman2014}). This is due to the poorer precision associated with the rapidly rotating stars, which makes it more difficult to identify long-period, low-amplitude binary stars. 

\subsection{Main-Sequence Binary Frequency}
We find 226 single and 54 binary member and likely member stars on the  main sequence of NGC 7789, restricting our analysis to $G \geq 14.0$ and $0.8 \leq$ \gming\, $\leq 1.0$. We incorporate the 60\% detection percentage to find a MS binary frequency of 31\% $\pm$ 3\% for binaries with periods less than 10$^{4}$ days. This is consistent with the binary frequency of 32\% found by \cite{Gim1998} in their RV study of giant stars in NGC 7789.

The P $<$ 10$^{4}$ d binary frequency found for other WOCS clusters include: 24\% $\pm$ 3\% in M35 (150 Myr; \citealt{Leiner2015}), 22\% $\pm$ 3\% in NGC 6819 (2.5 Gyr; \citealt{Milliman2014}), 34\% $\pm$ 3\% in M67 (4 Gyr; Geller et al. 2020, in prep.), and 29\% $\pm$ 3\% in NGC 188 (7 Gyr; \citealt{Geller2012}). NGC 7789's value is consistent with these clusters.

\section{Photometric Variables}
We have cross-referenced our primary sample to the photometric variables found by \cite{Jahn1995}, \cite{MK1999}, and \cite{Zhang2003}, and list the results in Table~\ref{ngc7789rv:tab:phot.var}. Along with the photometric types and periods determined by these sources, we also list the orbital period for spectroscopic binaries, the \PRV, \PPM, and any additional comments we have for the star.  

\clearpage
\movetabledown=3in
\begin{rotatetable}
\begin{deluxetable*}{rcccccccccccp{45mm}}
\tabletypesize{\footnotesize}
\tablewidth{0pt}
\centering
\tablecaption{NGC 7789 Photometric Variables \label{ngc7789rv:tab:phot.var}}
\tablehead{\colhead{WOCS} & \colhead{Orbital} & \multicolumn{3}{c}{\underline{\cite{Jahn1995}}} & \multicolumn{3}{c}{\underline{\cite{MK1999}}} & \multicolumn{3}{c}{\underline{\cite{Zhang2003}}}& \colhead{\PPM} & \colhead{Comments}\\
\colhead{ID} & \colhead{Period (d)} & \colhead{ID} & \colhead{Type} & \colhead{Period (d)}  & \colhead{ID} & \colhead{Type} & \colhead{Period (d)} & \colhead{ID} & \colhead{Type} & \colhead{Period (d)} & \colhead{(\%)} & \colhead{}}
\startdata
 7029 &  \nodata & \nodata & \nodata      & \nodata & \nodata & \nodata      & \nodata & v12     & EW      & 0.3917  & 0       & VRR\\
11016 &  2.27414\tablenotemark{*}  & \nodata & \nodata      & \nodata & V39     & EA           & 2.1077  & \nodata & \nodata & \nodata & 0       & BNM, SB1, RR, BSS Cand.\\
13004 &  \nodata & V1      & EW           & 1.19    & V1      & EW           & 1.1614  & v1      & \nodata & \nodata & 100     & BU, RR\\
14014 & 2.36601 & \nodata & \nodata      & \nodata & V37     & EA           & 1.2007  & v10     & \nodata & \nodata & 96      & BM, SB2\\
17028 & 3.553964 & \nodata & \nodata      & \nodata & \nodata & \nodata      & \nodata & v30     & EA      & \nodata & 100      & BM, SB2\\
20007 &  \nodata & V13     & \nodata      & \nodata & \nodata & \nodata      & \nodata & \nodata & \nodata & \nodata & 98      & BU\\
25008 &  \nodata & V10     & $\delta$ Scu & 0.0955  & V10     & $\delta$ Scu & 0.0868  & \nodata & \nodata & \nodata & 81      & SM, RR, BSS Cand.\\
33005 &  \nodata & V2      & EW           & 0.72    & V2      & EB           & 0.7165  & \nodata & \nodata & \nodata & \nodata & VRR\\
33010 &  \nodata & V6      & EW           & 0.884   & \nodata & \nodata      & \nodata & \nodata & \nodata & \nodata & 0       & SNM
\enddata
\tablecomments{EA: $\beta$ Persei type (semi-detached), EB: $\beta$ Lyrae type (detached), EW: W Ursae Majoris type (contact)}
\tablenotetext{*}{While this binary is considered a field star, we have a complete orbital solution for it and provide its parameters as follows: $P = 2.27414 \pm 0.00003$ d, $\gamma = -56.5 \pm 0.4$ \kms, $K = 17.5 \pm 0.5$ \kms, $e = 0.07 \pm 0.03$, $\omega = 90 \pm 30$ deg, $T_\circ = 55596.82 \pm 0.16$, $a \sin i = (0.547 \pm 0.015)\times10^6$ km, $f(m) = 1.26\text{e}{-3}\pm0.11\text{e}-3$ \Msolar, $\sigma = 1.598$ \kms, $N = 23$.}
\end{deluxetable*}
\end{rotatetable}

\clearpage

Of the nine photometric variables that overlap with our primary NGC 7789 sample, we find three to be PM and RV members or likely members. WOCS 14014 and WOCS 17028, noted as eclipsing binaries by \cite{MK1999} and \cite{Zhang2003}, respectively, are SB2's with orbital solutions. The orbital period we find for WOCS 17028 is almost twice the eclipse period found by \cite{MK1999}. \cite{Zhang2003} do not fit a period to WOCS 14014 because of their short observing window and incomplete phase coverage. Another eclipsing binary, WOCS 11016, is an SB1 BSS candidate that has \PPM~= 0\% and we consider it a field star. WOCS 13004 and WOCS 33005 are noted as W UMa's. WOCS 13004 has a very high proper-motion membership probability, and we also classify it as a BU RR with \vsini~= 110 \kms. WOCS 33005 is a VRR with high-noise proper motion membership information. WOCS 20007 is detected by \cite{Jahn1995}, but they did not have the time baseline to determine specifics about this system. We find it to be a binary which we classify as a BU. The one pulsating variable in this sample is a $\delta$ Scuti, WOCS 25008, which we find to be a single member of NGC 7789.

\section{Tidal Circularization}
\label{ngc7789rv:sec:circ}
We update the standard WOCS method to measure the tidal circularization period (CP) for NGC 7789. Like \cite{Meibom2005}, we fit the piecewise function 
\begin{equation}
	e(P) = 
	\begin{cases}
	0.0 & \text{if} ~P \leq P',\\
	\alpha~\left(1-e^{\beta(P'-P)}\right)^{\Gamma}  & \text{if} ~P > P'
	\end{cases}
\label{ngc7789rv:eq:pcirc}
\end{equation}
to the period-eccentricity distribution of MS binaries from 1 mag below the cluster turnoff  to our magnitude limit of G $\sim$ 15 (excluding proper-motion non-members), using the same cutoffs as in Section \ref{ngc7789rv:sec:bin.detect}. $\Gamma$ is set to 1.0, $\beta$ is set to 0.14, and $\alpha$ is set to 0.35, the average eccentricity for binary orbits over 50 days in the Pleiades, M35, Hyades, M67, and NGC 188. We minimize the total absolute deviation between the observed eccentricity and the model to determine the circularization function and P$'$. Specifically, we follow Geller et al. (2020, in prep.) and use a bootstrap technique to determine the circularization period and its uncertainties. CP is the period at which the best-fit circularization function equals 0.01, or $e$(CP) = 0.01.

\begin{figure*}
\includegraphics[width=\linewidth]{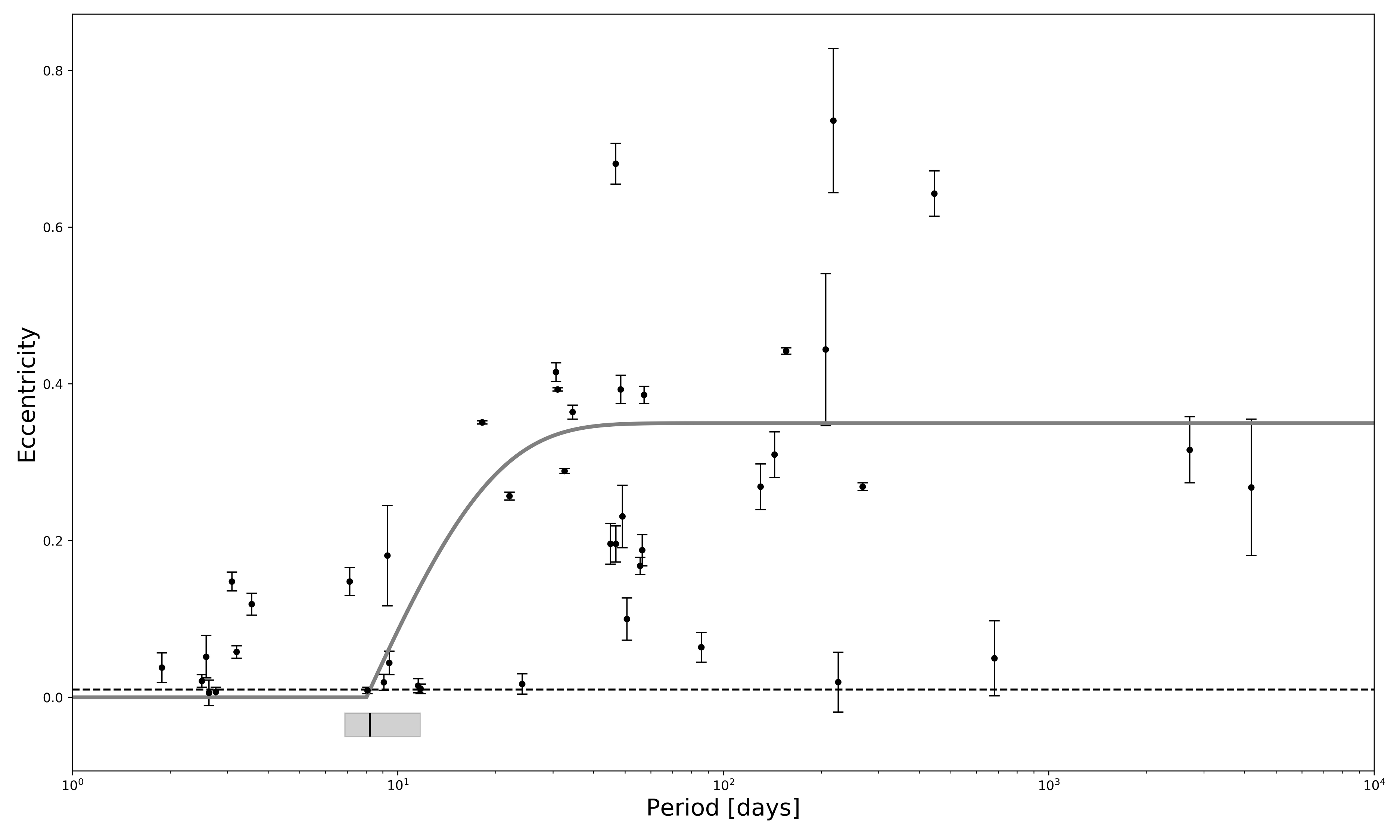}
\caption{The orbital eccentricity versus the orbital period for the MS binaries in NGC 7789. The large eccentricities for long period systems transition to circular orbits for short period binaries. We determine the circularization period to be \PcircCass$^{\cperrupCass}_{\cperrdnCass}\,$~days indicated by a black dash near the bottom of the plot, with the uncertainty overplotted as a light grey bar.  We also overplot the best-fit circularization function of \cite{Meibom2005} in grey.}
\label{ngc7789rv:fig:elogp}
\end{figure*}

We find a CP of~\PcircCass$^{\cperrupCass}_{\cperrdnCass}\,$~days, and plot the period-eccentricity distribution along with the best-fit circularization function in Figure~\ref{ngc7789rv:fig:elogp}.

\begin{figure*}
\includegraphics[width=\linewidth]{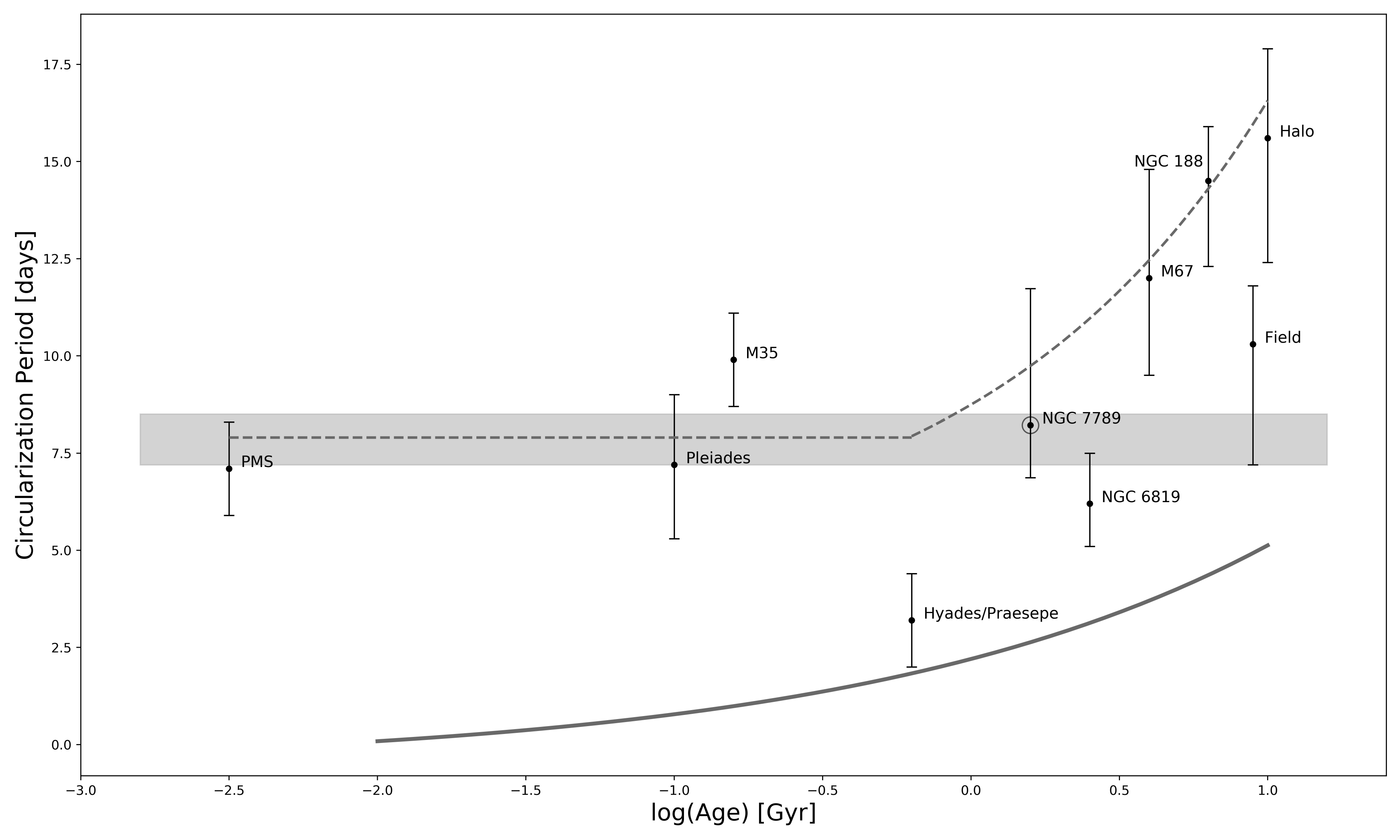}
\caption{Tidal circularization period as a function of age for the eight clusters studied by \cite{Meibom2005} including the update to M35 from \cite{Leiner2015}, the addition of NGC 6819 from \cite{Milliman2014} and our result for NGC 7789. The \cite{ZahnBouchet1989} prediction that tidal circularization is only significant during the PMS phase is plotted in the broad gray line. The theoretical model from the BSE algorithm from \cite{Hurley2002} and the ad hoc tidal energy dissipation rate from \cite{Geller2013} are also plotted as the solid gray curve and the dashed line, respectively.}
\label{ngc7789rv:fig:cp}
\end{figure*}

\cite{Meibom2005} studied eight main-sequence binary populations: the pre-main-sequence, Pleiades, M35, Hyades, M67, NGC 188, the field, and the halo to investigate tidal dissipation mechanisms. Since then the tidal circularization of NGC 6819 was measured by \cite{Milliman2014}, \cite{Leiner2015} updated the value for M35, and Geller et al. (2020, in prep.) updated the value for M67. With this work we incorporate the results for NGC 7789 and plot all the data in Figure~\ref{ngc7789rv:fig:cp}. (But note that only the NGC 7789 and M67 data are derived with the new bootstrap algorithm. This approach finds larger uncertainties than found by \citealt{Meibom2005}).

Our tidal circularization period for NGC 7789 is consistent with the trends of the prior clusters. It is worth reminding that the tidal circularization periods of both NGC 6819 and NGC 7789 are derived from main-sequence stars that have moved off of the cluster ZAMS, and are of slightly higher mass than the other clusters. Tidal circularization depends sensitively on stellar radius and convection zone structure. Future studies should re-determine the tidal circularization periods for these clusters with ZAMS stars.

We also include the theoretical predictions that tidal circularization is only significant during the pre-main-sequence (PMS) phase (\citealt{ZahnBouchet1989}), the algorithm from the Binary-Star Evolution (BSE) code (\citealt{Hurley2002}) and an ad hoc rate from \cite{Geller2013} that combines the PMS circularization from \cite{Kroupa1995} and the BSE from \citealt{Hurley2002} with an artificially increased convective damping term tuned to match the distribution of tidal circularization period with cluster age.
Still, no self-consistent theory is able to explain the relationship between cluster age and tidal circularization.

\section{Blue Stragglers}
\label{ngc7789rv:sec:bs}
\begin{figure*}[p!]
\includegraphics[width=1.0\linewidth]{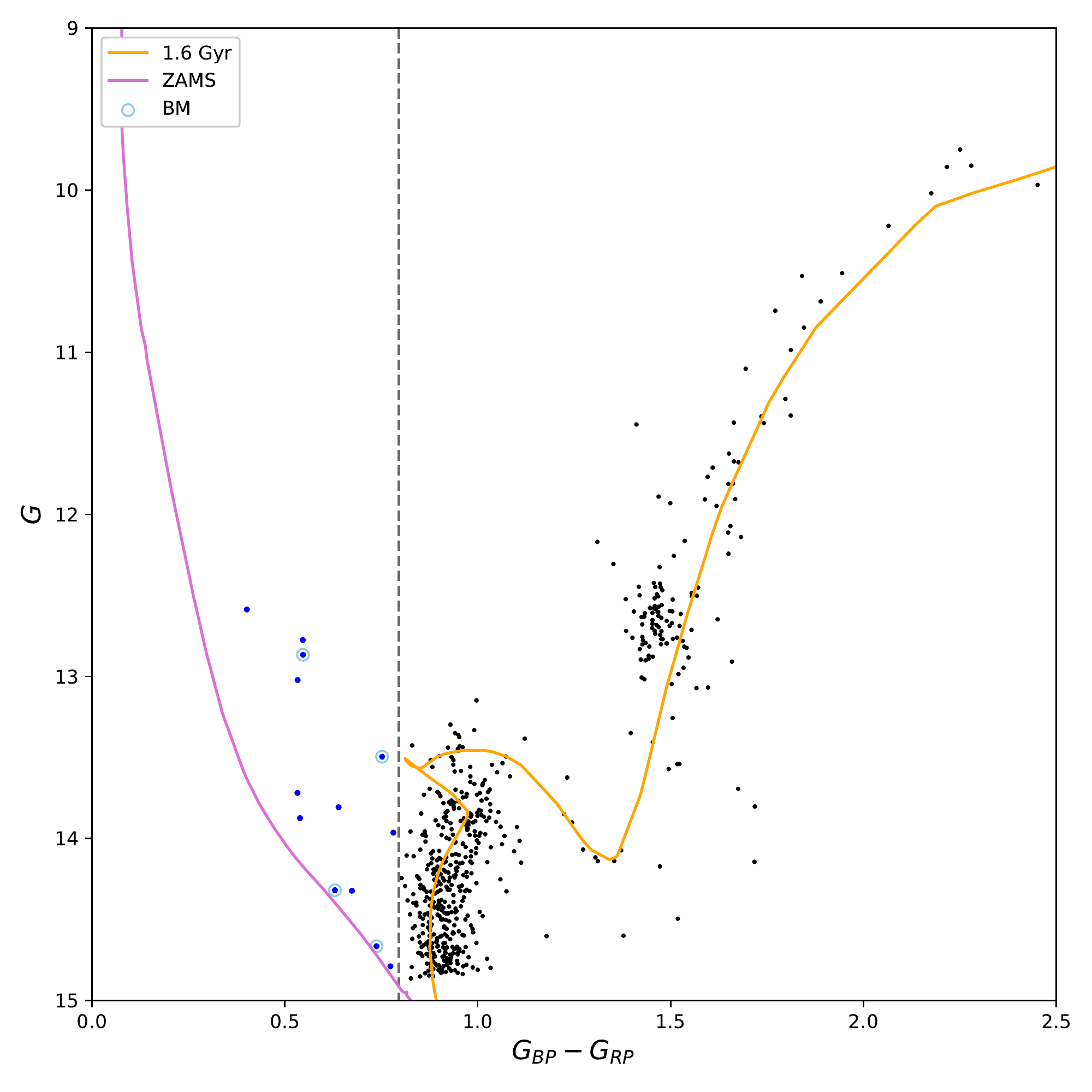}
\caption{The \gaia$\,$ CMD of NGC 7789 highlighting the BSS that fall to the blue of the dashed line. BSS non-velocity-variable RV members with \PPM~$\geq$ 50\% are indicated by blue dots. Binary BSS, all of which have orbital solutions, are marked with an additional circle. The curves are the same as in Figure \ref{ngc7789rv:fig:cmd.member}.}
\label{ngc7789rv:fig:cmd.bs}
\end{figure*}

We define BSS as being to the blue of the dashed line in Figure~\ref{ngc7789rv:fig:cmd.bs}, which we place based on single-star and equal-mass binary isochrones. We list all the BSS in our primary NGC 7789 sample in Table~\ref{ngc7789rv:tab:bsinfo} along with the number of WIYN observations that satisfy our quality thresholds, \gaia~ photometry, \PRV, \PPM, our RV membership classifications, as well as any additional comments.

\startlongtable
\begin{deluxetable*}{rcccccccp{40mm}}
\tabletypesize{\footnotesize}
\tablecaption{NGC 7789 Blue Straggler Candidates \label{ngc7789rv:tab:bsinfo}}
\tablehead{\colhead{WOCS ID} & \colhead{$N$} & \colhead{$G$} & \colhead{\gming} & \colhead{\PRV} & \colhead{\PPM} & \colhead{Class} & \colhead{Comment}}
\tablewidth{0pt}
\startdata
\multicolumn{8}{l}{\textit{RV and PM members:}} \\
  5004 & 14 & 12.78 &  0.55 &      95 &      99 &       SM &  \nodata \\ 
  5011 & 34 & 12.87 &  0.55 &      95 &      99 &       BM &       RR \\
  9027 &  7 & 13.02 &  0.53 &      94 &      99 &       SM &       RR \\
 10010 &  9 & 13.72 &  0.53 &      95 &      96 &       SM &  \nodata \\
 15015 &  9 & 13.87 &  0.54 &      95 &      99 &       SM &  \nodata \\
 16020 & 18 & 13.81 &  0.64 &      95 &      99 &       SM &  \nodata \\
 20009 & 42 & 14.32 &  0.63 &      62 &      99 &       BM &  \nodata \\
 25008\tablenotemark{*} & 13 & 13.96 &  0.78 &      94 &      81 &       SM &       RR \\
 25024\tablenotemark{*} & 11 & 14.79 &  0.77 &      94 &      99 &       SM &       RR \\
 27010 & 14 & 14.32 &  0.67 &      94 &      99 &       SM &  \nodata \\
 36011 & 13 & 14.66 &  0.74 &      94 &      98 &       BM &  \nodata \\
\multicolumn{8}{l}{\textit{RV members with high-noise PM information:}} \\
  4009\tablenotemark{*} & 15 & 12.59 &  0.40 &      92 & \nodata &       SM &  \nodata \\
 10011\tablenotemark{*} & 22 & 13.50 &  0.75 &      96 & \nodata &       BM &  \nodata \\
\multicolumn{8}{l}{\textit{RV members and PM non-members:}} \\
  8022 & 31 & 13.35 &  0.65 &      93 &      29 &       BNM &  \nodata \\
 11015 & 13 & 13.35 &  0.53 &      70 &       0 &       SNM &      Hot \\
 11016 & 25 & 13.79 &  0.65 &      65 &       0 &       BNM &       RR \\
 13016 & 14 & 13.81 &  0.77 &      87 &       0 &       SNM &  \nodata \\
 20012 & 12 & 14.37 &  0.69 &      95 &       0 &       SNM &  \nodata \\
 24034 & 12 & 14.44 &  0.75 &      95 &       0 &       SNM &  \nodata \\
 37005 & 12 & 14.89 &  0.79 &      86 &      17 &       BNM &  \nodata \\
\multicolumn{8}{l}{\textit{RV non-members and PM members:}} \\
  1014 & 18 & 11.98 &  0.42 &       5 &      99 &       SNM &       RR \\
  3024 &  7 & 12.30 &  0.56 &       0 &      94 &       SNM &       RR \\
\enddata
\tablenotetext{*}{These are the BSS marked as borderline as discussed in Section \ref{ngc7789rv:sec:bs}, whether by proximity to our cutoff or by astrometric noise.}
\end{deluxetable*}

Eleven BSS have \PRV~$\geq$ 50\% and \PPM~$\geq$ 50\%, and two more BSS  have \PRV~$\geq$ 50\% but high-noise PM information. We consider these 13 stars to be BSS cluster members. Seven other BSS with \PRV~$\geq$ 50\% all have \PPM~$<$ 50\% and we treat them as field stars. For the member BSS, four have completed orbital solutions: WOCS 5011, WOCS 10011, WOCS 20009, and WOCS 36011. All of these stars are SB1s. WOCS 5011 and WOCS 20009 are both long-period SB1s with periods of 2710\,d and 4190\,d, respectively. Such periods are consistent with mass transfer from asymptotic branch giants (e.g., \citealt{GellerNature2011} and \citealt{Gosnell2019}, albeit will less massive turnoffs).

WOCS 36011, on the other hand, has a shorter period of 217\,d and a much higher eccentricity with $e \sim 0.74$. Blue stragglers with periods on the order of $10^2$ days are likely the result of mass transfer from an RGB star, as in the case of WOCS 5379 in the open cluster NGC 188 (\citealt{Gosnell2014, MathieuEco2015, Gosnell2019}). Interestingly, the CMD location of WOCS 36011 in NGC 7789 relative to the ZAMS turnoff is very similar to that of WOCS 5379 in NGC 188 (although the primary star of WOCS 36011 is of higher mass).

Our analysis of the mass function of WOCS 36011 is inconclusive regarding the presence of a white dwarf companion from an RGB donor star, as we can only place a lower limit on the companion stellar mass of $\sim$0.05 \Msolar. The very high eccentricity may be challenging to a mass-transfer origin modeling. This blue straggler, like its counterpart in NGC 188, is the subject of continued observation.

WOCS 10011 has a similarly low period and significant eccentricity, with a period of 517\,d and $e \sim 0.37$. It may be the case that this binary is not a true blue straggler, but rather on the cluster blue hook. For the purposes of this work, we consider it to be a blue straggler candidate and it remains the subject of continued study.

The long-period orbits we find for two of the NGC 7789 blue stragglers are in line with the BSS orbital periods found for the intermediate-aged open cluster NGC 6819 (2.5 Gyr) and the old open cluster NGC 188 (7 Gyr). For NGC 6819, the three SB1 BSS with orbital solutions have periods of 762$\,$d, 1144$\,$d, and 3900$\,$d (\citealt{Milliman2014}). In NGC 188 all but one of the SB1 BSS (13/14) have a period on the order of $10^3$ days (\citealt{Geller2012}).

With the four detected velocity-variable BSS we find a BSS binary frequency of 31\% $\pm$ 15\% (4/13). This is consistent with the 40\% $\pm$ 16\% found for NGC 6819 (\citealt{Milliman2014}) and less than the 80\% $\pm$ 20\% BSS binary fraction found for NGC 188 (\citealt{MathieuNature2009}) and the 79\% $\pm$ 24\% BSS binary fraction found in M67 (\citealt{Geller2015}). If we exclude the borderline BSS candidates WOCS 10011, WOCS 25008, and WOCS 25024 the BSS binary frequency is then 30\% $\pm$ 17\%, still consistent with NGC 6819. Excluding also the other BSS candidate with high-noise astrometry, WOCS 4009, the binary frequency is then 33\% $\pm$ 19\%.

\section{Summary}
We present the results of a complete and extensive time-series radial-velocity survey of 1206 stars in the open cluster NGC 7789. The stellar sample extends to a radius of $\sim$18$'$ (10 pc in projection) from the cluster center and covers a $G$ mag range from $\sim$ 10 to 15. This magnitude range includes evolved MS stars to 1 mag below the turnoff, giants, red-clump stars, blue stragglers, and sub-subgiant candidates. 

We use these radial-velocity measurements and \gaia\, data to derive a sample of 564 three-dimensional cluster members.
We identify 105 velocity-variable cluster members, and present spectroscopic orbital solutions for 83 binaries. We calculate an incompleteness-corrected MS binary fraction of 31\% $\pm$ 3\% for binaries with periods under 10,000$\,$days, consistent with other open clusters studied by WOCS. We find a tidal circularization period of \PcircCass$^{\cperrupCass}_{\cperrdnCass}\,$~days, consistent with the ad hoc tidal dissipation rate of \cite{Geller2013}. 

We detect 31\% $\pm$ 15\% of BSS as velocity variable, consistent with the frequency of velocity-variable BSS in the similarly-aged cluster NGC 6819. We also identify two BSS, WOCS 10011 and WOCS 36011, whose low periods and significant eccentricities may be the result of mass transfer from an RGB star.

We also find two sub-subgiants and two red stragglers in our sample. We note that the amplitude of the velocity variability of the sub-subgiant WOCS 20035 is lower than that of other sub-subgiants, which typically have short orbital periods. Neither of the red stragglers are velocity variables. We look forward to our expanded observations of NGC 7789 that will shed more light on these alternative stellar evolution products.

\acknowledgments{
The University of Wisconsin-Madison authors acknowledge funding support from NSF AST-1714506, the Graduate School, the Office of the Vice Chancellor for Research and Graduate Education at the University of Wisconsin-Madison, and the Wisconsin Alumni Research Foundation.

This work has made use of data from the European Space Agency (ESA) mission
{\it Gaia} (\url{https://www.cosmos.esa.int/gaia}), processed by the {\it Gaia}
Data Processing and Analysis Consortium (DPAC,
\url{https://www.cosmos.esa.int/web/gaia/dpac/consortium}). Funding for the DPAC
has been provided by national institutions, in particular the institutions
participating in the {\it Gaia} Multilateral Agreement.

This work has also made use of Astropy (\url{http://www.astropy.org}), a community-developed core Python package for Astronomy \citep{astropy2013, astropy2018}, as well as MATLAB (\citealt{MATLAB:2019a}).

This work was conducted at the University of Wisconsin-Madison which is located on occupied ancestral land of the Ho-chunk Nation, and observations for this work were conducted on the traditional lands of the Tohono O'odham Nation. We respect the inherent sovereignty of these nations, along with the other eleven First Nations in Wisconsin as well as the southwestern tribes of Arizona. We honor with gratitude these lands and the peoples who have stewarded them, and who continue to steward them, throughout the generations. 
}

\bibliographystyle{aasjournal}
\bibliography{ngc7789.version2.bbl}

\begin{thebibliography}{}
\expandafter\ifx\csname natexlab\endcsname\relax\def\natexlab#1{#1}\fi
\providecommand{\url}[1]{\href{#1}{#1}}
\providecommand{\dodoi}[1]{doi:~\href{http://doi.org/#1}{\nolinkurl{#1}}}
\providecommand{\doeprint}[1]{\href{http://ascl.net/#1}{\nolinkurl{http://ascl.net/#1}}}
\providecommand{\doarXiv}[1]{\href{https://arxiv.org/abs/#1}{\nolinkurl{https://arxiv.org/abs/#1}}}

\bibitem[{MAT(2019)}]{MATLAB:2019a}
 2019, {MATLAB version 9.6 (R2019a)}, The Mathworks, Inc., Natick,
  Massachusetts

\bibitem[{{Astropy Collaboration} {et~al.}(2013){Astropy Collaboration},
  {Robitaille}, {Tollerud}, {Greenfield}, {Droettboom}, {Bray}, {Aldcroft},
  {Davis}, {Ginsburg}, {Price-Whelan}, {Kerzendorf}, {Conley}, {Crighton},
  {Barbary}, {Muna}, {Ferguson}, {Grollier}, {Parikh}, {Nair}, {Unther},
  {Deil}, {Woillez}, {Conseil}, {Kramer}, {Turner}, {Singer}, {Fox}, {Weaver},
  {Zabalza}, {Edwards}, {Azalee Bostroem}, {Burke}, {Casey}, {Crawford},
  {Dencheva}, {Ely}, {Jenness}, {Labrie}, {Lim}, {Pierfederici}, {Pontzen},
  {Ptak}, {Refsdal}, {Servillat}, \& {Streicher}}]{astropy2013}
{Astropy Collaboration}, {Robitaille}, T.~P., {Tollerud}, E.~J., {et~al.} 2013,
  \aap, 558, A33, \dodoi{10.1051/0004-6361/201322068}

\bibitem[{{Astropy Collaboration} {et~al.}(2018){Astropy Collaboration},
  {Price-Whelan}, {Sip{\H{o}}cz}, {G{\"u}nther}, {Lim}, {Crawford}, {Conseil},
  {Shupe}, {Craig}, {Dencheva}, {Ginsburg}, {Vand erPlas}, {Bradley},
  {P{\'e}rez-Su{\'a}rez}, {de Val-Borro}, {Aldcroft}, {Cruz}, {Robitaille},
  {Tollerud}, {Ardelean}, {Babej}, {Bach}, {Bachetti}, {Bakanov}, {Bamford},
  {Barentsen}, {Barmby}, {Baumbach}, {Berry}, {Biscani}, {Boquien}, {Bostroem},
  {Bouma}, {Brammer}, {Bray}, {Breytenbach}, {Buddelmeijer}, {Burke},
  {Calderone}, {Cano Rodr{\'\i}guez}, {Cara}, {Cardoso}, {Cheedella}, {Copin},
  {Corrales}, {Crichton}, {D'Avella}, {Deil}, {Depagne}, {Dietrich}, {Donath},
  {Droettboom}, {Earl}, {Erben}, {Fabbro}, {Ferreira}, {Finethy}, {Fox},
  {Garrison}, {Gibbons}, {Goldstein}, {Gommers}, {Greco}, {Greenfield},
  {Groener}, {Grollier}, {Hagen}, {Hirst}, {Homeier}, {Horton}, {Hosseinzadeh},
  {Hu}, {Hunkeler}, {Ivezi{\'c}}, {Jain}, {Jenness}, {Kanarek}, {Kendrew},
  {Kern}, {Kerzendorf}, {Khvalko}, {King}, {Kirkby}, {Kulkarni}, {Kumar},
  {Lee}, {Lenz}, {Littlefair}, {Ma}, {Macleod}, {Mastropietro}, {McCully},
  {Montagnac}, {Morris}, {Mueller}, {Mumford}, {Muna}, {Murphy}, {Nelson},
  {Nguyen}, {Ninan}, {N{\"o}the}, {Ogaz}, {Oh}, {Parejko}, {Parley}, {Pascual},
  {Patil}, {Patil}, {Plunkett}, {Prochaska}, {Rastogi}, {Reddy Janga},
  {Sabater}, {Sakurikar}, {Seifert}, {Sherbert}, {Sherwood-Taylor}, {Shih},
  {Sick}, {Silbiger}, {Singanamalla}, {Singer}, {Sladen}, {Sooley},
  {Sornarajah}, {Streicher}, {Teuben}, {Thomas}, {Tremblay}, {Turner},
  {Terr{\'o}n}, {van Kerkwijk}, {de la Vega}, {Watkins}, {Weaver}, {Whitmore},
  {Woillez}, {Zabalza}, \& {Astropy Contributors}}]{astropy2018}
{Astropy Collaboration}, {Price-Whelan}, A.~M., {Sip{\H{o}}cz}, B.~M., {et~al.}
  2018, \aj, 156, 123, \dodoi{10.3847/1538-3881/aabc4f}

\bibitem[{{Barden} {et~al.}(1994){Barden}, {Armandroff}, {Muller},
  {et~al.}}]{Barden1994}
{Barden}, S.~C., {Armandroff}, T., {Muller}, G., {et~al.} 1994, in Proc. SPIE,
  Vol. 2198, Instrumentation in Astronomy VIII, ed. D.~L. {Crawford} \& E.~R.
  {Craine}, 87--97

\bibitem[{{Burbidge} \& {Sandage}(1958)}]{Burbidge1958}
{Burbidge}, E.~M., \& {Sandage}, A. 1958, \apj, 128, 174,
  \dodoi{10.1086/146535}

\bibitem[{{Cantat-Gaudin} {et~al.}(2018){Cantat-Gaudin}, {Jordi}, {Vallenari},
  {Bragaglia}, {Balaguer-N{\'u}{\~n}ez}, {Soubiran}, {Bossini}, {Moitinho},
  {Castro-Ginard}, {Krone-Martins}, {Casamiquela}, {Sordo}, \&
  {Carrera}}]{CG2018}
{Cantat-Gaudin}, T., {Jordi}, C., {Vallenari}, A., {et~al.} 2018, \aap, 618,
  A93, \dodoi{10.1051/0004-6361/201833476}

\bibitem[{{Casamiquela} {et~al.}(2016){Casamiquela}, {Carrera}, {Jordi},
  {Balaguer-N{\'u}{\~n}ez}, {Pancino}, {Hidalgo},
  {Mart{\'{\i}}nez-V{\'a}zquez}, {Murabito}, {del Pino}, {Aparicio},
  {Blanco-Cuaresma}, \& {Gallart}}]{Casamiquela2016}
{Casamiquela}, L., {Carrera}, R., {Jordi}, C., {et~al.} 2016, \mnras, 458,
  3150, \dodoi{10.1093/mnras/stw518}

\bibitem[{{Deliyannis} {et~al.}(2019){Deliyannis}, {Anthony-Twarog},
  {Lee-Brown}, \& {Twarog}}]{Deliyannis2019}
{Deliyannis}, C.~P., {Anthony-Twarog}, B.~J., {Lee-Brown}, D.~B., \& {Twarog},
  B.~A. 2019, \aj, 158, 163, \dodoi{10.3847/1538-3881/ab3fad}

\bibitem[{{Dotter}(2016)}]{Dotter2016}
{Dotter}, A. 2016, \apjs, 222, 8, \dodoi{10.3847/0067-0049/222/1/8}

\bibitem[{{Fletcher} {et~al.}(1982){Fletcher}, {Harris}, {McClure}, \&
  {Scarfe}}]{Fletcher1982}
{Fletcher}, J.~M., {Harris}, H.~C., {McClure}, R.~D., \& {Scarfe}, C.~D. 1982,
  \pasp, 94, 1017, \dodoi{10.1086/131102}

\bibitem[{{Gaia Collaboration} {et~al.}(2016){Gaia Collaboration}, {Prusti},
  {de Bruijne}, {Brown}, {Vallenari}, {Babusiaux}, {Bailer-Jones}, {Bastian},
  {Biermann}, {Evans}, {Eyer}, {Jansen}, {Jordi}, {Klioner}, {Lammers},
  {Lindegren}, {Luri}, {Mignard}, {Milligan}, {Panem}, {Poinsignon},
  {Pourbaix}, {Randich}, {Sarri}, {Sartoretti}, {Siddiqui}, {Soubiran},
  {Valette}, {van Leeuwen}, {Walton}, {Aerts}, {Arenou}, {Cropper}, {Drimmel},
  {H{\o}g}, {Katz}, {Lattanzi}, {O'Mullane}, {Grebel}, {Holland}, {Huc},
  {Passot}, {Bramante}, {Cacciari}, {Casta{\~n}eda}, {Chaoul}, {Cheek}, {De
  Angeli}, {Fabricius}, {Guerra}, {Hern{\'a}ndez}, {Jean-Antoine-Piccolo},
  {Masana}, {Messineo}, {Mowlavi}, {Nienartowicz}, {Ord{\'o}{\~n}ez-Blanco},
  {Panuzzo}, {Portell}, {Richards}, {Riello}, {Seabroke}, {Tanga},
  {Th{\'e}venin}, {Torra}, {Els}, {Gracia-Abril}, {Comoretto},
  {Garcia-Reinaldos}, {Lock}, {Mercier}, {Altmann}, {Andrae}, {Astraatmadja},
  {Bellas-Velidis}, {Benson}, {Berthier}, {Blomme}, {Busso}, {Carry},
  {Cellino}, {Clementini}, {Cowell}, {Creevey}, {Cuypers}, {Davidson}, {De
  Ridder}, {de Torres}, {Delchambre}, {Dell'Oro}, {Ducourant}, {Fr{\'e}mat},
  {Garc{\'\i}a-Torres}, {Gosset}, {Halbwachs}, {Hambly}, {Harrison}, {Hauser},
  {Hestroffer}, {Hodgkin}, {Huckle}, {Hutton}, {Jasniewicz}, {Jordan},
  {Kontizas}, {Korn}, {Lanzafame}, {Manteiga}, {Moitinho}, {Muinonen},
  {Osinde}, {Pancino}, {Pauwels}, {Petit}, {Recio-Blanco}, {Robin}, {Sarro},
  {Siopis}, {Smith}, {Smith}, {Sozzetti}, {Thuillot}, {van Reeven}, {Viala},
  {Abbas}, {Abreu Aramburu}, {Accart}, {Aguado}, {Allan}, {Allasia},
  {Altavilla}, {{\'A}lvarez}, {Alves}, {Anderson}, {Andrei}, {Anglada Varela},
  {Antiche}, {Antoja}, {Ant{\'o}n}, {Arcay}, {Atzei}, {Ayache}, {Bach},
  {Baker}, {Balaguer-N{\'u}{\~n}ez}, {Barache}, {Barata}, {Barbier}, {Barblan},
  {Baroni}, {Barrado y Navascu{\'e}s}, {Barros}, {Barstow}, {Becciani},
  {Bellazzini}, {Bellei}, {Bello Garc{\'\i}a}, {Belokurov}, {Bendjoya},
  {Berihuete}, {Bianchi}, {Bienaym{\'e}}, {Billebaud}, {Blagorodnova},
  {Blanco-Cuaresma}, {Boch}, {Bombrun}, {Borrachero}, {Bouquillon}, {Bourda},
  {Bouy}, {Bragaglia}, {Breddels}, {Brouillet}, {Br{\"u}semeister},
  {Bucciarelli}, {Budnik}, {Burgess}, {Burgon}, {Burlacu}, {Busonero}, {Buzzi},
  {Caffau}, {Cambras}, {Campbell}, {Cancelliere}, {Cantat-Gaudin}, {Carlucci},
  {Carrasco}, {Castellani}, {Charlot}, {Charnas}, {Charvet}, {Chassat},
  {Chiavassa}, {Clotet}, {Cocozza}, {Collins}, {Collins}, {Costigan}, {Crifo},
  {Cross}, {Crosta}, {Crowley}, {Dafonte}, {Damerdji}, {Dapergolas}, {David},
  {David}, {De Cat}, {de Felice}, {de Laverny}, {De Luise}, {De March}, {de
  Martino}, {de Souza}, {Debosscher}, {del Pozo}, {Delbo}, {Delgado},
  {Delgado}, {di Marco}, {Di Matteo}, {Diakite}, {Distefano}, {Dolding}, {Dos
  Anjos}, {Drazinos}, {Dur{\'a}n}, {Dzigan}, {Ecale}, {Edvardsson}, {Enke},
  {Erdmann}, {Escolar}, {Espina}, {Evans}, {Eynard Bontemps}, {Fabre},
  {Fabrizio}, {Faigler}, {Falc{\~a}o}, {Farr{\`a}s Casas}, {Faye}, {Federici},
  {Fedorets}, {Fern{\'a}ndez-Hern{\'a}ndez}, {Fernique}, {Fienga}, {Figueras},
  {Filippi}, {Findeisen}, {Fonti}, {Fouesneau}, {Fraile}, {Fraser}, {Fuchs},
  {Furnell}, {Gai}, {Galleti}, {Galluccio}, {Garabato}, {Garc{\'\i}a-Sedano},
  {Gar{\'e}}, {Garofalo}, {Garralda}, {Gavras}, {Gerssen}, {Geyer}, {Gilmore},
  {Girona}, {Giuffrida}, {Gomes}, {Gonz{\'a}lez-Marcos},
  {Gonz{\'a}lez-N{\'u}{\~n}ez}, {Gonz{\'a}lez-Vidal}, {Granvik}, {Guerrier},
  {Guillout}, {Guiraud}, {G{\'u}rpide}, {Guti{\'e}rrez-S{\'a}nchez}, {Guy},
  {Haigron}, {Hatzidimitriou}, {Haywood}, {Heiter}, {Helmi}, {Hobbs},
  {Hofmann}, {Holl}, {Holland }, {Hunt}, {Hypki}, {Icardi}, {Irwin}, {Jevardat
  de Fombelle}, {Jofr{\'e}}, {Jonker}, {Jorissen}, {Julbe}, {Karampelas},
  {Kochoska}, {Kohley}, {Kolenberg}, {Kontizas}, {Koposov}, {Kordopatis},
  {Koubsky}, {Kowalczyk}, {Krone-Martins}, {Kudryashova}, {Kull}, {Bachchan},
  {Lacoste-Seris}, {Lanza}, {Lavigne}, {Le Poncin-Lafitte}, {Lebreton},
  {Lebzelter}, {Leccia}, {Leclerc}, {Lecoeur-Taibi}, {Lemaitre}, {Lenhardt},
  {Leroux}, {Liao}, {Licata}, {Lindstr{\o}m}, {Lister}, {Livanou}, {Lobel},
  {L{\"o}ffler}, {L{\'o}pez}, {Lopez-Lozano}, {Lorenz}, {Loureiro},
  {MacDonald}, {Magalh{\~a}es Fernandes}, {Managau}, {Mann}, {Mantelet},
  {Marchal}, {Marchant}, {Marconi}, {Marie}, {Marinoni}, {Marrese},
  {Marschalk{\'o}}, {Marshall}, {Mart{\'\i}n-Fleitas}, {Martino}, {Mary},
  {Matijevi{\v{c}}}, {Mazeh}, {McMillan}, {Messina}, {Mestre}, {Michalik},
  {Millar}, {Miranda}, {Molina}, {Molinaro}, {Molinaro}, {Moln{\'a}r},
  {Moniez}, {Montegriffo}, {Monteiro}, {Mor}, {Mora}, {Morbidelli}, {Morel},
  {Morgenthaler}, {Morley}, {Morris}, {Mulone}, {Muraveva}, {Musella},
  {Narbonne}, {Nelemans}, {Nicastro}, {Noval}, {Ord{\'e}novic},
  {Ordieres-Mer{\'e}}, {Osborne}, {Pagani}, {Pagano}, {Pailler}, {Palacin},
  {Palaversa}, {Parsons}, {Paulsen}, {Pecoraro}, {Pedrosa}, {Pentik{\"a}inen},
  {Pereira}, {Pichon}, {Piersimoni}, {Pineau}, {Plachy}, {Plum}, {Poujoulet},
  {Pr{\v{s}}a}, {Pulone}, {Ragaini}, {Rago}, {Rambaux}, {Ramos-Lerate},
  {Ranalli}, {Rauw}, {Read}, {Regibo}, {Renk}, {Reyl{\'e}}, {Ribeiro},
  {Rimoldini}, {Ripepi}, {Riva}, {Rixon}, {Roelens}, {Romero-G{\'o}mez},
  {Rowell}, {Royer}, {Rudolph}, {Ruiz-Dern}, {Sadowski}, {Sagrist{\`a}
  Sell{\'e}s}, {Sahlmann}, {Salgado}, {Salguero}, {Sarasso}, {Savietto},
  {Schnorhk}, {Schultheis}, {Sciacca}, {Segol}, {Segovia}, {Segransan},
  {Serpell}, {Shih}, {Smareglia}, {Smart}, {Smith}, {Solano}, {Solitro},
  {Sordo}, {Soria Nieto}, {Souchay}, {Spagna}, {Spoto}, {Stampa}, {Steele},
  {Steidelm{\"u}ller}, {Stephenson}, {Stoev}, {Suess}, {S{\"u}veges}, {Surdej},
  {Szabados}, {Szegedi-Elek}, {Tapiador}, {Taris}, {Tauran}, {Taylor},
  {Teixeira}, {Terrett}, {Tingley}, {Trager}, {Turon}, {Ulla}, {Utrilla},
  {Valentini}, {van Elteren}, {Van Hemelryck}, {van Leeuwen}, {Varadi},
  {Vecchiato}, {Veljanoski}, {Via}, {Vicente}, {Vogt}, {Voss}, {Votruba},
  {Voutsinas}, {Walmsley}, {Weiler}, {Weingrill}, {Werner}, {Wevers},
  {Whitehead}, {Wyrzykowski}, {Yoldas}, {{\v{Z}}erjal}, {Zucker}, {Zurbach},
  {Zwitter}, {Alecu}, {Allen}, {Allende Prieto}, {Amorim},
  {Anglada-Escud{\'e}}, {Arsenijevic}, {Azaz}, {Balm}, {Beck}, {Bernstein},
  {Bigot}, {Bijaoui}, {Blasco}, {Bonfigli}, {Bono}, {Boudreault}, {Bressan},
  {Brown}, {Brunet}, {Bunclark}, {Buonanno}, {Butkevich}, {Carret}, {Carrion},
  {Chemin}, {Ch{\'e}reau}, {Corcione}, {Darmigny}, {de Boer}, {de Teodoro}, {de
  Zeeuw}, {Delle Luche}, {Domingues}, {Dubath}, {Fodor}, {Fr{\'e}zouls},
  {Fries}, {Fustes}, {Fyfe}, {Gallardo}, {Gallegos}, {Gardiol}, {Gebran},
  {Gomboc}, {G{\'o}mez}, {Grux}, {Gueguen}, {Heyrovsky}, {Hoar}, {Iannicola},
  {Isasi Parache}, {Janotto}, {Joliet}, {Jonckheere}, {Keil}, {Kim},
  {Klagyivik}, {Klar}, {Knude}, {Kochukhov}, {Kolka}, {Kos}, {Kutka}, {Lainey},
  {LeBouquin}, {Liu}, {Loreggia}, {Makarov}, {Marseille}, {Martayan},
  {Martinez-Rubi}, {Massart}, {Meynadier}, {Mignot}, {Munari}, {Nguyen},
  {Nordlander}, {Ocvirk}, {O'Flaherty}, {Olias Sanz}, {Ortiz}, {Osorio},
  {Oszkiewicz}, {Ouzounis}, {Palmer}, {Park}, {Pasquato}, {Peltzer}, {Peralta},
  {P{\'e}turaud}, {Pieniluoma}, {Pigozzi}, {Poels}, {Prat}, {Prod'homme},
  {Raison}, {Rebordao}, {Risquez}, {Rocca-Volmerange}, {Rosen}, {Ruiz-Fuertes},
  {Russo}, {Sembay}, {Serraller Vizcaino}, {Short}, {Siebert}, {Silva},
  {Sinachopoulos}, {Slezak}, {Soffel}, {Sosnowska}, {Strai{\v{z}}ys}, {ter
  Linden}, {Terrell}, {Theil}, {Tiede}, {Troisi}, {Tsalmantza}, {Tur},
  {Vaccari}, {Vachier}, {Valles}, {Van Hamme}, {Veltz}, {Virtanen}, {Wallut},
  {Wichmann}, {Wilkinson}, {Ziaeepour}, \& {Zschocke}}]{GAIA2016}
{Gaia Collaboration}, {Prusti}, T., {de Bruijne}, J.~H.~J., {et~al.} 2016,
  \aap, 595, A1, \dodoi{10.1051/0004-6361/201629272}

\bibitem[{{Gaia Collaboration} {et~al.}(2018a){Gaia Collaboration},
  {Babusiaux}, {van Leeuwen}, {Barstow}, {Jordi}, {Vallenari}, {Bossini},
  {Bressan}, {Cantat-Gaudin}, {van Leeuwen}, \& et~al.}]{GaiaCollab2018a}
{Gaia Collaboration}, {Babusiaux}, C., {van Leeuwen}, F., {et~al.} 2018a, \aap,
  616, A10, \dodoi{10.1051/0004-6361/201832843}

\bibitem[{{Gaia Collaboration} {et~al.}(2018b){Gaia Collaboration}, {Brown},
  {Vallenari}, {Prusti}, {de Bruijne}, {Babusiaux}, {Bailer-Jones}, {Biermann},
  {Evans}, {Eyer}, \& et~al.}]{GaiaCollab2018b}
{Gaia Collaboration}, {Brown}, A.~G.~A., {Vallenari}, A., {et~al.} 2018b, \aap,
  616, A1, \dodoi{10.1051/0004-6361/201833051}

\bibitem[{{Gao}(2018)}]{Gao2018}
{Gao}, X.-h. 2018, \pasp, 130, 124101, \dodoi{10.1088/1538-3873/aae0d2}

\bibitem[{{Geller} {et~al.}(2013){Geller}, {Hurley}, \& {Mathieu}}]{Geller2013}
{Geller}, A.~M., {Hurley}, J.~R., \& {Mathieu}, R.~D. 2013, \aj, 145, 8,
  \dodoi{10.1088/0004-6256/145/1/8}

\bibitem[{{Geller} {et~al.}(2015){Geller}, {Latham}, \& {Mathieu}}]{Geller2015}
{Geller}, A.~M., {Latham}, D.~W., \& {Mathieu}, R.~D. 2015, \aj, 150, 97,
  \dodoi{10.1088/0004-6256/150/3/97}

\bibitem[{{Geller} {et~al.}(2017b){Geller}, {Leiner}, {Chatterjee}, {Leigh},
  {Mathieu}, \& {Sills}}]{Geller2017b}
{Geller}, A.~M., {Leiner}, E.~M., {Chatterjee}, S., {et~al.} 2017b, \apj, 842,
  1, \dodoi{10.3847/1538-4357/aa72ef}

\bibitem[{{Geller} \& {Mathieu}(2011)}]{GellerNature2011}
{Geller}, A.~M., \& {Mathieu}, R.~D. 2011, \nat, 478, 356,
  \dodoi{10.1038/nature10512}

\bibitem[{{Geller} \& {Mathieu}(2012)}]{Geller2012}
---. 2012, \aj, 144, 54, \dodoi{10.1088/0004-6256/144/2/54}

\bibitem[{{Geller} {et~al.}(2010){Geller}, {Mathieu}, {Braden},
  {et~al.}}]{Geller2010}
{Geller}, A.~M., {Mathieu}, R.~D., {Braden}, E.~K., {et~al.} 2010, \aj, 139,
  1383, \dodoi{10.1088/0004-6256/139/4/1383}

\bibitem[{{Geller} {et~al.}(2008){Geller}, {Mathieu}, {Harris}, \&
  {McClure}}]{Geller2008}
{Geller}, A.~M., {Mathieu}, R.~D., {Harris}, H.~C., \& {McClure}, R.~D. 2008,
  \aj, 135, 2264, \dodoi{10.1088/0004-6256/135/6/2264}

\bibitem[{{Geller} {et~al.}(2017a){Geller}, {Leiner}, {Bellini}, {Gleisinger},
  {Haggard}, {Kamann}, {Leigh}, {Mathieu}, {Sills}, {Watkins}, \&
  {Zurek}}]{Geller2017a}
{Geller}, A.~M., {Leiner}, E.~M., {Bellini}, A., {et~al.} 2017a, \apj, 840, 66,
  \dodoi{10.3847/1538-4357/aa6af3}

\bibitem[{{Gim} {et~al.}(1998{\natexlab{a}}){Gim}, {Hesser}, {McClure}, \&
  {Stetson}}]{Gim1998a}
{Gim}, M., {Hesser}, J.~E., {McClure}, R.~D., \& {Stetson}, P.~B.
  1998{\natexlab{a}}, \pasp, 110, 1172, \dodoi{10.1086/316241}

\bibitem[{{Gim} {et~al.}(1998{\natexlab{b}}){Gim}, {Vandenberg}, {Stetson},
  {Hesser}, \& {Zurek}}]{Gim1998}
{Gim}, M., {Vandenberg}, D.~A., {Stetson}, P.~B., {Hesser}, J.~E., \& {Zurek},
  D.~R. 1998{\natexlab{b}}, \pasp, 110, 1318, \dodoi{10.1086/316266}

\bibitem[{{Gosnell} {et~al.}(2019){Gosnell}, {Leiner}, {Mathieu}, {Geller},
  {Knigge}, {Sills}, \& {Leigh}}]{Gosnell2019}
{Gosnell}, N.~M., {Leiner}, E.~M., {Mathieu}, R.~D., {et~al.} 2019, \apj, 885,
  45, \dodoi{10.3847/1538-4357/ab4273}

\bibitem[{{Gosnell} {et~al.}(2015){Gosnell}, {Mathieu}, {Geller}, {Sills},
  {Leigh}, \& {Knigge}}]{Gosnell2015}
{Gosnell}, N.~M., {Mathieu}, R.~D., {Geller}, A.~M., {et~al.} 2015, \apj, 814,
  163, \dodoi{10.1088/0004-637X/814/2/163}

\bibitem[{{Gosnell} {et~al.}(2014){Gosnell}, {Mathieu}, {Geller},
  {et~al.}}]{Gosnell2014}
---. 2014, \apjl, 783, L8, \dodoi{10.1088/2041-8205/783/1/L8}

\bibitem[{{Hurley} {et~al.}(2002){Hurley}, {Tout}, \& {Pols}}]{Hurley2002}
{Hurley}, J.~R., {Tout}, C.~A., \& {Pols}, O.~R. 2002, \mnras, 329, 897,
  \dodoi{10.1046/j.1365-8711.2002.05038.x}

\bibitem[{{Jacobson} {et~al.}(2011){Jacobson}, {Pilachowski}, \&
  {Friel}}]{Jacobson2011}
{Jacobson}, H.~R., {Pilachowski}, C.~A., \& {Friel}, E.~D. 2011, \aj, 142, 59,
  \dodoi{10.1088/0004-6256/142/2/59}

\bibitem[{{Jahn} {et~al.}(1995){Jahn}, {Kaluzny}, \& {Rucinski}}]{Jahn1995}
{Jahn}, K., {Kaluzny}, J., \& {Rucinski}, S.~M. 1995, \aap, 295, 101

\bibitem[{{Kroupa}(1995)}]{Kroupa1995}
{Kroupa}, P. 1995, \mnras, 277, 1507

\bibitem[{{Kurucz}(1993)}]{Kurucz1993}
{Kurucz}, R.~L. 1993, in Light Curve Modeling of Eclipsing Binary Stars, ed.
  E.~F. Milone (Springer-Verlag), 93--101

\bibitem[{{Latham} {et~al.}(1985){Latham}, {Stefanik}, \&
  {Carney}}]{Latham1985}
{Latham}, D.~W., {Stefanik}, R.~P., \& {Carney}, B.~W. 1985, in Stellar Radial
  Velocities, ed. A.~G.~D. {Philip} \& D.~W. {Latham}, 381--384

\bibitem[{{Leiner} {et~al.}(2017){Leiner}, {Mathieu}, \& {Geller}}]{Leiner2017}
{Leiner}, E., {Mathieu}, R.~D., \& {Geller}, A.~M. 2017, \apj, 840, 67,
  \dodoi{10.3847/1538-4357/aa6aff}

\bibitem[{{Leiner} {et~al.}(2019){Leiner}, {Mathieu}, {Vanderburg}, {Gosnell},
  \& {Smith}}]{Leiner2019BL}
{Leiner}, E., {Mathieu}, R.~D., {Vanderburg}, A., {Gosnell}, N.~M., \& {Smith},
  J.~C. 2019, \apj, 881, 47, \dodoi{10.3847/1538-4357/ab2bf8}

\bibitem[{{Leiner} {et~al.}(2015){Leiner}, {Mathieu}, {Gosnell}, \&
  {Geller}}]{Leiner2015}
{Leiner}, E.~M., {Mathieu}, R.~D., {Gosnell}, N.~M., \& {Geller}, A.~M. 2015,
  \aj, 150, 10, \dodoi{10.1088/0004-6256/150/1/10}

\bibitem[{{Mathieu}(1983)}]{Mathieu1983}
{Mathieu}, R.~D. 1983, \apjl, 267, L97, \dodoi{10.1086/184011}

\bibitem[{{Mathieu}(2000)}]{Mathieu2000}
{Mathieu}, R.~D. 2000, in ASP Conf. Ser., Vol. 198, Stellar Clusters and
  Associations: Convection, Rotation, and Dynamos, ed. R.~{Pallavicini},
  G.~{Micela}, \& S.~{Sciortino}, 517

\bibitem[{{Mathieu} \& {Geller}(2009)}]{MathieuNature2009}
{Mathieu}, R.~D., \& {Geller}, A.~M. 2009, \nat, 462, 1032,
  \dodoi{10.1038/nature08568}

\bibitem[{{Mathieu} \& {Geller}(2015)}]{MathieuEco2015}
---. 2015, in Astrophysics and Space Science Library, Vol. 413, Ecology of Blue
  Straggler Stars, ed. H.~M.~J. Boffin, G.~Carraro, \& G.~Beccari
  (Springer-Verlag), 29--63

\bibitem[{{Mathieu} \& {Leiner}(2019)}]{MathieuLeiner2019ASEP}
{Mathieu}, R.~D., \& {Leiner}, E.~M. 2019, in The Impact of Binary Stars on
  Stellar Evolution, Cambridge Astrophysics (Cambridge University Press),
  261--276

\bibitem[{{McNamara} \& {Solomon}(1981)}]{McNamara1981}
{McNamara}, B.~J., \& {Solomon}, S. 1981, \aaps, 43, 337

\bibitem[{{Meibom} {et~al.}(2015){Meibom}, {Barnes}, {Platais}, {Gilliland},
  {Latham}, \& {Mathieu}}]{Meibom2015}
{Meibom}, S., {Barnes}, S.~A., {Platais}, I., {et~al.} 2015, \nat, 517, 589,
  \dodoi{10.1038/nature14118}

\bibitem[{{Meibom} {et~al.}(2009){Meibom}, {Grundahl}, {Clausen},
  {et~al.}}]{Meibom2009}
{Meibom}, S., {Grundahl}, F., {Clausen}, J.~V., {et~al.} 2009, \aj, 137, 5086,
  \dodoi{10.1088/0004-6256/137/6/5086}

\bibitem[{{Meibom} \& {Mathieu}(2005)}]{Meibom2005}
{Meibom}, S., \& {Mathieu}, R.~D. 2005, \apj, 620, 970, \dodoi{10.1086/427082}

\bibitem[{{Milliman} {et~al.}(2016){Milliman}, {Leiner}, {Mathieu},
  {Tofflemire}, \& {Platais}}]{Milliman2016}
{Milliman}, K.~E., {Leiner}, E., {Mathieu}, R.~D., {Tofflemire}, B.~M., \&
  {Platais}, I. 2016, \aj, 151, 152, \dodoi{10.3847/0004-6256/151/6/152}

\bibitem[{{Milliman} {et~al.}(2014){Milliman}, {Mathieu}, {Geller},
  {et~al.}}]{Milliman2014}
{Milliman}, K.~E., {Mathieu}, R.~D., {Geller}, A.~M., {et~al.} 2014, \aj, 148,
  38, \dodoi{10.1088/0004-6256/148/2/38}

\bibitem[{{Milliman} {et~al.}(2015){Milliman}, {Mathieu}, \&
  {Schuler}}]{Milliman2015}
{Milliman}, K.~E., {Mathieu}, R.~D., \& {Schuler}, S.~C. 2015, \aj, 150, 84,
  \dodoi{10.1088/0004-6256/150/3/84}

\bibitem[{{Mochejska} \& {Kaluzny}(1999)}]{MK1999}
{Mochejska}, B.~J., \& {Kaluzny}, J. 1999, \actaa, 49, 351

\bibitem[{{Overbeek} {et~al.}(2015){Overbeek}, {Friel}, {Jacobson},
  {et~al.}}]{Overbeek2015}
{Overbeek}, J.~C., {Friel}, E.~D., {Jacobson}, H.~R., {et~al.} 2015, \aj, 149,
  15, \dodoi{10.1088/0004-6256/149/1/15}

\bibitem[{{Paczy{\'n}ski}(1971)}]{Paczy1971}
{Paczy{\'n}ski}, B. 1971, \araa, 9, 183,
  \dodoi{10.1146/annurev.aa.09.090171.001151}

\bibitem[{{Pancino} {et~al.}(2010){Pancino}, {Carrera}, {Rossetti}, \&
  {Gallart}}]{Pancino2010}
{Pancino}, E., {Carrera}, R., {Rossetti}, E., \& {Gallart}, C. 2010, \aap, 511,
  A56, \dodoi{10.1051/0004-6361/200912965}

\bibitem[{{Platais} {et~al.}(2013){Platais}, {Gosnell}, {Meibom},
  {et~al.}}]{Platais2013}
{Platais}, I., {Gosnell}, N.~M., {Meibom}, S., {et~al.} 2013, \aj, 146, 43,
  \dodoi{10.1088/0004-6256/146/2/43}

\bibitem[{{Platais} {et~al.}(2003){Platais}, {Kozhurina-Platais}, {Mathieu},
  {Girard}, \& {van Altena}}]{Platais2003}
{Platais}, I., {Kozhurina-Platais}, V., {Mathieu}, R.~D., {Girard}, T.~M., \&
  {van Altena}, W.~F. 2003, \aj, 126, 2922, \dodoi{10.1086/379677}

\bibitem[{{Raghavan} {et~al.}(2010){Raghavan}, {McAlister}, {Henry},
  {et~al.}}]{R2010}
{Raghavan}, D., {McAlister}, H.~A., {Henry}, T.~J., {et~al.} 2010, \apjs, 190,
  1, \dodoi{10.1088/0067-0049/190/1/1}

\bibitem[{{Rasio}(1995)}]{Rasio1995}
{Rasio}, F.~A. 1995, \apjl, 444, L41, \dodoi{10.1086/187855}

\bibitem[{{Sandquist} {et~al.}(2018){Sandquist}, {Mathieu}, {Quinn}, {Pollack},
  {Latham}, {Brown}, {Esselstein}, {Aigrain}, {Parviainen}, {Vanderburg},
  {Stello}, {Somers}, {Pinsonneault}, {Tayar}, {Orosz}, {Bedin}, {Libralato},
  {Malavolta}, \& {Nardiello}}]{Sandquist2018}
{Sandquist}, E.~L., {Mathieu}, R.~D., {Quinn}, S.~N., {et~al.} 2018, \aj, 155,
  152, \dodoi{10.3847/1538-3881/aab0ff}

\bibitem[{{Sarajedini} {et~al.}(1999){Sarajedini}, {von Hippel},
  {Kozhurina-Platais}, \& {Demarque}}]{Sarajedini1999}
{Sarajedini}, A., {von Hippel}, T., {Kozhurina-Platais}, V., \& {Demarque}, P.
  1999, \aj, 118, 2894, \dodoi{10.1086/301149}

\bibitem[{{Taylor}(2005)}]{Taylor2005}
{Taylor}, M.~B. 2005, in Astronomical Society of the Pacific Conference Series,
  Vol. 347, Astronomical Data Analysis Software and Systems XIV, ed.
  P.~{Shopbell}, M.~{Britton}, \& R.~{Ebert}, 29

\bibitem[{{Thompson} {et~al.}(2014){Thompson}, {Frinchaboy}, {Kinemuchi},
  {Sarajedini}, \& {Cohen}}]{Thompson2014}
{Thompson}, B., {Frinchaboy}, P., {Kinemuchi}, K., {Sarajedini}, A., \&
  {Cohen}, R. 2014, \aj, 148, 85, \dodoi{10.1088/0004-6256/148/5/85}

\bibitem[{{Tofflemire} {et~al.}(2014){Tofflemire}, {Gosnell}, {Mathieu}, \&
  {Platais}}]{Tofflemire2014}
{Tofflemire}, B.~M., {Gosnell}, N.~M., {Mathieu}, R.~D., \& {Platais}, I. 2014,
  \aj, 148, 61, \dodoi{10.1088/0004-6256/148/4/61}

\bibitem[{{van Dokkum}(2001)}]{Dokkum2001}
{van Dokkum}, P.~G. 2001, \pasp, 113, 1420, \dodoi{10.1086/323894}

\bibitem[{{Webbink}(1976)}]{Webbink1976}
{Webbink}, R.~F. 1976, \apj, 209, 829, \dodoi{10.1086/154781}

\bibitem[{{Wu} {et~al.}(2009){Wu}, {Du}, {Ma}, \& {Zhou}}]{Wu2009}
{Wu}, Z.-Y., {Du}, C.-H., {Ma}, J., \& {Zhou}, X. 2009, Chinese Physics
  Letters, 26, 029701, \dodoi{10.1088/0256-307X/26/2/029701}

\bibitem[{{Wu} {et~al.}(2007){Wu}, {Zhou}, {Ma}, {Jiang}, {Chen}, \&
  {Wu}}]{Wu2007}
{Wu}, Z.-Y., {Zhou}, X., {Ma}, J., {et~al.} 2007, \aj, 133, 2061,
  \dodoi{10.1086/512189}

\bibitem[{{Yang} {et~al.}(2013){Yang}, {Sarajedini}, {Deliyannis},
  {et~al.}}]{Yang2013}
{Yang}, S.-C., {Sarajedini}, A., {Deliyannis}, C.~P., {et~al.} 2013, \apj, 762,
  3, \dodoi{10.1088/0004-637X/762/1/3}

\bibitem[{{Zahn} \& {Bouchet}(1989)}]{ZahnBouchet1989}
{Zahn}, J.-P., \& {Bouchet}, L. 1989, \aap, 223, 112

\bibitem[{{Zhang} {et~al.}(2003){Zhang}, {Deng}, {Xin}, \& {Zhou}}]{Zhang2003}
{Zhang}, X.-B., {Deng}, L.-C., {Xin}, Y., \& {Zhou}, X. 2003, \cjaa, 3, 151

\bibitem[{{Zucker} \& {Mazeh}(1994)}]{Zucker1994}
{Zucker}, S., \& {Mazeh}, T. 1994, \apj, 420, 806, \dodoi{10.1086/173605}

\end{thebibliography}

\end{document}